\documentstyle[11pt,epsf]{article}
\begin{document}

\begin{center}
{\bf Los Alamos Electronic Archives: physics/9808031 v2\\}
{\sl La verdad os har\'a libres\\
Universidad de Guanajuato\\
IFUG, Le\'on, Guanajuato, M\'exico}
\end{center}

\bigskip
\bigskip

\begin{center}
{\large \bf MEC\'ANICA CU\'ANTICA I.}\end{center}

\begin{center} {\large \bf MC I} \end{center}

\bigskip

\begin{center}
{\large \bf Haret C. Rosu}\end{center}
\begin{center} e-mail: rosu@ifug3.ugto.mx\\
fax: 0052-47187611\\
phone: 0052-47183089  \end{center}
%\begin{center} {\bf IFUG (M\'exico) \& IGSS (Rumania)} \end{center}
%Julio de 1998}
%\end{center}

\bigskip
\bigskip

%%%%%%%%%%%%%%
\vskip 2ex
\centerline{
\epsfxsize=280pt
\epsfbox{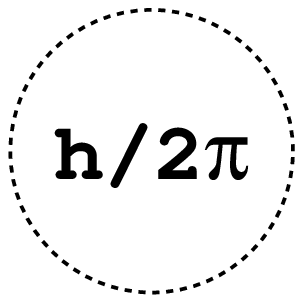}}
\vskip 4ex
%\begin{center}
%{\small{Fig. 1}\\
%}
%\end{center}
%%%%%%%%%%%%%%%%

\vspace{5.5cm}
\begin{center} A cargo del Prof. Haret Rosu para el
beneficio de los estudiantes \end{center}
\begin{center} presentes y futuros del IFUG y otros lugares.\end{center}
\begin{center} Primer curso de mec\'anica cu\'antica
en castellano publicado en Internet. \end{center}

\begin{center} Copyright \copyright 1998. All rights are reserved.
{\cal H}.{\cal C}. {\cal R}{\cal O}{\cal S}{\cal U}
\end{center}
\vspace{2cm}
\centerline{\bf Julio de 1998}

\begin{abstract}
%\begin{center}
This is the first Internet course on elementary quantum mechanics written in 
Spanish (``castellano") for the benefit of Spanish speaking students. I thank
my eight Mexican students at the Institute of Physics, University of Guanajuato,
Leon, (IFUG), for the collaboration that made this possible. The topics 
included refer to the postulates of quantum mechanics, one-dimensional barriers 
and wells, angular momentum, WKB method, harmonic oscillator, hydrogen atom, 
quantum scattering, and partial waves.
%\end{center}
\end{abstract}

%\end{center}

\newpage

\centerline{\'INDICE DE CONTENIDO}

\vspace{0.5cm}

\noindent

{\bf 0. Introducci\'on General}
- Haret C. Rosu\\

{\bf 1. Los Postulados de la MC}
- Mart\'{\i}n Gilberto Castro Esparza\\

{\bf 2. Potenciales Barreras y Pozos}
- Juan Manuel Rodr\'{\i}guez Vizca\'{\i}no\\

{\bf 3. El Momento Angular}
- Teodoro C\'ordova Fraga\\

{\bf 4. El M\'etodo WKB}
-  Luis Antonio Garc\'{\i}a Trujillo\\

{\bf 5. El Oscilador Arm\'onico}
-  Jos\'e Torres Arenas\\

{\bf 6. El \'Atomo de Hidr\'ogeno}
-  Edgar Alvarado Anell\\

{\bf 7. La Dispersi\'on en la MC}
- Daniel Jim\'enez \'Alvarez\\

{\bf 8. Las Ondas Parciales}
-   Pedro Basilio Espinoza Padilla\\

Incluye tambi\'en alrededor de 25 problemas con soluciones.

\newpage

{\sl

\section*{0. Introduction}  %%%%%%%   0

The energy quanta occured in 1900 in works of Max Planck (Nobel prize, 1918)
on the black body electromagnetic radiation. Planck's ``quanta of light" have
been used
by Einstein (Nobel prize, 1921) to explain the photoelectric effect, but
the first ``quantization" of a quantity having units of action (the angular
momentum) belongs to Niels Bohr (Nobel Prize, 1922). This
opened the road to the universality of quanta, since the action is the
basic functional to describe any type of motion. However, only in the
1920's the formalism of quantum mechanics has been developed in a systematic
manner and the remarkable works of that decade contributed in a decisive way
to the rising of quantum mechanics at the level of fundamental theory of the
universe from the mankind standpoint and one of the most successful from
the point of view of technology. Moreover, it is quite probable that many
of the cosmological misteries may be disentangled by means of various
quantization procedures of the gravitational field leading to our progress in
understanding the origins of the universe. On the other hand, in recent years,
there is a strong surge of activity in the information aspect of
quantum mechanics, that has not been very much exploited in the past,
aiming at a very attractive ``quantum computer" technology.

At the philosophical level, the famous paradoxes of quantum mechanics
(showing the difficulties of the `quantum' thinking) are
actively pursued ever since they have been first posed. Perhaps the
famous of them is the EPR paradox (Einstein, Podolsky, Rosen, 1935) on the
existence of {\em elements of physical reality}, or in EPR words:
``If, without in any way disturbing a system, we can predict with certainty
(i.e., with probability equal to unity) the value of a physical quantity, then
there exists an element of physical reality corresponding to this physical
quantity."
Another famous paradox is that of Schr\"odinger's cat which is related to
the fundamental quantum property of entanglement and the way we
understand and detect it.
What one should emphasize is that all these delicate points are the sourse
of many interesting experiments (such as the so-called ``teleportation"
of quantum states) pushing up the technology.

Here, we present eight elementary topics in nonrelativistic quantum mechanics
from a course in Spanish (``castellano")
on quantum mechanics that I taught in
the Institute of Physics, University of Guanajuato (IFUG), Le\'on, Mexico,
during January-June semester of 1998. The responsability of the idiom
belongs mostly to the eight students, respectively.

\hfill Haret C. Rosu

\newpage

\section*{0. Introducci\'on General}   %%%%%%% 0

Los cuantos de energ\'{\i}a surgieron en 1900 como consecuencia de los
trabajos de Max Planck (premio Nobel 1918)
sobre el problema de la radiaci\'on del cuerpo
negro. Los ``cuantos de luz" de Planck fueron usados por Einstein (premio
Nobel 1921) para
explicar el efecto fotoel\'ectrico, pero la primera ``cuantificaci\'on" de
una cantidad que tiene las unidades de una acci\'on (el momento angular)
se debe a Niels Bohr (premio Nobel 1922). Eso abri\'o el camino de la
universalidad de los
cuantos ya que la acci\'on es la funci\'on b\'asica para describir
cualquier tipo de movimiento. A\'un as\'{\i}, solo los a\~nos
veinte se consideran como el inicio del formalismo cu\'antico que
levant\'o a la mec\'anica cu\'antica al nivel de teor\'{\i}a fundamental del
universo y una de las m\'as exitosas en cuanto a la tecnolog\'{\i}a.
En verdad, es muy probable que muchos de los misterios
cosmol\'ogicos est\'an por ejemplo detras de las diferentes maneras de
cuantificar el campo gravitacional y tales avances pueden contribuir al
entendimiento de los origines del universo.
Por otro lado, el aspecto inform\'atico de la mec\'anica cu\'antica, que no se
aprovech\'o en el pasado, se est\'a desarrollando de una manera muy activa
en los \'ultimos a\~nos con el prop\'osito de investigar la posibilidad de la
construcci\'on de las llamadas ``computadoras cu\'anticas".
%\bigskip
En la parte filos\'ofica cabe mencionar que en la mec\'anica cu\'antica
hay paradojas famosas que todav\'{\i}a se mantienen en pol\'emica y que
reflejan las dificultades de la l\'ogica que impone la manera de pensar
cu\'antica (y/o probabil\'{\i}stica). Una de las m\'as c\'elebres es la de
Einstein (que nunca acepto por completo la MC), Podolsky y Rosen (EPR, 1935)
sobre si hay o no
``elementos verdaderos de la realidad f\'{\i}sica" (seg\'un Einstein la MC
prohibe la existencia independiente del acto de medici\'on de los sistemas
f\'{\i}sicos). Otra de igual
celebridad es la del ``gato de Schr\"odinger". Lo que se debe subrayar es
que todos estos puntos te\'oricos delicados generan experimentos muy
interesantes (como son por ejemplo los de la llamada ``teletransportaci\'on"
de estados cu\'anticos) que impulsan a la tecnolog\'{\i}a.
%\bigskip
Lo que sigue son algunos temas introductivos en la mec\'anica cu\'antica
norelativista que sirvieron como base para el curso de maestr\'{\i}a
de mec\'anica cu\'antica I en el IFUG durante el semestre Enero-Junio de
1998. Este curso fue impartido por mi a los
estudiantes enlistados, los cuales se encargaron de los temas
correspondientes. La responsabilidad del idioma pertenece en gran parte
a cada uno de los estudiantes.

\hfill Haret C. Rosu}

\newpage
{\sl

\section*{1. LOS POSTULADOS DE LA MC} %%%%% 1
%\end{center}
%\author{Martin Gilberto Castro Esparza}
%\date{}
%\maketitle

%\section*{}
Los siguientes 6 postulados se pueden considerar como la base de la teor\'{\i}a
y los experimentos de la mec\'anica cu\'antica.
\begin{enumerate}
\item[{\bf P1.}-]
A cualquier cantidad f\'{\i}sica L le corresponde un operador Hermitiano 
$\hat{L}$.
\end{enumerate}
\begin{enumerate}
\item[{\bf P2.}-]
A cualquier estado (f\'{\i}sico) estacionario de un sistema f\'{\i}sico le
corresponde una
funci\'on de onda normalizada $\psi$ ($\parallel\psi\parallel^2=1$).
\end{enumerate}
\begin{enumerate}
\item[{\bf P3.}-]
La cantidad f\'{\i}sica L puede tomar solo los valores propios del operador 
$\hat{L}$.
\end{enumerate}
\begin{enumerate}
\item[{\bf P4.}-]
Lo que se mide es siempre el valor promedio $\overline{L}$ de la cantidad L
en el estado $\psi$, la cual en teor\'{\i}a es el elemento de matriz diagonal

 $<f\mid\hat{L}\mid{f}>=\overline{L}$.

\end{enumerate}
\begin{enumerate}
\item[{\bf P5.}-]
Los elementos de matriz de los operadores coordenada cartesiana y momento
$\widehat{x_{i}}$ y $\widehat{p_{k}}$, calculados entre las funciones de 
onda f y g satisfacen a las ecuaciones de Hamilton de la mec\'anica cl\'asica
 en la forma:\\
$\frac{d}{dt}<f\mid\widehat{p_{i}}\mid{g}>=-<f\mid\frac{\partial\widehat{H}}
{\partial\widehat{x_{i}}}\mid{g}>$,
$\frac{d}{dt}<f\mid\widehat{x_{i}}\mid{g}>=<f\mid\frac{\partial\widehat{H}}
{\partial\widehat{p_{i}}}\mid{g}>$

donde $\widehat{H}$ es el operador Hamiltoniano.
\end{enumerate}
\begin{enumerate}
\item[{\bf P6.}-]
Los operadores $\widehat{p_{i}}$ y $\widehat{x_{k}}$ tienen los siguientes 
conmutadores:
\begin{center}

$[\widehat{p_{i}},\widehat{x_{k}}]=-i\hbar\delta_{ik}$,\\
$[\widehat{p_{i}},\widehat{p_{k}}]=0$\\
$[\widehat{x_{i}},\widehat{x_{k}}]=0$,\hspace{5mm}

$\hbar=h/2\pi=1.0546\times10^{-27}$ erg.seg.
\end{center}
\end{enumerate}
\newpage
\begin{enumerate}
\item[1.-]
La correspondencia de una cantidad f\'{\i}sica L que tiene un an\'alogo
cl\'asico
$L(x_{i},p_{k})$ se hace sustituyendo $x_{i}$, $p_{k}$ por $\widehat{x_{i}}$,
$\widehat{p_{k}}$. La funci\'on L se supone que se puede desarrollar en serie 
de potencias. Si la funci\'on no contiene productos $x_{k}p_{k}$, el operador 
$\hat{L}$ es directamente hermitiano.\\
Ejemplo:\\
\begin{center}

$T=(\sum_{i}^3p_{i}^2)/2m$ $\longrightarrow$
$\widehat{T}=(\sum_{i}^3\widehat{p}^2)/2m$.\\
\end{center}

Si L contiene productos $x_{i}p_{i}$, $\hat{L}$ no es hermitiano, en tal caso
L se sustituye por $\hat\Lambda$ la parte hermitica de $\hat{L}$ 
($\hat\Lambda$ es un operador autoadjunto).\\
Ejemplo:\\
\begin{center}

$w(x_{i},p_{i})=\sum_{i}p_{i}x_{i}$ $\longrightarrow$             
$\widehat{w}=1/2\sum_{i}^3(\widehat{p_{i}}\widehat{x_{i}}+\widehat
{x_{i}}\widehat{p_{i}})$.\\
\end{center}

Resulta tambi\'en que el tiempo no es un operador sino un par\'ametro.
\end{enumerate}
\begin{enumerate}
\item[2.-](Probabilidad en el espectro discreto y continuo)
Si $\psi_{n}$ es funci\'on propia del operador $\hat{L}$, entonces:\\

$\overline{L}=<n\mid\hat{L}\mid{n}>=<n\mid\lambda_{n}\mid{n}>=
\lambda_{n}<n\mid{n}>=\delta_{nn}\lambda_{n}=\lambda_{n}$.\\

Tambi\'en se puede demostrar que $\overline{L}^k=(\lambda_{n})^k$.

Si la funci\'on $\phi$ no es funci\'on propia de $\hat{L}$ se usa el desarrollo en un sistema completo de $\hat{L}$, entonces:\\
Sean las siguientes definiciones:\\
\begin{center}

 $\hat{L}\psi_{n}=\lambda_{n}\psi_{n}$,\hspace{10mm} 
 $\phi=\sum_{n}a_{n}\psi_{n}$\\
\end{center}
combinando estas dos definiciones obtenemos lo 
siguiente:\\

\begin{center}
 $\hat{L}\phi=\sum_{n}\lambda_{n}a_{n}\psi_{n}$.\\
\end{center}
Con las definiciones pasadas ya podremos calcular los elementos de matriz del operador L. Entonces:\\
\begin{center}

$<\phi\mid\hat{L}\mid{\phi}>=\sum_{n,m}a_{m}^{\ast}a_{n}\lambda_{n}<m\mid{n}>=\sum_{m}\mid{a_{m}}\mid^2\lambda_{m}$,\\
\end{center}
lo cual nos dice que el resultado del experimento es $\lambda_{m}$ con la 
probabilidad $\mid{a_{m}}\mid^2$.\\
Si el espectro es discreto: de acuerdo con el postulado 4 eso significa que 
$\mid{a_{m}}\mid^2$, o sea, los coeficientes del desarrollo en un sistema completo determinan las probabilidades de observar el valor propio $\lambda_{n}$.\\
Si el espectro es continuo: usando la siguiente definici\'on \\

\begin{center}
$\phi(\tau)=\int{a}(\lambda)\psi{(\tau,\lambda)}d\lambda$,\\
\end{center}
se calcular\'an los elementos de matriz para el espectro continuo\\
\begin{center}
$\overline{L}=<\phi\mid{\hat{L}}\mid{\phi}>$\\
\end{center}
\begin{center}
$=\int{d}\tau\int{a}^\ast(\lambda)\psi^\ast(\tau,\lambda)d\lambda\int\mu{a}
(\mu)\psi(\tau,\mu)d\mu$\\
\end{center}
\begin{center}
$=\int\int{a}^{\ast} a(\mu)\mu\int\psi^\ast(\tau,\lambda)\psi(tau,\mu)d\lambda
 {d}\mu{d}\tau$\\
\end{center}
\begin{center}
$=\int\int{a}^\ast(\lambda){a}(\mu)\mu\delta(\lambda-\mu){d}\lambda{d}\mu$\\
\end{center}
\begin{center}
$=\int{a}^\ast(\lambda)a(\lambda)\lambda{d}\lambda$\\
\end{center}
\begin{center}
$=\int\mid{a}(\lambda)\mid^2\lambda{d}\lambda$.\\
\end{center}
En el caso continuo se dice que $\mid{a}(\lambda)\mid^2$ es la densidad de 
probabilidad de observar el valor de $\lambda$ del espectro continuo. Tambi\'en vale\\
\begin{center}
 $\overline{L}=<\phi\mid\hat{L}\mid\phi>$.
\end{center}
\end{enumerate} 
\begin{enumerate}
\item[3.-]
Definici\'on de la derivada con respecto a un operador:\\

\begin{center}
$\frac{\partial{F(\hat{L})}}{\partial\hat{L}}={\rm lim}_{\epsilon\rightarrow
\infty}
\frac{F(\hat{L}+\epsilon\hat{I})-F(\hat{L})}{\epsilon}.$
\end{center}
\end{enumerate}
\begin{enumerate}
\item[4.-](Representaci\'on del momento)
Cu\'al es la forma concreta de $\widehat{p_{1}}$, $\widehat{p_{2}}$ y 
$\widehat{p_{3}}$, si los argumentos de las funciones de onda son coordenada 
cartesiana $x_{i}$.\\
Vamos a considerar el siguiente conmutador:\\
\begin{center}
$[\widehat{p_{i}}, \widehat{x_{i}}^2]= \widehat{p_{i}}\widehat{x_{i}}^2-\widehat{x_{i}}^2
\widehat{p_{i}}$\\
\end{center}
\begin{center}
$= \widehat{p_{i}}\widehat{x_{i}}\widehat{x_{i}}-\widehat{x_{i}}\widehat{p_{i}}
\widehat{x_{i}}+\widehat{x_{i}}\widehat{p_{i}}\widehat{x_{i}}-\widehat{x_{i}}
\widehat{x_{i}}
\widehat{p_{i}}$\\
\end{center}
\begin{center}
$=(\widehat{p_{i}}\widehat{x_{i}}-\widehat{x_{i}}\widehat{p_{i}})\widehat{x_{i}}
+\widehat{x_{i}}(\widehat{p_{i}}\widehat{x_{i}}-\widehat{x_{i}}\widehat{p_{i}})$\\
\end{center}
\begin{center}
$=[\widehat{p_{i}},
\widehat{x_{i}}]\widehat{x_{i}}+\widehat{x_{i}}[\widehat{p_{i}}, \widehat{x_{i}}]$\\
\end{center}
\begin{center}
$=-i\hbar\widehat{x_{i}}-i\hbar\widehat{x_{i}}=-2i\hbar\widehat{x_{i}}$.\\
\end{center}
En general se tiene:\\

\begin{center}
$\widehat{p_{i}}\widehat{x_{i}}^n-\widehat{x_{i}}^n\widehat{p_{i}}=
-ni\hbar\widehat{x_{i}}^{n-1}.$\\
\end{center}
 Entonces para todas las funciones anal\'{\i}ticas se tiene lo siguiente:\\
\begin{center}
$\widehat{p_{i}}\psi(x)-\psi(x)\widehat{p_{i}}=-i\hbar\frac{\partial\psi}
{\partial{x_{i}}}$.\\
\end{center}
Ahora sea $\widehat{p_{i}}\phi=f(x_{1},x_{2},x_{3})$ la acci\'on de $\widehat{p_{i}}$
sobre $\phi(x_{1},x_{2},x_{3})=1$.
Entonces:

 $\widehat{p_{i}}\psi=-i\hbar\frac{\partial\psi}{\partial{x_{1}}}+f_{1}\psi$ y hay 
relaciones an\'alogas para $x_{2}$ y $x_{3}$.\\
Del conmutador $[\widehat{p_{i}},\widehat{p_{k}}]=0$ se obtiene 
$\nabla\times\vec{f}=0$,
por lo tanto,\\
 $f_{i}=\nabla_{i}F$.\\
La forma m\'as general de $\widehat{P_{i}}$ es:\\
$\widehat{p_{i}}=-i\hbar\frac{\partial}{\partial{x_{i}}}+\frac{\partial{F}}
{\partial{x_{i}}}$, donde F es cualquier funci\'on.
La funci\'on F se puede eliminar utilizando una transformaci\'on unitaria
$\widehat{U}^\dagger=\exp^{\frac{i}{\hbar}F}$.\\

\begin{center}
$\widehat{p_{i}}=\widehat{U}^\dagger(-i\hbar\frac{\partial}{\partial{x_{i}}}+
\frac{\partial{F}}{\partial{x_{i}}})\widehat{U}$\\
\end{center}
\begin{center}

$=\exp^{\frac{i}{\hbar}F}(-i\hbar
\frac{\partial}{\partial{x_{i}}}+\frac{\partial{F}}{\partial{x_{i}}})
\exp^{\frac{-i}{\hbar}F}$\\
\end{center}
\begin{center}
$=-i\hbar\frac{\partial}{\partial{x_{i}}}$\\
\end{center}
resultando que \hspace{10mm} $\widehat{p_{i}}=-i\hbar\frac{\partial}{\partial{x_{i}}}$ 
$\longrightarrow$ $\widehat{p}=-i\hbar\nabla$.\\

\end{enumerate}
\begin{enumerate}
\item[5.-](C\'alculo de la constante de normalizaci\'on)
Cualquier funci\'on de onda $\psi(x)$ $\in$ $L^2$ de variable $x$ se puede escribir como:\\
\begin{center}
$\psi(x)=\int\delta(x-\xi)\psi(\xi)d\xi$\\
\end{center}
y considerar la expresi\'on como desarrollo de $\psi$ en las funciones propias del operador coordenada $\hat{x}\delta(x-\xi)=\xi(x-\xi)$.
Entonces, $\mid\psi(x)\mid^2$ es la densidad de probabilidad de la coordenada 
en el estado $\psi(x)$. De aqu\'{\i} resulta la interpretaci\'on de la norma\\
\begin{center}
$\parallel\psi(x)\parallel^2=\int\mid\psi(x)\mid^2 dx=1$.\\
\end{center}
El sistema descrito por la funci\'on $\psi(x)$ debe encontrarse en alg\'un 
lugar del eje real.\\
Las funciones propias del operador momento son:\\
$-i\hbar\frac{\partial\psi}{\partial{x_{i}}}=p_{i}\psi$, integr\'andola se
obtiene $\psi(x_{i})=A\exp^{\frac{i}{\hbar}p_{i}x_{i}}$, $x$ y $p$ tienen
espectro 
continuo y entonces se tiene que hacer la normalizaci\'on con la funci\'on
 delta.\\
C\'omo se obtiene la constante de normalizaci\'on?\\
se puede obtener utilizando las siguientes transformadas de Fourier:\\
$f(k)=\int{g(x)}\exp^{-ikx}dx$,\hspace{3mm}$g(x)=\frac{1}{2\pi}\int{f(k)}\exp
^{ikx}dk.$\\
Tambi\'en se obtiene de la siguiente manera:\\
Sea la funci\'on de onda no normalizada de la part\'{\i}cula libre\\
$\phi_{p}(x)=A\exp^{\frac{ipx}{\hbar}}$ y la f\'ormula
\begin{center}
$\delta(x-x^{'})=\frac{1}{2\pi}\int_{-\infty}^{\infty}\exp^{ik(x-x^{'})}dx$\\
\end{center}
se ve que \\

\begin{center}
$\int_{-\infty}^{\infty}\phi_{p^{'}}^{\ast}(x)\phi_{p}(x)dx$\\
\end{center}
\begin{center}
$=\int_{-\infty}^{\infty}A^{\ast}\exp^{\frac{-ip^{'}x}{\hbar}}A\exp^{\frac{ipx}
{\hbar}}dx$\\
\end{center}
\begin{center}
$=\int_{-\infty}^{\infty}\mid{A}\mid^2\exp^{\frac{ix(p-p^{'})}{\hbar}}dx$\\
\end{center}
\begin{center}
$=\mid{A}\mid^2\hbar\int_{-\infty}^{\infty}\exp^{\frac{ix(p-p^{'})}{\hbar}}
d\frac{x}{\hbar}$\\
\end{center}
\begin{center}
$=2\pi\hbar\mid{A}\mid^2\delta(p-p^{'})$\\
\end{center}
entonces la constante de normalizaci\'on es:
\begin{center}              
$A=\frac{1}{\sqrt{2\pi\hbar}}$.\\
\end{center}
Tambi\'en resulta que las funciones propias del operador momento forman un
sistema completo (en el sentido del caso continuo) para las funciones 
$L^2$.\\

\begin{center}
$\psi(x)=\frac{1}{\sqrt{2\pi\hbar}}\int{a(p)}\exp^{\frac{ipx}{\hbar}}dp$\\
\end{center}
\begin{center}
$a(p)=\frac{1}{\sqrt{2\pi\hbar}}\int\psi(x)\exp^{\frac{-ipx}{\hbar}}dx$.\\
\end{center}
Estas f\'ormulas establecen la conexi\'on entre las representaciones x y p.

\end{enumerate}
\begin{enumerate}
\item[6.-]
Representaci\'on p: La forma explic\'{\i}ta de los operadores $\hat{p_{i}}$ y
$\hat{x_{k}}$ se puede obtener de las relaciones de conmutaci\'on, pero 
tambi\'en usando los n\'ucleos\\
\begin{center}

$x(p,\beta)=U^{\dagger}xU=\frac{1}{2\pi\hbar}\int\exp^{\frac{-ipx}{\hbar}}x
\exp^{\frac{i\beta{x}}{\hbar}}dx$\\
\end{center}
\begin{center}
$=\frac{1}{2\pi{\hbar}}\int\exp^{\frac{-ipx}{\hbar}}(-i\hbar\frac{\partial}
{\partial\beta}\exp^{\frac{i\beta{x}}{\hbar}})$.\\
\end{center}
La integral pasada tiene la forma siguiente:\\

$M(\lambda,\lambda^{'})=\int{U^{\dagger}}(\lambda,x)\widehat{M}U(\lambda^{'},x)
dx$, y usando $\hat{x}f=\int{x}(x,\xi)f(\xi)d\xi$.\\

Entonces la acci\'on de $\hat{x}$ sobre a(p) $\in$ $L^2$ es:\\
\begin{center}
$\hat{x}a(p)=\int{x}(p,\beta)a(\beta)d\beta$\\
\end{center}
\begin{center}
$=\int(\frac{1}{2\pi\hbar}\int
\exp^{\frac{-ipx}{\hbar}}(-i\hbar\frac{\partial}{\partial\beta}\exp^{\frac
{i\beta{x}}{\hbar}})dx)a(\beta)d\beta$\\
\end{center}
\begin{center}
$=\frac{-i}{2\pi}\int\int\exp^{\frac{-ipx}{\hbar}}\frac{\partial}{\partial\beta}
\exp^{\frac{i\beta{x}}{\hbar}}a(\beta)dxd\beta$\\
\end{center}
\begin{center}
$=\frac{-i\hbar}{2\pi}\int\int\exp^{\frac{-ipx}{\hbar}}\frac
{\partial}{\partial\beta}\exp^{\frac{i\beta{x}}{\hbar}}a(\beta)d\frac{x}{\hbar}
d\beta$\\
\end{center}
\begin{center}
$=\frac{-i\hbar}{2\pi}\int\int\exp^{\frac{ix(\beta-p)}{\hbar}}\frac{\partial}
{\partial\beta}a(\beta)d\frac{x}{\hbar}d\beta$\\
\end{center}
\begin{center}
$=-i\hbar\int\frac{\partial{a(p)}}{\partial\beta}\delta(\beta-p)d\beta
=-i\hbar\frac{\partial{a(p)}}
{\partial{p}}$,\\

\end{center}
 donde \hspace{15mm}$\delta(\beta-p)=\frac{1}{2\pi}\int\exp^{\frac{ix(\beta-p)}
{\hbar}}d\frac{x}{\hbar}$.\\

El operador momento en la representaci\'on p se caracteriza por el n\'ucleo:\\

\begin{center}
$p(p,\beta)=\widehat{U}^{\dagger}p\widehat{U}$\\
\end{center}
\begin{center}
$=\frac{1}{2\pi\hbar}\int\exp^{\frac{-ipx}{\hbar}}
(-i\hbar\frac{\partial}{\partial{x}})\exp^{\frac{i\beta{x}}{\hbar}}dx$\\
\end{center}
\begin{center}
$=\frac{1}{2\pi\hbar}\int\exp^{\frac{-ipx}{\hbar}}\beta\exp^{\frac{i\beta{x}}{\hbar}}dx
=\beta\delta(p-\beta)$ \\
\end{center}
resultando que $\hat{p}a(p)=pa(p)$.\\

Lo que pasa con $\hat{x}$ y $\hat{p}$ es que aunque son herm\'{\i}ticos sobre 
todas
f(x) $\in$ $L^2$ no son herm\'{\i}ticos sobre las funciones propias.\\
Si $\hat{p}a(p)=p_{o}a(p)$ y $\hat{x}=\hat{x}^\dagger$ $\hat{p}=
\hat{p}^\dagger$, entonces:\\
\begin{center}
$<a\mid\hat{p}\hat{x}\mid{a}>-<a\mid\hat{x}\hat{p}\mid{a}>=-i\hbar<a\mid{a}>$\\
\end{center}
\begin{center}
$p_{o}[<a\mid\hat{x}\mid{a}>-<a\mid\hat{x}\mid{a}>]=-i\hbar<a\mid{a}>$\\
\end{center}
\begin{center}
$p_{o}[<a\mid\hat{x}\mid{a}>-<a\mid\hat{x}\mid{a}>]=0$
\end{center}
La parte izquierda es cero, mientras tanto la derecha esta indefinida, lo que
es un contradicci\'on.
\end{enumerate}

\begin{enumerate}
\item[7.-](Representaciones de Schr\"{o}dinger y Heisenberg) \\
Las ecuaciones de movimiento dadas por el postulado 5 tienen varias interpretaciones, por el hecho de que en la expresi\'on $\frac{d}{dt}<f\mid\hat{L}\mid{f}>$ uno puede considerar la dependencia del tiempo atribuida completamente a las funciones de onda 
o completamente a los operadores.\\
\begin{itemize}
\item
Para un operador dependiente del tiempo $\widehat{O}=\widehat{O(t)}$ tenemos:\\

\begin{center}
$\hat{p_{i}}=-\frac{\partial\widehat{H}}{\partial\hat{x_{i}}}$,\hspace{5mm}
$\hat{x_{i}}=\frac{\partial\widehat{H}}{\partial\hat{p_{i}}}$\\
\end{center}
\begin{center}
$[\hat{p},f]=\hat{p}f-f\hat{p}=-i\hbar\frac{\partial{f}}{\partial\hat{x_{i}}}$\\
\end{center}
\begin{center}
$[\hat{x},f]=\hat{x}f-f\hat{x}=-i\hbar\frac{\partial{f}}{\partial\hat{p_{i}}}$\\
\end{center}
se obtienen las ecuaciones de movimiento de Heisenberg:\\

\begin{center}
$\hat{p_{i}}=\frac{-i}{\hbar}[\hat{p},\widehat{H}]$,\hspace{5mm}
$\hat{x_{i}}=\frac{-i}{\hbar}[\hat{x},\widehat{H}]$.
\end{center}

\end{itemize}
\begin{itemize}
\item
Si las funciones dependen del tiempo, todav\'{\i}a se puede usar\\
$\hat{p_{i}}=\frac{-i}{\hbar}[\hat{p_{i}},\widehat{H}]$, porque es consecuencia
solo de las relaciones de conmutaci\'on y entonces no dependen de la representaci\'on.\\

\begin{center}
$\frac{d}{dt}<f\mid\hat{p_{i}}\mid{g}>=\frac{-i}{\hbar}<f\mid[\hat{p},{\widehat{H}}]
\mid{g}>$.\\
\end{center}

Si ahora $\hat{p_{i}}$ y ${\widehat{H}}$ no dependen del tiempo y teniendo en cuenta su hermiticidad se obtiene:\\

\begin{center}
$(\frac{\partial{f}}{\partial{t}},\hat{p_{i}}g)+(\hat{p_{i}}f,\frac
{\partial{g}}{\partial{t}})$\\
\end{center}
\begin{center}
$=\frac{-i}{\hbar}(f,\hat{p_{i}}\hat{H}g)+\frac
{i}{\hbar}(f,\hat{H}\hat{p_{i}}g)$\\
\end{center}
\begin{center}
$=\frac{-i}{\hbar}(\hat{p}f,\hat{H}g)+\frac{i}{\hbar}(\hat{H}f,\hat{p_{i}}g)$\\
\end{center}

\begin{center}
$(\frac{\partial{f}}{\partial{t}}+\frac{i}{\hbar}\hat{H}f,\hat{p_{i}}g)+
(\hat{p_{i}}f,\frac{\partial{g}}{\partial{t}}-\frac{i}{\hbar}\hat{H}g)=0$\\
\end{center}
La \'ultima relaci\'on se cumple para cualquier pareja de funciones $f(x)$ y
$g(x)$ al momento inicial si cada una satisface la ecuaci\'on\\
\begin{center}
$i\hbar\frac{\partial\psi}{\partial{t}}=H\psi$.\\
\end{center}
Esta es la ecuaci\'on de Schr\"{o}dinger y la descripci\'on del sistema por
operadores independientes del tiempo se llama
representaci\'on de Schr\"{o}dinger.
\end{itemize}
En las dos representaciones la evoluci\'on temporal del sistema se caracteriza
por el operador $\widehat{H}$, el cual se obtiene de la funci\'on de Hamilton de
la mec\'anica cl\'asica.\\
Ejemplo: $\widehat{H}$ de una part\'{\i}cula en potencial $U(x_{1},x_{2},x_{3})$ 
es:\\

$\widehat{H}=\frac{\hat{p^2}}{2m}+U(x_{1},x_{2},x_{3})$, y en la
representaci\'on x es:\\
\begin{center}
$\widehat{H}=-\frac{\hbar^{2}}{2m}\nabla+U(x_{1},x_{2},x_{3})$.
\end{center}         
\end{enumerate} 
\begin{enumerate}
\item[8.-]
El postulado 5 vale en las representaciones de Schr\"{o}dinger y de Heisenberg.
Por eso, el valor promedio de cualquier observable coincide en las dos representaciones, y entonces, hay una transformada unitaria que pasa de una representaci\'on a otra.
Tal transformaci\'on es del
tipo $\hat{s}^\dagger=\exp^{\frac{-i\hat{H}t}{\hbar}}$. Para pasar a la
representaci\'on de Schr\"{o}dinger hay que usar la transformada de
Heisenberg $\psi=\hat{s^{\dagger}}f$ con $f$ y $\hat{L}$, y para pasar a la
representaci\'on de Heisenberg se usar\'a la transformaci\'on de
 Schr\"{o}dinger $\hat{\Lambda}=\hat{s^{\dagger}}\hat{L}\hat{s}$ con $\psi$ y 
$\hat{\Lambda}$.
Ahora se obtendr\'a la ecuaci\'on de Schr\"{o}dinger: como en la
transformaci\'on $\psi=\hat{s^{\dagger}}f$ la funci\'on $f$ no depende del
tiempo, derivaremos la transformaci\'on con respecto al tiempo obteni\'endose
lo sig.:\\

$\frac{\partial{\psi}}{\partial{t}}=\frac{\partial{S^{\dagger}}}{\partial{t}}f=
\frac{\partial}{\partial{t}}(\exp^{\frac{-i\widehat{H}t}{\hbar}})f=\frac{-i}
{\hbar}\widehat{H}\exp^{\frac{-i\widehat{H}t}{\hbar}}f=\frac{-i}{\hbar}\widehat{H}\hat{s^{\dagger}}f=\frac{-i}{\hbar}\widehat{H}\psi$.\\

por lo tanto, tenemos:\\
\begin{center}
$\widehat{H}\psi-i\hbar\frac{\partial\psi}{\partial{t}}$.\\
\end{center}
Enseguida calcularemos la ecuaci\'on de Heisenberg: poniendo la transformaci\'on de Schr\"{o}dinger de la siguiente manera $\hat{s}\hat{\Lambda}\hat
{s^{\dagger}}=\hat{L}$ y derivandola con respecto al tiempo se obtiene la 
ecuaci\'on de Heisenberg\\

\begin{center}
$\frac{\partial\hat{L}}{\partial{t}}=\frac{\partial\hat{s}}{\partial{t}}
\hat{\Lambda}\hat{s^{\dagger}}+\hat{s}\hat{\Lambda}\frac{\partial\hat
{s^{\dagger}}}{\partial{t}}=\frac{i}{\hbar}\widehat{H}\exp^{\frac{i\widehat{H}t}{\hbar}}\hat\Lambda\hat{s^{\dagger}}-\frac{i}{\hbar}\hat{s}\hat\lambda
\exp^{\frac{-i\hat{H}t}{\hbar}}\widehat{H}$\\
\end{center}
\begin{center}
$=\frac{i}{\hbar}(\widehat{H}\hat{s}\hat{\Lambda}
\hat{s^{\dagger}}-\hat{s}\hat{\Lambda}\hat{s^{\dagger}}\widehat{H})=\frac{i}{\hbar}(\widehat{H}\hat{L}-\hat{L}\widehat{H})=\frac{i}{\hbar}[\widehat{H},\hat{L}]$.\\
\end{center}
Por lo tanto,tenemos:\\
\begin{center}
$\frac{\partial\hat{L}}{\partial{t}}=\frac{i}{\hbar}[\widehat{H},\hat{L}]$.\\
\end{center}
Tambi\'en la ecuaci\'on de Heisenberg se puede escribir de la sig. manera:\\
\begin{center}
$\frac{\partial\hat{L}}{\partial{t}}=\frac{i}{\hbar}\hat{s}[\widehat{H},\hat
{\Lambda}]\hat{s^{\dagger}}$.\\
\end{center}
A $\hat{L}$ se le conoce como la integral de movimiento si $\frac{d}{dt}
<\psi\mid\hat{L}\mid\psi>=0$ y est\'a caracterizada por los siguentes 
conmutadores:\\
\begin{center}
$[\widehat{H},\hat{L}]=0$,\hspace{6mm} $[\widehat{H},\hat\Lambda]=0$.

\end{center}
\end{enumerate} 
\begin{enumerate}
\item[9.-]
Los estados de un sistema descrito por las funciones propias de $\widehat{H}$ se llaman estados estacionarios del sistema, y al conjunto de valores propios 
correspondientes se les llaman espectro de energ\'{\i}a (espectro energ\'etico)
 del sistema. En tal caso la ecuaci\'on de Schr\"{o}dinger es :\\
\begin{center}
$i\hbar\frac{\partial\psi_{n}}{\partial{t}}=E_{n}\psi_{n}=\widehat{H}\psi_{n}$.\\
\end{center}
Y su soluci\'on es: \hspace{11mm}$\psi_{n}(x,t)=\exp^{\frac{-iE_{n}t}{\hbar}}\phi_{n}(x)$.\\
\begin{itemize}
\item
La probabilidad es la siguiente:\\
\begin{center}
$\delta(x)=\mid\psi_{n}(x,t)\mid^2=\mid\exp^{\frac{-iE_{n}t}{\hbar}}\phi_{n}(x)
\mid^2$\\
\end{center}
\begin{center}
$=\exp^{\frac{iE_{n}t}{\hbar}}\exp^{\frac{-iE_{n}t}{\hbar}}\mid\phi_{n}(x)
\mid^2=\mid\phi_{n}(x)\mid^2$.\\
\end{center}
Resultando que la probabilidad es constante en el tiempo.
\end{itemize}
\begin{itemize}
\item
En los estados estacionarios, el valor promedio de cualquier conmutador de tipo $[\widehat{H},\hat{A}]$ es cero, donde $\hat{A}$ es cualquier operador:\\
\begin{center}
$<n\mid\widehat{H}\hat{A}-\hat{A}\widehat{H}\mid{n}>=<n\mid\widehat{H}\hat{A}\mid{n}>-
<n\mid\hat{A}\widehat{H}\mid{n}>$\\
\end{center}
\begin{center}
$=<n\mid{E_{n}}\hat{A}\mid{n}>-<n\mid\hat{A}E_{n}\mid{n}>$\\
\end{center}
\begin{center}
$=E_{n}<n\mid\hat{A}\mid{n}>-E_{n}<n\mid\hat{A}\mid{n}>=0$.\\
\end{center}
\end{itemize}
\begin{itemize}
\item
Teorema del virial en mec\'anica cu\'antica.- Si $\widehat{H}$ es el operador 
Hamiltoniano de una part\'{\i}cula en un campo $U(r)$ y usando\\ 
$\hat{A}=1/2\sum_{i=1}^3(\hat{p_{i}}\hat{x_{i}}-\hat{x_{i}}
\hat{p_{i}})$ se obtiene lo siguiente:\\
\begin{center}
$<\psi\mid[\hat{A},\widehat{H}]\mid\psi>=0=<\psi\mid\hat{A}\widehat{H}-\widehat{H}\hat{A}
\mid\psi>$\\
\end{center}
\begin{center}
$=\sum_{i=1}^3<\psi\mid\hat{p_{i}}\hat{x_{i}}\widehat{H}-\widehat{H}\hat{p_{i}}\hat{x_{i}}\mid\psi>$\\
\end{center}
\begin{center}
$=\sum_{i=1}^3<\psi\mid[\widehat{H},\hat{x_{i}}]\hat{p_{i}}+
\hat{x_{i}}[\widehat{H},\hat{p_{i}}]\mid\psi>$.\\
\end{center}
usando varias veces los conmutadores y $\hat{p_{i}}=-i\hbar\Delta$,
$\hat{H}=\widehat{T}+U(r)$, se tiene entonces:\\
\begin{center}
$<\psi\mid[\hat{A},\widehat{H}]\mid\psi>=0$\\
\end{center}
\begin{center}
$=-i\hbar(2<\psi\mid\widehat{T}\mid\psi>-<\psi\mid\vec{r}\cdot\nabla{U(r)}\mid
\psi>)$.\\
\end{center}
Que es el teorema del virial. Si el potencial es $U(r)=U_{o}r^{n}$, entonces 
tenemos el teorema del virial como en mec\'anica cl\'asica, s\'olo
que para valores promedios\\
\begin{center}
$\overline{T}=\frac{n}{2}\overline{U}$.
\end{center}
\end{itemize}
\begin{itemize}
\item
Para un Hamiltoniano $\widehat{H}=-\frac{\hbar^2}{2m}\nabla+U(r)$ y
$[\vec{r},H]=\frac{-i\hbar}{m}\vec{p}$, y calculando los elementos de matriz se tiene:\\
\begin{center}
$(E_{k}-E_{n})<n\mid\vec{r}\mid{k}>=\frac{i\hbar}{m}<n\mid\hat{p}\mid{k}>$.
\end{center}
\end{itemize}  
\end{enumerate} 
\begin{enumerate}
\item[10.-](Densidad de corriente de probabilidad)
La siguiente integral :\\
\begin{center}
$\int\mid{\psi_{n}}(x)\mid^2dx=1$,\\
\end{center}
es la normalizaci\'on de una funci\'on propia de un espectro discreto en la representaci\'on de coordenada, y ocurre como una condici\'on de movimiento en una regi\'on finita. Por eso, los estados del espectro discreto se llaman estados
ligados.\\
Para las funciones propias del espectro continuo $\psi_{\lambda}(x)$ no se 
puede dar de manera directa una interpretaci\'on de probabilidad.\\
Supongamos una funci\'on dada $\phi$ $\in$ $L^2$, la cual la escribimos como
combinaci\'on lineal de funciones propias en el continuo:\\
\begin{center}
$\phi=\int{a(\lambda)}\psi_{\lambda}(x)dx.$\\
\end{center}
Se dice que $\phi$ corresponde a un movimiento infinito.
En muchos casos, la funci\'on $a(\lambda)$ es diferente de cero s\'olo en una 
vecindad de un punto $\lambda=\lambda_{o}$. En este caso $\phi$ se le conoce
como paquete de onda(s).\\
Vamos a calcular el cambio en el tiempo de la probabilidad de encontrar el sistema en el volumen $\Omega$.\\
\begin{center}
$P=\int_{\Omega}\mid\psi(x,t)\mid^2dx=\int_{\Omega}\psi^{\ast}(x,t)
\psi(x,t)dx$.\\
\end{center}
Derivando la integral con respecto al tiempo tenemos:\\
\begin{center}
$\frac{dP}{dt}=\int_{\Omega}(\psi\frac{\partial{\psi^{\ast}}}{\partial{t}}+
\psi^{\ast}\frac{\partial{\psi}}{\partial{t}})dx$.\\
\end{center}
Utilizando la ecuaci\'on de Schr\"{o}dinger del lado derecho de la integral se 
tiene lo siguiente:\\
\begin{center}
$\frac{dP}{dt}=\frac{i}{\hbar}\int_{\Omega}(\psi\hat{H}\psi^{\ast}-\psi^{\ast}
\hat{H}\psi)dx$.\\
\end{center}
Usando la identidad $f\nabla{g}-g\nabla{f}=div[(f) grad{(g)}-(g) 
grad{(f)}]$
y la ecuaci\'on de Schr\"{o}dinger de la forma siguiente:\\
\begin{center}
$\hat{H}\psi=\frac{\hbar^2}{2m}\nabla{\psi}$.\\
\end{center}
Sustituyendo lo anterior en la integral se obtiene:\\

\begin{center}
$\frac{dP}{dt}=\frac{i}{\hbar}\int_{\Omega}[\psi(-\frac{\hbar^2}{2m}\nabla{\psi
^{\ast}})-\psi^{\ast}(\frac{-\hbar^2}{2m}\nabla{\psi})]dx$\\
\end{center}
\begin{center}
$=-\int_{\Omega}\frac{i\hbar}{2m}(\psi\nabla{\psi^{\ast}}-\psi^{\ast}\nabla
\psi)dx$\\
\end{center}
\begin{center}
$=-\int_{\Omega}div\frac{i\hbar}{2m}(\psi\nabla{\psi^{\ast}}-\psi^{\ast}
\nabla{\psi})dx$.\\
\end{center}
Usando el teorema de la divergencia para transformar la integral de volumen en
una de superficie, entonces tenemos lo siguiente:\\
\begin{center}
$\frac{dP}{dt}=-\oint\frac{i\hbar}{2m}(\psi\nabla{\psi^{\ast}}-\psi^{\ast}
\nabla{\psi})dx$.\\
\end{center}
A la cantidad $\vec J(\psi)=\frac{i\hbar}{2m}(\psi\nabla{\psi^{\ast}}-
\psi^{\ast}\nabla{\psi})$ se le conoce como densidad de corriente de
probabilidad y de inmediato se obtiene una ecuaci\'on de continuidad,\\
\begin{center}
$\frac{d\rho}{dt}+div(\vec J)=0$.\\
\end{center}
\begin{itemize}
\item
Si $\psi(x)=AR(x)$, donde $R(x)$ es funci\'on real,
entonces: $\vec J(\psi)=0$.\\
\end{itemize}
\begin{itemize}
\item
Para las funciones propias del impulso $\psi(x)=\frac{1}{(2\pi{\hbar})^3/2}
\exp^{\frac{i\vec{p}\vec{x}}{\hbar}}$ se obtiene:\\

\begin{center}
$J(\psi)=\frac{i\hbar}{2m}(\frac{1}{(2\pi{\hbar})^3/2}\exp^{\frac
{i\vec{p}\vec{x}}{\hbar}}(\frac{i\vec{p}}{\hbar(2\pi{\hbar})^3/2}\exp
^{\frac{-i\vec{p}\vec{x}}{\hbar}})$\\
\end{center}
\begin{center}
$-(\frac{1}{(2\pi{\hbar})^3/2}\exp^{\frac{-i\vec
{p}\vec{x}}{\hbar}}\frac{i\vec{p}}{\hbar(2\pi{\hbar})^3/2}
\exp^{\frac{i\hbar
\vec{p}\vec{x}}{\hbar}}))$\\
\end{center}
\begin{center}
$=\frac{i\hbar}{2m}(-\frac{2i\vec{p}}{\hbar(2\pi{\hbar})^3})=\frac{\vec{p}}
{m(2\pi{\hbar})^3}$,\\
\end{center}
 lo cual nos dice que no depende de la coordenada la densidad de probabilidad.
\end{itemize}
\end{enumerate} 
\begin{enumerate}
\item[11.-](Operador de transporte espacial)
Si $\widehat{H}$ es invariante ante translaciones de cualquier vector $\vec{a}$,\\
\begin{center}
$\widehat{H}(\vec{r}+\vec{a})=\widehat{H}\vec{(r)}$.\\
\end{center}
Entonces, hay un $\widehat{T}(\vec{a})$ unitario $\widehat{T}^{\dagger}(\vec
{a})\widehat{H}(\vec{r})\widehat{T}(\vec{a})=\widehat{H}(\vec{r}+\vec{a})$.\\
Por la conmutaci\'on de las translaciones\\
\begin{center}
 $\widehat{T}(\vec{a})\widehat{T}(\vec{b})=
\widehat{T}(\vec{b})\widehat{T}(\vec{a})=\widehat{T}(\vec{a}+\vec{b})$,\\
\end{center}
 resulta que $\widehat{T}$ tiene la forma $\widehat{T}=\exp^{i\hat{k}a}$,
 donde, $\hat{k}=\frac{\hat{p}}{\hbar}$.\\
En el caso infinitesimal:\\
\begin{center}
$\widehat{T}(\delta\vec{a})\widehat{H}\widehat{T}(\delta\vec{a})\approx(\hat{I}+i\hat{k}
\delta\vec{a})\widehat{H}(\hat{I}-i\hat{k}\delta\vec{a})$,\\
\end{center}
\begin{center}
$\widehat{H}(\vec{r})+i[\hat{K},\widehat{H}]\delta\vec{a}=\widehat{H}(\vec{r})+(\nabla\widehat{H})\delta\vec{a}$.\\
\end{center}
Tambi\'en $[\hat{p},\widehat{H}]=0$, donde $\hat{p}$ es una integral de movimiento.
El sistema tiene funciones de onda de la forma $\psi(\vec{p},\vec{r})=\frac
{1}{(2\pi\hbar)^3/2}\exp^{\frac{i\vec{p}\vec{r}}{\hbar}}$ y la transformada 
unitaria hace que $\exp^{\frac{i\vec{p}\vec{a}}{\hbar}}\psi(\vec{r})=\psi(\vec{r}+\vec{a})$.\\
El operador de transporte espacial $\widehat{T}^\dagger
=\exp^{\frac{-i\vec{p}\vec{a}}
{\hbar}}$ es an\'alogo al\\
 $\hat{s}^\dagger=\exp^{\frac{-i\hat{H}t}{\hbar}}$ el operador de transporte 
temporal.
\end{enumerate}   
\begin{enumerate}
\item[12.-]Ejemplo:
Si $\widehat{H}$ es invariante con respecto a una traslaci\'on discreta (por ejemplo
en una red cristalina) $\widehat{H}(\vec{r}+\vec{a})=\widehat{H}(\vec{r})$ donde
$\vec{a}=\sum_{i}\vec{a_{i}}n_{i}$, $n_{i}$ $\in$ $N$ y $a_{i}$ son los 
vectores b\'aricos.\\
Entonces:\\
\begin{center}
$\widehat{H}(\vec{r})\psi(\vec{r})=E\psi(\vec{r})$,\\
\end{center}
\begin{center}
$\widehat{H}(\vec{r}+\vec{a})\psi(\vec{r}+\vec{a})=E\psi(\vec{r}+\vec{a})=\hat{H}
(\vec{r})\psi(\vec{r}+\vec{a})$.\\
\end{center}
Resultando que $\psi(\vec{r})$ y $\psi(\vec{r}+\vec{a})$ son funciones de onda para el mismo valor propio de $\widehat{H}$.
La relaci\'on entre $\psi(\vec{r})$ y $\psi(\vec{r}+\vec{a})$ se puede buscar 
en la forma $\psi(\vec{r}+\vec{a})=\hat{c}(\vec{a})\psi(\vec{r})$ donde
$\hat{c}(\vec{a})$ es una matriz gxg (g es el grado de degeneraci\'on del nivel E). Dos matrices tipo c, $\hat{c}(\vec{a})$ y $\hat{c}(\vec{b})$ conmutan y entonces son diagonalizables en el mismo tiempo.\\
Adem\'as para los elementos diagonales hay de tipo \\
$c_{ii}(\vec{a})c_{ii}(\vec{b})=c_{ii}(\vec{a}+\vec{b})$, donde i=1,2,....,,g.
La soluci\'on de esta ecuaci\'on es del tipo $c_{ii}(a)=\exp^{ik_{i}a}$ resulta 
que $\psi_{k}(\vec{r})=U_{k}(\vec{r})\exp^{i\vec{k}\vec{a}}$ donde $\vec{k}$
es un vector real cualquiera, y la
funci\'on $U_{k}(\vec{r})$ es peri\'odica con
periodo $\vec{a}$, entonces: $U_{k}(\vec{r}+\vec{a})=U_{k}(\vec{r})$.\\
La aseveraci\'on de que las funciones propias de un $\hat{H}$ peri\'odico 
cristalino $\hat{H}(\vec{r}+\vec{a})=\hat{H}(\vec{r})$ se pueden escribir 
$\psi_{k}(\vec{r})=U_{k}(\vec{r})\exp{i\vec{k}\vec{a}}$ con
$U_{k}(\vec{r}+\vec{a})=U_{k}(\vec{r})$ se llama teorema de Bloch.\\
En el caso continuo, $U_{k}$ debe ser constante, porque la constante es la 
\'unica funci\'on peri\'odica para cualquier $\vec{a}$.\\
El vector $\vec{p}=\hbar\vec{k}$ se llama cuasi-impulso (analog\'{\i}a con el
caso continuo). El vector $\vec{k}$ no est\'a determinado de manera un\'{\i}voca, 
porque se le puede juntar cualquier vector $\vec{g}$ para el cual $ga=2\pi{n}$
donde n $\in$ N.\\
El vector $\vec{g}$ se puede escribir $\vec{g}=\sum_{i=1}^{3}\vec{b_{i}}m_{i}$
donde $m_{i}$ son n\'umeros enteros y $b_{i}$ est\'an dados por\\
\begin{center}
$\vec{b_{i}}=2\pi\frac{\hat{a_{j}}\times\vec{a_{k}}}{\vec{a_{i}}(\vec{a_{j}}\times\vec
{a_{k}})}$ \\
\end{center}
si $i\neq{j}\neq{k}$ son los vectores b\'aricos de la red cristalina.\\

\end{enumerate}
Citas:

\noindent 1. Acetatos del Prof. H. Rosu.

\noindent 2. E. Farhi, J. Goldstone, S. Gutmann, ``How probability arises in
quantum mechanics",

\noindent Annals of Physics {\bf 192}, 368-382 (1989)

\noindent 3. N.K. Tyagi en Am. J. Phys. {\bf 31}, 624 (1963) da una 
demostraci\'on muy corta del principio de incertidumbre de Heisenberg, que 
dice que la medici\'on simult\'anea de dos operadores hermitianos que no 
conmutan produce una incertidumbre relacionada con el valor del conmutador.

\noindent
Notas:

\noindent
1. Para ``la creaci\'on de la MC...", Werner Heisenberg ha sido galardonado
con el premio Nobel en 1932 (y lo recibi\'o en 1933). El
art\'{\i}culo ``Zur Quantenmechanik. II", Zf. f. Physik {\bf 35}, 557-615
(1926) (recibido el 16 de Noviembre de 1925) de M. Born, W. Heisenberg y
P. Jordan, se le conoce como ``el trabajo de las tres gentes" y est\'a
considerado como el que abri\'o de verdad los grandes horizontes de la MC.

\noindent
2. Para ``la interpretaci\'on estad\'{\i}stica de la funci\'on de onda"
Max Born recibi\'o el premio Nobel en 1954.
\newpage
\section*{P r o b l e m a s\\} % de mec\'anica cu\'antica:\\}
{\bf Problema 1.1}:\\

Considerar dos operadores A y B los cuales por hipot\'esis, conmutan.
Entonces se derivar\'a la relaci\'on:\\

$\exp^{A}\exp^{B}=\exp^{A+B}\exp^{1/2[A,B]}$,\hspace{10mm}
(f\'ormula de Glauber).\\

Definiremos un operador F(t), una funci\'on de variable real t, por:\\
$F(t)=\exp^{At}\exp^{Bt}$.\\
tenemos:\\
$\frac{dF}{dt}=A\exp^{At}\exp^{Bt}+\exp^{At}B\exp^{Bt}=(A+\exp^{At}B\exp^{-At})
F(t)$.\\

Ahora aplicando la f\'ormula $[A,F(B)]=[A,B]F^{'}(B)$, tenemos que \\
$[\exp^{At},B]=t[A.B]\exp^{At}$, por lo tanto:
$\exp^{At}B=B\exp^{At}+t[A,B]\exp^{At}$\\
multiplicando ambos lados de la ecuaci\'on pasada por $\exp^{-At}$
y sustituyendo en la primera ecuaci\'on, obtenemos:\\

$\frac{dF}{dt}=(A+B+t[A,B])F(t)$.\\

Los operadores $A$ y $B$ y $[A,B]$ conmutan por hipot\'esis. Por lo tanto, podemos
integrar la ecuaci\'on diferencial como
si $A+B$ y $[A,B]$ fueran n\'umeros.\\
Entonces tenemos:\\

$F(t)=F(0)\exp^{(A+B)t+1/2[A,B]t^2}$.\\

Poniendo $t=0$, se ve que $F(0)=1$, y :\\

$F(t)=\exp^{(A+B)t+1/2[A,B]t^2}$.\\

Entonces poniendo $t=1$, obtenemos el resultado deseado.\\

\bigskip

\noindent
{\bf Problema 1.2}:\\

Se calcular\'a el conmutador $[X,D_{x}]$. Para hacerlo, tomaremos una funci\'on arbitraria $\psi(\vec{r})$:\\

$[X,D_{x}]\psi(\vec{r})=(x\frac{\partial}{\partial{x}}-\frac{\partial}{\partial{x}}x)\psi(\vec{r})=x\frac{\partial}{\partial{x}}\psi(\vec{r})-\frac{\partial}{\partial{x}}[x\psi(\vec{r})]\\
=x\frac{\partial}{\partial{x}}\psi(\vec{r})-\psi(\vec{r})-x\frac{\partial}{\partial{x}}\psi(\vec{r})=-\psi(\vec{r})$.\\
Entonces si es v\'alido para toda $\psi(\vec{r})$, se puede deducir que:\\

$[X,D_{x}]=-1$.         

\bigskip
 
%\newpage
\noindent
{\bf Problema 1.3}:\\

Se probar\'a que la traza es invariante ante un cambio de base ortonormal discreta.\\

La suma de los elementos de la diagonal de la matriz la cual representa un operador A en una base arbitraria no depende de la base.\\
Se derivar\'a esta propiedad para el caso del cambio de una base ortonormal dicreta $[\mid{u_{i}}>]$ a otra base ortonormal dicreta $[\mid{t_{k}}>]$. Tenemos:\\
$\sum_{i}<u_{i}\mid{A}\mid{u_{i}}>=\sum_{i}<u_{i}\mid[\sum_{k}\mid{t_{k}}><t_{k}\mid]A\mid{u_{i}}>$\\

(donde se ha usado la relaci\'on de cerradura para el
estado $t_{k}$). El lado derecho de la relaci\'on pasada es igual a:\\

$\sum_{i,j}<u_{i}\mid{t_{k}}><t_{k}\mid{A}\mid{u_{i}}>=\sum_{i,j}<t_{k}\mid{A}\mid{u_{i}}><u_{i}\mid{t_{k}}>$,\\

(es posible el cambio de orden del producto de dos n\'umeros). Entonces, podemos 
reemplazar $\sum_{i}\mid{u_{i}}><u_{i}\mid$ por uno (relaci\'on de cerradura para el estado $\mid{u_{i}}>$), y se obtiene finalmente:\\

$\sum_{i}<u_{i}\mid{A}\mid{u_{i}}>=\sum_{k}<t_{k}\mid{A}\mid{t_{k}}>$.
Por lo tanto, se ha mostrado la propiedad de invariancia para la traza.

%\end{document}

\newpage
%%%%%%%%%%%%%%%%%%%%%%%%%%%%%%%%%%%%
%\documentstyle[12pt]{article}
\newcommand{\aple}{\mbox{${}_{\textstyle\sim}^{\textstyle<}$}}
\newcommand{\apge}{\mbox{${}_{\textstyle\sim}^{\textstyle>}$}}
\newcommand{\slsh}[1]{\mbox{$\displaystyle {#1}\!\!\!{/}$}}
\newcommand{\lpr}{\mbox{$ \displaystyle O_L $}}
\newcommand{\rpr}{\mbox{$ \displaystyle O_R $}}
\newcommand{\GeV}{\mbox{$\rm  \, GeV $}}
%\begin{document}
%\date{}
%\title
\section*{\bf 2. POTENCIALES BARRERAS y POZOS}      %%%%%%%%%%%%  2
%\end{center}
%\maketitle

\section*{Comportamiento de una Funci\'on de Onda Estacionaria $\psi(x)$}

\subsection*{Regiones de Energ\'{\i}a Potencial Constante}

\qquad En el caso de un potencial cuadrado, $V(x)$ es una funci\'on constante $V(x)=V$en cierta regi\'on del espacio. En tal regi\'on, la ecuaci\'on de Schr\"odinger puede ser escrita:
\begin{equation}
\frac{d^2}{dx^2} \psi(x) + \frac{2m}{\hbar^2} (E-V)\psi(x) = 0
\end{equation}
 
Distinguiremos entre varios casos:

{\bf (i) $E>V$}

Introduzcamos la constante positiva $k$, definida por
\begin{equation}
k = \frac{\sqrt{2m(E-V)}}{\hbar}
\end{equation}

\noindent
La soluci\'on de la ecuaci\'on (1) puede ser entonces escrita:
\begin{equation}
\psi(x) = Ae^{ikx} + A'e^{-ikx}
\end{equation}

\noindent
donde $A$ y $A'$ son constantes complejas.

%\newpage
% .................................................

{\bf (ii) $E<V$}

Esta condici\'on corresponde a regiones del espacio las cuales
estar\'{\i}an prohibidas para la part\'{\i}cula por las leyes de
la mec\'anica cl\'asica. En este caso, introducimos la constante
positiva $q$ definida por:

\begin{equation}
 q = \frac{\sqrt{2m(V-E)}}{\hbar} 
\end{equation}
y la soluci\'on de (1) puede ser escrita:
\begin{equation}
\psi(x) = Be^{q x} + B'e^{-q x}
\end{equation}
donde $B$ y $B'$ son constantes complejas.

{\bf (iii) $E = V$}

\noindent
En este caso especial, $\psi(x)$ es una funci\'on lineal de $x$.

\noindent
\subsection*{Comportamiento de $\psi(x)$ en
una discontinuidad de energ\'{\i}a potencial.}

\qquad Podr\'{\i}a pensarse
que en el punto $x=x_1$, donde el potencial $V(x)$ es discontinuo,
la funci\'on de onda $\psi(x)$ se
comportar\'{\i}a extra\~namente,llegando a ser por s\'{\i} misma
discontinua, por ejemplo. Este no es el caso: $\psi(x)$ y
$\frac{d\psi}{dx}$ son continuas, y es s\'olo la segunda derivada la que
es discontinua en $x=x_1$

\noindent
\subsection*{Visi\'on general del c\'alculo}

\qquad El procedimiento para determinar el estado estacionario en
un ``potencial cuadrado'' es por lo tanto el siguiente: en todas
las regiones donde $V(x)$ es constante, escribimos $\psi(x)$ en
cualquiera de las dos formas (3) o (5) seg\'un sea aplicable; entonces
pegamos estas funciones por requerimientos de continuidad de $\psi(x)$ y
de $\frac{d\psi}{dx}$ en los puntos donde $V(x)$ es discontinuo.

\noindent
\section*{Examinaci\'on de ciertos casos simples}
\qquad Llevemos a cabo el c\'alculo cuantitativo de los estados
estacionarios, hecho de acuerdo al m\'etodo descrito arriba.

% ========================================================================
\subsection*{Potencial escal\'on}

%%%%%%%%%%%%%%
\vskip 2ex
\centerline{
\epsfxsize=280pt
\epsfbox{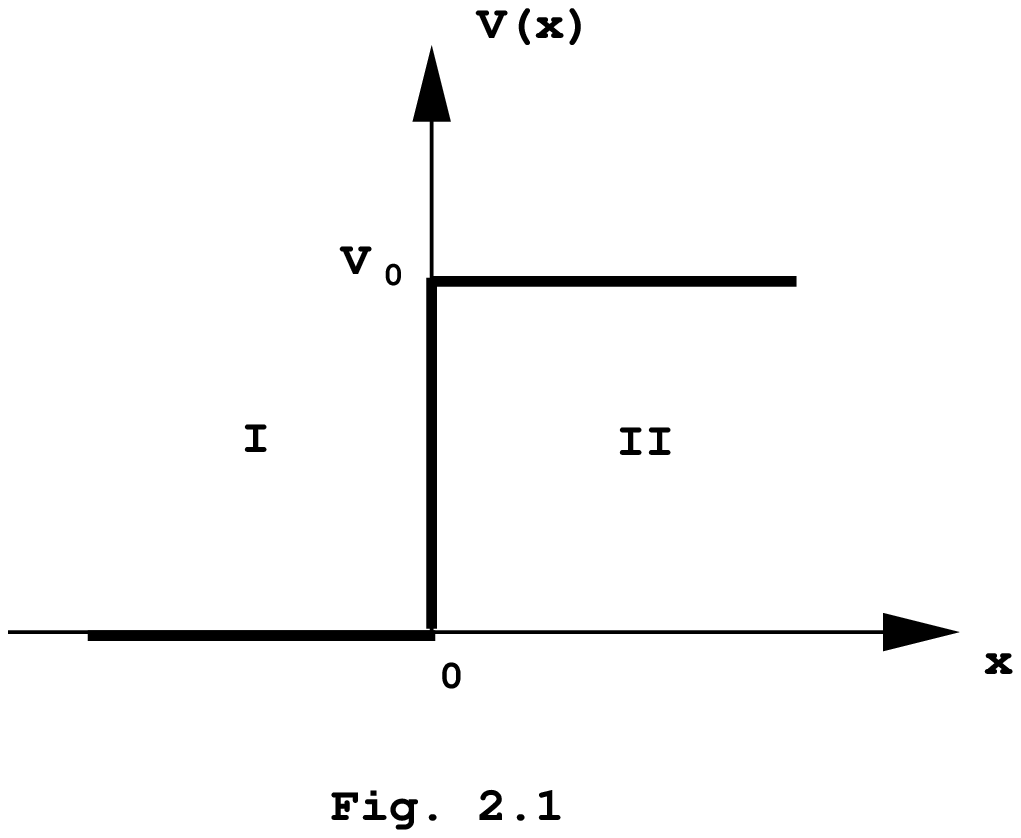}}
\vskip 4ex
%\begin{center}
%{\small{Fig. 1}\\
%}
%\end{center}
%%%%%%%%%%%%%%%%

{\bf a. Caso donde $E>V_0$; {\it reflexi\'on parcial}}

Pongamos la ec. (2) como:

\begin{eqnarray}
k_1 &=& \frac{\sqrt{2mE}}{\hbar}  \\
\nonumber\\
k_2 &=& \frac{\sqrt{2m(E-V_0)}}{\hbar}
\end{eqnarray}

La soluci\'on de la ec. (1) tiene la forma de la ec. (3) en las regiones $I (x<0)$ y $II (x>0)$:
\begin{eqnarray}
\psi_I &=& A_1e^{ik_1x} + A_1'e^{-ik_1x}  \nonumber
\nonumber\\
\psi_{II} &=& A_2e^{ik_2x} + A_2'e^{-ik_2x} \nonumber
\end{eqnarray}
% .......................................................................
En la regi\'on I la ec. (1) toma la forma:
\begin{eqnarray}
\psi''(x) + \frac{2mE}{\hbar^2}\psi(x) = \psi''(x) + k^2\psi(x) = 0 \nonumber
\end{eqnarray}

\noindent
y en la region II:
\begin{eqnarray}
\psi''(x) - \frac{2m}{\hbar^2} [V_0-E]\phi(x) = \psi''(x) - q^2\psi(x) = 0 \nonumber
\end{eqnarray}
% ......................................................................
Si nos limitamos al caso de una part\'{\i}cula incidente viniendo desde $x=-\infty$, debemos elegir $A_2'=0$ y podemos determinar los radios $A_1'/A_1$ y $A_2/A_1$. Las condiciones de pegado entonces dan:
\begin{itemize}
\item
$\psi_I = \psi_{II}$,\qquad en $x=0:$
\begin{equation}
A_1+A_1' = A_2
\end{equation}
\item
$\psi'_I = \psi'_{II}$,\qquad en $x=0:$
\begin{equation}
A_1ik_1 - A_1'ik_1 = A_2ik_2
\end{equation}
\end{itemize}

\noindent
Sustituyendo $A_1$ y $A_1'$ de (8) en (9):
\begin{eqnarray}
A_1' &=& \frac{A_2(k_1 - k_2)}{2k_1} \\
\nonumber\\
A_1 &=& \frac{A_2(k_1 + k_2)}{2k_1}
\end{eqnarray}
\noindent
La igualaci\'on de la constante $A_2$ en (10) y (11) resulta:
\begin{equation}
\frac{A_1'}{A_1} = \frac{k_1 - k_2}{k_1 + k_2},
\end{equation}
y un despeje en (11) nos da:
\begin{equation}
\frac{A_2}{A_1} = \frac{2k_1}{k_1+k_2}
\end{equation}

$\psi(x)$ es la superposici\'on de dos ondas. La primera
(el t\'ermino en $A_1$) corresponde a una part\'{\i}cula incidente, con
momento $p = \hbar k_1$, propag\'andose de izquierda a derecha. La
segunda (el t\'ermino en $A_1'$) corresponde a una part\'{\i}cula reflejada,
con momento $-\hbar k_1$, propag\'andose en la direcci\'on opuesta. Ya que
hemos elegido $A_2' = 0$, $\psi_{II}(x)$ consiste de una sola onda, la cual
est\'a asociada con una part\'{\i}cula transmitida. (En la siguiente
p\'agina se muestra c\'omo es posible, usando el concepto de una corriente de
probabilidad, definir el coeficiente de transmisi\'on T y el coeficiente de
reflexi\'on R del potencial escal\'on). Estos coeficientes dan la probabilidad
de que la part\'{\i}cula, llegando de $x
=-\infty$, pase el potencial escal\'on en $x=0$ o se regrese. As\'{\i}
encontramos:
\begin{equation}
R = | \frac{A_1'}{A_1}|^2
\end{equation}

\noindent
y, para $T$:
\begin{equation}
T = \frac{k_2}{k_1}| \frac{A_2}{A_1}|^2
\end{equation}

Tomando en cuenta a (12) y (13), tenemos:
\begin{eqnarray}
R &=& 1- \frac{4 k_1 k_2}{(k_1 + k_2)^2} \\
\nonumber\\
T &=& \frac{4 k_1 k_2}{(k_1 + k_2)^2}
\end{eqnarray}
  
Es f\'acil verificar que $R+T=1$: es cierto que la part\'{\i}cula ser\'a transmitida o reflejada. Contrariamente a las predicciones de la mec\'anica cl\'asica, la part\'{\i}cula incidente tiene una probabilidad no nula de regresarse.

Finalmente es f\'acil verificar, usando (6) y (7) y (17), que, si $E \gg V_0$, $T \simeq 1$: cuando la energ\'{\i}a de la part\'{\i}cula es suficientemente grande comparada con la altura del potencial escal\'on, la part\'{\i}cula salva este escal\'on como
 si no existiera.

\bigskip

%\newpage
Considerando la soluci\'on en la regi\'on I:
\begin{eqnarray}
\psi_I = A_1e^{ik_1x} + A_1'e^{-ik_1x}  \nonumber
\end{eqnarray}

\begin{equation}
j = -\frac{i\hbar}{2m}(\phi^* \bigtriangledown \phi - \phi \bigtriangledown \phi^*)
\end{equation}

con $A_1 e^{ik_1x}$ y su conjugado $A_1^* e^{-ik_1x}$:
\begin{eqnarray}
j &=& -\frac{i\hbar}{2m}[(A_1^* e^{-ik_1x})(A_1 i k_1 e^{ik_1x})-(A_1 e^{ik_1x})(-A_1^* i k_1 e^{-ik_1x})] \nonumber \\
\nonumber \\
j &=& \frac{\hbar k_1}{m}|A_1|^2 \nonumber
\end{eqnarray}

Ahora con $A'_1 e^{-ik_1x}$ y su conjugado $A_1^* e^{ik_1x}$ resulta:

\noindent
\begin{eqnarray}
j = -\frac{\hbar k_1}{m}|A'_1|^2 \nonumber
\end{eqnarray}

Deseamos en seguida verificar la proporci\'on de corriente que se refleja con respecto a la corriente que incide (m\'as precisamente, queremos verificar la probabilidad de que la part\'{\i}cula se regrese):
\begin{eqnarray}
R = \frac{|j(\phi_-)|}{|j(\phi_+)|} = \frac{| -\frac{\hbar k_1}{m}|A'_1|^2|}{| \frac{\hbar k_1}{m}|A_1|^2|} = |\frac{A'_1}{A_1}|^2
\end{eqnarray}

En forma similar, la proporci\'on de lo que se transmite con respecto a lo
que incide (o sea la probabilidad de que la part\'{\i}cula se transmita) es,
tomando ahora en cuenta la soluci\'on de la regi\'on II:
\begin{eqnarray}
T = \frac{|\frac{\hbar k_2}{m}|A_2|^2|}{| \frac{\hbar k_1}{m}|A_1|^2|} = \frac{k_2}{k_1}|\frac{A_2}{A_1}|^2
\end{eqnarray}

% ...........................................................................
%\newpage
{\bf a. Caso donde $E<V_0$; {\it reflexi\'on total}}

En este caso tenemos:
\begin{eqnarray}
k_1 &=& \frac{\sqrt{2mE}}{\hbar}  \\ 
\nonumber\\
q_2 &=& \frac{\sqrt{2m(V_0-E)}}{\hbar}
\end{eqnarray}
En la regi\'on $I (x<0)$, la soluci\'on de la ec. (1) [dada como $\psi(x)'' + k_1^2\psi(x) = 0$] tiene la forma de la ec. (3):

\begin{equation}
\psi_I = A_1e^{ik_1x} + A_1'e^{-ik_1x} 
\end{equation}
 
\noindent
Y, en la regi\'on $II (x>0)$, la misma ec. (1) [ahora dada como $\psi(x)'' -  q_2^2\psi(x) = 0$] tiene la forma de la ec. (5):
\begin{equation}
\psi_{II} = B_2e^{q_2x} + B_2'e^{-q_2x}
\end{equation}

\noindent
Para que la soluci\'on permanezca limitada cuando $x \rightarrow + \infty$, es necesario que:
\begin{equation}
B_2 = 0
\end{equation}
Las condiciones de pegado en $x=0$ dan en este caso:

\begin{itemize}
\item
$\psi_I = \psi_{II}$,\qquad en $x=0:$
\begin{equation}
A_1 + A_1' = B_2'
\end{equation}
\item
$\psi'_I = \psi'_{II}$,\qquad en $x=0:$
\begin{equation}
A_1 ik_1 - A_1' ik_1 = - B_2' q_2
\end{equation}
\end{itemize}

\noindent
Sustituyendo $A_1$ y $A_1'$ de (26) en (27):
\begin{eqnarray}
A_1' &=& \frac{B_2'(i k_1 + q_2)}{2i k_1} \\
\nonumber\\
A_1 &=& \frac{B_2'(i k_1 -  q_2)}{2i k1}
\end{eqnarray}
\noindent
La igualaci\'on de la constante $B_2'$ en (28) y (29) resulta:
\begin{equation}
\frac{A_1'}{A_1} = \frac{i k_1 + q_2}{i k_1 - q_2} = \frac{k_1 - iq_2}{k_1 + iq_2}, 
\end{equation}
y un despeje en (29) nos da:
\begin{equation}
\frac{B_2'}{A_1} = \frac{2i k_1}{i k_1 - q_2} =\frac{2 k_1}{k_1 - iq_2} 
\end{equation}

\noindent
El coeficiente de reflexi\'on $R$ es entonces:
\begin{equation}
R = | \frac{A_1'}{A_1}|^2 = | \frac{k_1 - i q_2}{k_1 + i q_2}|^2 = \frac{k_1^2 + q_2^2}{k_1^2 + q_2^2} = 1   
\end{equation}
\noindent
Como en la mec\'anica cl\'asica, la part\'{\i}cula es siempre reflejada
(reflexi\'on total). Sin embargo, hay una diferencia importante: debido a
la existencia de una onda evanescente $e^{-q_2x}$, la part\'{\i}cula tiene
una probabilidad no nula de estar 
presente en la regi\'on del espacio la cual, cl\'asicamente, le
ser\'{\i}a prohibida. Esta probabilidad decrece exponencialmente
con $x$ y llega a ser despreciable cuando $x$ es m\'as grande que
el ``rango'' $1/q_{2}$ de la onda evanescente. Notemos tambi\'en que el
coeficiente $A_1'/A_1$ es complejo. Un cierto cambio de fase aparece a causa
de la  reflexi\'on, el cual, f\'{\i}sicamente, es debido al hecho de que la
part\'{\i}cula es retardada cuando penetra la regi\'on $x>0$. No hay
analog\'{\i}a en la mec\'anica cl\'asica.

% =============================================================================
\subsection*{Potencial Barrera}

%%%%%%%%%%%%%%
\vskip 2ex
\centerline{
\epsfxsize=280pt
\epsfbox{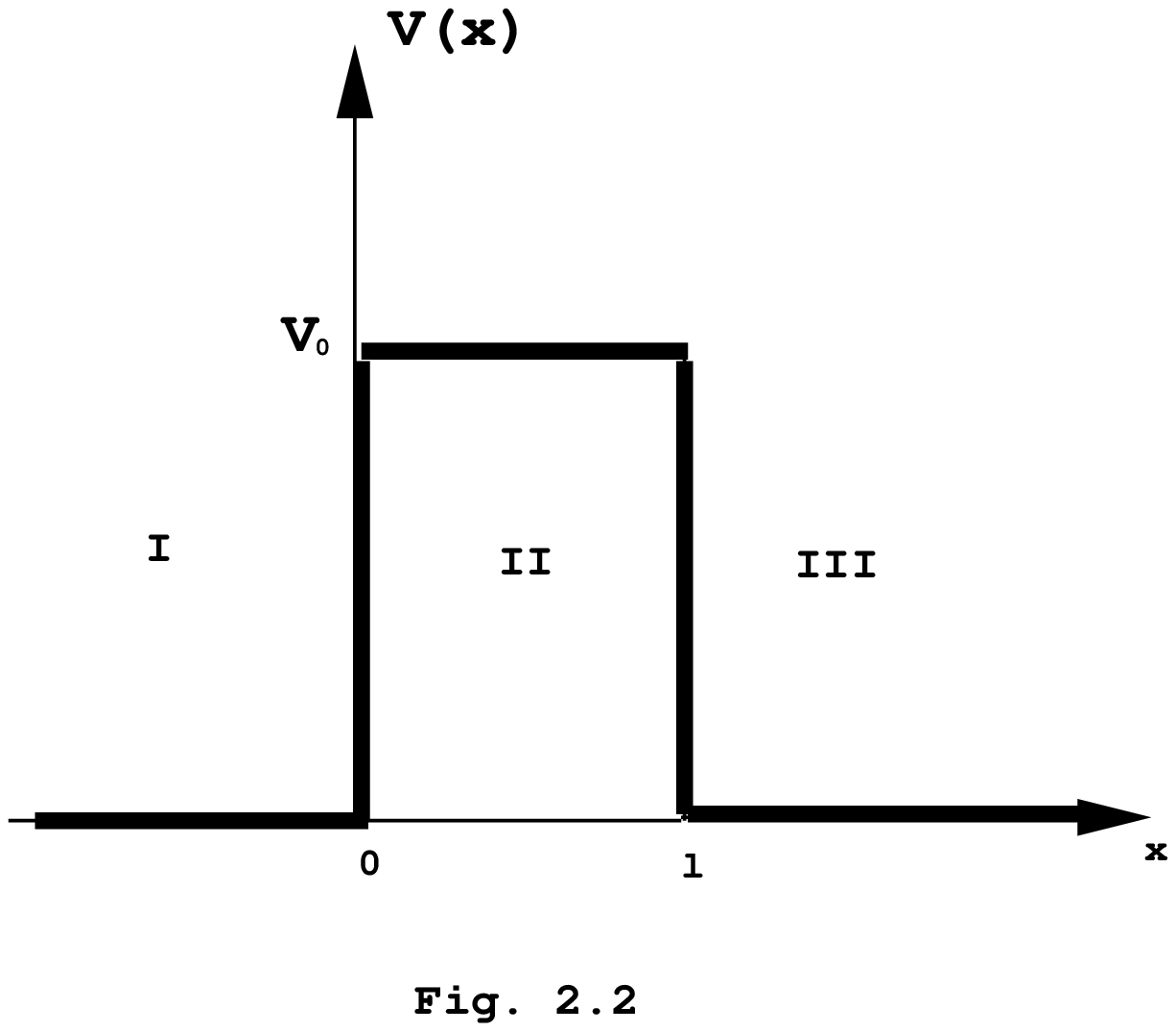}}
\vskip 4ex
%\begin{center}
%{\small{Fig. 1}\\
%}
%\end{center}
%%%%%%%%%%%%%%%%

{\bf a. Caso donde $E>V_0$}; {\it resonancias}

Pongamos aqu\'{\i} la ec. (2) como:
\begin{eqnarray}
k_1 &=& \frac{\sqrt{2mE}}{\hbar}  \\
\nonumber\\
k_2 &=& \frac{\sqrt{2m(E-V_0)}}{\hbar}
\end{eqnarray}

La soluci\'on de la ec. (1) tiene la forma de la ec. (3) en las 
regiones $I (x<0)$, $II (0<x<a$) y $III (x>a):$
\begin{eqnarray}
\psi_I &=& A_1e^{ik_1x} + A_1'e^{-ik_1x} \nonumber
\nonumber\\
\psi_{II} &=& A_2e^{ik_2x} + A_2'e^{-ik_2x}\nonumber
\nonumber\\
\psi_{III} &=& A_3e^{ik_1x} + A_3'e^{-ik_1x}\nonumber   
\end{eqnarray}

Si nos limitamos al caso de una part\'{\i}cula incidente viniendo 
desde $x=-\infty$, debemos elegir $A_3'=0$. 
% ......................................................

\begin{itemize}
\item
$\psi_I = \psi_{II}$,\qquad en $x=0:$
\begin{equation}
A_1 + A_1' = A_2 + A_2'
\end{equation}
\item
$\psi'_I = \psi'_{II}$,\qquad en $x=0:$
\begin{equation}
A_1ik_1 - A_1'ik_1 = A_2ik_2 - A_2'ik_2
\end{equation}
\item
$\psi_{II} = \psi_{III}$,\qquad en $x=a:$
\begin{equation}
A_2e^{ik_2a} + A_2'e^{-ik_2a} = A_3e^{ik_1a} 
\end{equation}
\item
$\psi'_{II} = \psi'_{III}$,\qquad en $x=a:$ 
\begin{equation}
A_2ik_2e^{ik_2a} - A_2'ik_2e^{-ik_2a} = A_3ik_1e^{ik_1a} 
\end{equation}
\end{itemize}

\noindent
Las condiciones de continuidad en $x=a$ dan entonces a $A_2$ y $A_2'$ en funci\'on de $A_3$, y aquellas en $x=0$ dan a $A_1$ y $A_1'$ en funci\'on de $A_2$ y $A_2'$ (y, consecuentemente, en funci\'on de $A_3$). Este proceso es mostrado enseguida:

\noindent 
Sustituyendo $A_2'$ de la ec. (37) en (38):
\begin{equation}
A_2 = \frac{A_3e^{ik_1a}(k_2+k_1)}{2k_2e^{ik_2a}}
\end{equation}

\noindent
Sustituyendo $A_2$ de la ec. (37) en (38):
\begin{equation}
A_2' = \frac{A_3e^{ik_1a}(k_2-k_1)}{2k_2e^{-ik_2a}}
\end{equation}
\noindent
Sustituyendo $A_1$ de la ec. (35) en (36):
\begin{equation}
A_1' = \frac{A_2(k_2-k_1)-A_2'(k_2+k_1}{-2k_1}
\end{equation}

\noindent
Sustituyendo $A_1'$ de la ec. (35) en (36):
\begin{equation}
A_1 = \frac{A_2(k_2+k_1)-A_2'(k_2-k_1}{2k_1}
\end{equation}

\noindent
Ahora, sustituyendo en (41) las ecuaciones (39) y (40):
\begin{equation}
A_1' = i \frac{(k_2^2 - k_1^2)}{2 k_1 k_2} (\sin k_2a) e^{ik_1a}A_3
\end{equation}

\noindent
Y, finalmente, sustituyendo en (42) las ecuaciones (39) y (40):
\begin{equation}
A_1 = [\cos k_2a - i\frac{k_1^2 + k_2^2}{2 k_1 k_2} \sin k_2a] e^{ik_1a}A_3
\end{equation}
$A_1'/A_1$ y $A_3/A_1$ [relaciones que salen de igualar las ecuaciones (43) y (44), y del despeje de la ec. (44), respectivamente] nos capacita para calcular el coeficiente de reflecci\'on $R$ y el de transmisi\'on $T$ de la barrera, los cuales aqu\'{\i} 
son iguales a:
\begin{equation}
R = |A_1'/A_1|^2 = \frac{(k_1^2 - k_2^2)^2\sin^2 k_2a}{4k_1^2k_2^2 + (k_1^2-k_2^2)^2 \sin^2 k_2a},
\end{equation}
\begin{equation}
T=|A_3/A_1|^2=\frac{4k_1^2k_2^2}{4k_1^2 k_2^2 + (k_1^2 - k_2^2)^2 \sin^2 k_2a},
\end{equation}

\noindent 
Entonces es  f\'acil de verificar que $R + T = 1$.

% .......................................................................

{\bf b. Caso donde $E<V_0$; {\it efecto t\'unel}}

\qquad Ahora tendr\'{\i}amos a las ecuaciones (2) y (4) dispuestas:
\begin{eqnarray}
k_1 &=& \frac{\sqrt{2mE}}{\hbar}  \\
\nonumber\\
q_2 &=& \frac{\sqrt{2m(V_0 - E)}} {\hbar}
\end{eqnarray}

La soluci\'on de la ec. (1) tiene la forma de la ec. (3) en las regiones $I (x<0)$ y $III (x>a)$ y la forma de la ec. (5) en la regi\'on $II (0<x<a$):
\begin{eqnarray}
\psi_I &=& A_1e^{ik_1x} + A_1'e^{-ik_1x}\nonumber
\nonumber\\
\psi_{II} &=& B_2e^{q_2x} + B_2'e^{-q_2x}\nonumber
\nonumber\\
\psi_{III} &=& A_3e^{ik_1x} + A_3'e^{-ik_1x}\nonumber   
\end{eqnarray}

Las condiciones de pegado en $x=0$ y $x=a$ nos capacita para calcular el 
coeficiente de transmisi\'on de la barrera. De hecho, no es necesario realizar 
otra vez el c\'alculo: todo lo que debemos hacer es sustituir, en la ecuaci\'on 
obtenida en el primer caso de esta misma secci\'on, $k_2$ por $-i q_2$.

% ======================================================================
\subsection*{Estados Ligados; Pozo de Potencial}
{\bf a. Pozo de profundidad finita}

%%%%%%%%%%%%%%
\vskip 2ex
\centerline{
\epsfxsize=280pt
\epsfbox{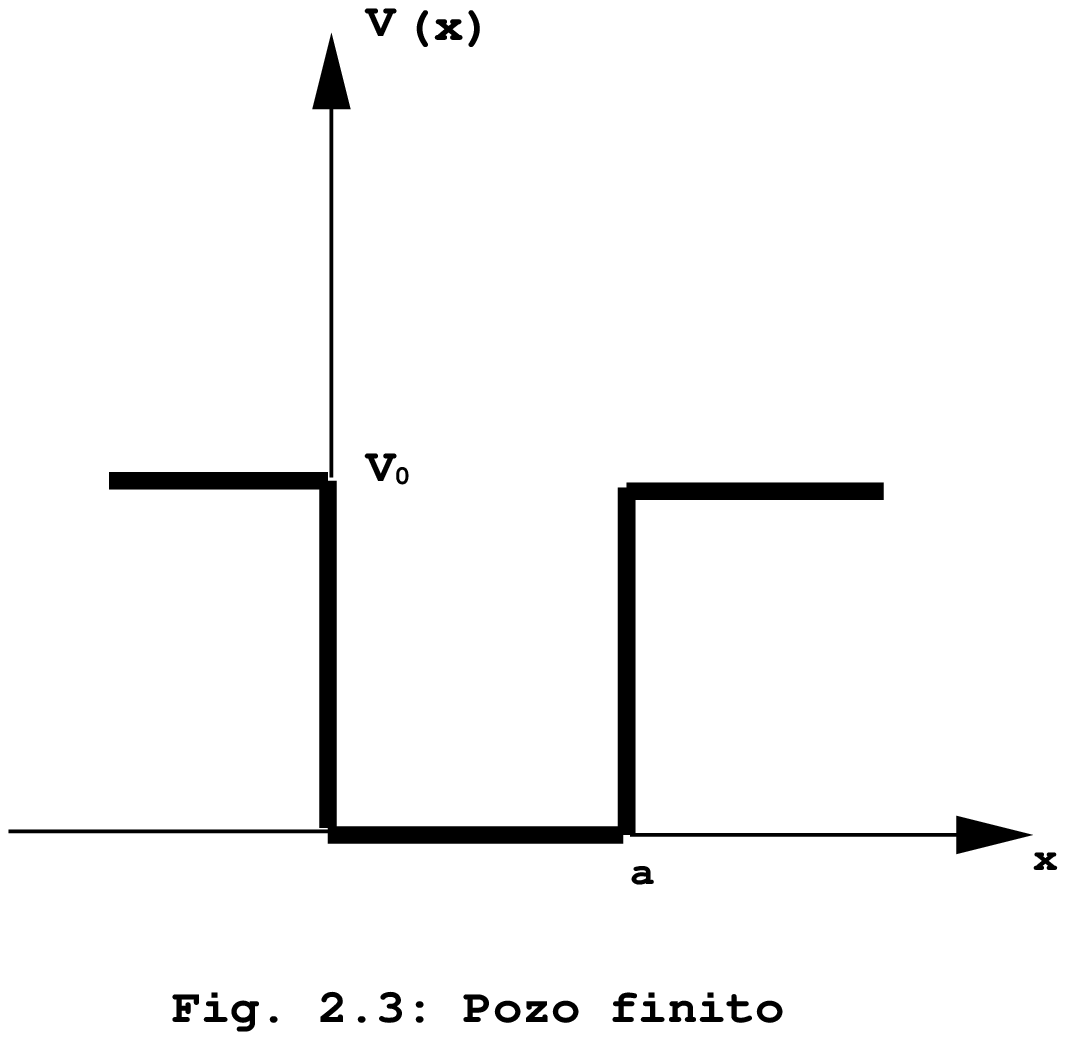}}
\vskip 4ex
%\begin{center}
%{\small{Fig. 1}\\
%}
%\end{center}
%%%%%%%%%%%%%%%%

En esta parte nos limitaremos s\'olo a tratar el caso $0<E<V_0$ 
(el caso $E>V_0$ es exactamente igual al calculado en la secci\'on precedente, 
``barrera de potencial''.

Para las regiones exteriores I $(x<0)$ y III $(x>a)$ usamos la ec. (4):

\begin{equation}
q = \frac{\sqrt{2m(V_0-E)}}{\hbar}
\end{equation}

Para la regi\'on central  II $(0<x<a)$ usamos la ec. (2):

\begin{equation}
k = \frac{\sqrt{2m(E)}}{\hbar}
\end{equation}

La soluci\'on de la ec. (1) tiene la forma de la ec. (5) en las regiones
exteriores y la forma de la ec. (3) en la regi\'on central:
\begin{eqnarray}
\psi_I &=& B_1e^{q x} + B_1'e^{-q x}\nonumber
\nonumber\\
\psi_{II} &=& A_2e^{ikx} + A_2'e^{-ikx}\nonumber
\nonumber\\
\psi_{III} &=& B_3e^{q x} + B_3'e^{-q x} \nonumber   
\end{eqnarray}

En la regi\'on $(0<x<a)$ la ec. (1) toma la forma:
\begin{equation}
\psi(x)'' + \frac{2mE}{\hbar^2}\psi(x) = \psi(x)'' + k^2\psi(x) = 0
\end{equation}

\noindent
y en las regiones exteriores:
\begin{equation}
\psi(x)'' - \frac{2m}{\hbar^2} [V_0-E]\phi(x) = \psi(x)'' - q^2\psi(x) = 0
\end{equation}

Ya que $\psi$ debe ser limitada en la regi\'on I, debemos tener:
\begin{equation}
B_1'=0
\end{equation}
Las condiciones de pegado dan:

$\psi_I = \psi_{II}$,\qquad en $x=0:$
\begin{equation}
B_1 = A_2 + A'_2
\end{equation}

$\psi'_I = \psi'_{II}$,\qquad en $x=0:$
\begin{equation}
B_1 q = A_2ik - A'_2ik
\end{equation}

$\psi_{II} = \psi_{III}$,\qquad en $x=a:$
\begin{equation}
A_2e^{ika} + A'_2e^{-ika} = B_3e^{q a} + B'_3e^{-q a}
\end{equation}

$\psi'_{II} = \psi'_{III}$,\qquad en $x=a:$
\begin{equation}
A_2ike^{ika} - A'_2ike^{-ika} = B_3q e^{q a} - B'_3q e^{-q a}
\end{equation}

Sustituyendo la constante $A_2$ y la constante $A'_2$ de la ec. (54) en la ec. (55) obtenemos, respectivamente:
%                .........................................................
\begin{eqnarray}
A'_2 &=& \frac{B_1(q-ik)}{-2ik}\nonumber
\nonumber\\
A_2 &=& \frac{B_1(q+ik)}{2ik}
\end{eqnarray}

Sustituyendo la constante $A_2$ y la constante $A'_2$ de la ec. (56) en la ec. (57) obtenemos, respectivamente:
\begin{eqnarray}
B'_3e^{-q a}(ik + q) + B_3e^{q a}(ik-q) + A'_2e^{-ika}(-2ik) &=& 0\nonumber
\nonumber\\
2ikA_2e^{ika} + B'_3e^{-q a}(-ik+q) + B_3E^{q a}(-ik-q) &=& 0
\end{eqnarray}

\noindent
Igualando $B'_3$ de las ecuaciones (59) y tomando en cuenta las ecuaciones (58):
\begin{equation}
\frac{B_3}{B_1} = \frac{e^{-q a}}{4ikq}[e^{ika}(q+ik)^2 - e^{-ika}(q - ik)^2]
\end{equation}

Pero $\psi(x)$ debe tambi\'en  estar limitada en la regi\'on III. Por lo tanto, 
es necesario que $B_3=0$, esto es:
\begin{equation}
[\frac{q - ik}{q + ik}]^2 = \frac{e^{ika}}{e^{-ika}} = e^{2ika}
\end{equation}

Ya que $q$ y $k$ dependen de $E$, la ec. (1) puede ser satisfecha para ciertos 
valores de $E$. Imponiendo un l\'{\i}mite sobre $\psi(x)$ en todas las regiones 
del espacio as\'{\i} se ocasiona la cuantizaci\'on de la energ\'{\i}a. M\'as 
precisamente dos casos son posibles:

{\bf (i) si:}
\begin{equation}
\frac{q - ik}{q + ik} = - e^{ika}
\end{equation}

\noindent
Si en esta expresi\'on igualamos en ambos miembros la parte real y la imaginaria, nos resulta:
\begin{equation}
\tan(\frac{ka}{2}) =\frac{q}{k} 
\end{equation}
Poniendo:
\begin{equation}
k_0 = \sqrt{\frac{2mV_0}{\hbar}} = \sqrt{k^2 + q^2}
\end{equation}
Entonces obtenemos:
\begin{equation}
\frac{1}{\cos^2(\frac{ka}{2})} = 1 + \tan^2(\frac{ka}{2}) = \frac{k^2 + q^2}{k^2} = (\frac{k_0}{k})^2
\end{equation}

La ec.(63) es as\'{\i} equivalente al sistema de ecuaciones:

% &&&&&&&&&&&&&&&&&&&&&&&&&&&&&&&&&&&&&&&&&&&&&&&&&&&&&&&&&&&&&&&&
% \[
% \left\{
% \begin{array}{ll}
% |\cos(ka/2)|= k/k_0&\mbox{}\\
% \tan(ka/2)>0&\mbox{}
% \end{array}
% \right.
% \]
% &&&&&&&&&&&&&&&&&&&&&&&&&&&&&&&&&&&&&&&&&&&&&&&&&&&&&&&&&&&&&&&
\begin{eqnarray}
|\cos(\frac{ka}{2})| &=& \frac{k}{k_0}
\nonumber\\
\tan(\frac{ka}{2}) &>& 0   
\end{eqnarray}

Los niveles de energ\'{\i}a est\'an determinados por la intersecci\'on de 
una l\'{\i}nea recta teniendo una inclinaci\'on $\frac{1}{k_0}$, con arcos 
senoidales (l\'{\i}neas largas interrumpidas en la figura 2.4). As\'{\i} 
obtenemos un cierto n\'umero de niveles
de energ\'{\i}a, cuyas funciones de onda son pares. Esto es m\'as claro si 
sustituimos (62) en (58) y (60); Es f\'acil verificar que $B'_3 = B_1$ y 
que $A_2 = A'_2$, as\'{\i} que $\psi(-x) =\psi(x)$.

{\bf (ii) si:}
\begin{equation}
\frac{q - ik}{q + ik} = e^{ika}
\end{equation}
Un c\'alculo del mismo tipo nos lleva a:
\begin{eqnarray}
|\sin(\frac{ka}{2})| &=& \frac{k}{k_0}
\nonumber\\
\tan(\frac{ka}{2}) &<& 0   
\end{eqnarray}
 
Los niveles de energ\'{\i}a est\'an entonces determinados por la intersecci\'on 
de la misma l\'{\i}nea recta como antes con otros arcos senoidales (l\'{\i}neas 
cortas en la figura 2.4). Los niveles as\'{\i} obtenidos caen entre aquellos 
encontrados en (i). Puede ser f\'acilmente mostrado que las correspondientes 
funciones de onda son impares.

%%%%%%%%%%%%%%
\vskip 2ex
\centerline{
\epsfxsize=280pt
\epsfbox{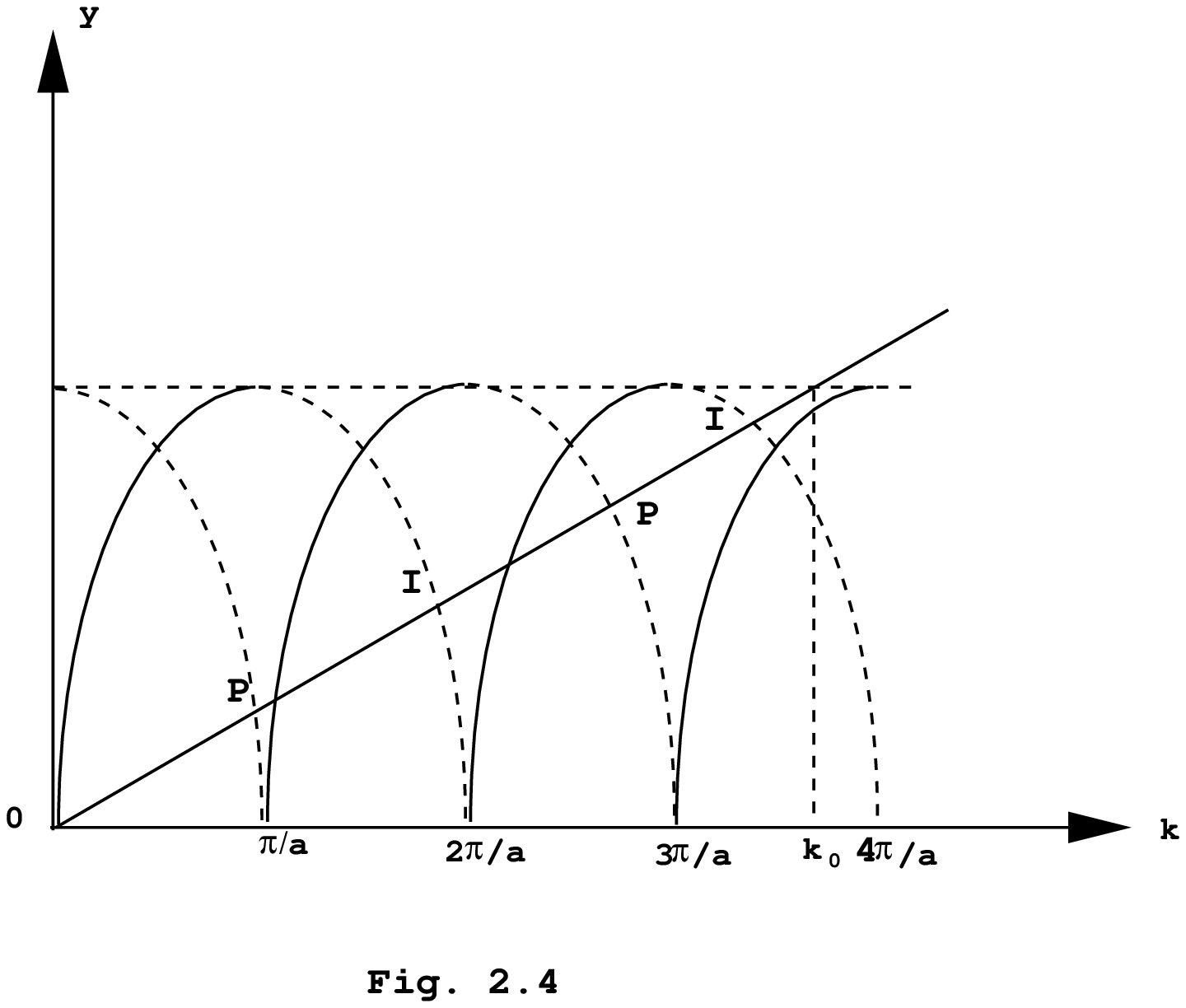}}
\vskip 4ex
%\begin{center}
%{\small{Fig. 1}\\
%}
%\end{center}
%%%%%%%%%%%%%%%%

{\bf a. Pozo de profundidad infinita}

\noindent
Asumiendo que $V(x)$ es cero para $0<x<a$ e infinito en cualquier 
parte m\'as. Pongamos:
\begin{equation}
k = \sqrt{\frac{2mE}{\hbar^2}}
\end{equation}
$\psi(x)$ debe ser cero fuera del intervalo $[0,a]$, y continuo en $x=0$, 
tambi\'en como en $x=a$.

\noindent
Ahora para $0 \leq x \leq a$:
\begin{equation}
\psi(x) = Ae^{ikx} + A'e^{-ikx}
\end{equation}
Ya que $\psi(0)=0$, puede ser deducido que $A' = -A$, lo cual nos lleva a:
\begin{equation}
\psi(x) = 2iA\sin(kx)
\end{equation}
Adem\'as $\psi(a)=0$, as\'{\i} esto:
\begin{equation}
k = \frac{n\pi}{a}
\end{equation}
donde $n$ es un arbitrario entero positivo. Si normalizamos la
funci\'on (71), tomando (72) en cuenta, entonces obtenemos las funciones
de onda estacionarias:
\begin{equation}
\psi_n(x) = \sqrt{\frac{2}{a}}\sin(\frac{n\pi x}{a})
\end{equation}
con energ\'{\i}as:
\begin{equation}
E_n = \frac{n^2\pi^2\hbar^2}{2ma^2}
\end{equation}
La cuantizaci\'on de los niveles de energ\'{\i}a es as\'{\i}, en este caso, 
particularmente simple.
% ----------------------------------------------------------

\newpage
\section*{P r o b l e m a s}

\vspace*{4mm}

\subsection*{Problema 2.1: El potencial Delta}

Supongamos que tenemos el potencial de la forma:

\begin{eqnarray}
V(x) = -V_0 \delta(x);\qquad  V_0 > 0; \qquad x \in \Re. \nonumber 
\end{eqnarray}
La correspondiente funci\'on de onda $\psi(x)$ se supone que es suave.
 
%{\bf Problema}.
% 
a) Busca los estados ligados ($E<0$) los cuales est\'an localizados en este potencial.

b) Calcula la dispersi\'on de una onda plana llegando en este potencial y encuentra el {\it  coeficiente de reflexi\'on}
\begin{eqnarray}
R = \frac{|\psi_{ref}|^2}{|\psi_{lleg}|^2}|_{x=0} \nonumber
\end{eqnarray}
donde $\psi_{ref}$, $\psi_{lleg}$ es la onda reflejada y la de llegada,
respectivamente.

\noindent
{\it Sugerencia}: Para evaluar el comportamiento de $\psi(x)$ en x=0, integra la ecuaci\'on de Schr\"odinger en el intervalo ($-\varepsilon,+\varepsilon$) y considera el l\'{\i}mite $\varepsilon$ $\rightarrow$ $0$.

{\bf Soluci\'on.} a) La ecuaci\'on de Schr\"odinger est\'a dada por
\begin{equation}
\frac{d^2}{dx^2} \psi(x) + \frac{2m}{\hbar^2} (E+V_0 \delta(x))\psi(x) = 0
\end{equation}
Lejos del origen tenemos una ecuaci\'on diferencial de la forma
\begin{equation}
\frac{d^2}{dx^2} \psi(x) = - \frac{2mE}{\hbar^2}\psi(x).
\end{equation}
Las funciones de onda son por lo tanto de la forma
\begin{equation}
\psi(x) = Ae^{-q x} + Be^{q x} \qquad si \qquad x>0 \qquad o\qquad x<0,
\end{equation}
con $q = \sqrt{-2mE/ \hbar^2}$ $ \in\Re.$ Como $|\psi|^2$ debe ser integrable, all\'{\i} no puede haber una parte increment\'andose exponencialmente. Adem\'as la funci\'on de onda debe ser continua en el origen. De aqu\'{\i},
\begin{eqnarray}
\psi(x) &=& Ae^{q x}; \qquad (x<0), \nonumber
\nonumber\\
\psi(x) &=& Ae^{-q x}; \qquad (x>0).
\end{eqnarray} 
Integrando la ecuaci\'on de Schr\"odinger desde $-\varepsilon$ a $+\varepsilon$, obtenemos
\begin{equation}
-\frac{\hbar^2}{2m}[\psi'(\varepsilon)-\psi'(-\varepsilon)] - V_0\psi(0) = E\int^{+\varepsilon} _{-\varepsilon} \psi(x)dx \approx 2\varepsilon E\psi(0)
\end{equation}
Insertando ahora el resultado (78) y tomando el l\'{\i}mite $\varepsilon \rightarrow \infty$, tenemos
\begin{equation}
-\frac{\hbar^2}{2m}(-q A-q A) - V_0 A = 0
\end{equation}
o $E = -m(V_0^2 / 2\hbar^2)$. Claramente hay s\'olo un eigenvalor de energ\'{\i}a. La constante de normalizaci\'on es encontrada que es $A = \sqrt{mV_0/ \hbar^2}$.
% .........
\begin{eqnarray}
\nonumber
\end{eqnarray} 
% .........
b) La funci\'on de onda de una onda plana es descrita por
\begin{equation}
\psi(x) = A e^{ikx}, \qquad k^2 = \frac{2mE}{\hbar^2} 
\end{equation}
Se mueve de izquierda a derecha y es reflejada en el potencial. Si $B$ o $C$ es la amplitud de la onda reflejada o transmitida, respectivamente, tenemos
\begin{eqnarray}
\psi(x) &=& Ae^{ikx} + Be^{-ikx}; \qquad (x<0), \nonumber
\nonumber\\
\psi(x) &=& Ce^{ikx}; \qquad \qquad \qquad (x>0).
\end{eqnarray} 

Las condiciones de continuidad y la relaci\'on $\psi'(\varepsilon)-\psi'(-\varepsilon) = - f\psi(0)$ con $f = 2mV_0 / \hbar^2$ da
\begin{eqnarray}
A + B &=& C \qquad  \qquad  \qquad B = -\frac{f}{f+2ik}A,     \nonumber
\nonumber\\
ik(C - A + B) &=& -fC \qquad  \qquad  C = \frac{2ik}{f+2ik}A.
\end{eqnarray} 
El coeficiente de reflexi\'on deseado es por lo tanto
\begin{eqnarray}
R = \frac{|\psi_{ref}|^2}{|\psi_{lleg}|^2}|_{x=0} = \frac{|B|^2}{|A|^2} = \frac{m^2V_0^2}{m^2V_0^2 + \hbar^4k^2}.
\end{eqnarray}
Si el potencial es extremadamente fuerte ($V_0 \rightarrow \infty$) $R \rightarrow 1$, o sea, la onda entera es reflejada.

El {\it coeficiente de transmisi\'on} es, por otro lado,
\begin{eqnarray}
T = \frac{|\psi_{trans}|^2}{|\psi_{lleg}|^2}|_{x=0} = \frac{|C|^2}{|A|^2} = \frac{\hbar^4 k^2}{m^2V_0^2 + \hbar^4k^2}.
\end{eqnarray} 
Si el potencial es muy fuerte, ($V_0 \rightarrow \infty$) $T \rightarrow 0$, o sea, la onda transmitida se desvanece.

Obviamente, $R + T = 1$ como se esperaba.

% ...................................................................................
\newpage
\subsection*{Problema 2.2:
Part\'{\i}cula en un Pozo de Potencial finito Unidimensional}

%{\bf Problema.}
Resuelve la ecuaci\'on unidimensional de
Schr\"odinger para un pozo de potencial finito descrito por el siguiente
potencial 
\[
V(x) = \left\{
\begin{array}{ll}
-V_0&\mbox{si $|x| \leq a$}\\
0&\mbox{si $|x|>a$}
\end{array}
\right.
\]

Considera s\'olo estados ligados ($E<0$).

%%%%%%%%%%%%%%
\vskip 2ex
\centerline{
\epsfxsize=280pt
\epsfbox{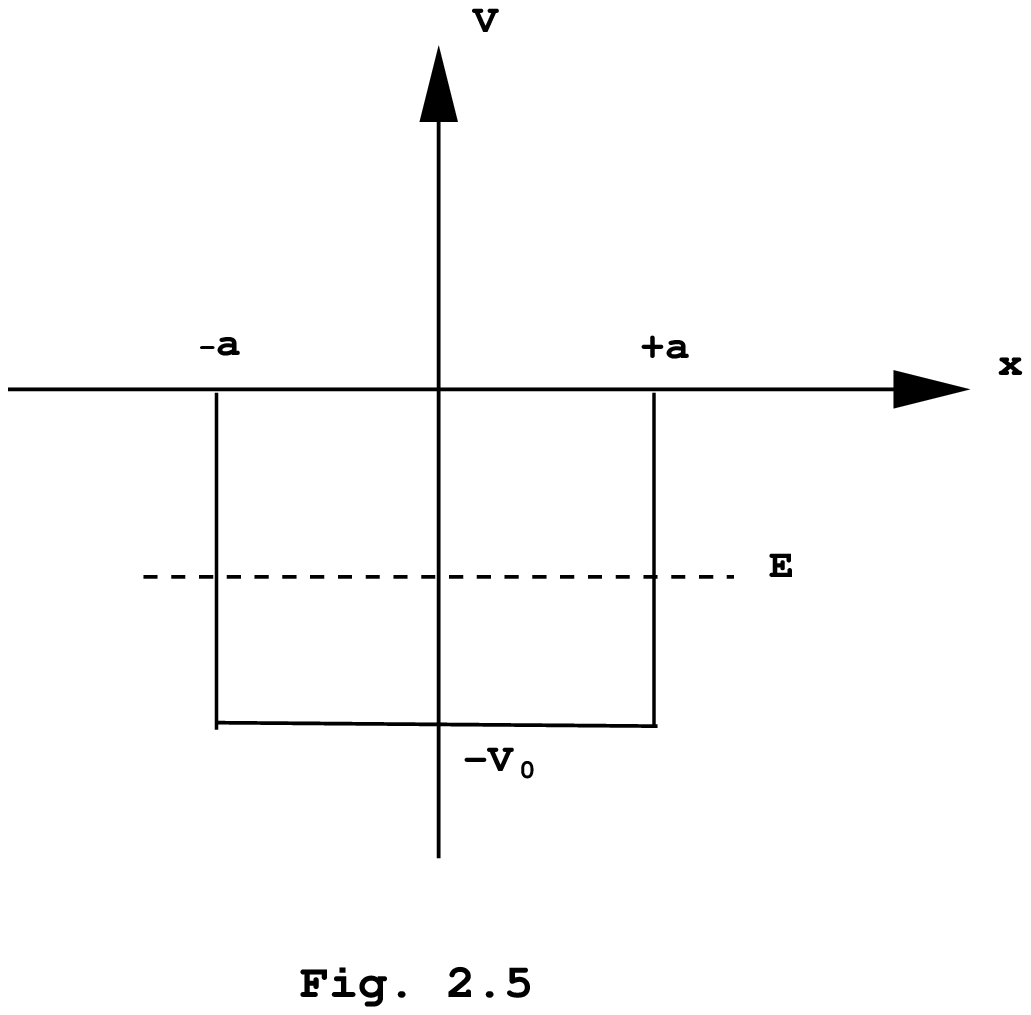}}
\vskip 4ex
%\begin{center}
%{\small{Fig. 1}\\
%}
%\end{center}
%%%%%%%%%%%%%%%%

%\vspace*{75mm}

\noindent
{\bf Soluci\'on.} a) La funci\'on de onda
para $|x|<a$ y $|x|>a$. La correspondiente ecuaci\'on de
Schr\"odinger est\'a dada por
\begin{equation}
-\frac{\hbar^2}{2m}\psi''(x) + V(x)\psi(x) = E\psi(x)
\end{equation}
Definimos
\begin{equation}
q^2 = -\frac{2mE}{\hbar^2}, \qquad  k^2 = \frac{2m(E+V_0)}{\hbar^2}
\end{equation}
%
%\newpage

\noindent
y obtenemos:
\begin{eqnarray}
{\rm 1)~ si~~~x<-a:}\qquad \psi''_1(x) - q^2 \psi_1 &=& 0,\ \psi_1 = A_1e^{q x} + B_1e^{-q x};\nonumber
\nonumber\\
{\rm 2)~ si~-a\leq x\leq a:}~\psi''_2(x) + k^2 \psi_2 &=& 0, \ \psi_2 = A_2 \cos(kx) + B_2 \sin(kx); \nonumber
\nonumber\\
{\rm 3)~ si~~~x>a:~~}\qquad \psi''_3(x) - q^2 \psi_3 &=& 0,\ \psi_3 = B_3 e^{q x} + B_3 e^{-q x}. \nonumber   
\end{eqnarray}

b) Formulaci\'on de las condiciones de frontera. La normalizaci\'on de los 
estados ligados requiere que la soluci\'on se haga cero en el infinito. Esto 
significa que $B_1=A_3=0$. Adem\'as, $\psi(x)$ debe ser continuamente 
diferenciable. Todas las soluciones
particulares son fijadas de tal forma que $\psi$ tambi\'en como su primera 
derivada $\psi'$ son suaves en ese valor de x correspondiendo a la frontera 
entre el interior y el exterior. La segunda derivada $\psi''$ contiene el 
salto requerido por el particular potencial tipo caja de esta ecuaci\'on de 
Schr\"odinger. Todo esto junto nos lleva a
\begin{eqnarray}
\psi_1(-a) &=& \psi_2(-a),\qquad  \psi_2(a) = \psi_3(a), \nonumber
\nonumber\\
\psi'_1(-a) &=& \psi'_2(-a),\qquad  \psi'_2(a) = \psi'_3(a).
\end{eqnarray} 

c) Las ecuaciones de eigenvalores. De (88) obtenemos cuatro ecuaciones lineales y homog\'eneas para los coeficientes $A_1$, $A_2$, $B_2$ y $B_3$:
\begin{eqnarray}
A_1 e^{-qa} &=& A_2\cos(ka) - B_2\sin(ka), \nonumber
\nonumber\\
qA_1 e^{-qa} &=& A_2k\sin(ka) + B_2k\cos(ka),  \nonumber
\nonumber\\
B_3 e^{-qa} &=& A_2\cos(ka)  + B_2\sin(ka), \nonumber
\nonumber\\
-qB_3 e^{-qa} &=& -A_2k\sin(ka) + B_2k\cos(ka).
\end{eqnarray} 

%\newpage

\noindent
Por adici\'on y sustracci\'on obtenemos un sistema de ecuaciones m\'as
f\'acil de resolver:
\begin{eqnarray}
 (A_1+B_3) e^{-qa} &=& 2A_2\cos(ka) \nonumber
\nonumber\\
q(A_1+B_3) e^{-qa} &=& 2A_2k\sin(ka) \nonumber
\nonumber\\
(A_1-B_3)  e^{-qa} &=& -2B_2\sin(ka) \nonumber
\nonumber\\
q(A_1-B_3) e^{-qa} &=&  2B_2k\cos(ka).
\end{eqnarray} 
Asumiendo que $A_1+B_3 \neq 0$ y $A_2 \neq 0$, Las primeras dos ecuaciones resultan
\begin{equation}
q = k\tan(ka).
\end{equation}
Insertando esta en una de las \'ultimas dos ecuaciones da
\begin{equation}
A_1 = B_3; \qquad B_2 = 0.
\end{equation}
De aqu\'{\i}, como resultado, tenemos una soluci\'on sim\'etrica
con $\psi(x) = \psi(-x)$. Entonces hablamos de una {\it paridad positiva}.

Casi id\'entico c\'alculo nos lleva para $A_1 - B_3 \neq 0$ y para $B_2 \neq 0$
a
\begin{equation}
q = -k\cot(ka) \qquad y \qquad A_1 = -B_3; \qquad A_2 = 0. 
\end{equation}
La funci\'on de onda as\'{\i} obtenida es antisim\'etrica, correspondiendo a
una paridad {\it negativa}.

d) Soluci\'on cualitativa del problema de eigenvalores. La ecuaci\'on que
conecta $q$ y $k$, la cual ya hemos obtenido, son condiciones para el
eigenvalor de energ\'{\i}a. Usando la forma corta
\begin{equation}
\xi = ka, \qquad \eta = qa,
\end{equation}
obtenemos de la definici\'on (87)
\begin{equation}
\xi^2 + \eta^2 = \frac{2mV_0a^2}{\hbar^2} = r^2.
\end{equation}
Por otro lado, usando (91) y (93) obtenemos las ecuaciones
\begin{eqnarray}
\eta = \xi \tan(\xi), \qquad \eta = -\xi\cot(\xi). \nonumber
\end{eqnarray}
Por lo tanto los valores de energ\'{\i}a deseados pueden ser obtenidos
construyendo la intersecci\'on de esas dos curvas con el c\'{\i}rculo
definido por (95), en el plano $\xi$-$\eta$ (ver figura 2.6).

%%%%%%%%%%%%%%
\vskip 2ex
\centerline{
\epsfxsize=280pt
\epsfbox{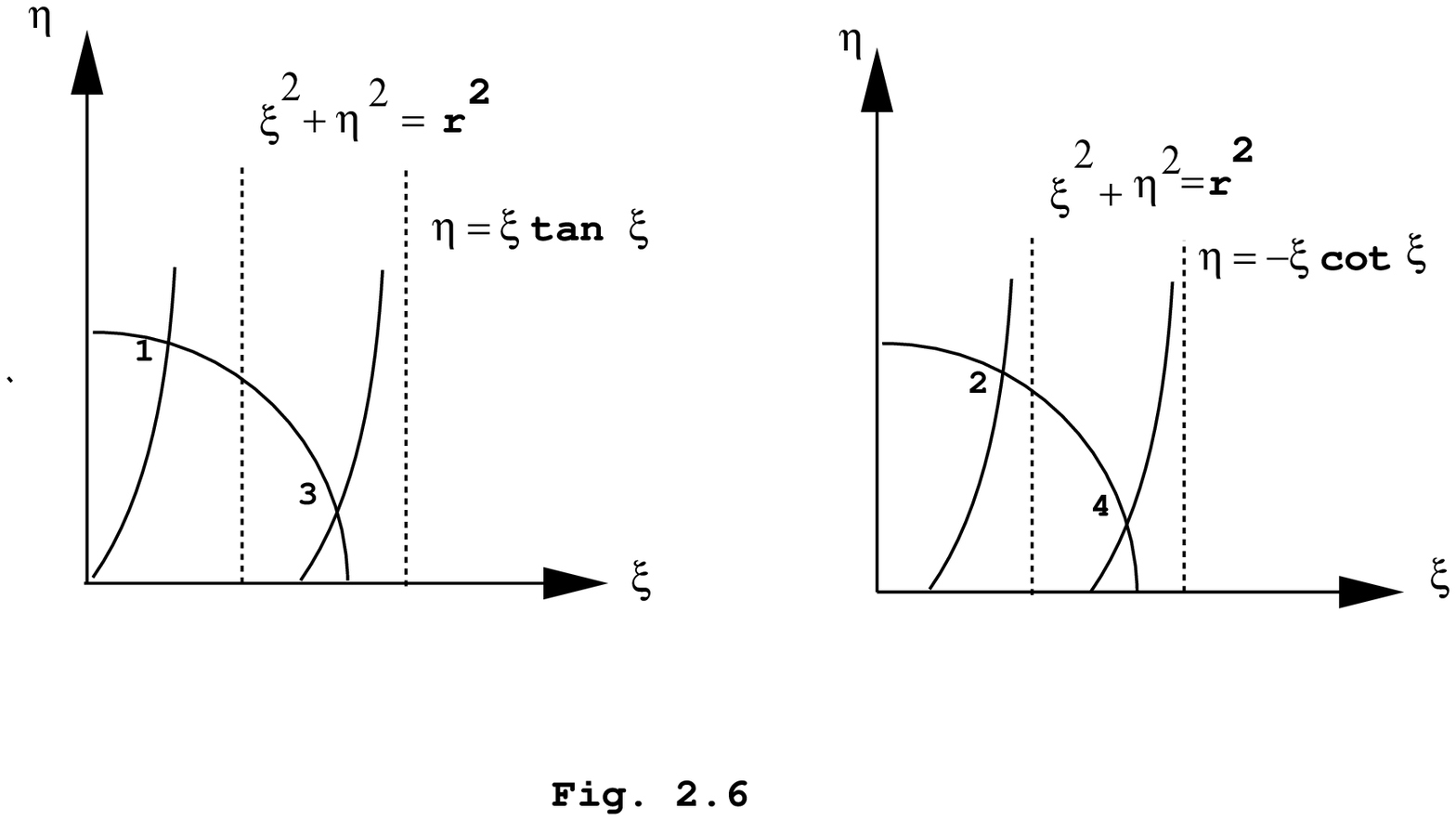}}
\vskip 4ex
%\begin{center}
%{\small{Fig. 1}\\
%}
%\end{center}
%%%%%%%%%%%%%%%%

%\vspace*{70mm}

Al menos una soluci\'on existe para valores arbitrarios del
par\'ametro $V_0$, en el caso de paridad positiva, porque la funci\'on
tangente intersecta el origen. Para paridad negativa, el radio del
c\'{\i}rculo necesita ser m\'as grande que un valor m\'{\i}nimo de tal forma
que las dos curvas puedan intersectarse. El potencial deber tener una cierta
profundidad en conexi\'on con un tama\~no dado $a$ y una masa dada $m$,
para permitir una soluci\'on con paridad negativa. El n\'umero de niveles
de energ\'{\i}a se incrementa con $V_0$, $a$ y la masa $m$. Para el caso en
que $mVa^2 \rightarrow \infty$, las intersecciones son encontradas en
\begin{eqnarray}
\tan(ka) &=& \infty \qquad {\rm correspondiendo~ a} \qquad  ka=\frac{2n-1}{2}\pi, \nonumber
\nonumber\\
-\cot(ka) &=& \infty \qquad {\rm correspondiendo~ a} \qquad ka = n \pi, 
\end{eqnarray} 
donde $n=1,2,3, \, \ldots $

\noindent
o combinado:
\begin{equation}
k(2a) = n \pi.
\end{equation}
Para el espectro de energ\'{\i}a esto significa que
\begin{equation}
E_n = \frac{\hbar^2}{2m}(\frac{n \pi}{2a})^2 - V_0.
\end{equation}
Ampliando el pozo de potencial y/o la masa de la part\'{\i}cula $m$, la 
diferencia entre dos eigenvalores de energ\'{\i}a vecinos decrecer\'a. El 
estado m\'as bajo ($n=1$) no est\'a localizado en $-V_0$, sino un poco m\'as 
arriba. Esta diferencia es llamada la {\it energ\'{\i}a de punto cero}.

e) La forma de la funci\'on de onda es mostrada en la figura 2.7 para la 
soluci\'on discutida.

%%%%%%%%%%%%%%
\vskip 2ex
\centerline{
\epsfxsize=280pt
\epsfbox{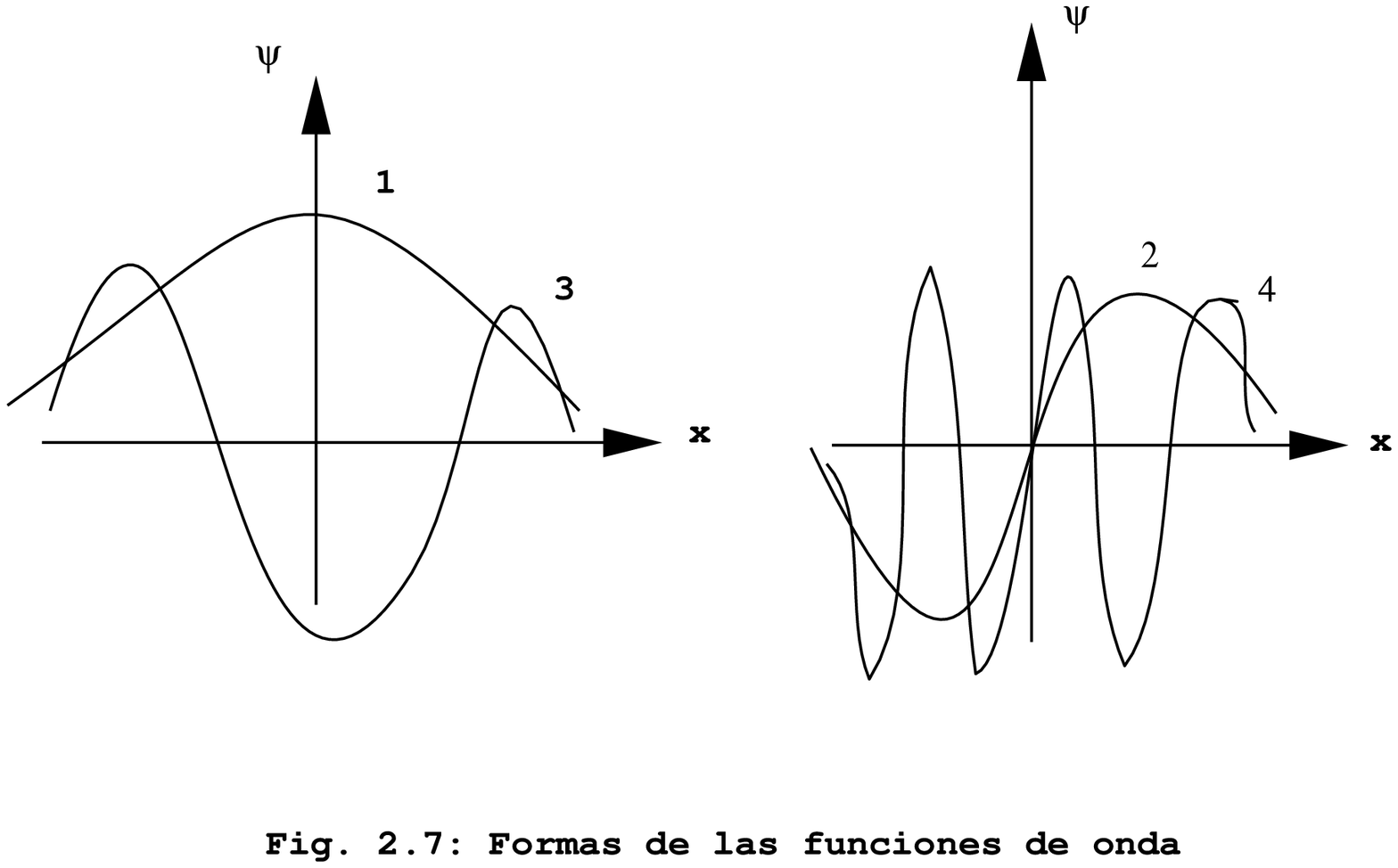}}
\vskip 4ex
%\begin{center}
%{\small{Fig. 1}\\
%}
%\end{center}
%%%%%%%%%%%%%%%%

%\newpage

\subsection*{Problema 2.3:
Part\'{\i}cula en un Pozo de Potencial infinito Unidimensional}

%{\bf Problema.}
Resuelve la ecuaci\'on unidimensional de
Schr\"odinger para una part\'{\i}cula que se encuentra en un pozo de
potencial infinito descrito como sigue:
\[
V(x) = \left\{
\begin{array}{ll}
0&\mbox{si $x'<x<x'+2a$}\\
\infty&\mbox{si $x'\geq x~~{\rm o}~~x\geq x'+2a$}
\end{array}
\right.
\]
Tenemos que la soluci\'on en forma general es
\begin{equation}
\psi(x) = A\sin(kx) + B\cos(kx)
\end{equation}
donde
\begin{equation}
k = \sqrt{\frac{2mE}{\hbar^2}}
\end{equation}
Como $\psi$ debe cumplir que $\psi(x') = \psi(x'+2a) = 0$, se tiene:
\begin{eqnarray}
A~\sin(kx')~~~ +~~~ B~\cos(kx') = 0 \\
A\sin[k(x'+2a)] + B\cos[k(x'+2a)] = 0
\end{eqnarray} 
Multiplicando (101) por $\sin[k(x'+2a)]$ y (102) por $\sin(kx')$ y, enseguida, restando el segundo resultado del primero obtenemos:
\begin{equation}
B[~~\cos(kx') \sin[k(x'+2a)] - \cos[k(x'+2a)]\sin(kx')~~] = 0
\end{equation}
o a trav\'es de una identidad trigonom\'etrica:
\begin{equation}
B \sin(2ak) = 0
\end{equation}
Multiplicando (101) por $\cos[k(x'+2a)]$ y rest\'andole la multiplicaci\'on
de (102) por $\cos(kx')$, se tiene:
\begin{equation}
A[~~\sin(kx') \cos[k(x'+2a)] - \sin[k(x'+2a)]\cos(kx')~~] = 0
\end{equation}
o a trav\'es de la misma identidad trigonom\'etrica:
\begin{equation}
A \sin[k(-2ak)] = A \sin[k(2ak)] =  0
\end{equation}

Como se descarta la soluci\'on trivial $\psi=0$, entonces por las ecuaciones
(104) y (106) se tiene que $\sin(2ak)=0$ y esto s\'olo ocurre
si $2ak = n \pi$, siendo $n$ un entero. Seg\'un lo anterior $k=n \pi/2a$ y
como $k^2=2mE/\hbar^2$ entonces tenemos que los eigenvalores est\'an dados por
la expresi\'on:
\begin{equation}
E = \frac{\hbar^2\pi^2n^2}{8a^2m}
\end{equation}
La energ\'{\i}a est\'a cuantizada ya que s\'olo se le permiten ciertos
valores; para cada $k_n = n\pi/2a$ le corresponde la
energ\'{\i}a $E_n=[n^2/2m][\pi\hbar/2a]^2$.

La soluci\'on en la forma general queda como:
\begin{equation}
\psi_n = a\sin(\frac{n\pi x}{2a}) + B\cos(\frac{n\pi x}{2a}).
\end{equation}
Normalizando:
\begin{equation}
1 = \int_{x'} ^{x'+2a} \psi \psi^* dx = a(A^2+B^2)
\end{equation}
lo que nos lleva a
\begin{equation}
A = \pm \sqrt{1/a - B^2}
\end{equation}
Sustituyendo este valor de $A$ en (101) se obtiene:
\begin{equation}
B = \mp\frac{1}{\sqrt{a}}\sin(\frac{n\pi x'}{2a})
\end{equation}
Sustituyendo este valor de $B$ en (110) se tiene:
\begin{equation}
A = \pm\frac{1}{\sqrt{a}}\cos(\frac{n\pi x'}{2a})
\end{equation}
Tomando los signos superiores de $A$ y $B$, y sustituyendo sus valores en
(108) se tiene:
\begin{equation}
\psi_n =\frac{1}{\sqrt{a}}\sin(\frac{n\pi}{2a})(x-x')
\end{equation}
Utilizando los signos inferiores de A y B se tiene:
\begin{equation}
\psi_n =-\frac{1}{\sqrt{a}}\sin(\frac{n\pi}{2a})(x-x').
\end{equation}

%\end{document}
\newpage
%%%%%%%%%%%%%%%%%%%%%%%%%%%%%%%%%%%%%%%%%%%%%%%%%%%%%%%
%\documentstyle[12pt]{article}
%\newcommand{\aple}{\mbox{${}_{\textstyle\sim}^{\textstyle<}$}}
%\newcommand{\apge}{\mbox{${}_{\textstyle\sim}^{\textstyle>}$}}
%\newcommand{\slsh}[1]{\mbox{$\displaystyle {#1}\!\!\!{/}$}}
%\newcommand{\lpr}{\mbox{$ \displaystyle O_L $}}
%\newcommand{\rpr}{\mbox{$ \displaystyle O_R $}}
%\newcommand{\GeV}{\mbox{$\rm  \, GeV $}}
%\baselineskip 25.1pt plus 0.2pt minus 0.1pt
%\begin{document}
%\baselineskip 20pt plus 0.2pt minus 0.1pt
%%%%%%%%%%%%%%%%%%%%%%%%%%%%%%%%%%%%%%%%%%%%%%%%%%%%%%%%%%%%%%
%\title
%\setcounter{equation}\\
%\section*
\begin{center}
{\large 3. MOMENTO ANGULAR EN LA MC}
\end{center}
%\setcounter{equation}
%\author{Teodoro C\'ordova Fraga}
%\date{}
%\maketitle

%\vspace*{-20pt}
\section*{Introducci\'on}
De la {\it Mec\'anica Cl\'asica} se sabe que, el {\it momento angular} 
$\bf{l}$ de las part\'{\i}culas macrosc\'opicas est\'a dado por
\setcounter{equation}{0}
\begin{equation}
{\bf l=r} \times {\bf p},
\label{1}
\end{equation}
donde $\bf r$ y $\bf p$ son respectivamente el vector de posici\'on  y
el momento lineal.

 Sin embargo, en {\it Mec\'anica Cu\'antica}, el operador momento angular 
(OMA), en general no es un operador que se exprese solamente por el operador 
coordenada $\hat{x}_j$ y el operador momento lineal $\hat{p}_k$, los cuales,
s\'olo act\'uan sobre las funciones propias (FP) de coordenadas. Por lo tanto,
es muy importante establecer, antes que nada, las relaciones de conmutaci\'on
para las  componentes del OMA, es decir, en {\it Mec\'anica Cu\'antica}
$\bf l$ se representa por el operador 
\begin{equation}
{\bf l}=-i\hbar {\bf r} \times \nabla,
\label{2}
\end{equation}
cuyas componentes son operadores que satisfacen las siguientes reglas de 
conmutaci\'on

\begin{equation}
[l_x,l_y]=il_z, \qquad  [l_y,l_z]=il_x, \qquad  [l_z,l_x]=il_y,
\label{3}
\end{equation}
adem\'as, cada uno de ellos conmuta con el cuadrado del OMA, esto es

\begin{equation}
l^2=l^2_x+l^2_y+l^2_z, \qquad [l_i,l^2]=0, \qquad i=1,2,3.
\label{4}
\end{equation}
Estas relaciones, adem\'as de ser v\'alidas para el OMA, tambi\'en se
cumplen para el operador esp\'{\i}n (OS), el cual, carece de an\'alogo en la
{\it mec\'anica cl\'asica}.

Estas relaciones de conmutacion son b\'asicas para obtener el espectro
de dichos operadores, as\'{\i} como sus representaciones diferenciales.

\section*{Momento Angular Orbital}
Para un punto cualquiera de un espacio espacio fijo (EP), se puede
tener una funci\'on $\psi(x,y,z)$, para el cual, consideramos dos sistemas 
cartesianos $\Sigma$ y $\Sigma '$, donde $\Sigma '$ se obtiene de rotar el 
eje $z$.

En general un EP se refiere a un sistema de coordenadas 
diferentes~a~$\Sigma$~y~${\Sigma}'$.

Ahora bien, comparemos los valores de $\psi$ en dos puntos del EP
con las mismas coordenadas (x,y,z) en $\Sigma$ y ${\Sigma}'$, esto es
equivalente a rotar un vector
\begin{equation}
\psi(x',y',z') = R \psi(x,y,z)
\label{5}
\end{equation}
donde $R$ es la matriz de rotaci\'on en {\sc R}$^3$
\begin{equation}
\left(\begin{array}{c}
x' \\ y' \\ z' 
\end{array}\right)
=
\left(\begin{array}{ccc}
\cos \phi & -\sin \phi & 0 \\
\sin \phi &  \cos \phi & 0 \\
0         &  0         & z 
\end{array}\right)
\left(\begin{array}{c}
x \\ y \\ z
\end{array}\right),
\label{6}
\end{equation}
entonces
\begin{equation}
R \psi (x,y,z) =
\psi(x \cos \phi -y \sin \phi ,
     x \sin \phi  +y \cos \phi , z).
\label{7}
\end{equation}

Por otro lado es importante recordar que las funciones de onda (FO) no
dependen del sistema de coordenadas y que la transformaci\'on a rotaciones
de las FP se hace por medio de operadores unitarios, luego entonces, para
establecer la forma del operador unitario (OU) $U^\dagger(\phi)$ que lleva
$\psi$ a $\psi '$, siempre se considera una rotaci\'on infinitesimal $d\phi$,
manteniendo solamente los t\'erminos lineales en $d\phi$ cuando se expande
$\psi '$ en series de Taylor alrededor del punto~$x$
\begin{eqnarray}\vspace*{-20pt}
\psi(x',y',z') & \approx &  \psi(x+yd\phi, xd\phi+y, z), \nonumber\\
	       & \approx & \psi(x,y,z) 
+ d\phi\left(y \frac{\partial \psi}{\partial x} 
      - x \frac{\partial \psi}{\partial y}\right), \nonumber\\
               & \approx & (1-id\phi l_z)\psi(x,y,z),
\label{8} 
\end{eqnarray}

\noindent
donde hemos empleado la notaci\'on\footnote{La demostraci\'on de (8) 
se presenta como el problema 3.1}
\begin{equation}
l_z = \hbar^{-1}(\hat{x}\hat{p}_y -\hat{y}\hat{p}_x ),
\label{9}
\end{equation}
la cual, como se ver\'a m\'as adelante, corresponde al operador de 
proyecci\'on en~$z$ del momento angular de acuerdo con la definici\'on 
en (2) y dividido~por~$\hbar$, tal que, para la rotaci\'on del \'angulo $\phi$
finito, se pueda representar como una exponencial, es decir
\begin{equation}
\psi(x', y', z) = e^{il_z \phi}\psi(x,y,z),
\label{10}
\end{equation}
donde
\begin{equation}
\hat{U}^\dagger(\phi)=e^{il_z\phi}.
\label{11}
\end{equation}
Para reafirmar el concepto de rotaci\'on, consideremosla de una manera m\'as
general con la ayuda del vector $\hat{\vec{A}}$ que act\'ua sobre $\psi$ 
asumiendo que $\hat{A}_x$,  $\hat{A}_y$ $\hat{A}_z$ tienen la misma forma
en $\Sigma$ y $\Sigma '$, es decir, que los valores promedio de
$\hat{\vec{A}}$ calculados en  $\Sigma$ y $\Sigma '$ deben ser iguales
cuando se ven desde el EF, esto es
\begin{eqnarray}
&&\int \psi^*(\vec{r}')
(\hat{A}_x\hat{\imath}' + \hat{A}_y\hat{\jmath}' + \hat{A}_z\hat{k}')
\psi^*(\vec{r}')\,d\vec{r} \nonumber\\
&& \qquad =\int \psi^*(\vec{r})
(\hat{A}_x\hat{\imath} + \hat{A}_y\hat{\jmath} + \hat{A}_z\hat{k})
\psi^*(\vec{r})\,d\vec{r},
\label{12}
\end{eqnarray}
donde

\begin{equation}
\hat{\imath}' = \hat{\imath}\cos\phi + \hat{\jmath}\sin\phi, \qquad
\hat{\jmath}' = \hat{\imath}\sin\phi + \hat{\jmath}\cos\phi, \qquad
\hat{k}' = \hat{k}.
\label{13}
\end{equation}

Luego entonces, si combinamos (10), (12) y (13) tenemos
\begin{eqnarray}
e^{il_z\phi} \hat{A}_x e^{-il_z\phi}&=& \hat{A}_x\cos\phi -\hat{A}_y\sin\phi,
\nonumber\\
e^{il_z\phi} \hat{A}_y e^{-il_z\phi}&=& \hat{A}_x\sin\phi -\hat{A}_y\cos\phi,
\nonumber\\
e^{il_z\phi} \hat{A}_z e^{-il_z\phi}&=& \hat{A}_z.
\label{14}
\end{eqnarray}

Nuevamente consideremos rotaciones infinitesimales y desarrollando las
partes de la izquierda en (14) se pueden determinar las relaciones de 
conmutaci\'on de $\hat{A}_x$, $\hat{A}_y$ y $\hat{A}_z$ con $\hat{l}_z$
\begin{equation}
[l_z,A_x]=iA_y, \qquad  [l_z,A_y]=-iA_x, \qquad  [l_z,A_z]= 0,
\label{15}
\end{equation}
y de manera similar para $l_x$ y $l_y$. 
\\

 Las condiciones b\'asicas para obtener tales relaciones de conmutaci\'on son

\begin{itemize}

\item[$\star$]
La FP se transforma como en (7) cuando $\Sigma \rightarrow \Sigma '$

\item[$\star$]
Las componentes $\hat{A}_x$, $\hat{A}_y$, $\hat{A}_z$ tienen la misma forma en
$\Sigma$ y $\Sigma '$

\item[$\star$] 
Los vectores de los valores promedio de $\hat{A}$ en $\Sigma$ y $\Sigma '$
coinciden (son iguales) para un observador en el EF.

\end{itemize}

Tambi\'en se puede usar otra representaci\'on en la cual la FO $\psi(x,y,z)$
no cambia cuando $\Sigma \rightarrow \Sigma '$ y lo operadores vectoriales se 
transforman como vectores. Para pasar a tal representaci\'on cuando $\phi$
rota alrededor de $z$ se usa el operador $\hat{U}(\phi)$, es decir

\begin{equation}
e^{il_z\phi} \psi'(x,y,z) = \psi(x,y,z),
\label{16}
\end{equation}
entonces
\begin{equation}
e^{-il_z\phi} \hat{\vec{A}} e^{il_z\phi} = \hat{\vec{A}}.
\label{17}
\end{equation}
y utilizando las relaciones dadas en (14)
\begin{eqnarray}
\hat{A}_x' & = & \hat{A}_x\cos\phi + \hat{A}_y\sin\phi 
             =   e^{-il_z\phi} \hat{A}_x e^{il_z\phi}, \nonumber\\ 
\hat{A}_y' & = & -\hat{A}_x\sin\phi + \hat{A}_y\cos\phi 
             =  e^{-il_z\phi} \hat{A}_y e^{il_z\phi}, \nonumber\\ 
\hat{A}_z' & = & e^{-il_z\phi} \hat{A}_z e^{il_z\phi}.
\label{18}
\end{eqnarray}

Dado que las trasformaciones de la nueva representaci\'on se hacen con
operadores unitarios, las relaciones de conmutaci\'on no se cambian.

\subsection*{Observaciones}

\begin{itemize}

\item[$\star$]
El operador $\hat{A}^2$ es invariante ante rotaciones, esto es
\begin{equation}
e^{-il_z\phi} \hat{A}^2 e^{il_z\phi} = \hat{A}'^2 = \hat{A}^2
\label{19}
\end{equation}

\item[$\star$]
Resulta que 
\begin{equation}
[\hat{l}_i, \hat{A}^2] = 0,
\label{20}
\end{equation}

\item[$\star$]
Si el operador Hamiltoniano es de la forma
\begin{equation}
\hat{H} = \frac{1}{2m}\hat{p}^2 + U(\vec{r}),
\label{21}
\end{equation}
entonces se mantiene invariante ante rotaciones dadas en cualquier eje
que pasa por el origen de coordenadas
\begin{equation}
[\hat{l}_i, \hat{H}] = 0
\label{22}
\end{equation}
donde las $\hat{l}_i$ son integrales de movimiento.
\end{itemize}

\subsection*{Definici\'on}
Si $\hat{A}_i$ son las componentes de un operador vectorial que act\'ua sobre 
una FO de coordenadas y si hay operadores $\hat{l}_i$ que cumplen con las
siguientes relaciones de conmutaciones.
\begin{equation}
[\hat{l}_i, \hat{A}_j] = i\varepsilon_{ijk}\hat{A}_k, \qquad
[\hat{l}_i, \hat{l}_j] = i\varepsilon_{ijk}\hat{l}_k.
\label{23}
\end{equation}

Luego entonces, las $\hat{l}_i$ se llaman {\it componentes del operador
momento angular orbital}, y podemos concluir de (20) y (23) que 
\begin{equation}
[\hat{l}_i, \hat{l}^2]=0.
\label{24}
\end{equation}
 
Por lo tanto, las tres componentes asociadas con las componentes de un
momento angular cl\'asico arbitrario satisfacen las relaciones de conmutaci\'on
M\'as a\'un, puede mostrarse que el origen de estas relaciones conducen a 
propiedades geom\'etricas de rotaci\'on en un espacio tridimensional. Esto
es porque adoptamos un punto de vista m\'as general y definiremos un momento 
angular $\bf J$ como cualquier conjunto de tres observables $J_x$, $J_y$ y
$J_z$ los cuales cumplen las relaciones de conmutaci\'on, es decir
\begin{equation}
[J_i, J_j] = i\varepsilon_{ijk}J_k.
\label{25}
\end{equation}

Entonces introducimos el operador
\begin{equation}
{\bf J}^2 = J^2_x + J^2_y + J^2_z,
\label{26}
\end{equation}
el (escalar) cuadrado del momento angular $\bf J$. Este operador es 
hermitiano, dado que $J_x$, $J_y$ y $J_z$ son hermitianos, y asumiremos que es
un observable, ahora mostremos que $\bf J^2$ conmuta con las tres componentes
de $\bf J$
\begin{equation}
[{\bf J}^2, {\bf J}]=0.
\label{27}
\end{equation}

Por el hecho de que $\bf J^2$ conmuta con cada una de sus componentes,
entonces hay un sistema completo de FP
\begin{equation}
{\bf J^2}\psi_{\gamma \mu} = \gamma \psi_{\gamma \mu}, \qquad
    J^2_i\psi_{\gamma \mu} = \mu    \psi_{\gamma \mu}.
\label{28}
\end{equation}
Los operadores $J_i$ y $J_k$ $(i \neq k)$ no conmutan, es decir, 
no tienen FP en com\'un.

En lugar de usar las componentes $J_x$ y $J_y$ del momento angular $\bf J$,
es m\'as conveniente introducir las siguientes combinaciones lineales
\begin{equation}
J_+ = J_x + iJ_y, \qquad J_{-}= J_x - iJ_y.
\label{29}
\end{equation}
Los cuales no son herm\'{\i}ticos, tal como lo operadores $a$ y $a^\dagger$
del oscilador arm\'onico, solamente son adjuntos.

Ahora, si hacemos las operaciones indicadas en la izquierda concluimos que
\begin{equation}
[J_z,   J_{\pm}] = \pm J_{\pm}, \qquad 
[J_{+}, J_{-}]   =     2J_z, 
\label{30}
\end{equation}
\begin{equation}
[J^2, J_{+}]   =  [J^2, J_{-}] = [J^2, J_{z}] = 0.
\label{31}
\end{equation}
\begin{equation}
J_z(J_{\pm}\psi_{\gamma\mu}) = \{J_{\pm}J_z + [J_z, J_{\pm}]\}\psi_{\gamma\mu}
=(\mu \pm 1) J_{\pm} \psi_{\gamma\mu}.
\label{32}
\end{equation}

Luego entonces las $\psi_{\gamma\mu}$ son FP de los valores de $J_z$
(autovalores) $\mu \pm 1$, es decir
\begin{eqnarray}
J_{\pm} \psi_{\gamma\mu - 1} &=& \alpha_\mu \psi_{\gamma\mu}, \nonumber\\
J_{-}\psi_{\gamma\mu}         =&& \beta_\mu \psi_{\gamma\mu-1}.
\label{33}
\end{eqnarray}
pero
\begin{equation}
\alpha^*_{\mu} = J_{+}\psi_{\gamma \mu_1} , \psi_{\gamma \mu} 
               = \psi_{\gamma \mu_1} J_{-} \psi_{\gamma \mu} = \beta_\mu,
\label{34}
\end{equation}
tal que, al tomar una fase del tipo $e^{ia}$ (con $a$ real) para la funci\'on
$\psi_{\gamma \mu}$ se puede hacer $\alpha_\mu$ real e igual a $\beta _\mu$, 
esto quiere decir
\begin{equation}
J_{+}\psi_{\gamma \mu - 1} = \alpha \mu \psi_{\gamma \mu},
J_{-}\psi_{\gamma \mu    } = \alpha \mu \psi_{\gamma \mu - 1},
\label{35}
\end{equation}
y por lo tanto
\begin{eqnarray}
\gamma &=& \psi_{\gamma \mu}, [J_x^2 + J_y^2 + J_z^2] \psi_{\gamma \mu}
= \mu^2 + a + b, \nonumber\\
a & = & (\psi_{\gamma \mu}, J^2 \psi_{\gamma \mu}) = 
       (J_x\psi_{\gamma \mu}, J_x \psi_{\gamma \mu}) \geq 0, \nonumber\\
b & = &  (\psi_{\gamma \mu}, J^2 \psi_{\gamma \mu}) = 
        (J_y\psi_{\gamma \mu}, J_y \psi_{\gamma \mu}) \geq 0.
\label{36}
\end{eqnarray}

La norma de cualquier funci\'on es no negativa, esto implica que
\begin{equation}
\gamma \geq \mu^2,
\label{37}
\end{equation}
para $\gamma$ fijo, el valor de $\mu$ tiene l\'{\i}mite superior e inferior
(es decir, tiene valores en un intervalo finito). 

Sean $\Lambda$ y $\lambda$ esos l\'{\i}mites (superior e inferior de $\mu$)
para un $\gamma$ dado
\begin{equation}
J_{+}\psi_{\gamma \Lambda} = 0, \qquad J_{-}\psi_{\gamma \lambda} = 0.
\label{38}
\end{equation}

Ahora, utilizando las siguientes igualdades operatoriales
\begin{eqnarray}
J_{-}J_{+} &=& {\bf J^2} - j^2_z + J_z = {\bf J^2} - J_z(J_z-1),\nonumber\\
J_{+}J_{-} &=& {\bf J^2} - j^2_z + J_z = {\bf J^2} - J_z(J_z+1),
\label{39}
\end{eqnarray}
actu\'ando sobre $\psi _ {\gamma \Lambda}$ y $\psi_{\gamma \lambda}$ se obtiene
\begin{eqnarray}
\gamma - \Lambda^2 - \Lambda &=& 0, \nonumber \\
\gamma - \lambda^2 + \lambda &=& 0, \nonumber \\
(\lambda - \lambda + 1) (\lambda + \lambda) &=& 0.
\label{40}
\end{eqnarray}

La condici\'on 
\begin{equation}
\Lambda \geq \lambda \rightarrow \Lambda = -\lambda = J \rightarrow
\gamma = J(J+1).
\label{41}
\end{equation} 

Para un $\gamma$ dado (fijo) la proyecci\'on del momento $\mu$ toma
$2J+1$ valores que difieren por una unidad de $\bf J$ a $\bf -J$. Por eso,
la diferencia $\Lambda -\lambda = 2J$ debe ser un n\'umero entero y 
consecuentemente los autovalores $J_z$ tomados por $m$ son enteros, esto es
\begin{equation}
m=k, \qquad k=0, \pm 1, \pm 2, \, \ldots \, ,
\label{42}
\end{equation}
o semienteros
\begin{equation}
m=k + {1 \over 2}, \qquad k=0, \pm 1, \pm 2, \, \ldots \, .
\label{43}
\end{equation}

Para un estado dado $\gamma = J(J+1)$, es degenerado con grado $g=2J+1$
respecto a los autovalores de la posici\'on del momento $m$ (esto es porque
$J_i,~J_k$ conmutan con $J^2$ pero no conmutan entre ellos).
 
Por ``estado de momento angular $J$''  se entiende en la mayor\'{\i}a de
los casos, un estado con $\gamma = J(J+1)$ en el cual, el valor m\'aximo de la 
proyecci\'on es $J$. Tales estados se denotan por 
$\psi_{jm}$ o $|jm\rangle$.

Vamos a encontrar los elementos de matriz de $J_x,~J_y$ de la representaci\'on
en la cual, $J^2$ y $J_z$ son diagonales, luego entonces de (35) y (39) se
tiene que
\begin{eqnarray}
J_{-}J_{+} \psi_{jm-1} = \alpha_mJ_{-}\psi_{jm} = \alpha_m\psi_{jm-1},
\nonumber\\
J(J+1)-(m-1)^2-(m-1) = \alpha_m^2,\nonumber\\
\alpha_m = \sqrt{(J+m)(J-m+1)}.
\label{44}
\end{eqnarray}

Ahora, combinando (44) con (35) se obtiene
\begin{equation}
J_+\psi_{jm-1} = \sqrt{(J+m)(J-m+1)}\psi_{jm},
\label{45}
\end{equation}
resulta que el elemento de matriz de $J_{+}$ es
\begin{equation}
\langle jm | J_+ | jm-1 \rangle = \sqrt{(J+m)(J-m+1)} \delta_{nm},
\label{46}
\end{equation}
y de manera an\'aloga
\begin{equation}
\langle jn | J_{-} | jm \rangle = -\sqrt{(J+m)(J-m+1)} \delta_{nm-1},
\label{47}
\end{equation}
por \'ultimo, de la definici\'on en (29) $J^{+},\ J_{-}$ se obtiene 
f\'acilmente
\begin{eqnarray}
\langle jm | J_x | jm-1 \rangle &=& {1 \over 2}\sqrt{(J+m)(J-m+1)} ,
\nonumber\\
\langle jm | J_y | jm-1 \rangle &=& {-i \over 2}\sqrt{(J+m)(J-m+1)} ,
\label{48}
\end{eqnarray}

\subsection*{A Modo de Conclusi\'on}
\begin{itemize}
\item[$\alpha$] ({\it Propiedades de los Eigenvalores de $\bf J$ y $J_z$})\\
Si $j(j+1)\hbar^2$ y $j_z$ son los eigenvalores de $\bf J$ y $J_z$
asociados con los eigenvectores $|kjm\rangle$, entonces $j$ y $m$ satisfacen
la desigualdad 
\[
-j \leq m \leq j.
\]

\item[$\beta$] ({\it Propiedades del Vector $J_{-}|kjm\rangle$})\\
Sea $|kjm\rangle$ un eigenvector de  $\bf J^2$ y $j_m$ con los eigenvalores
$j(j+1)\hbar^2$ y $m\hbar$
\begin{itemize}
\item{(i)}
Si $m=-j$, $J_{-}|kj-j\rangle=0$.
\item{(ii)}
Si $m>-j$, $J_{-}|kjm \rangle$ es un eigenvector no nulo de $J^2$ y $J_z$
con los eigenvalores $j(j+1)\hbar^2$ y $(m-1)\hbar$.
\end{itemize}

\item[$\gamma$] ({\it Propiedades del Vector $J_+|kjm\rangle$})\\
Sea $|kjm\rangle$ un eigenvector de $\bf J^2$ y $J_z$ con los eigenvalores
$j(j+1)\hbar$ y $m\hbar$
\begin{itemize}
\item[$\star$]
Si $m=j,\quad J_+|kjm\rangle =0.$
\item[$\star$]
Si $m<j$, $J+=|kjm\rangle$ es un vector no nulo de $\bf J^2$ y $j_z$
con los eigenvalores $j(j+1)\hbar^2$ y $(m+1)\hbar$
\end{itemize}

\item[$\delta$]  Luego entonces
\begin{eqnarray}
J_z|kjm\rangle &=& m\hbar|kjm\rangle,\nonumber\\
J_+|kjm\rangle &=& m\hbar\sqrt{j(j+1) - m(m+1)}|kjm+1\rangle, \nonumber\\
J_-|kjm\rangle &=& m\hbar\sqrt{j(j+1) - m(m-1)}|kjm+1\rangle. \nonumber
\end{eqnarray}
\end{itemize}

\section*{Aplicaci\'on del Momento Angular Orbital}
Hemos considerado las propiedades del momento angular derivadas \'unicamente 
de las relaciones de conmutaci\'on, ahora retomaremos el momento 
angular~$\bf L$ de una part\'{\i}cula sin giro y veremos como dicha teor\'{\i}a
desarrollada en la secci\'on anterior se aplica a un caso particular, esto es
\begin{equation}
[\hat{l}_i, \hat{p}_j] = i\varepsilon_{ijk}\hat{p}_k.
\label{49}
\end{equation} 
Por lo tanto, $\hat{l}_z$ y $\hat{p}_j$ tienen un sistema com\'un de funciones
propias. Por otro lado, el Hamiltoniano de una part\'{\i}cula libre
\[
\hat{H} = \left(\frac{\hat{\vec{p}}}{\sqrt{2m}}\right)^2,
\]
por el hecho de ser el cuadrado de un operador vectorial tiene el mismo
sistema de FP que $\hat{l}$ y $\hat{l}_z$. Adem\'as, esto implica que la
part\'{\i}cula libre se puede encontrar en un estado con $E$, $l$, $m$
bien determinados.

\subsection*{Eigenvalores y Eigenfunciones de $\bf L^2$ y $\bf L$}
Es m\'as conveniente trabajar con coordenadas esf\'ericas (o polares),
dado que, como veremos, varios operadores del momento angular actu\'an solamente
sobre los \'angulos variables $\theta,\ \phi$ y no en la variable $r$.
En lugar de caracterizar el vector $r$ por sus componentes cartesianas
$x,\ y,\ z$ llamaremos el punto correspondiente $M$ en el espacio
($\bf OM=r$) por sus coordenadas esf\'ericas, esto es
\begin{equation}
x=r\cos\phi\sin\theta, \qquad  
y=r\sin\phi\sin\theta, \qquad
z=r\cos\theta,
\label{50}
\end{equation}
con
\[
r \geq 0, \qquad 
0 \leq \theta \leq \pi, \qquad 
0 \leq \phi   \leq 2\pi.
\]

Sean $\Phi(r, \theta,\phi)$ y $\Phi'(r, \theta,\phi)$ las FO de una 
part\'{\i}cula en $\Sigma$ y $\Sigma'$ en la cual la rotaci\'on 
infinitesimal est\'a dada por~$\delta\alpha$ alrededor de~$z$
\begin{eqnarray}
\Phi'(r, \theta,\phi) &=& \Phi(r, \theta,\phi+\delta\alpha),\nonumber\\
&=& \Phi(r, \theta,\phi) + \delta\alpha\frac{\partial \Phi}{\partial\phi},
\label{51}
\end{eqnarray}
\'o bien
\begin{equation}
\Phi'(r, \theta,\phi) = (1+i\hat{l}_z\delta\alpha)\Phi(r, \theta,\phi).
\label{52}
\end{equation}

Luego entonces, resulta que
\begin{equation}
\frac{\partial \Phi}{\partial \phi} = i\hat{l_z}\Phi, \qquad 
\hat{l}_z = -i{\partial \over \partial \phi}.
\label{53}
\end{equation}

Para una rotaci\'on infinitesimal en $x$
\begin{eqnarray}
\Phi'(r, \theta,\phi) &=& \Phi+\delta\alpha
\left(\frac{\partial \Phi}{\partial \theta}
      \frac{\partial \theta}{\partial \alpha} +
      \frac{\partial \Phi}{\partial \theta}
      \frac{\partial \phi}{\partial \alpha}
\right), \nonumber\\
&=& (1+i\hat{l}_x\delta\alpha)\Phi(r, \theta,\phi),
\label{54}
\end{eqnarray}
pero en tal rotaci\'on
\begin{equation}
z' = z+y\delta\alpha; \qquad
z' = z+y\delta\alpha; \qquad x' = x
\label{55}
\end{equation}
y de (50) se obtiene
\begin{eqnarray}
r\cos(\theta + d\theta) &=& 
r\cos\theta + r\sin\theta\sin\phi\delta\alpha, \nonumber\\
r\sin\phi\sin\theta + d\theta) &=& 
r\sin\theta\sin\phi + r\sin\theta\sin\phi -r\cos\theta\delta\alpha,
\label{56}
\end{eqnarray}
es decir
\begin{equation}
\sin\theta d\theta = \sin\theta\sin\phi \, \delta\alpha \rightarrow
{d\theta \over d\alpha} = -\sin\phi,
\label{57}
\end{equation}
y
\begin{eqnarray}
\cos\theta\sin\phi \, d\theta + \sin\theta\cos\phi \, d\phi 
&=& -\cos\theta \, \delta\alpha,\nonumber\\
\cos\phi\sin\theta{d\phi \over d\alpha} &=& -\cos\theta - 
\cos\theta\sin\phi{d\theta \over d\alpha},
\label{58}
\end{eqnarray}
ahora, sustituyendo (57) en (56)
\begin{equation}
\frac{d\phi}{d\alpha} = -\cot\theta\cos\phi,
\label{59}
\end{equation}
tal que al sustituir (56) y (58) en (51) y comparando las partes de la 
derecha en (51) se obtiene
\begin{equation}
\hat{l}_x = i\left(
\sin\phi {\partial \over \partial \theta} + 
\cot \theta\cos\phi {\partial \over \partial \phi}
\right).
\label{60}
\end{equation}

En el caso de la rotaci\'on en $y$, el resultado es similar, tal que
\begin{equation}
\hat{l}_y = i\left(
\sin\phi {\partial \over \partial \theta} + 
\cot \theta\cos\phi {\partial \over \partial \phi}
\right).
\label{61}
\end{equation}

Usando $\hat{l}_x, \ \hat{l}_y$ tambi\'en se puede obtener 
$\hat{l}_\pm, \ \hat{l}^2$, esto es
\begin{eqnarray}
\hat{l}_\pm &=& \exp(\pm i\phi)
\left(
\pm{\partial \over \partial\theta} + i\cot\theta{\partial \over \partial\phi}
\right), \nonumber\\
\hat{l}^2 &=& \hat{l}_{-}\hat{l}_{+} + \hat{l}^2 + \hat{l}_z, \nonumber\\
&=&-\left[
{1 \over \sin^2\theta}
{\partial^2 \over \partial \phi^2} +
{1 \over \sin^2\theta}
{\partial^2 \over \partial \theta}
\bigg(
\sin\theta{\partial \over \partial\theta} 
\bigg)\right].
\label{62}
\end{eqnarray}
de (62) se ve que $\hat{l}^2$ es id\'entico hasta una constante al operador 
de Laplace en la parte angular, esto es
\begin{equation}
\nabla^2 f = {1 \over r^2}{\partial \over \partial r}
\left( r^2{\partial f \over \partial r} \right) + 
{1 \over r^2}
\left[
{1 \over \sin\theta}
{\partial \over \partial\theta} 
\left(
\sin\theta{\partial f \over \partial\theta}
\right) + 
{1 \over \sin^2\theta}{\partial^2 \over \partial \phi^2}
\right].
\label{63}
\end{equation}

\subsection*{Funciones Propias de $l_z$}
%%%%%%%%%%%%%%%%%%%%%%%%%%%%%%%%%%%%%%%%%
\begin{eqnarray}
\hat{l}_z\Phi_m = m\Phi = -i{\partial \Phi_m \over \partial \phi},
\nonumber\\
\Phi_m = {1 \over \sqrt{2\pi}}e^{im\phi}.
\label{64}
\end{eqnarray}

\subsection*{Condiciones de Hermiticidad de $\hat{l}_z$}
%%%%%%%%%%%%%%%%%%%%%%%%%%%%%%%%%%%%%%%%%%%%%%%%%%%%%%%%%%%
\begin{equation}
\int_0^{2\pi} f^*\hat{l}_zg\,d\phi = 
\left( \int_0^{2\pi} g^*\hat{l}_zf\,d\phi \right)^* +
f^*g(2\pi) - f^*g(0).
\label{65}
\end{equation}

Entonces $\hat{l}_z$ es herm\'{\i}tico en la clase de funciones para las
cuales
\begin{equation}
f^*g(2\pi) = f^*g(0).
\label{66}
\end{equation}

Las funciones propias de $\Phi_m$ de $\hat{l}_m$ pertenecen a 
$L^2(0, 2\pi)$ y cumplen con (66), as\'{\i} como para cualquier funci\'on que 
se pueda desarrollar en $\Phi_m(\phi)$, esto es
\begin{eqnarray}
F(\phi) &=& \sum^k a_ke^{ik\phi}, \qquad k = 0, \pm 1, \pm 2, \, \ldots \, ,
\nonumber\\
G(\phi) &=& \sum^k b_ke^{ik\phi},\qquad  
k = \pm 1/2,\pm 3/2, \pm 5/2 \, \ldots \, ,
\label{67}
\end{eqnarray}
solo $m$ enteros \'o $m$ semi-enteros, pero no para combinaciones de
$F(\phi), \ G(\phi)$.

Las elecciones apropiadas de $m$ est\'an basadas en el experimento de FP 
comunes de $\hat{l}_z$ y $\hat{l}^2$.

\subsection*{Arm\'onicos Esf\'ericos}
En la representaci\'on $\{\bf \vec{r}\}$, las eigenfunciones asociadas
con los eigenvalores $l(l+1)\hbar^2$, de $\bf L^2$ y $m\hbar$ de $l_z$ son
las soluciones de la ec.\ diferencial parcial
\begin{eqnarray}
-\left({\partial^2 \over \partial \theta^2} + {1 \over \tan\theta}
{\partial \over \partial \theta} + {1 \over \sin^2\theta} 
{\partial^2 \over \partial \phi^2}\right)\psi(r, \theta, \phi) &=&
l(l+1)\psi(r, \theta, \phi), \nonumber\\
-i{\partial \over \partial \phi}\psi(r, \theta, \phi) &=& 
m\hbar\psi(r, \theta, \phi). 
\label{68}
\end{eqnarray}

Considerando que los resultados generales obtenidos son aplicables al 
momento angular, sabemos que $l$ es un entero o un semientero y que para
$l,m$ fijos, pueden tomarse solamente los valores 
\[
-l, -l+1, \, \dots \, ,l-1, l.
\]

En (68), $r$ no aparece en el operador diferencial, as\'{\i} que puede
con\-si\-de\-rar\-se como un par\'ametro y tomar en cuenta s\'olo la dependencia
en $\theta,\ \phi$ de $\psi$. Luego entonces podemos denotar por
$Y_{lm}(\theta, \phi)$ como una eigenfunci\'on com\'un de $\bf L^2$ y $l_z$
la cual corresponde a los eigenvalores de $l(l+1)\hbar^2, m\hbar $, esto es
\begin{eqnarray}
{\bf L^2}Y_{lm}(\theta, \phi) &=& l(l+1)\hbar^2Y_{lm}(\theta,\phi), \nonumber\\
 l_z Y_{lm}(\theta, \phi)     &=& m\hbar Y_{lm}(\theta,\phi).
\label{69}
\end{eqnarray}

Para ser completamente rigurosos, tenemos que introducir un \'{\i}ndice 
adicional con el objeto de distinguir entre varias soluciones de (69), 
las cuales correspondan al mismo par de valores $l,\ m$. En efecto, como se 
ver\'a m\'as adelante, estas ecs.\ tienen solamente una soluci\'on (en un
factor constante) para cada par de valores permitidos de $l,\ m$; esto es 
porque los sub\'{\i}ndices $l,\ m$ son suficientes.

La ec.\ (69) di\'o a $\theta,\ \phi$ dependencia de las eigenfunci\'ones
de $\bf L^2$ y $l_z$. 
Una de las soluciones de $Y_{lm}(\theta,\, \phi)$ de estas
ecs.\ han sido encontradas de la siguiente manera\footnote{Demostraci\'on
en el problema 3.4}
\begin{equation}
\psi_{lm}(r, \theta,\phi) = f(r)\psi_{lm}(\theta,\phi),
\label{70}
\end{equation}
donde $f(r)$ es una funci\'on de $r$ la cual aparece como una constante de
integraci\'on para las ecuaciones diferenciales parciales de (68). El hecho
de que $f(r)$ sea arbitraria muestra que $\bf L^2$ y 
$l_z$ no forman un conjunto
completo de observables comunes\footnote{Por definici\'on, el operador 
herm\'{\i}tico A es un observable si este sistema ortogonal de vectores
forma una base en el espacio de estados} en el espacio 
$\varepsilon_r$\footnote{Cada estado cu\'antico de la part\'{\i}cula es
caracterizado por un estado vectorial perteneciente a un espacio abstracto
$\varepsilon_r$}
o funciones de $\vec{r}$ (o de $r, \theta, \phi$).

Con el objeto de normalizar $\psi_{lm}(r, \theta, \phi)$, es conveniente 
normalizar $Y_{lm}(\theta, \phi) $ y $f(r)$ separadamente (como se muestra).
Entonces, debemos tomar un diferencial de \'angulo s\'olido
\begin{eqnarray}
\int_0^{2\pi}d\phi\int_0^{\pi}\sin\theta|\psi_{lm}(\theta, \phi)|^2d\theta 
&=& 1, 
\label{71} \\
\int_0^\infty r^2|f(r)|^2dr &=& 1.
\label{72}
\end{eqnarray}

\subsection*{Valores de $l,\ m$}

\noindent
$\alpha:\ $  {\it $l,\ m$ deben ser enteros}\\
Usando $l_z = {\hbar \over i}{\partial \over \partial \phi}$, podemos escribir
(69) como sigue
\begin{equation}
{\hbar \over i}{\partial \over \partial \phi}Y_{lm}(\theta, \phi) = m\hbar 
Y_{lm}(\theta, \phi),
\label{73}
\end{equation}
la cual muestra que 
\begin{equation}
Y_{lm}(\theta, \phi) = F_{lm}(\theta, \phi)e^{im\phi}.
\label{74}
\end{equation}

Si permitimos que $0\leq \phi < 2\pi$, entonces podemos cubrir todo el 
espacio ya que la funci\'on debe ser continua en todas partes, tal que
\begin{equation}
Y_{lm}(\theta, \phi=0) = Y_{lm}(\theta, \phi=2\pi),
\label{75}
\end{equation}
lo que implica 
\begin{equation}
e^{im\pi} = 1.
\label{76}
\end{equation}

Seg\'un se vi\'o, $m$ es un  entero o un semientero, la aplicaci\'on al
momento angular orbital, muestra que $m$ debe ser un entero. ($e^{2im\pi}$
ser\'a igual $-1$ si $m$ fuera semientero).

\noindent
$\beta$: Todo valor entero (positivo o cero) de $l$ puede ser encontrado 
escogiendo un valor entero de $l$, se sabe de la teor\'{\i}a general que 
$Y_{lm}(\theta, \phi)$ debe cumplirse, esto es
\begin{equation}
L_+Y_{lm}(\theta, \phi)=0,
\label{77}
\end{equation}
la cual, al combinar $L_+ = \hbar e^{i\phi}$ y (62)
\begin{equation}
\left({d \over d\theta} - l\cot\theta \right) F_{ll}(\theta) = 0.
\label{78}
\end{equation}
Esta ec.\ de primer orden puede ser integrada inmediatamente si notamos~que
\begin{equation}
\cot\theta d\theta = \frac{d(\sin\theta)}{\sin\theta},
\label{79}
\end{equation}
su soluci\'on general es
\begin{equation}
F_{ll}=c_l(\sin\theta)^l,
\label{80}
\end{equation}
donde $c_l$ es una constante de normalizaci\'on.

Consecuentemente, para cualquier valor positivo o cero de $l$, existe una 
funci\'on $Y_{ll}(\theta, \phi)$ la cual es igual (con un factor constante)
\begin{equation}
Y^{ll}(\theta, \phi) = c_l(\sin\theta)^le^{il\phi}.
\label{81}
\end{equation}

A trav\'es de la acci\'on repetida de $L_{-}$, construimos 
$Y_{ll-1}(\theta, \phi), \, \dots \, , Y_{lm}(\theta, \phi),$  
$\dots\, , Y_{l-l}(\theta, \phi)$. Luego entonces, vemos que la correspondencia
para el par de eigenvalores $l(l+1)\hbar, m\hbar $ (donde $l$ es un entero 
positivo arbitrario o cero y $m$ es otro entero tal que $l\leq m \leq l$ ),
de (78) una y solamente una eigenfunci\'on $Y_{lm}(\theta, \phi)$, puede ser 
ambiguamente calculada de (78). A las eigenfunciones $Y_{lm}(\theta, \phi)$
se les conocen como arm\'onicos esf\'ericos.

\subsection*{Propiedades de los Arm\'onicos Esf\'ericos}

\noindent
$\alpha\ $ {\it Relaci\'ones de Recurrencia}\\
Seg\'un los resultados generales podemos tener
\begin{equation}
l_\pm Y_{lm}(\theta,\phi)=\hbar\sqrt{l(l+1)-m(m\pm 1)}Y_{lm\pm1}(\theta, \phi).
\label{82}
\end{equation}
Usando la expresi\'on (62) para 
$l_\pm $ y el hecho de que $Y_{lm}(\theta,\phi)$
es el producto de una funci\'on dependiente s\'olo de $\theta$ y $e^{\pm i\phi}$,
obtenemos
\begin{equation}
e^{\pm i\phi}\left(\frac{\partial}{\partial \theta} - m\cot\theta \right)
Y_{lm}(\theta,\phi) = \sqrt{l(l+1)-m(m\pm 1)}Y_{lm\pm 1}(\theta, \phi)
\label{83}
\end{equation}

\noindent
$\beta \ $ {\it Ortonormalizaci\'on y Relaci\'on de Cerradura}\\
La Ec.\ (68) determina solamente los arm\'onicos esf\'ericos con un factor 
cons\-tan\-te. Ahora eligiremos este factor tal como ortonormalizaci\'on de 
$Y_{lm}(\theta, \phi)$ (como funci\'on de variable angular $\theta,\ \phi$)
\begin{equation}
\int^{2\pi}_0d\phi\int^\pi_0\sin\theta\,d\theta Y^*_{lm}(\theta, \phi)
Y_{lm}(\theta, \phi) - \delta_{l'l}\delta_{m'm}.
\label{84}
\end{equation}
Adem\'as, cualquier funci\'on de $\theta,\ \phi$ pueden ser expresadas en
t\'erminos de los arm\'onicos esf\'ericos, esto es
\begin{equation}
f(\theta, \phi) = \sum^\infty_{l=0}\sum^l_{m= -l}c_{lm}Y_{lm}(\theta,\phi),
\label{85}
\end{equation}
donde
\begin{equation}
c_{lm}=\int^{2\pi}_0d\phi\int^\pi_0\sin\theta\,d\theta\, Y^*_{lm}(\theta, \phi)
f(\theta, \phi).
\label{86}
\end{equation}

Los arm\'onicos esf\'ericos constituyen una base ortonormal en el espacio
$\varepsilon_{\Omega}$ de funciones de $\theta,\ \phi$. Este hecho se 
expresa por la relaci\'on de cerradura
\begin{eqnarray}
 \sum^\infty_{l=0}\sum^l_{m=l}Y_{lm}(\theta,\phi)Y^*_{lm}(\theta',\phi)
&=&\delta(\cos\theta-\cos\theta' )\delta(\phi, \phi),\nonumber\\
&=&\frac{1}{\sin\theta}\delta(\theta-\theta')\delta(\phi, \phi).
\label{87} 
\end{eqnarray}

\noindent
Es $\delta(\cos\theta-\cos\theta' )$ y no $\delta(\theta-\theta')$ los 
cuales entran en el lado derecho de la relaci\'on de cerradura porque la 
integral sobre la variable $\theta$ se efect\'ua usando el elemento diferencial
$\sin\theta\,d\theta = -d(\cos\theta)$.

\subsection*{Operador de Paridad $\cal P$}
El comportamiento de ${\cal P}$ en tres dimensiones es esencialmente igual
que en una dimensi\'on, es decir, al aplicarlo sobre una funci\'on 
$\psi(x,y,z)$ de coordenadas cartesianas s\'olo le cambi\'o el signo, esto es
\begin{equation}
{\cal P}\psi(x,y,z) = \psi(-x,-y,-z).
\label{88}
\end{equation}
$\cal P$ tiene las propiedades de un operador herm\'{\i}tico, adem\'as es
un operador unitario y de proyecci\'on.

El operador ${\cal P}^2$ es un operador identidad
\begin{eqnarray}
\langle \bf{r} | \cal{P} | \bf{r'} \rangle =
\langle \bf{r} | \bf{-r'} \rangle  = \delta({\bf{r} + \bf{-r'} }), \nonumber\\
\cal{P}^2 |\bf{r}\rangle = {\cal P}({\cal P}|\bf{r}\rangle  = 
\cal{P}|-\bf{r}\rangle  = |\bf{r}\rangle,
\label{89}
\end{eqnarray}
entonces
\begin{equation}
{\cal P}^2 = \hat{1}, 
\label{90}
\end{equation}
cuyos valores propios son $\cal{P}=\pm1$. 
Adem\'as se tiene que las FP se llaman pares si $\cal{P} = 1$ e impares si
$\cal{P}=-1$. En mec\'anica cu\'antica no relativista, el operador $\hat{H}$
en un sistema cerrado es invariante ante transformaciones unitarias directas
\begin{equation}
{\cal P}\hat{H}{\cal P} = {\cal P}^{-1}\hat{H}{\cal P} = \hat{H}.
\label{91}
\end{equation}
Entonces $\hat{H}$ conmuta con $\cal P$ y consecuentemente la paridad del
estado es una integral de movimiento. Tambi\'en se cumple para $\hat{l}$
\begin{equation}
[{\cal P}, \hat{l}_i] = 0, \qquad [{\cal P}, \hat{l}_\pm].
\label{92}
\end{equation}
Si $\hat{H}$ es par y uno de sus eigenestados $|\Phi_n \rangle$ el cual tiene
paridad definida $({\cal P}|\Phi_n \rangle )$, no colinear a $|\psi_n \rangle$,
se ha encontrado y puede inferirse que el eigenvalor correspondiente es
degenerado con un grado de degeneraci\'on $n^2$, dado que $\cal P$ conmuta
con $\hat{H}$, $({\cal P}|\Phi_n \rangle )$ es un eigenvector  de $\hat{H}$
con el mismo eigenvalor como $|\Phi_n \rangle )$. Si $\psi$ es FP de 
${\cal P},\ \hat{l}$ y $\hat{l}_z$ de (92) resulta que las paridades de los
estados diferentes s\'olo en $\hat{l}_z$ coinciden. Queda as\'{\i} determinado 
la paridad de una part\'{\i}cula de momento angular $\hat{l}$.

En coordenadas esf\'ericas, para este operador se considera la siguiente
sustituci\'on
\begin{equation}
r\rightarrow r, \qquad \theta \rightarrow \pi-\theta \qquad
\phi \rightarrow \pi+\phi.
\label{93}
\end{equation}

Consecuentemente, si usamos una base est\'andar para el espacio de funciones 
de onda de una part\'{\i}cula sin giro, la parte radial de la funci\'on base
$\psi_{klm}(\vec{r})$ es modificada por el operado paridad. La transformaci\'on
s\'olo se da en los arm\'onicos esf\'ericos, como se ver\'a.

De (93) podemos notar que 
\begin{equation}
\sin\theta \rightarrow \sin\theta, \qquad \cos\theta \rightarrow
-\cos\theta \qquad e^{im\phi} \rightarrow (-1)^me^{im\phi},
\label{94}
\end{equation}
bajo estas condiciones, la funci\'on $Y_{ll}(\theta, \phi)$ es transformada en
\begin{equation}
Y_{ll}(\phi - \theta, \pi + \phi) = (-1)^l Y_{ll}(\theta, \phi),
\label{95}
\end{equation}
de (95) podemos ver que la paridad de los arm\'onicos  esf\'ericos va como
$(-1)^l$. Por otro lado
\begin{equation}
{\partial \over \partial\theta} \rightarrow -{\partial \over \partial\theta},
\qquad {\partial \over \partial\phi} \rightarrow {\partial \over \partial\phi}.
\label{96}
\end{equation}

Relacionando (95) y (96) mostramos que $\hat{l}_\pm$ permanece sin cambio (lo
cual implica que los operadores $\hat{l}_\pm$ son pares). Consecuentemente
podemos calcular $Y_{lm}(\theta, \phi)$,
\begin{equation}
Y_{lm}(\phi - \theta, \pi + \phi) = (-1)^l Y_{lm}(\theta, \phi).
\label{97}
\end{equation}
Por lo tanto, los arm\'onicos esf\'ericos son funciones cuya paridad est\'a
bien definidad e independiente de $m$, par si $l$ es par e impar si $l$
es impar.

\section*{El Operador Esp\'{\i}n}
%%%%%%%%%%%%%%%%%%%%%%%%%%%
Algunas part\'{\i}culas, adem\'as de su momento angular tienen un 
{\it momento propio\/}, el cual, se le conoce como {\it esp\'{\i}n\/} y 
denominaremos
como $\hat{S}$. Este operador no est\'a relacionado con rotaciones normales
espaciales, a\'un as\'{\i}, cumple con las mismas relaciones de conmutaci\'on
que tienen el momento angular,~esto~es
\begin{equation}
[\hat{S}_i,\hat{S}_j]= i\varepsilon_{ijk}\hat{S}_k.
\label{98}
\end{equation}
As\'{\i} como, las siguientes propiedades

\begin{itemize}

\item[(1).]
Para el esp\'{\i}n, valen todas las f\'ormulas de (23) a (48) del momento angular,
las cuales son similares a (98)

\item[(2).] 
El espectro de la proyecci\'on del esp\'{\i}n, es una secuencia de 
n\'umeros enteros \'o semienteros que difieren por una unidad.

\item[(3).]
Los valores propios de $\hat{S}^2$ son 
\begin{equation}
\hat{S}^2\psi=S(S+1)\psi_s.
\label{99}
\end{equation}

\item[(4).]
Para un $S$ dado, la componente $S_z$ s\'olo puede tomar $2S+1$ valores, 
de $-S$ a $+S$.
 
\item[(5).]
Las FP de las part\'{\i}culas con esp\'{\i}n, adem\'as de depender 
de $\vec{r}$~\'o
$\vec{p}$, dependen de una variable discreta (propia del esp\'{\i}n)~$\sigma$,
la cual denota la proyecci\'on del esp\'{\i}n en $z$.

\item[(6).]
La FP $\psi(\vec{r}, \sigma)$ de una part\'{\i}cula con esp\'{\i}n se puede 
desarrollar en FP con proyecciones dadas del esp\'{\i}n $S_z$, esto es
\begin{equation}
\psi(\vec{r}, \sigma) = \sum_{\sigma = -S}^S \psi_\sigma(\vec{r})\chi(\sigma),
\label{100}
\end{equation}
donde $\psi_\sigma(\vec{r})$ es la parte orbital y $\chi(\sigma)$ es la parte
espinorial.

\item[(7).]
Las funci\'ones de esp\'{\i}n (espinores) $\chi(\sigma_i)$ son ortogonales para 
cualquier par de $\sigma_i \ne \sigma_k$. Las funciones 
$\psi_\sigma(\vec{r})\chi(\sigma)$ se les conoce como las componentes de las 
FO de una part\'{\i}cula con esp\'{\i}n.

\item[(8).]
La funci\'on $\psi_\sigma(\vec{r})$ se llama parte orbital de la FO \'o s\'olo 
orbital.

\item[(9)]
La normalizaci\'on se hace como sigue
\begin{equation}
\sum_{\sigma = -S}^S ||\psi_\sigma(\vec{r})|| = 1.
\label{101}
\end{equation}

\end{itemize}

Las relaciones de conmutaci\'on permiten establecer la forma concreta de los
operadores (matrices) de esp\'{\i}n que act\'uan en el espacio de las FP del
operador proyecci\'on del esp\'{\i}n.

Muchas part\'{\i}culas elementales tales como el electr\'on, el neutr\'on,
el prot\'on, etc.\ tienen esp\'{\i}n $1/2$, por eso la proyecci\'on toma s\'olo 
dos valores, es decir $S_z = \pm 1/2$ (en unidades $\hbar$).

Por otro lado, las matrices $S_x,\S_y,\ S_z$ en el espacio de las FP de
$\hat{S}^2,\ \hat{S}_z$~son
\begin{eqnarray}
S_x = {1 \over 2}\left(\begin{array}{cc}
 0 & 1  \\
 1 & 0   
\end{array}\right), \qquad
&&S_y = {1 \over 2}\left(\begin{array}{cc}
 0 & -i  \\
 i &  0   
\end{array}\right), \nonumber\\
\nonumber\\
S_z = {1 \over 2}\left(\begin{array}{cc}
 1 &  0  \\
 0 & -1  
\end{array}\right), \qquad
&&S^2 = {3 \over 4}\left(\begin{array}{cc}
 1 &  0  \\
 0 &  1  
\end{array}\right).
\label{102}
\end{eqnarray}

\subsection*{Definici\'on de las Matrices de Pauli}
Las matrices
\begin{equation}
\sigma_i = 2S_i
\label{103}
\end{equation}
se llaman matrices de Pauli, las cuales son matrices hermitianas, 
tienen la misma ec.\ caracter\'{\i}stica
\begin{equation}
\lambda^2 - 1 = 0,
\label{104}
\end{equation}
por consiguiente, los eigenvalores de $\sigma_x,\ \sigma_y$ y  $\sigma_z$ son
\begin{equation}
\lambda = \pm 1.
\label{105}
\end{equation}
Por lo tanto, son consistentes con el hecho de que $S_x,\ S_y$ y $S_z$ sean
iguales~a~$ \pm 1$.

Adem\'as 
\begin{equation}
\sigma_i^2 = \hat{I}, \qquad  \sigma_k\sigma_j = -\sigma_j\sigma_k
= i\sigma_z,\qquad \sigma_j\sigma_k = i\sum_l \varepsilon_{jkl}\sigma_l. +
\delta_{jk}I
\label{106}
\end{equation}

En el caso para el cual un sistema con esp\'{\i}n tiene simetr\'{\i}a esf\'erica 
(esf\'erico sim\'etrico)
\begin{equation}
\psi_1(r, +\textstyle{1\over 2}), \qquad
\psi_1(r, -\textstyle{1\over 2})~,
\label{107}
\end{equation}
son diferentes soluciones por la proyecci\'on $S_z$.

El valor de la probabilidad de una u otra de las proyecciones est\'a 
determinada por el cuadrado de $||\psi_{1,2}||^2$ de tal modo que
\begin{equation}
||\psi_1||^2 + ||\psi_2||^2 = 1.
\label{109}
\end{equation}
Como las  FP de $S_z$ tiene dos componentes, entonces
\begin{equation}
\chi_1= \left(\begin{array}{c}  1  \\ 0  
\end{array}\right), \qquad
\chi_2= \left(\begin{array}{c}  0  \\ 1  
\end{array}\right),
\label{109}
\end{equation}
tal que, las FP de una part\'{\i}cula de esp\'{\i}n $1/2$ se puede escribir
como
\begin{equation}
\psi = \psi_1\chi_1 + \psi_2\chi_2 = 
\left(\begin{array}{c}  \psi_1 \\ \psi_2 \end{array}\right).
\label{110}
\end{equation}

A continuaci\'on las orbitas van a ser sustituidas por n\'umeros dado que
solamente la parte del esp\'{\i}n es importante.

\section*{Las Transformaciones a las Rotaciones}
%%%%%%%%%%%%%%%%%%%%%%%%%%%%%%%%%%%%%%%%%%%%%%%%%
Sea $\psi$ la FO de un sistema con esp\'{\i}n en $\Sigma$. Encontremos la
probabilidad de la proyecci\'on del esp\'{\i}n en una direcci\'on arbitraria 
en el espacio tridimensional (3D) que la toma como eje $z'$ de $\Sigma '$.
Como ya se vi\'o, se tienen dos m\'etodos para su soluci\'on

\begin{itemize}

\item[$\alpha$)]
$\psi$ no cambia cuando $\Sigma \rightarrow \Sigma'$ y el operador
$\hat{\Lambda}$ cambia como un vector. Debemos encontrar las FP de la 
proyecci\'on $S'_z$ y desarrollamos $\psi$ en esas FP. Los cuadradados de
los m\'odulos de los coeficientes dan el resultado.
\begin{eqnarray}
\hat{S}_x' = \hat{S}_x\cos\phi + \hat{S}_y\sin\phi &=& 
e^{-il\phi}\hat{S}_x e^{il\phi},\nonumber\\
\hat{S}_y' = -\hat{S}_x\sin\phi + \hat{S}_y\cos\phi &=& 
e^{-il\phi}\hat{S}_y e^{il\phi},\nonumber\\
\hat{S}_z' = -\hat{S}_z = e^{il\phi}\hat{S}_z,
\label{111}
\end{eqnarray}
con rotaciones infinitesimales y de las relaciones de conmutaci\'on se puede
encontrar
\begin{equation}
\hat{L} =\hat{S}_z,
\label{112}
\end{equation}
donde $\hat{L}$ es el generador.

\item[$\beta$)]
La segunda representaci\'on es:\\
\noindent
$\hat{S}$ no se cambia a la $\Sigma \rightarrow \Sigma'$ y las componentes
de $\psi$ se cambian.
La transformaci\'on a esta representaci\'on se hace con una 
transformaci\'on unitaria
\begin{eqnarray}
\hat{V}^\dagger\hat{S}'\hat{V} &=& \hat{\Lambda}, \nonumber\\
\left(\begin{array}{c}  \psi_1'  \\  \psi_2'  \end{array}\right) &=&
\hat{V}^\dagger
\left(\begin{array}{c}  \psi_1  \\  \psi_2  \end{array}\right), 
\label{113}
\end{eqnarray}
de (111) y (113) resulta que 
\begin{eqnarray}
\hat{V}^\dagger e^{-i\hat{S}_z\phi}\hat{S} e^{i\hat{S}_z\phi}
\hat{V}&=&\hat{S},\nonumber\\
\hat{V}^\dagger &=& e^{i\hat{S}_z\phi},
\label{114}
\end{eqnarray}
de (114) se obtiene
\begin{equation}
\left(\begin{array}{c}  \psi_1'  \\  \psi_2'  \end{array}\right) =
e^{i\hat{S}_z\phi}
\left(\begin{array}{c}  \psi_1  \\  \psi_2  \end{array}\right), 
\label{115}
\end{equation}
Usando la forma concreta de $\hat{S}_z$ y las propiedades de las matrices de
Pauli se obtiene la forma concreta $\hat{V}^\dagger_z$, tal que
\begin{equation}
\hat{V}^\dagger_z(\phi) = \left(\begin{array}{cc}  
e^{{i \over 2}\phi}  & 0 \\  
0 &  e^{{-i \over 2}\phi}  \end{array}\right).
\label{116}
\end{equation}

\end{itemize}

\section*{Un Resultado de Euler}
%%%%%%%%%%%%%%%%%%%%%%%%%%%%%%%%%
Cualquier sistema de referencia $\Sigma'$ de orientaci\'on arbitraria con 
respecto a $\Sigma$ puede ser alcanzado con solo tres rotaciones, primero
alrededor del eje $z$, en seguida una rotaci\'on del \'angulo $\theta$   
sobre el nuevo eje de coordenadas $x'$ y por \'ultimo el \'angulo $\psi_a$
en el nuevo eje $z'$.

Los param\'etros $(\varphi, \theta, \psi_a)$ se les llama \'angulos de Euler
\begin{equation}
\hat{V}^\dagger(\varphi, \theta, \psi_a) = 
\hat{V}^\dagger_{z'}(\psi_a)\hat{V}^\dagger_{x'}(\theta)
\hat{V}^\dagger_{z}(\varphi).
\label{117}
\end{equation}

Las matrices $\hat{V}^\dagger_z$ son del tipo de (116), en cuanto a 
$\hat{V}^\dagger_x$ es de la forma
\begin{equation}
\hat{V}^\dagger_x(\varphi) = \left(\begin{array}{cc}  
\cos{\theta \over 2}  & i\sin{\theta \over 2} \\  
i\sin{\theta \over 2} &  \cos{\theta \over 2}  
\end{array}\right),
\label{118}
\end{equation}
de tal modo que 
\begin{equation}
\hat{V}^\dagger(\varphi, \theta, \psi_a) = 
\left(\begin{array}{cc}  
e^{i{\varphi + \psi_a \over 2}}\cos{\theta \over 2}   & 
ie^{i{\psi_a - \varphi \over 2}}\sin{\theta \over 2}  \\  
ie^{i{\varphi - \psi_a \over 2}}\sin{\theta \over 2}  &
e^{-i{\varphi + \psi_a \over 2}}\cos{\theta \over 2} 
\end{array}\right).
\label{119}
\end{equation}

Entonces por la rotaci\'on de $\Sigma$, las componentes de la funci\'on 
espinorial se cambian como sigue
\begin{eqnarray}
\psi'_1 &=& \psi_1 e^{i{\varphi + \psi_a \over 2}}\cos{\theta \over 2} +
       i\psi_2 e^{i{\psi_a - \varphi \over 2}}\sin{\theta \over 2},\nonumber\\ 
\psi'_2 &=&i\psi_1 e^{i{\varphi - \psi_a \over 2}}\sin{\theta \over 2} +
\psi_2 e^{-i{\varphi + \psi_a \over 2}}\cos{\theta \over 2}.
\label{120}
\end{eqnarray}
De (120) se puede ver que para una rotaci\'on en $E_3$ le corresponde una 
transformaci\'on lineal en $E_2$---espacio Euclidiano bidimensional 
(2D)---las dos componentes de la funci\'on espinorial. La rotaci\'on en
$E_3$ no implica una rotaci\'on en $E_2$, la cual significa
\begin{equation}
\langle \Phi' | \psi' \rangle = 
\langle \Phi | \psi \rangle = 
\Phi^*_1\psi_1 + \Phi^*_2\psi_2.
\label{121}
\end{equation}

De (119) se encuentra que (121) no se cumple, sin embargo, hay una invariancia
en las transformaciones (119) en el espacio $E_2$ de las funciones espinoriales,
el cual es
\begin{equation}
\{\Phi | \psi \}= 
\psi_1\Phi_2 - \psi_2\Phi_1.
\label{122}
\end{equation} 

Las transformaciones lineales que dejan invariantes tales formas
bilineales se llaman binarias.

Una transformaci\'on f\'{\i}sica con dos componentes para la cual una rotaci\'on
del sistema de coordenadas es una transformaci\'on binaria se llama 
{\it esp\'{\i}n de primer orden\/} o solamente {\it esp\'{\i}n}.

\subsection*{ Funciones de Onda Espinoriales de
%\hspace*{1pt}
un Sistema con 2 Fermiones}
Las funciones propias de $_i\hat{s}^2\ _i\hat{s}_z$---$i = 1,2$---tienen
la forma siguiente
\begin{equation}
i| + \rangle = \left(\begin{array}{c}
1 \\ 0
\end{array}\right)_i, \qquad 
i| - \rangle = \left(\begin{array}{c}
0 \\ 1
\end{array}\right)_i.
\label{123}
\end{equation}

Una variable---o mejor dicho, un operador---corriente en un sistema de dos 
fermiones es el esp\'{\i}n total
\begin{equation}
\hat{S} = _1\hat{S} + _2\hat{S}
\label{124}
\end{equation}

Las funciones propias espinoriales  de $\hat{s}^2\ \hat{s}_z $
son los kets $|\hat{S}, \sigma  \rangle$, las cuales son combinaciones 
lineales de las FP de $_i\hat{s}^2\ _i\hat{s}_z$, esto es
\begin{eqnarray}
| + +  \rangle = 
\left(\begin{array}{c} 1 \\ 0 \end{array}\right)_1
\left(\begin{array}{c} 1 \\ 0 \end{array}\right)_1, && \qquad 
| + -  \rangle = 
\left(\begin{array}{c} 1 \\ 0 \end{array}\right)_1
\left(\begin{array}{c} 0 \\ 1 \end{array}\right)_2, \nonumber\\
| - +  \rangle = 
\left(\begin{array}{c} 0 \\ 1 \end{array}\right)_2
\left(\begin{array}{c} 1 \\ 0 \end{array}\right)_1, && \qquad 
| - -  \rangle = 
\left(\begin{array}{c} 0 \\ 1 \end{array}\right)_2
\left(\begin{array}{c} 0 \\ 1 \end{array}\right)_2.
\label{125}
\end{eqnarray}

Las funciones de (125) son ortonormalizadas. En $E_n$ el estado 
$| ++ \rangle$ es  $S_z = 1$ y al mismo tiempo es funci\'on propia del 
operador 
\begin{equation}
\hat{S} = _1\hat{s}^2 + 2(_1\hat{s})(_2\hat{s})+_2\hat{s}^2.
\label{126}
\end{equation}
Como se puede ver de 
\begin{eqnarray}
\label{127}
\hspace*{-30pt}
\hat{S}^2 &=& | + +  \rangle = \textstyle{3 \over 2} | + +  \rangle +
2(_1\hat{s}_x \cdot _2\hat{s}_x + _1\hat{s}_y \cdot _2\hat{s}_y +
  _1\hat{s}_z \cdot _2\hat{s}_z)| + +  \rangle , \hspace*{-30pt}\\
\label{128}\hspace*{-30pt}
\hat{S}^2 &=& | + +  \rangle = 2 | + +  \rangle  =1(1+1) | + +  \rangle .
\hspace*{-30pt}
\end{eqnarray} 

Si se introduce el operador 
\begin{equation}
\hat{S}_{-} = _1\hat{s}_{-} + _2\hat{s}_{-},
\label{129}
\end{equation}
se obtiene que
\begin{equation}
[\hat{S}_{-} ,\hat{S}^2] = 0.
\label{130}
\end{equation}
Entonces $(\hat{S}_{-})^k|1,1\rangle$ se puede escribir en funci\'on de las
FP del operador $\hat{S}^2$, esto es
\begin{equation}
\hat{S}_{-}|1,1\rangle = \hat{S}_{-}|+ +\rangle =
\sqrt{2}|+ -\rangle + \sqrt{2}|- +\rangle.
\label{131}
\end{equation}
Resulta que $S_z = 0$ en el estado $\hat{S}_{-}|1,1\rangle $. Por otro lado,
de la condici\'on de normalizaci\'on tenemos
\begin{eqnarray}
|1,0\rangle  = \textstyle{1\over\sqrt{2}}(|+ -\rangle + |- +\rangle)\\
\label{132}
\hat{S}_{-}|1,0\rangle =|- -\rangle + |- -\rangle = \alpha|1, -1\rangle.
\label{133} 
\end{eqnarray}

De la coordenada de normalizaci\'on 
\begin{equation}
|1, -1\rangle  = |-, -\rangle. 
\label{134}
\end{equation}

S\'olo hay una combinaci\'on lineal independiente  m\'as de funciones de (125)
diferentes de $|1, 1\rangle,\ |1, 0\rangle$ y $|1, -1\rangle$, esto es
\begin{eqnarray}
\psi_4 = \textstyle{1\over\sqrt{2}}(|+ -\rangle - |- +\rangle), \label{135}\\
\hat{S}_z \psi_4 = 0, \qquad \hat{S}^2 \psi_4.
\label{136}
\end{eqnarray}
Por consiguiente
\begin{equation}
\psi_4 =|0, 0\rangle.
 \label{137}
\end{equation}
$\psi_4$ describe el estado de un sistema de dos fermiones con el esp\'{\i}n 
total
igual a cero. Este tipo de estado se llama {\it singlet\/}. Consecuentemente
el estado de dos fermiones de esp\'{\i}n total igual a uno se llama {\it triplet\/}
y tiene un grado de degeneraci\'on $g=3$.

\section*{Momento Angular Total}
%%%%%%%%%%%%%%%%%%%%%%%%%%%%%%%%%
Se define como la suma del momento angular orbital m\'as el spin, esto  es
\begin{equation}
\hat{J} = \hat{l}+\hat{S},
\label{138}
\end{equation}
donde, $\hat{l}$ y $\hat{S}$ como hemos visto, act\'uan en espacios
diferentes, pero el cuadrado de $\hat{l}$ y $\hat{S}$ conmutan con
$\hat{J}$, es decir
\begin{equation}
[\hat{J}_i, \hat{J}_j] = i\varepsilon_{ijk}\hat{J}_k, \qquad
[\hat{J}_i, \hat{l}^2] = 0, \qquad [\hat{J}_i, \hat{S}^2] = 0,
\label{139}
\end{equation}
(139) que  $\hat{l}^2$ y $\hat{S}^2$ tienen un sistema de FP con $\hat{J}^2$,
y $\hat{J}_z$.

Encontramos el espectro de las proyecciones de  $\hat{J}_z$ para un fermi\'on.
el estado de proyecci\'on de m\'aximo de $\hat{J}_z$ se puede escribir como
\begin{eqnarray}
\bar{\psi} &=& \psi_{ll}
\left(\begin{array}{c} 1 \\ 0 \end{array}\right) =
 |l,l,+ \rangle \\
\label{140}
\hat{\jmath}_z\psi &=& (l + \textstyle{1 \over 2})\bar{\psi}, \rightarrow
j= l + \textstyle{1 \over 2}.
\label{141}
\end{eqnarray}

Si introducimos el operador $\hat{J}_{-}$ definido por
\begin{equation}
\hat{J}_{-}=\hat{l}_{-}+\hat{S}_{-} = \hat{l}_{-}+
\left(\begin{array}{cc} 0 & 0 \\ 1 & 0 \end{array}\right).
\label{142}
\end{equation}

De la normalizaci\'on $\alpha = \sqrt{(J+M)(J-M+1)}$ se obtiene
\begin{equation}
\hat{J}_{-}\psi_{ll}\left(\begin{array}{c} 1 \\ 0 \end{array}\right) =
\sqrt{2l}|l,l-1, +\rangle + |l,l-1, -\rangle,
\label{143}
\end{equation}
tal que el valor de la proyecci\'on de $\hat{j}_{-}$ en 
$\hat{j}_{-}\bar{\psi}$ ser\'a
\begin{equation}
\hat{\jmath}_z = (l-1) + \textstyle{1 \over 2} = l - \textstyle{1 \over 2},
\label{144}
\end{equation}
resulta que $\hat{\jmath}_{-}$ disminuye por una unidad a $\hat{J}_z$.

En el caso general tenemos que
\begin{equation}
\hat{\jmath}_{-}^k = \hat{l}_{-}^k + k\hat{l}_{-}^{k-1}\hat{S}_{-},
\label{145}
\end{equation}
se observa que (145) se obtiene del desarrollo binomial si se considera que
$\hat{s}^2_{-}$  y todas las potencias superiores de $\hat{s}$ son cero.
\begin{equation}
\hat{\jmath}_{-}^k |l,l,+\rangle = \hat{l}_{-}^k |l,l,+\rangle + 
k\hat{l}_{-}^{k-1} |l,l,-\rangle.
\label{146}
\end{equation}

Sabemos que 
\[
(\hat{l}_{-})^k\psi_{l,l} =
\textstyle{\sqrt{\frac{k!(2l)!}{(2l-k)!}}\psi_{l,l-k}},
\]
tal que al usarla se obtiene
\begin{equation}
\hat{\jmath}_{-}^k | l,l,+\rangle = 
\textstyle{\sqrt{\frac{k!(2l)!}{(2l-k)!}}}| l,l-k,+\rangle +
\textstyle{\sqrt{\frac{(k+1)!(2l)!}{(2l-k+1)!}}} k| l,l-k+1,-\rangle,
\label{147}
\end{equation}
con la notaci\'on $m = l-k$
\begin{equation}
\hat{\jmath}_{-}^{l-m} | l,l,+\rangle = 
\textstyle{\sqrt{\frac{(l-m)!(2l)!}{(l+m)!}}}| l,m,+\rangle +
\textstyle{\sqrt{\frac{(l-m-1)!(2l)!}{(2l+m+1)!}}} (l-m)| l,m+1,-\rangle.
\label{148}
\end{equation}

\noindent
Los valores propios de la proyecci\'on del momento angular total es la 
secuencia de n\'umeros que difieren por la unidad desde 
$j=l+{1\over 2}$ hasta $j=l-{1\over 2}$.
Todos estos estados pertenecen a la misma funci\'on propia de $\hat{J}$
como $|l,l,+\rangle$ porque $[\hat{J}_1,\hat{J}^2]=C$
\begin{eqnarray}
\hat{J}^2|l,l,+\rangle &=& 
(\hat{l}^2 + 2\hat{l}\hat{S} + \hat{S}^2)|l,l,+\rangle, \nonumber\\ &=& 
[l(l+1) + 2l\textstyle{1\over 2} + \textstyle{3\over 4}]|l,l,+\rangle
\label{149}
\end{eqnarray}
donde $j(j+1) = (l+{1 \over 2})(l + {3 \over 2})$.

En la derecha de (149) una contribuci\'on diferente de cero da solamente
$j=\hat{l}_z\hat{S}_z$. Entonces las FP obtenidas corresponden a 
$j=l+{1 \over 2}$, $m_j=m+{1 \over 2}$. Las FP son de forma
\begin{equation} 
\Phi |l+{1 \over 2}, m+{1 \over 2} \rangle = 
\sqrt{l+m+1 \over 2l+1}|l, m, + \rangle +
\sqrt{l-m \over 2l+1}|l, m+1, - \rangle.
\label{150}
\end{equation}

El n\'umero total de estados lineales independientes es
\begin{equation}
N=(2l+1)(2s + 1) = 4l+2.
\label{151}
\end{equation}

El sistema de FP constituido de tal manera contiene
$2j+1 + 2l+1$ estados
\begin{equation}
|l-\textstyle{1 \over 2}, m-\textstyle{1 \over 2} \rangle = 
\sqrt{l-m \over  2l+1} | l, m, + \rangle -
\sqrt{l+m+1 \over  2l+1} | l, m+1, - \rangle.
\label{152}
\end{equation}

Si dos subsistemas est\'an interaccionando de tal manera que el
momento angular de cada $\hat{J}_i$ se conserva, entonces las FP del
operador momento angular total
\begin{equation}
\hat{J} =\hat{\jmath}_1 + \hat{\jmath}_2,
\label{153}
\end{equation}
se puede encontrar como lo hicimos anteriormente. Para valores propios de 
$\hat{\jmath}_1$ y  $\hat{\jmath}_2$  hay 
\[
(2j_1+1)(2j_2+1),
\]
FP ortonormalizadas de la proyecci\'on del momento angular total $\hat{J}_z$.
 La FP que corresponde al valor m\'aximo de la proyecci\'on $\hat{J}_z$ es 
decir, 
\[
M_J = j_1 + j_2,
\]
se puede construir de manera \'unica y por lo tanto $J = j_1 + j_2$ es el
valor m\'aximo del momento angular total del sistema. Aplicando el operador
$\hat{J} = \hat{\jmath}_1 + \hat{\jmath}_2$ de manera repetida a la funci\'on
\begin{equation}
|j_1 + j_2, j_1 + j_2, j_1 + j_2 \rangle = |j_1 , j_1\rangle \cdot
|j_2 , j_2\rangle,
\label{154}
\end{equation}
se obtienen todas las $2(j_1 + j_2) + 1$ funciones ortogonales de la FP de
$\hat{J} = j_1 + j_2$ con varios $M$ 
\[
-(j_1 + j_2) \leq M \leq (j_1 + j_2).
\]
Por ejemplo las FP para $M = j_1 + j_2-1$ es:
\begin{equation}
|j_1 + j_2, j_1 + j_2-1, j_1 , j_2 \rangle =
\sqrt{j_1 \over j_1 + j_2}|j_1,j_1-1, j_2, j_2 \rangle +
\sqrt{j_2 \over j_1 + j_2}|j_1,j_1, j_2, j_2-1 \rangle .
\label{155}
\end{equation}

Aplicando en seguida varias veces el operador $\hat{J}_{-}$ se puede obtener
las $2(j_1 + j_2-1) -1$ funciones de $J = j_1 + j_2-1$.

Se puede demostrar que 
\[
|j_1 - j_2| \leq J \leq j_1 + j_2
\]
tal que
\begin{equation}
\sum_{{\rm min} \ J}^{{\rm max} \ J}(2J +1) = (2J_1 +1)(2J_2 +1),
\label{156}
\end{equation}
 Consecuentemente
\begin{equation}
|J,M,j_1,j_2\rangle = 
\sum_{m_1+m_2 = M} (j_1, m, j_2|J +) | j_1, m_1, j_2, m_2 \rangle
\label{156}
\end{equation}

\bigskip

\noindent
Citas: 1. Acetatos del Prof. H. Rosu

\noindent
Referencia bibliogr\'afica:

\noindent
1. H.A. Buchdahl, ``Remark concerning the eigenvalues of orbital angular
momentum",

\noindent
Am. J. Phys. {\bf 30}, 829-831 (1962)

 \newpage
\section*{P r o b l e m a s}
\subsection*{Problema No. 3.1}
Mostrar que si $\psi' = R\psi$, entonces el operador $R$ se puede 
representar como un operador exponencial\\
\noindent
{\bf Soluci\'on}\\
Para mostrarlo, expandemos $\psi'(\vec{r})$ en serie de Taylor en el punto
$x' = x+dx$ y considerando s\'olo las primeras potencias tenemos
\begin{eqnarray}
\psi(x',y',z') &=& \psi(x,y,z) + (x'-x)\frac{\partial}{\partial x'}
\psi(x',y',z')\bigg|_{\vec{r'} = \vec{r}}  \nonumber \\
&& + (y'-y)\frac{\partial}{\partial y'} \psi(x',y',z')
\bigg|_{\vec{r'} = \vec{r}} \nonumber \\ 
&& + (z'-z)\frac{\partial}{\partial z'} \psi(x',y',z')
\bigg|_{\vec{r'} = \vec{r}}\, , \nonumber 
\end{eqnarray}
ahora considerando el hecho de que
\begin{eqnarray}
\frac{\partial}{\partial x'_i}\psi(\vec{r}')\bigg|_{\vec{r}'} &=& 
\frac{\partial}{\partial x_i}\psi(\vec{r}), \nonumber\\ 
x' = x -y d \phi, \qquad y' &=& y + d \phi, \qquad z' = z, \nonumber
\end{eqnarray}
tal que, esto reduce la serie de tres dimensiones a solamente dos 
\begin{eqnarray}
\psi(\vec{r}') &=& \psi(\vec{r}) + 
  (x-yd\phi-x)\frac{\partial\psi(\vec{r})}{\partial x} 
+ (y+xd\phi -y)\frac{\partial\psi(\vec{r})}{\partial y'}, \nonumber\\
&=& \psi(\vec{r})- d\phi y\frac{\partial\psi(\vec{r})}{\partial x}  
	         - d\phi x\frac{\partial\psi(\vec{r})}{\partial y},\nonumber\\ 
&=& \left[1 - d\phi\left(x\frac{\partial}{\partial y} 
            - y\frac{\partial}{\partial x}\right)\right]\psi(\vec{r})\nonumber
\end{eqnarray}
considerando que $i\hat{l}_z = 
 \left(x\frac{\partial}{\partial y} - y\frac{\partial}{\partial x}\right)$
entonces tenemos que
$
R = \left[1 - d\phi\left(x\frac{\partial}{\partial y} 
            - y\frac{\partial}{\partial x}\right)\right]\psi(\vec{r})
$
entonces $R$ puede escribirse como una exponencial
\[
R=e^{i\hat{l}_zd\phi}.
\]

%\newpage
\subsection*{Problema No. 3.2}
Mostrar que de las expresiones dadas en (14) se puede llegar a (15) \\
{\bf{Soluci\'on}}\\
Nuevamente, consideremos s\'olo los t\'erminos lineales en la expansi\'on
de la serie de Taylor y dado que tenemos rotaciones infinitesimales, entonces
\[
e^{i\hat{l}_zd\phi} = 1 + i\hat{l}_zd\phi + 
\textstyle{1 \over 2!}(i\hat{l}_zd\phi)^2 + \ldots\, , \nonumber
\]
entonces tenemos que 
\begin{eqnarray}
(1 + i\hat{l}_zd\phi)\hat{A}_x(1-i\hat{l}_zd\phi) &=& 
\hat{A}_x -\hat{A}_xd\phi,\nonumber\\
(\hat{A}_x + i\hat{l}_zd\phi\hat{A}_x)(1-i\hat{l}_zd\phi) &=& 
\hat{A}_x -\hat{A}_xd\phi,\nonumber\\
\hat{A}_x -\hat{A}_xi\hat{l}_zd\phi+ i\hat{l}_zd\phi\hat{A}_x +
\hat{l}_zd\phi\hat{A}_x \hat{l}_zd\phi &=&\hat{A}_x -\hat{A}_xd\phi,\nonumber\\
i(\hat{l}_z\hat{A}_x -\hat{A}_x\hat{l}_z)d\phi &=&-\hat{A}_yd\phi.\nonumber
\end{eqnarray} 
Luego entonces, concluimos que
\[
[ \hat{l}_z, \hat{A}_x]  = i\hat{A}_y. \nonumber
\]
entonces tenemos que 
\begin{eqnarray}
(1 + i\hat{l}_zd\phi)\hat{A}_y(1-i\hat{l}_zd\phi) &=& 
\hat{A}_xd\phi -\hat{A}_y,\nonumber\\
(\hat{A}_y + i\hat{l}_zd\phi\hat{A}_y)(1-i\hat{l}_zd\phi) &=& 
\hat{A}_xd\phi -\hat{A}_y,\nonumber\\
\hat{A}_y -\hat{A}_yi\hat{l}_zd\phi+ i\hat{l}_zd\phi\hat{A}_y +
\hat{l}_zd\phi\hat{A}_y \hat{l}_zd\phi &=&\hat{A}_xd\phi -\hat{A}_y,\nonumber\\
i(\hat{l}_z\hat{A}_y -\hat{A}_y\hat{l}_z)d\phi &=&-\hat{A}_xd\phi.\nonumber
\end{eqnarray} 

%\newpage
\subsection*{Problema No. 3.3}
Determine la precesi\'on del esp\'{\i}n en un campo magn\'etico homog\'eneo.\\

\noindent
{\bf Soluci\'on}\\
Si el cuerpo cargado se mueve en un campo magn\'etico homog\'eneo
circular alrededor de la direcci\'on del campo con una frecuencia
\[
\omega= 2\omega_L = {-eB \over mc}.
\]
Aqu\'{\i}, la carga del electr\'on es $-e$. Esto se sigue del hecho de que
la fuerza de Lorentz equilibra la fuerza centr\'{\i}fuga.
\[
{eBv \over c} = my\omega^2,
\]
entonces
\[
\omega = - {eB \over mc}.
\]
Por lo tanto
\[
\omega_L = {eB \over 2mc}
\]
la cual, se conoce  con el nombre de {\it frecuencia de Larmor}.

%\newpage
\subsection*{Problema No. 3.4}
Resolver la ec.\ de Laplace usando coordenadas esf\'ericas\\
{\bf Soluci\'on}\\
Asumiendo que podemos tener 
\[
U(r, \theta, \phi) = R(r)\Theta(\theta, \phi),
\]
de este modo vemos que
\[
{r \over R(r)} {\partial^2 \over \partial r^2} [rR(r)] = {1 \over \Theta} 
{\bf L^2}\Theta = l(l+1)
\]
  
%\end{document}

\newpage
%%%%%%%%%%%%%%%%%%%%%%%%%%%%%%%%%%%%%%%
%\documentclass[a4paper,12pt]{article}
%\usepackage{latexsym}
%\pagestyle{empty}
%\begin{document}
%\begin{center}
%\setcounter{equation}
%%%%%%%%%%%%%%%%%%%%%%%%%%%%%%%%%%%%%%%%
%\documentclass[a4paper,12pt]{article}
%\usepackage{latexsym}
%\pagestyle{empty}
%\begin{document}
\begin{center}
{\huge{4. El M\'etodo WKB}}
%\section*{4. EL METODO WKB}
\end{center}
\setcounter{equation}{0}
\hspace{0.6cm}Para estar en condiciones de estudiar los efectos de 
potenciales m\'as rea\-lis\-tas, que los correspondientes a 
{\em barreras
y pozos de potencial}, es necesario encontrar m\'etodos que permitan 
resolver la ecuaci\'on de Schr\"odinger para dichos potenciales. 

En general no es posible construir soluciones exactas para tales casos, y 
lo que se hace, es recurrir a m\'etodos de aproximaci\'on que 
proporcionen una soluci\'on suficientemente buena y simple,
como para poder estudiar el comportamineto del sistema con 
ella.

M\'etodos como estos hay muchos, pero nos concentraremos en el m\'etodo 
desarrollado simult\'aneamente por 
{\em G. Wentzel, M. A. Kramers y L.
Bri\-llouin} en 1926; y de cuyos apellidos deriva el acr\'onimo {\em WKB}.
%\textit{\textbf{WKB}}.

Es importante mencionar que el m\'etodo WKB, s\'olo es aplicable a la 
ecuaci\'on de Schr\"odinger  1-dimensional.

Para resolver la ecuaci\'on de Schr\"odinger
\begin{equation}
-\frac{\hbar^2}{2m}\frac{\partial^2\psi}{\partial{y}^2}+u(y)\psi=E\psi
\end{equation}
supongamos que el potencial tiene la forma:
\begin{equation}
u(y)=u_0f\Big(\frac{y}{a}\Big)
\end{equation}

Y hacemos los cambios de variables:
\begin{equation}
\xi^2=\frac{\hbar^2}{2mu_0a^2}
\end{equation}
\begin{equation}
\eta=\frac{E}{u_0}
\end{equation}
\begin{equation}
x=\frac{y}{a}
\end{equation}
de la ecuaci\'on $(5)$ obtenemos:
\begin{equation}
\frac{\partial}{\partial{x}}=
\frac{\partial{y}}{\partial{x}}\frac{\partial}{\partial{y}}=
a\frac{\partial}{\partial{y}}
\end{equation}
\begin{equation}
\frac{\partial^2}{\partial{x}^2}
=\frac{\partial}{\partial{x}}\Big(a\frac{\partial}{\partial{y}}\Big)
=\Big(a\frac{\partial}{\partial{x}}\Big)\Big(a\frac{\partial}{\partial{x}}
\Big)=a^2\frac{\partial^2}{\partial{y}^2}
\end{equation}
y la ecuaci\'on de Schr\"odinger se escribe:
\begin{equation}
-\xi^2\frac{\partial^2\psi}{\partial{x}^2}+f(x)\psi=\eta\psi
\end{equation}
multiplic\'andola por $-1/\xi^2$ y definiendo $r(x)=\eta-f(x)$, es posible 
escribirla en la forma:
\begin{equation}
\frac{\partial^2\psi}{\partial{x}^2}+\frac{1}{\xi^2}r(x)\psi=0
\end{equation}
para resolver esta \'ultima, proponemos la siguiente soluci\'on:
\begin{equation}
\psi(x)=\exp{\Bigg[\frac{i}{\xi}\int_a^x{q(x)dx}\Bigg]}
\end{equation}

En general: 
$\int_a^xq(x)dx=Q(x)\vert_a^x=
Q(x)-Q(a)\ni\frac{\partial{Q(x)}}{\partial{x}}=\frac{dQ(x)}{dx}=q(x)$. 
Esto de acuerdo con el teorema fundamental del c\'alculo.

Por lo que:
%\begin{displaymath}
$$
\frac{\partial^2{\psi}}{\partial{x}^2}
=\frac{\partial}{\partial{x}}\bigg(\frac{\partial{\psi}}{\partial{x}}\bigg)
=\frac{\partial}{\partial{x}}\Bigg\{\frac{i}{\xi}q(x)
\exp\Bigg[{\frac{i}{\xi}\int_a^xq(x)dx\Bigg]}\Bigg\}
$$
%\end{displaymath}
%\begin{displaymath}
$$
\Longrightarrow\frac{\partial^2{\psi}}{\partial{x}^2}
=\frac{i}{\xi}\Bigg\{\frac{i}{\xi}q^2(x)\exp\Bigg[\frac{i}{\xi}
\int_a^xq(x)dx\Bigg]+\frac{\partial{q(x)}}{\partial{x}}
\exp\Bigg[\frac{i}{\xi}\int_a^xq(x)dx\Bigg]\Bigg\}
$$
%\end{displaymath}
Factorizando $\psi$ tenemos:
\begin{equation}
\frac{\partial^2{\psi}}{\partial{x}^2}=\Bigg[-\frac{1}{\xi^2}q^2(x)+\frac{i}{\xi}\frac{\partial{q(x)}}{\partial{x}}\Bigg]\psi
\end{equation}

Olvid\'andonos de la dependencia en $x$, la ecuaci\'on de Schr\"odinger 
se escribe ahora:
\begin{equation}
\Bigg[-\frac{1}{\xi^2}q^2+\frac{i}{\xi}\frac{\partial{q}}{\partial{x}}+\frac{1}{\xi^2}r\Bigg]\psi=0
\end{equation}

En general $\psi\neq0$ por lo que:
\begin{equation}
i\xi\frac{dq}{dx}+r-q^2=0
\end{equation}
que es una ecuaci\'on diferencial lineal tipo Riccati, cuya soluci\'on se 
busca como una serie de potencias de $\xi$; suponiendo que $\xi$ es muy 
peque\~na.

Dicha serie proponemos que tiene la forma:
\begin{equation}
q(x)=\sum^\infty_{n=0}(-i\xi)^nq_n(x)
\end{equation}
Sustituimos \'esta en la Riccati para obtener:
\begin{equation}
i\xi\sum_{n=0}^\infty(-i\xi)^n\frac{dq_n}{dx}+r(x)-\sum_{\mu=0}^\infty(-i\xi)^{\mu}q_{\mu}\sum_{\nu=0}^\infty(-i\xi)^{\nu}q_{\nu}=0
\end{equation}
Rearreglando los t\'erminos de la ecuaci\'on $(15)$ tenemos:
\begin{equation}
\sum_{n=0}^\infty(-1)^n(i\xi)^{n+1}\frac{dq_n}{dx}+r(x)-\sum_{\mu=0}^\infty\sum_{\nu=0}^\infty(-i\xi)^{\mu+\nu}q_{\mu}q_{\nu}=0
\end{equation}
Las series dobles cumplen con:
\begin{displaymath}
\sum_{\mu=0}^\infty\sum_{\nu=0}^\infty{a_{\mu\nu}}=\sum_{n=0}^\infty\sum_{m=0}^n{a_{m,n-m}}
\end{displaymath}
Donde: $\mu=n-m\quad\nu=m$\\\\
De esta forma:
\begin{equation}
\sum_{n=0}^\infty(-1)^n(i\xi)^{n+1}\frac{dq_n}{dx}+r(x)-\sum_{n=0}^\infty\sum_{m=0}^n(-i\xi)^{n-m+m}q_{m}q_{n-m}=0
\end{equation}

Veamos por separado unos cuantos t\'erminos de cada una de las series 
contenidas en la ecuaci\'on (17): 
\begin{equation}
\sum_{n=0}^\infty(-1)^n(i\xi)^{n+1}\frac{dq_n}{dx}=i\xi\frac{dq_0}{dx}+\xi^2\frac{dq_1}{dx}-i\xi^3\frac{dq_2}{dx}+\dots
\end{equation}
\begin{equation}
\sum_{n=0}^\infty\sum_{m=0}^n(-i\xi)^{n}q_{m}q_{n-m}=q^2_0-i2{\xi}q_0q_1+\dots
\end{equation}
Para que ambas series contengan en su primer t\'ermino, de sus respectivos 
desarrollos a $i\xi$, debemos escribirlas:
$$
%\begin{displaymath}
\sum_{n=1}^\infty(-1)^{n-1}(i\xi)^n\frac{dq_{n-1}}{dx}+r(x)-q_0^2-
\sum_{n=1}^\infty\sum_{m=0}^n(-i\xi)^nq_mq{n-m}=0
%\end{displaymath}
$$
De la cual obtenemos:
\begin{equation}
\sum_{n=1}^\infty\Bigg[-(-i\xi)^n\frac{dq_{n-1}}{dx}-
\sum_{m=0}^n(-i\xi)^nq_mq_{n-m}\Bigg]+\Bigg[r(x)-q_0^2\Bigg]=0
\end{equation}

Para que se satisfaga la igualdad anterior, debemos exigir que:
\begin{equation}
r(x)-q_0^2=0 \quad\Rightarrow\quad q_0=\pm\sqrt{r(x)}
\end{equation}
$$
%\begin{displaymath}
-(-i\xi)^n\frac{dq_{n-1}}{dx}-\sum_{m=0}^n(-i\xi)^nq_mq_{n-m}=0 \quad 
%\end{displaymath}
$$
\begin{equation}
\Rightarrow\quad\quad\frac{dq_{n-1}}{dx}=-\sum_{m=0}^{n}q_mq_{n-m}
\quad\quad{n\geq1}
\end{equation}
A la cual llamaremos {\bf relaci\'on de recurrencia}.
Recordando que definimos 
$r(x)=\eta-f(x),\quad\eta=\frac{E}{u_0}\quad\&\quad{f(x)=\frac{u}{u_0}}$ 
; obtenemos con ayuda de la ecuaci\'on $(21)$ que:
\begin{equation}
q_0=\pm\sqrt{\eta-f(x)}=\pm\sqrt{\frac{E}{u_0}-\frac{u}{u_0}}=
\pm\sqrt{\frac{2m(E-u)}{2mu_0}}
\end{equation}

Esta \'ultima es la conexi\'on cl\'asica para el momento de una 
part{\'\i}cula de energ{\'\i}a $E$ en el potencial $u$, en unidades 
de $\sqrt{2mu _0}$. Por ello:
$$
%\begin{displaymath}
q_0=p(x)=\sqrt{\eta-f(x)}$$ \hspace{1mm} {(\bf que no es operador)}
%\end{displaymath}
%$$
Si aproximamos hasta segundo orden, tenemos lo siguiente:
$$
%\begin{displaymath}
q(x)=q_0-i{\xi}q_1-\xi^2q_2
%\end{displaymath}
$$
y empleando la ecuaci\'on de recurrencia calculamos $q_1$ y $q_2$:
$$
%\begin{displaymath}
\frac{dq_0}{dx}=-2q_0q_1
\quad \Rightarrow \quad 
q_1=-\frac{1}{2}\frac{\frac{dq_0}{dx}}{q_0}=
-\frac{1}{2}\frac{d}{dx}(\ln\vert{q_0}\vert)
%\end{displaymath}
$$
\begin{equation}
\Rightarrow \quad q_1=-\frac{1}{2}\frac{d}{dx}(\ln\vert p(x)\vert)
\end{equation}
\begin{equation}
\frac{dq_1}{dx}=-2q_0q_2-q_1^2 \quad\Rightarrow\quad 
q_2=-\frac{\frac{dq_1}{dx}-q_1^2}{2q_0}
\end{equation}

De la ecuaci\'on $(24)$, nos percatamos de que $q_1$ es la pendiente con 
el signo cambiado de $\ln\vert q_0\vert$; cuando $q_0$ es muy 
peque\~no, $q_1\ll0\quad\Rightarrow\quad -\xi{q_1}\gg0$, y en 
consecuencia la serie diverge. Lo que nos lleva a exigir la
siguiente {\bf condici\'on WKB}:
$$
%\begin{displaymath}
\vert q_0\vert\gg\vert -\xi{q_1}\vert=\xi\vert{q_1}\vert$$
%\hspace{1mm}

La condici\'on WKB no se satisface para puntos $x_k$ tales que:
$$
%\begin{displaymath}
q_0(x_k)=p(x_k)=0
%\end{displaymath}
$$
recordando que $q_0=p=\sqrt{\frac{2m(E-u)}{2mu_0}}$, la ecuaci\'on 
anterior nos conduce a:
\begin{equation}
E=u(x_k)
\end{equation}

En f{\'\i}sica cl\'asica puntos $x_k$ que satisfacen la ecuaci\'on 
(26), son llamados {\bf puntos de retorno}; ya que en
ellos la part{\'\i}cula invierte el sentido de su movimiento.

En base a lo anterior, podemos decir que $q_0$ es una soluci\'on 
cl\'asica del problema, y que $q_1$ y $q_2$ son 
respectivamente, la primer y segunda correcciones cu\'anticas del problema.

Para obtener las funciones de onda, s\'olo consideraremos la soluci\'on 
cl\'asica, y la primer correcci\'on cu\'antica del problema; y las 
ssutituimos en nuestra propuesta para $\psi$:
$$
%\begin{displaymath}
\psi=\exp\Bigg[\frac{i}{\xi}\int_a^x{q(x)dx}\Bigg]=
\exp\Bigg[\frac{i}{\xi}\int_a^x(q_0-i\xi{q_1})dx\Bigg]
%\end{displaymath}
$$
$$
%\begin{displaymath}
\Rightarrow\quad\psi=\exp\Bigg(\frac{i}{\xi}\int_a^xq_0dx\Bigg)\cdot
\exp\Bigg(\int_a^xq_1dx\Bigg)
%\end{displaymath}
$$

Para el segundo factor tenemos:
$$
%\begin{displaymath}
\exp\Bigg(\int_a^xq_1dx\Bigg)=\exp\Bigg[-\frac{1}{2}
\int_a^x\frac{d}{dx}(\ln \vert p(x)\vert)dx\Bigg]=
%\end{displaymath}
$$
$$
%\begin{displaymath}
\quad\quad\quad\quad\quad\quad\quad=\exp\Bigg[-\frac{1}{2}(\ln\vert 
p(x)\vert)\Big{\vert}_a^x\Bigg]=\frac{A}{\sqrt{p(x)}} 
%\end{displaymath}
$$
con $A$ una constante. Para el primer factor tenemos:
$$
%\begin{displaymath}
\exp\Bigg(\frac{i}{\xi}\int_a^xq_0dx\Bigg)=\exp\Bigg[\pm\frac{i}{\xi}
\int_a^xp(x)dx\Bigg]
%\end{displaymath}
$$
Y $\psi$ puede ser escrita como:
\begin{equation}
\psi^{\pm}=\frac{1}{\sqrt{p(x)}}\exp\Bigg[\pm\frac{i}{\xi}\int_a^xp(x)dx\Bigg]
\end{equation}
y se llaman %\textbf
{\bf las soluciones WKB de la ecuaci\'on de Schr\"odinger 
uno-dimensional}.

La soluci\'on general WKB en la regi\'on para la cual se cumple la 
condici\'on WKB, se escribe:
\begin{equation}
\psi=a_+\psi^++a_-\psi^-
\end{equation}

Como ya se dijo, no hay soluci\'on WKB en los puntos de retorno; lo que 
nos lleva a cuestionarnos como es que $\psi(x<x_k)$ pasa a 
$\psi(x>x_k)$, y para esto se hace necesaria la introducci\'on de las 
f\'ormulas de conexi\'on.

%pero antes veamos como hacer una estimaci\'on
%del error que se comete al resolver una ecuaci\'on diferencial
%ordinaria por el m\'etodo aproximativo WKB. %\end{document}
\bigskip

%%%%%%%%%%%%%%%%%%%%%%%%%%%%%%%%%%%%%%%%%%
%\documentclass[a4paper,12pt]{article}
%\usepackage{latexsym}
%\pagestyle{empty}
%\begin{document}
%\begin{center}
%\centerline{\huge
\subsection*{Las F\'ormulas De Conexi\'on}
%\end{center}

\hspace{0.6cm}Ya se dijo que las soluciones WKB, son singulares 
en los puntos cl\'asicos de retorno; no obstante estas soluciones son 
v\'alidas a la izquierda, y a la derecha de un punto cl\'asico de retorno 
$x_k$. Y por ello nos cuestionamos como es que $\psi(x<x_k)$ pasa a 
$\psi(x>x_k)$; es decir, debemos encontrar las f\'ormulas de conexi\'on.

De la teor{\'\i}a de ecuaciones diferenciales ordinarias, y con apoyo del 
an\'alisis de funciones de variable compleja, 
puede demostrarse que las f\'ormulas de conexi\'on existen y que son las 
siguientes: 
$$
%\begin{displaymath}
\psi_1(x)=
\frac{1}{\left[-r(x)\right]^{\frac{1}{4}}}
\exp\left(-\int_x^{x_k}\sqrt{-r(x)}dx\right)\rightarrow 
%\end{displaymath}
$$
\begin{equation}
\rightarrow\frac{2}{\left[r(x)\right]^{\frac{1}{4}}}
\cos\left(\int_{x_k}^x\sqrt{r(x)}dx-\frac{\pi}{4}\right)
\end{equation}
donde $\psi_1(x)$ s\'olo tiene un comportamiento exponencial decreciente para 
$x<x_k$. Lo que significa nuestra primer f\'ormula de conexi\'on, es que una 
funci\'on $\psi(x)$, que a la izquierda de un punto cl\'asico de retorno se 
comporte como una exponencial decreciente, pasa a la derecha del punto 
cl\'asico de retorno como un $coseno$ de fase $\phi=\frac{\pi}{4}$, y con
el doble de la amplitud de la exponencial.

Ahora, en el caso de una funci\'on $\psi(x)$ m\'as general; es decir, una 
funci\'on que tenga un comportamiento exponencial creciente y decreciente; 
la f\'ormula de conexi\'on correspondiente es:
$$
%\begin{displaymath}
\sin\left(\phi+\frac{\pi}{4}\right)
\frac{1}{\left[-r(x)\right]^{\frac{1}{4}}}
\exp\left(\int_x^{x_k}\sqrt{-r(x)}dx\right)\leftarrow
%\end{displaymath}
$$
\begin{equation}
\leftarrow\frac{1}{\left[r(x)\right]^{\frac{1}{4}}}
\cos\left(\int_{x_k}^x\sqrt{r(x)}dx+\phi\right)
\end{equation}
siempre que $\phi$ no tenga un valor muy cercano a $-\frac{\pi}{4}$; la 
raz\'on de ello es que si $\phi=-\frac{\pi}{4}$, el $seno$ se anula. Esta 
segunda f\'ormula de conexi\'on, significa que una funci\'on que se 
comporta como un $coseno$ a la derecha de un punto cl\'asico de retorno, 
pasa a la izquierda de \'el como una exponencial creciente con amplitud 
modulada por un $seno$.

Para ver los detalles de como son obtenidas estas f\'ormulas de 
conexi\'on, debe consultarse el libro: %\textbf
{\em Mathematical Methods of
Physics} by %\textbf
{\em J. Mathews \& R.L. Walker.}\\\\
%\textbf
\bigskip
%\newpage
%\documentclass[a4paper,12pt]{article}
%\usepackage{latexsym}
%\pagestyle{empty}
%\begin{document}
%\begin{center}
%\centerline{\huge
\subsection*{Estimaci\'on Del Error Introducido en la
Aproximaci\'on WKB}

%\end{center}
%\textbf{(Problema 1)}\\\\
\hspace{0.6cm}Hemos encontrado la soluci\'on a la ecuaci\'on de 
Schr\"odinger en cualquier regi\'on donde se satisfaga la 
condici\'on WKB; no obstante, las soluciones WKB divergen en los 
puntos cl\'asicos de retorno como ya se ha se\~nalado. 
Analizaremos un tanto superficialmente esta problem\'atica a fin de proponer 
las llamadas %\textbf
{\em f\'ormulas de conexi\'on} en una vecindad
pr\'oxima a los puntos cl\'asicos de retorno.

Supongamos que $x=x_k$, es un punto cl\'asico de retorno; es decir, es un 
punto tal que se cumple: $q_0(x_k)=p(x_k)=0\quad\Rightarrow\quad 
E=u(x_k)$. Ahora bien, a la izquierda de $x_k$; es decir para 
puntos del espacio 1-dimensional tales que $x<x_k$, supongamos que 
$E<u(x)$, de modo que en esta regi\'on la soluci\'on WKB es:
$$
%\begin{displaymath}
\psi(x)=\frac{a}{\left[\frac{u(x)-E}{u_0}\right]^\frac{1}{4}}\exp\left(-\frac{1}{\xi}\int_x^{x_k}\sqrt{\frac{u(x)-E}{u_0}}dx\right)\quad+
%\end{displaymath}
$$
\begin{equation}
\quad\quad+\quad\frac{b}{\left[\frac{u(x)-E}{u_0}\right]^\frac{1}{4}}\exp\left(\frac{1}{\xi}\int_x^{x_k}\sqrt{\frac{u(x)-E}{u_0}}dx\right)
\end{equation}
de igual forma a la derecha de $x_k$, es decir para puntos del espacio 
1-dimensioanl tales que $x>x_k$ supondremos que $E>u(x)$, en 
consecuencia la soluci\'on WKB en esta regi\'on es:
$$
%\begin{displaymath}
\psi(x)=\frac{c}{\left[\frac{E-u(x)}{u_0}\right]^\frac{1}{4}}
\exp\left(\frac{i}{\xi}\int_{x_k}^x\sqrt{\frac{E-u(x)}{u_0}}dx\right)\quad+
%\end{displaymath}
$$
\begin{equation}
\quad\quad\quad\quad+
\quad\frac{d}{\left[\frac{E-u(x)}{u_0}\right]^\frac{1}{4}}
\exp\left(-\frac{i}{\xi}\int_{x_k}^x\sqrt{\frac{E-u(x)}{u_0}}dx\right)
\end{equation}

Si $\psi(x)$ es una funci\'on real, lo ser\'a tanto a la derecha como a la 
izquierda de $x_k$, a esto le llamaremos %\textbf
{\it ``reality  condition''}, y establece que
si $a,b\in\Re$, entonces $c=d^*$.

Nuestro problema es conectar las aproximaciones a cada lado de $x_k$ a modo de 
que se refieran a la misma soluci\'on exacta; esto es encontrar 
$c$ y $d$ si conocemos $a$ y $b$, y viceversa. Para hacer dicha 
conexi\'on, debemos utilizar una soluci\'on aproximada, la cual 
sea v\'alida a lo largo de un camino que conecte las regiones a 
ambos lados de $x_k$, donde las soluciones WKB sean v\'alidas 
tambi\'en.

Lo m\'as com\'un es recurrir a un m\'etodo propuesto por %\textbf
{\em Zwann} y %\textbf
{\em Kemble}, el
cual consiste en evadir el eje real en las cercanias de $x_k$, 
mediante el recorrido de un camino que encierre a $x_k$ en el 
plano complejo; a lo largo de este camino las soluciones WKB seguir\'an siendo
v\'alidas. En esta exposici\'on recurriremos a dicho m\'etodo, pero s\'olo 
con la finalidad de obtener un medio de estimar errores en la 
aproximaci\'on WKB.

La estimaci\'on de errores es importante, a causa de que se desea obtener
soluciones aproximadas, en un amplio intervalo de puntos del espacio 
1-dimensional; y se debe estar preocupado en si el error se acumula, y si 
posteriormente traer\'a consigo corrimientos de fase.

Para esto definimos %\textbf
{\em las funciones WKB asociadas} como:
\begin{equation}
W_{\pm}=\frac{1}{\left[\frac{E-u(x)}{u_0}\right]^\frac{1}{4}}
\exp\left(\pm\frac{i}{\xi}\int_{x_k}^x\sqrt{\frac{E-u(x)}{u_0}}dx\right)
\end{equation}
a \'estas las consideraremos como funciones de variable compleja respecto de 
$x$, y emplearemos cortes de tipo rama para eludir discontinuidades en los 
ceros de $r(x)=\frac{E-u(x)}{u_0}$. Estas funciones satisfacen una 
ecuaci\'on diferencial, que puede obtenerse diferenci\'andolas respecto a 
$x$, para tener:
$$
%\begin{displaymath}
W_{\pm}'=\left(\pm\frac{i}{\xi}\sqrt{r}-\frac{1}{4}\frac{r'}{r}\right)W_{\pm}
%\end{displaymath}
$$
\begin{equation}
W_{\pm}''+\left[\frac{r}{\xi^2}+\frac{1}{4}\frac{r''}{r}-\frac{5}{16}\left(\frac{r'}{r}\right)^2\right]W_{\pm}=0
\end{equation}
nombramos a:
\begin{equation}
s(x)=\frac{1}{4}\frac{r''}{r}-\frac{5}{16}\left(\frac{r'}{r}\right)^2
\end{equation}
entonces las $W_{\pm}$ son soluciones exactas de:
\begin{equation}
W_{\pm}''+\left[\frac{1}{\xi^2}r(x)+s(x)\right]W_{\pm}=0
\end{equation}
mientras que s\'olo satisfacen aproximadamente a la ecuaci\'on de 
Schr\"odinger; la cual es regular en $x=x_k$, mientras que la 
ecuaci\'on satisfecha por las funciones WKB asociadas es singular 
en dicho punto.

Procedamos a definir funciones $\alpha_{\pm}(x)$ tales que cumplan con las 
dos relaciones siguientes: 
\begin{equation}
\psi(x)=\alpha_+(x)W_+(x)+\alpha_-(x)W_-(x)
\end{equation}
\begin{equation}
\psi'(x)=\alpha_+(x)W_+'(x)+\alpha_-(x)W_-'(x)
\end{equation}
donde $\psi(x)$ es una soluci\'on a la ecuaci\'on de Schr\"odinger. 
Resolviendo las ecuaciones anteriores para las $\alpha_{\pm}$; 
tenemos:
$$
%\begin{displaymath}
\alpha_+=\frac{\psi W_-'-\psi'W_-}{W_+W_-'-W_+'W_-}\qquad\qquad\alpha_-=-\frac{\psi W_+'-\psi'W_+}{W_+W_-'-W_+'W_-}
%\end{displaymath}
$$
siendo el denominador de \'estas el {\bf Wronskiano} de $W_+$ y $W_-$; no
es dif{\'\i}cil demostrar que \'este toma el valor 
$-\frac{2}{\xi}i$, as{\'\i} que las $\alpha_{\pm}$ se simplifican 
a:
\begin{equation}
\alpha_+=\frac{\xi}{2}i\left(\psi W_-'-\psi'W_-\right)
\end{equation}
\begin{equation}
\alpha_-=\frac{-\xi}{2}i\left(\psi W_+'-\psi'W_+\right)
\end{equation}
Tomando la derivada respecto a $x$ de las ecuaciones $(39)$ y $(40)$, tenemos:
\begin{equation}
\frac{d\alpha_{\pm}}{dx}=\frac{\xi}{2}i\left(\psi'W_{\mp}'+\psi W_{\mp}''-\psi''W_{\mp}-\psi'W_{\mp}'\right)
\end{equation}
dentro del par\'entesis el primer y cuarto t\'ermino se anulan; recordemos que:
$$
%\begin{displaymath}
\psi''+\frac{1}{\xi^2}r(x)\psi=0\quad\&\quad 
W_{\pm}''+\left[\frac{1}{\xi^2}r(x)+s(x)\right]W_{\pm}=0
%\end{displaymath}
$$
podemos escribir la ecuaci\'on $(41)$ como:
$$
%\begin{displaymath}
\frac{d\alpha_{\pm}}{dx}=\frac{\xi}{2}i\left[-\psi\left(\frac{r}{\xi^2}+s\right)W_{\mp}+\frac{r}{\xi^2}\psi W_{\mp}\right]
%\end{displaymath}
$$
\begin{equation}
\frac{d\alpha_{\pm}}{dx}=\mp\frac{\xi}{2}is(x)\psi(x)W_{\mp}(x)
\end{equation}
y en base a las ecuaciones $(33)$ y $(37)$:
\begin{equation}
\frac{d\alpha_{\pm}}{dx}=\mp\frac{\xi}{2}i\frac{s(x)}{\left[r(x)\right]^\frac{1}{2}}\left[\alpha_{\pm}+\alpha_{\mp}\exp\left(\mp\frac{2}{\xi}i\int_{x_k}^x\sqrt{r(x)}dx\right)\right]
\end{equation}

Las ecuaciones $(42)$ y $(43)$ son usadas para %\textbf
{\em estimar el error que se
comete en la aproximaci\'on WKB para un punto particular del espacio 
1-dimensional.} 

La raz\'on de que se considere a $\frac{d\alpha_{\pm}}{dx}$, como una 
estimaci\'on del error que se comete en la aproximaci\'on WKB, es que en las
ecuaciones $(31)$ y $(32)$ las constantes $a$, $b$ y $c$, $d$ 
respectivamente, tan s\'olo nos dan soluciones $\psi$ a\-pro\-xi\-ma\-das; 
mientras que las funciones $\alpha_{\pm}$ al introducirlas en las
ecuaciones $(37)$ y $(38)$, nos proporcionan soluciones $\psi$ exactas; y al 
tomar su derivada obtenemos la pendiente de la recta 
tangente a ellas, y \'esta nos dice cuanto es que se desv\'{\i}an las 
$\alpha_{\pm}$ de las constantes $a$, $b$, $c$ y $d$.
%\end{document}

\bigskip

\noindent
{\bf Nota}: Los art\'{\i}culos WKB originales son:\\

\noindent
G. Wentzel, ``Eine Verallgemeinerung der Wellenmechanik",

\noindent
Zeitschrift f\"ur Physik {\bf 38}, 518-529 (1926) [recibido 18 Junio 1926]\\

\noindent
L. Brillouin, ``La m\'ecanique ondulatoire de Schr\"odinger; une m\'ethode
g\'en\'erale de resolution par approximations successives",

\noindent
Compte Rendue Acad. Sci. (Paris) {\bf 183}, 24-26 (1926) 
[recibido 5 Julio 1926]\\

\noindent
H.A. Kramers, ``Wellenmechanik und halbzahlige Quantisierung",

\noindent
Zf. Physik {\bf 39}, 828-840 (1926) [recibido 9 Sept. 1926]\\

\noindent
H. Jeffreys, ``On certain approx. solutions of linear diff. eqs. of the
second order",

\noindent
Proc. Lond. Math. Soc. {\bf 23}, 428-436 (1925)

\newpage
\centerline{{\bf P r o b l e m a s}}
%%%%%%%%%%%%%%%%%%%%%%%%%%%%%%%%%%%%%%%
{\em Problema 4.1}\\

Veamos un ejemplo de como se usa el m\'etodo WKB en mec\'anica cu\'antica:
Consideremos una part{\'\i}cula de energ{\'\i}a $E$ que se mueve en un 
potencial $u(x)$, la correspondiente ecuaci\'on estacionaria de 
Schr\"odinger es:
\begin{equation}
\frac{d^2\psi}{dx^2}+\frac{2m}{\hbar^2}\left[E-u(x)\right]\psi=0
\end{equation}
y consideremos que $u(x)$ tiene la forma que se muestra en la 
figura 4.1.

%%%%%%%%%%%%%%
\vskip 2ex
\centerline{
\epsfxsize=280pt
\epsfbox{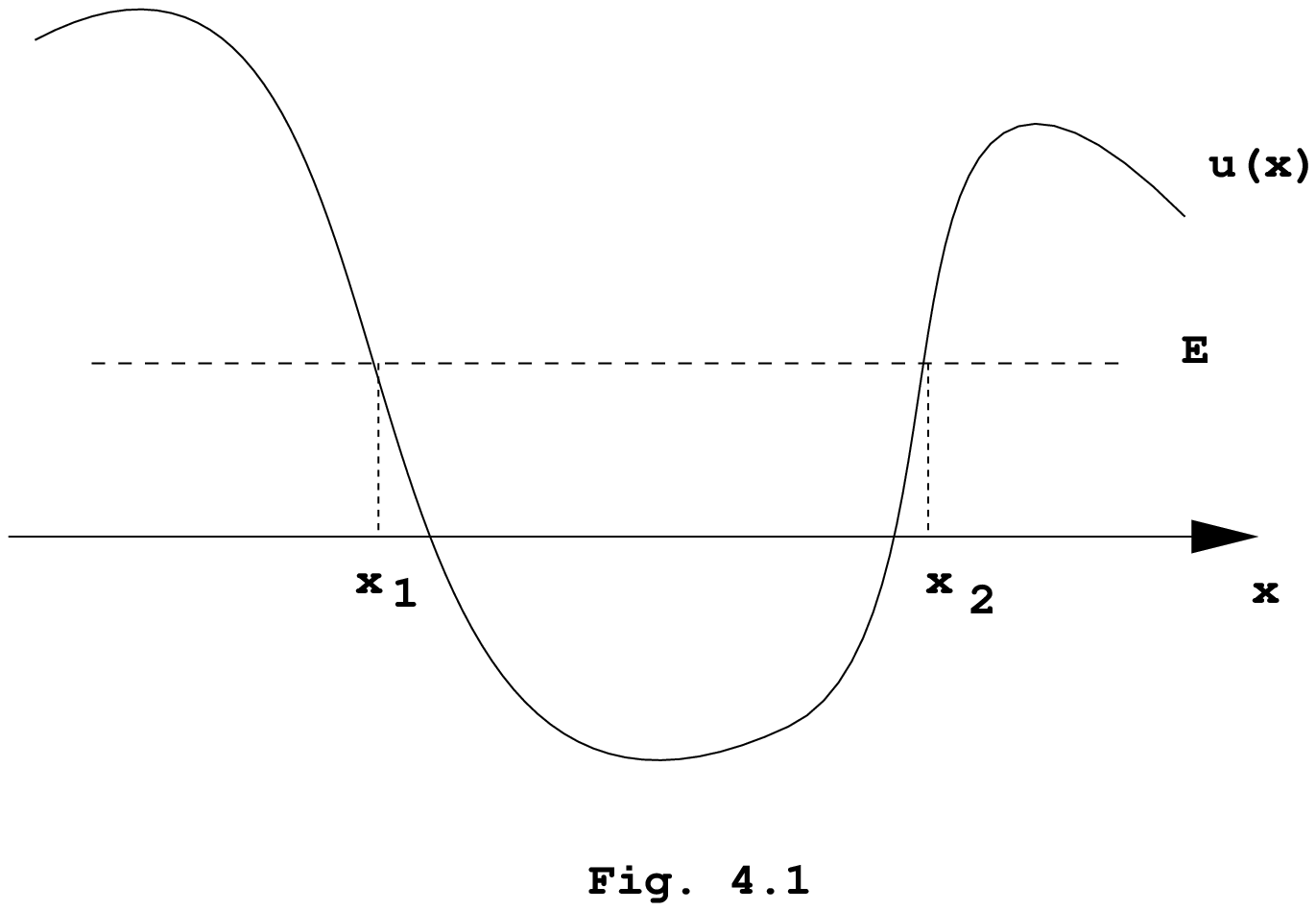}}
\vskip 4ex
%\begin{center}
%{\small{Fig. 1}\\
%}
%\end{center}
%%%%%%%%%%%%%%%%

Como podemos ver:
$$%\begin{displaymath}
r(x)=\frac{2m}{\hbar^2}\left[E-u(x)\right]\qquad
\left\{
\begin{array}{ll}
\mbox{es positiva para $a<x<b$}\\
\mbox{es negativa para $x<a, x>b$}
\end{array}
\right.
$$%\end{displaymath}

Si $\psi(x)$ corresponde a puntos tales que $x<a$, al pasar al
intervalo $a<x<b$, nuestra f\'ormula de conexi\'on es la ecuaci\'on $(29)$ y
nos dice que:
\begin{equation}
\psi(x)\approx\frac{A}{\left[E-u\right]^{\frac{1}{4}}}\cos\left(\int_a^x
\sqrt{\frac{2m}{\hbar^2}(E-u)}dx-\frac{\pi}{4}\right)
\end{equation}
donde $A$ es una constante arbitraria, 

Cuando $\psi(x)$ corresponde a $x>b$, al pasar al intervalo $a<x<b$ 
similarmente: 
\begin{equation}
\psi(x)\approx-\frac{B}{\left[E-u\right]^{\frac{1}{4}}}\cos\left(\int_x^b
\sqrt{\frac{2m}{\hbar^2}(E-u)}dx-\frac{\pi}{4}\right)
\end{equation}
donde $B$ es una constante arbitraria. La raz\'on de que nuestra f\'ormula
de conexi\'on sea nuevamente la ecuaci\'on (29), es que cuando la
part{\'\i}cula llega al segundo punto cl\'asico de retorno, $x=b$,
esta invierte la direcci\'on de su movimiento, y entonces es como si 
hubiera venido de derecha a izquierda; lo que equivale a ver la primer 
situaci\'on de izquierda a derecha en el punto $x=a$, en un espejo.

Estas dos expresiones independientemente de las constantes $A$ y $B$, deben de 
ser las mismas; as\'{\i} que:
$$
%\begin{displaymath}
\cos\left(\int_a^x\sqrt{\frac{2m}{\hbar^2}(E-u)}dx-\frac{\pi}{4}\right)
=-\cos\left(\int_x^b\sqrt{\frac{2m}{\hbar^2}(E-u)}dx-\frac{\pi}{4}\right)\\
$$%\end{displaymath}
\begin{equation}
\Rightarrow\cos\left(\int_a^x\sqrt{\frac{2m}{\hbar^2}(E-u)}dx-\frac{\pi}{4}\right)+\cos\left(\int_x^b\sqrt{\frac{2m}{\hbar^2}(E-u)}dx-\frac{\pi}{4}\right)=0
\end{equation}
Recordando que:
$$
%\begin{displaymath}
\cos A+\cos B= 2\cos\left(\frac{A+B}{2}\right)\cos\left(\frac{A-B}{2}\right)
$$%\end{displaymath}
la ecuaci\'on $(47)$ se escribe:
$$%\begin{displaymath}
2\cos\left[\frac{1}{2}\left(
\int_a^x\sqrt{\frac{2m}{\hbar^2}(E-u)}dx-\frac{\pi}{4}
+\int_x^b\sqrt{\frac{2m}{\hbar^2}(E-u)}dx-\frac{\pi}{4}\right)\right]\cdot
$$%\end{displaymath}
\begin{equation}
\cdot\cos\left[\frac{1}{2}\left(\int_a^x\sqrt{\frac{2m}{\hbar^2}(E-u)}dx
-\frac{\pi}{4}-\int_x^b\sqrt{\frac{2m}{\hbar^2}(E-u)}dx
+\frac{\pi}{4}\right)\right]=0
\end{equation}
lo que implica que los argumentos de estos $cosenos$ sean un m\'ultiplo entero 
de $\frac{\pi}{2}$; el argumento del segundo $coseno$ no nos lleva a alg\'un 
resultado interesante, por lo que s\'olo prestaremos atenci\'on al argumento 
del primer $coseno$, el cual por el contrario si nos lleva a un resultado de 
gran importancia; entonces:
$$%\begin{displaymath}
\frac{1}{2}\left(\int_a^x\sqrt{\frac{2m}{\hbar^2}(E-u)}dx
-\frac{\pi}{4}+\int_x^b\sqrt{\frac{2m}{\hbar^2}(E-u)}dx
-\frac{\pi}{4}\right)=\frac{n}{2}\pi\quad\mbox{para n impar}
$$%\end{displaymath}
$$%\begin{displaymath}
\Rightarrow\quad\quad\int_a^b\sqrt{\frac{2m}{\hbar^2}(E-u)}dx
-\frac{\pi}{2}=n\pi
$$%\end{displaymath}
$$%\begin{displaymath}
\Rightarrow\quad\quad\int_a^b\sqrt{\frac{2m}{\hbar^2}(E-u)}dx=
(n+\frac{1}{2})\pi
$$%\end{displaymath}
\begin{equation}
\Rightarrow\quad\quad\int_a^b\sqrt{2m(E-u)}dx=(n+\frac{1}{2})\pi\hbar
\end{equation}

Este resultado es muy similar a las reglas de cuantizaci\'on de 
%\textbf
{\em Bohr - Sommerfeld}.

Recordemos que el postulado de Bohr establece que el momento angular de un 
electr\'on, que se mueve en una \'orbita permitida en torno a un nucleo 
at\'omico, est\'a cuantizado y su valor es igual a: $L=n\hbar$, 
$n=1,2,3,\dots$. Y recordemos tambi\'en que las reglas de cuantizaci\'on de 
Wilson - Sommerfeld, establecen que toda coordenada de un sistema 
f{\'\i}sico que var{\'\i}e peri\'odicamente en el tiempo deber\'a satisfacer 
la condici\'on cu\'antica: $\oint p_qdq=n_q h$; donde $q$ es una 
coordenada peri\'odica, $p_q$ es el momento asociado con ella, $n_q$ es un 
n\'umero entero y $h$ es la constante de Planck. Y como podemos ver el 
resultado obtenido de la aproximaci\'on WKB es muy similar a estos dos.\\
%\textbf

\bigskip
\newpage

{\em Problema 4.2}\\

Estimemos el error que se comete en la soluci\'on WKB, en un
punto $x_1\neq x_k$, con $x_k$ un punto cl\'asico de retorno; para la
ecuaci\'on diferencial $y''+xy=0$. {\em La soluci\'on de este
problema en f{\'\i}sica, es importante para el estudio de campos uniformes; 
tales como el campo gravitacional \'o el campo el\'ectrico uniforme debido a 
placas planas con carga el\'ectrica.}\\
{\it Soluci\'on:}\\
Para esta ecuaci\'on diferencial tenemos que:
$$%\begin{displaymath}
\xi=1,\qquad r(x)=x\qquad\&\qquad s(x)=-\frac{5}{16}x^{-2}
$$%\end{displaymath}
$r(x)=x$ solamente tiene un cero en $x_k=0$, as{\'\i} que para $x\gg0$: 
\begin{equation} 
W_{\pm}=x^{-\frac{1}{4}}
\exp\left(\pm i\int_0^x\sqrt{x}dx\right)
=x^{-\frac{1}{4}}\exp\left(\pm\frac{2}{3}ix^{\frac{3}{2}}\right)
\end{equation}
Derivando las $W_{\pm}$ una y dos veces respecto a $x$, nos damos cuenta
de que satisfacen la siguiente ecuaci\'on diferencial:
\begin{equation}
W_{\pm}''+(x-\frac{5}{16}x^{-2})W_{\pm}=0
\end{equation}

La soluci\'on exacta $y(x)$ a esta ecuaci\'on diferencial, la escribiremos 
como una combinaci\'on l{\'\i}neal de las $W_{\pm}$, tal y 
como se indic\'o en la secci\'on correspondiente a la 
estimaci\'on de error en la aproximaci\'on WKB; si 
recordamos la combinaci\'on l{\'\i}neal se propuso de la 
forma:
$$%\begin{displaymath}
y(x)=\alpha_+(x)W_+(x)+\alpha_-(x)W_-(x) 
$$%\end{displaymath}

Para $x$ muy grandes una soluci\'on general de nuestra ecuaci\'on 
diferencial, est\'a descrita por la aproximaci\'on WKB como:
\begin{equation}
y(x)=Ax^{-\frac{1}{4}}
\cos\left(\frac{2}{3}x^{\frac{3}{2}}
+\delta\right)\qquad \mbox{cuando} \quad x\rightarrow\infty
\end{equation}
de modo que $\alpha_+\rightarrow\frac{A}{2}e^{i\delta}$ y 
$\alpha_-\rightarrow\frac{A}{2}e^{-i\delta}$ para 
$x\rightarrow\infty$. Deseamos calcular el error en esta 
soluci\'on WKB; el cual es medido por la desviaci\'on de 
$\alpha_+$ y $\alpha_-$ respecto de las constantes $A$. Para esto 
utilizamos la ecuaci\'on:
$$%\begin{displaymath}
\frac{d\alpha_{\pm}}{dx}
=\mp\frac{\xi}{2}i\frac{s(x)}{\sqrt{r(x)}}
\left[\alpha_{\pm}+\alpha_{\mp}
\exp\left(\mp2i\int_{x_k}^x\sqrt{r(x)}dx\right)\right]
$$%\end{displaymath}
y efectuando las sustituciones correspondientes:
\begin{equation}
\frac{d\alpha_{\pm}}{dx}
=\mp\frac{i}{2}\left(-\frac{5}{16}x^{-2}\right)
x^{-\frac{1}{2}}\left[\frac{A}{2}e^{\pm i\delta}
+\frac{A}{2}e^{\mp i\delta}\exp\left(\mp 2i\frac{2}{3}
x^{\frac{3}{2}}\right)\right]
\end{equation}
sabemos que $\Delta\alpha_{\pm}$, representa los cambios que sufren las  
$\alpha_{\pm}$ cuando $x$ va desde $x_1$ hasta $\infty$, y estos cambios se 
calculan mediante: 
$$%\begin{displaymath}
\frac{\Delta\alpha_{\pm}}{A/2}
=\frac{2}{A}\int_{x_1}^\infty\frac{d\alpha_{\pm}}{dx}dx
=\qquad\qquad\qquad\qquad\qquad\qquad\qquad\qquad\quad
$$%\end{displaymath}
\begin{equation}
=\pm i\frac{5}{32}e^{\pm i\delta}\left[\frac{2}{3}x_1^{-\frac{3}{2}}+e^{\mp 
2i\delta}\int_{x_1}^\infty x^{-\frac{5}{2}}\exp\left(\mp 
i\frac{4}{3}x^\frac{3}{2}\right)dx\right] 
\end{equation}
El segundo t\'ermino dentro del par\'entesis es menos importante que el 
primero, esto se debe a que la exponencial compleja, oscila 
entre $1$ y $-1$ y $x^{-\frac{5}{2}}<x^{-\frac{3}{2}}$. de este modo:
\begin{equation}
\frac{\Delta\alpha_{\pm}}{A/2}\approx\pm\frac{5}{48}ie^{\pm 
i\delta}x_1^{-\frac{3}{2}} \end{equation}
y como podemos ver el error que se introduce es realmente peque\~no, esto 
porque igualmente la exponencial compleja oscila entre $-1$ y $1$, y 
$x_1^{-\frac{3}{2}}$ ser\'a tambi\'en peque\~no.\\
%\textbf

{\em Problema 4.3}\\

?`Porqu\'e la ecuaci\'on diferencial que satisfacen las funciones WKB
asociadas, difiere de la ecuaci\'on de Schr\"odinger que es satisfecha por 
las funciones WKB; en la inclusi\'on de la funci\'on $s(x)$, si las 
funciones WKB y las asociadas WKB tienen la misma forma?\\

{\it Justificaci\'on:}\\

Recordemos que en el proceso de obtenci\'on de las soluciones WKB, nos
encontramos con una ecuaci\'on diferencial tipo Riccati; para la cual 
propusimos una soluci\'on en forma de serie de potencias de $-i\xi$, dicha 
serie es $q(x)=\sum_{n=0}^\infty(-i\xi)^nq_n(x)$. Pero recordemos 
tambi\'en que \'esta la aproximamos s\'olo hasta segundo orden, 
por lo que nuestras funciones $\psi^{\pm}(x)$ satisfacen la 
ecuaci\'on de Schr\"odinger %\%textbf
{\em s\'olo aproximadamente}. Por
otra parte se proponen las funciones WKB asociadas $W_{\pm}$, como 
funciones que tienen la misma forma que las funciones 
$\psi^{\pm}$; y para obtener la ecuaci\'on diferencial que \'estas 
satisfacen, simplemente las derivamos; y en consecuencia esta 
ecuaci\'on diferencial es satisfecha exactamente por ellas, y como 
vemos se introduce de manera natural la funci\'on s(x); y hace su 
aparici\'on para indicarnos que tanto se %\textbf
{\em ``desv\'{\i}an''} las
funciones $\psi^{\pm}$ de la soluci\'on exacta a la ecuaci\'on de 
Schr\"odinger 1 - dimensional.
%\end{document}

% References

% R. Landauer, ``Path concepts in HJ theory", AJP {\bf }, 363-367 (1952)

%\end{document}

\newpage
%%%%%%%%%%%%%%%%%%%%%%%%%%%%%%%%%%%%%
%\documentstyle[12pt]{article}
\baselineskip 24.1pt plus 0.2pt minus 0.1pt
\newcommand{\oa}{oscilador arm\'onico\hspace{0.2cm}}
\newcommand{\bc}{\begin{center}}
\newcommand{\ec}{\end{center}}
\newcommand{\ii}{\'{\i}}
\newcommand{\be}{\begin{equation}}
\newcommand{\ee}{\end{equation}}
\newcommand{\dd}{\dagger}
\newcommand{\ad}{a^{\dd}}
\newcommand{\m}{\mid}
%\begin{document}

%\setcounter{equation}
%\baselineskip 24.1pt plus 0.2pt minus 0.1pt
%%%%%%%%%%%%%%%%%%%%%%%%%%%%%%%%%%%%%%%%%%%%%%%%%%%
\section*{5. EL OSCILADOR  ARM\'ONICO}
%\author{ Jos\'e Torres Arenas}
%\date {}
%\maketitle
%\setcounter{equation}
\section*{Soluci\'on de la ecuaci\'on de Schr\"odinger}
\setcounter{equation}{0}
El \oa  es quiz\'a el modelo m\'as usado en la F\'{\i}sica, y su utilidad
va desde los campos de la F\'{\i}sica cl\'asica hasta la Electrodin\'amica
cu\'antica.\\
De la mec\'anica cl\'asica sabemos que muchos potenciales complicados, pueden
ser aproximados en la vecindad de sus puntos de equilibrio por un \oa
\be
V(x) \sim \frac{1}{2}V^{\prime\prime}(a)(x-a)^2
\ee

Esto en el caso unidimensional. Para este caso tenemos que la
funci\'on hamiltoniana cl\'asica, de una part\'{\i}cula con
masa {\em m}, oscilando con frecuencia $\omega$, toma la siguiente forma :
\be
H=\frac{p^2}{2m}+\frac{1}{2}m\omega^2x^2
\ee
y el correspondiente hamiltoniano cu\'antico en el espacio de configuraciones
es :
\be
\hat{H}=\frac{1}{2m}(-i\hbar\frac{d}{dx})^2+\frac{1}{2}m\omega^2x^2
\ee
\be
\hat{H}=-\frac{\hbar^2}{2m}\frac{d^2}{dx^2}+\frac{1}{2}m\omega^2x^2
\ee

Dado que el potencial es independiente del tiempo,
lo que determina las eigenfunciones $\Psi_n$ y sus correspondientes
eigenvalores $E_n$, es la ecuaci\'on de Schr\"odinger independiente del tiempo :
\be
\hat{H}\Psi_n=E_n\Psi_n
\ee

Considerando el hamiltoniano para el \oa , se tiene que la ecuaci\'on de
Schr\"odinger correspondiente es :
\be
\frac{d^2\Psi}{dx^2}+[\frac{2mE}{\hbar^2}
-\frac{m^2\omega^2}{\hbar^2}x^2]\Psi=0
\ee

Hemos suprimido los sub\'{\i}ndices de $E$ y $\Psi$ por comodidad.
Definiremos ahora las siguientes cantidades:
\be
k^2=\frac{2mE}{\hbar^2}
\ee
\be
\lambda=\frac{m\omega}{\hbar}
\ee

Con estas definiciones, nuestra ecuaci\'on de Schr\"odinger es:
\be
\frac{d^2\Psi}{dx^2}+[k^2-\lambda^2x^2]\Psi=0
\ee

A esta \'ultima ecuaci\'on se le
conoce como ``ecuaci\'on diferencial de Weber'' \\
Haremos enseguida la transformaci\'on:
\be
y=\lambda x^2
\ee

En general, en un cambio de variable, suponiendo que hacemos el cambio
de la variable $x$ a la variable $y$ , se tiene que los operadores
diferenciales toman la forma siguiente:
\be
\frac{d}{dx}=\frac{dy}{dx}\frac{d}{dy}
\ee
\be
\frac{d^2}{dx^2}=\frac{d}{dx}(\frac{dy}{dx}\frac{d}{dy})
=\frac{d^2y}{dx^2}\frac{d}{dy}+(\frac{dy}{dx})^2\frac{d^2}{dy^2}
\ee

Aplicando esto a la transformaci\'on propuesta obtenemos la siguiente
ecuaci\'on diferencial en la variable $y$ :
\be
y\frac{d^2\Psi}{dy^2}+\frac{1}{2}\frac{d\Psi}{dy}+[\frac{k^2}{4\lambda}-\frac{1}{4}y]\Psi=0
\ee

o bien, definiendo :
\be
\kappa=\frac{k^2}{2\lambda}=\frac{\bar k^2}{2m\omega}=\frac{E}{\hbar\omega}
\ee

Obtenemos entonces la ecuaci\'on:
\be
y\frac{d^2\Psi}{dy^2}+\frac{1}{2}\frac{d\Psi}{dy}
+[\frac{\kappa}{2}-\frac{1}{4}y]\Psi=0
\ee

Pasaremos enseguida a resolver esta ecuaci\'on, haciendo
primeramente el an\'alisis asint\'otico en el
l\'{\i}mite $y\rightarrow\infty$ , para hacer esto reescribimos la
ecuaci\'on anterior  como sigue :
\be
\frac{d^2\Psi}{dy^2}+\frac{1}{2y}\frac{d\Psi}{dy}
+[\frac{\kappa}{2y}-\frac{1}{4}]\Psi=0
\ee

Observamos que , en el
l\'{\i}mite $y\rightarrow\infty$, esta ecuaci\'on  se comporta as\'{\i}:
\be
\frac{d^2\Psi_{\infty}}{dy^2}-\frac{1}{4}\Psi_{\infty}=0
\ee

Esta ecuaci\'on tiene como soluci\'on:
\be
\Psi_{\infty}(y)=A\exp{\frac{y}{2}}+B\exp{\frac{-y}{2}}
\ee

Desechamos $\exp{\frac{y}{2}}$ dado que \'esta
diverge en el l\'{\i}mite $y\rightarrow\infty$, nos
quedamos entonces con la exponencial decreciente. Podemos sugerir
entonces que $\Psi$ tiene la siguiente forma:
\be
\Psi(y)=\exp{\frac{-y}{2}}\psi(y)
\ee

Sustituyendo esto en la ecuaci\'on diferencial para $y$ ( ec. $15$) se tiene:
\be
y\frac{d^2\psi}{dy^2}
+(\frac{1}{2}-y)\frac{d\psi}{dy}+(\frac{\kappa}{2}-\frac{1}{4})\psi=0
\ee

Lo que hemos obtenido es la ecuaci\'on hipergeom\'etrica confluente
\footnote{tambi\'en conocida como la ecuaci\'on diferencial de Kummer} :
\be
z\frac{d^2y}{dz^2}+(c-z)\frac{dy}{dz}-ay=0
\ee

La soluci\'on general a esta ecuaci\'on es :
\be
y(z)=A \hspace{.2cm} _1F_1(a;c,z)+
B \hspace{.2cm} z^{1-c} \hspace{.1cm}  _1F_1(a-c+1;2-c,z)
\ee

Con la funci\'on hipergeom\'etrica confluente :
\be
_1F_1(a;c,z)=\sum_{n=0}^{\infty}\frac{(a)_n x^n}{(c)_n n!}
\ee

Comparando entonces nuestra ecuaci\'on , con la ecuaci\'on hipergeom\'etrica 
confluente , se observa que la soluci\'on general a nuestra ecuaci\'on es :
\be
\psi(y)=A\hspace{.2cm} _1F_1(a;\frac{1}{2},y)+
B \hspace{.2cm} y^{\frac{1}{2}}
\hspace{.2cm} _1F_1(a+\frac{1}{2};\frac{3}{2},y)
\ee

donde
\be
a=-(\frac{\kappa}{2}-\frac{1}{4})
\ee

Si dejamos estas soluciones as\'{\i} como est\'an, entonces la condici\'on de
normalizaci\'on de la funci\'on de onda no se cumple, pues del comportamiento 
asint\'otico de la funci\'on hipergeom\'etrica 
confluente \footnote{ El comportamiento asint\'otico
para $\mid x \mid\rightarrow \infty$ es:
\bc
$_1F_1(a;c,z)\rightarrow
\frac{\Gamma(c)}{\Gamma(c-a)}e^{-ia\pi}x^{-a}
+\frac{\Gamma(c)}{\Gamma(a)}e^{x}x^{a-c}$
\ec
} se sigue que
( considerando \'unicamente el comportamiento exponencial, dado 
que es el dominante) :
\be
\Psi(y)=e^{\frac{-y}{2}}\psi(y)\rightarrow
\hspace{.3cm}const. \hspace{.2cm} e^{\frac{y}{2}}y^{a-\frac{1}{2}}
\ee

Esto  nos lleva a una divergencia en la integral de normalizaci\'on, lo cual es
f\'{\i}sicamente inaceptable. Lo que se hace entonces, es imponer la 
condici\'on de terminaci\'on de la serie \footnote{La condici\'on de 
truncamiento de la serie para la funci\'on hipergeom\'etrica 
confluente $_1F_1(a;c,z)$ es $a=-n$, con $n$ un entero 
no negativo ( esto es, incluye el cero ).} , esto es, la serie se corta y surge 
entonces un polinomio de grado $n$.\\
{\em Observamos entonces que, el hecho de pedir que la integral de 
normalizaci\'on sea finita (para tener significado f\'{\i}sico en t\'erminos 
de probabilidades), nos lleva al truncamiento de la serie, y esto a su vez es 
lo que da lugar a la cuantizaci\'on de la energ\'{\i}a.}\\
Consideremos enseguida los dos posibles casos :

$1)\hspace{.4cm} a=-n \hspace{.3cm}$ y $ B=0$
\be
\frac{\kappa}{2}-\frac{1}{4}=n
\ee

Con las eigenfunciones:
\be
\Psi_n(x)=D_n \exp{\frac{-\lambda x^2}{2}}
\hspace{.1cm} _1F_1(-n;\frac{1}{2},\lambda x^2)
\ee

y la energ\ii a:
\be
E_n=\hbar\omega(2n+\frac{1}{2})
\ee

$2)\hspace{.4cm} a+\frac{1}{2}=-n \hspace{.3cm}$ y $A=0$
\be
\frac{\kappa}{2}-\frac{1}{4}=n+\frac{1}{2}
\ee

Teniendo las eigenfunciones siguientes:
\be
\Psi_n(x)=D_n \exp{\frac{-\lambda x^2}{2}}
\hspace{.2cm}x \hspace{.2cm}_1F_1(-n;\frac{3}{2},\lambda x^2)
\ee

y para la energ\ii a se tiene:
\be
E_n=\hbar\omega[(2n+1)+\frac{1}{2}]
\ee

Los polinomios dados anteriormente por las funciones hipergeom\'etricas,
son conocidos como polinomios de Hermite, y est\'an definidos en
t\'erminos de la funci\'on hipergeom\'etrica como sigue :
\be
H_{2n}(\eta)=(-1)^n \frac{(2n)!}{n!}
\hspace{.2cm} _1F_1(-n;\frac{1}{2},\eta^2)
\ee
\be
H_{2n-1}(\eta)=(-1)^n \frac{2(2n+1)!}{n!}
\hspace{.2cm}\eta \hspace{.2cm} _1F_1(-n;\frac{3}{2},\eta^2)
\ee

Podemos finalmente combinar los resultados
obtenidos ( pues unos nos dan los valores pares y otros los
impares ) en una sola expresi\'on para los eigenvalores y las
eigenfunciones , obteni\'endose :
\be
\Psi_n (x)=D_n \exp{ \frac{-\lambda x^2}{2}} H_n (\sqrt{\lambda}x)
\ee
\be
E_n =(n+\frac{1}{2})\hbar\omega \hspace{1cm}n=0,1,2\ldots
\ee

El espectro de energ\ii a
del \oa es equidistante, esto es, existe la misma
diferencia $\hbar \omega$ entre cualesquiera dos estados . Otra
observaci\'on que podemos hacer, es acerca
del m\ii nimo valor de energ\ii a que toma el oscilador; lo sorprendente es
que \'este es distinto de cero; esto es un resultado puramente
mec\'anico cu\'antico, a este valor se le conoce
como {\em energ\ii a de punto cero} y el hecho de que sea distinta de
cero , es una caracter\ii stica de todos los potenciales
ligantes (aquellos que confinan a las part\ii culas) .\\

La constante de normalizaci\'on puede ser calculada, y tiene el valor:
\be
D_n = ( \sqrt{\frac{\lambda}{\pi}}\frac{1}{2^n n!})^{\frac{1}{2}}
\ee

Con lo cual obtenemos finalmente las eigenfunciones del \oa unidimensional, normalizadas :
\be
\Psi_n (x)= ( \sqrt{\frac{\lambda}{\pi}}\frac{1}{2^n n!})^{\frac{1}{2}}
\hspace{.2cm} \exp{ \frac{-\lambda x^2}{2}}
\hspace{.2cm} H_n( \sqrt{\lambda} x)
\ee

%\newpage

\section*{Operadores de creaci\'on ($\hat{a}^{\dagger}$) y
aniquilaci\'on ($\hat{a}$)}

Existe otra forma de tratar el \oa de una forma distinta a la
convencional de resolver la ecuaci\'on de Schr\"odinger, esta otra manera es
llamada el m\'etodo algebraico o m\'etodo de operadores, \'este es un
poderoso m\'etodo el cual es aplicado tambi\'en en otra clase de problemas
mec\'anico cu\'anticos.\\
Definiremos dos operadores no hermitianos $a$ y $a^{\dd}$ :
\be
a=\sqrt{\frac{m\omega}{2\hbar}}(x+\frac{ip}{m\omega})
\ee
\be
a^{\dd}=\sqrt{\frac{m\omega}{2\hbar}}(x-\frac{ip}{m\omega})
\ee

Estos operadores son conocidos
como \hspace{.1cm} {\em operador de aniquilaci\'on}
\hspace{.1cm}  y \hspace{.1cm} {\em operador de creaci\'on},
\hspace{.1cm}  respectivamente  (las razones para estos nombres se
ver\'an despu\'es ).\\
Vamos a calcular ahora el conmutador de estos dos operadores:
\be
[a,a^{\dd}]=\frac{m\omega}{2\hbar}[x
+\frac{ip}{m\omega},x-\frac{ip}{m\omega}]=\frac{1}{2\hbar}(-i[x,p]+i[p,x])=1
\ee

Donde hemos usado el conmutador:
\be
[x,p]=i\hbar
\ee

Esto es, tenemos que los operadores de creaci\'on y aniquilaci\'on satisfacen la relaci\'on de conmutaci\'on :
\be
[a,a^{\dd}]=1
\ee

Vamos a definir tambi\'en el llamado operador de n\'umero $\hat{N}$ como:
\be
\hat{N}=\ad a
\ee

Este\hspace{.1cm} operador es
hermitiano\hspace{.1cm} como lo podemos demostrar
f\'acilmente\hspace{.1cm} usando $(AB)^{\dd}=B^{\dd}A^{\dd}$ :
\be
\hat{N}^{\dd}=(\ad a)^{\dd}=\ad (\ad)^{\dd}=\ad a=\hat{N}
\ee

Ahora bien, considerando que :
\be
\ad a =\frac{m\omega}{2\hbar}(x^2+\frac{p^2}{m^2\omega^2})+\frac{i}{2\hbar}[x,p]=\frac{\hat{H}}{\hbar\omega}-\frac{1}{2}
\ee

observamos que el hamiltoniano est\'a dado de una manera
simple en t\'erminos del  operador de n\'umero :
\be
\hat{H}=\hbar\omega(\hat{N}+\frac{1}{2})
\ee

El operador de n\'umero recibe su nombre debido a que sus eigenvalores
son justo el \ii ndice de la funci\'on de onda sobre la que opera, esto es:
\be
\hat{N}\m n>=n\m n>
\ee

Donde hemos usado la notaci\'on:
\be
\m \Psi_n> \hspace{.2cm}= \hspace{.2cm}\m n>
\ee

Aplicando este hecho a $(47)$ tenemos :
\be
\hat{H}\m n>=\hbar\omega(n+\frac{1}{2})\m n>
\ee

Pero sabemos de la ecuaci\'on de Schr\"odinger que $\hat{H}\m n>=E\m n>$ de
lo cual se sigue que los valores de la energ\ii a est\'an dador por :
\be
E_n=\hbar\omega(n+\frac{1}{2})
\ee

El cual es id\'entico (como deb\ii a ser ) con el resultado $(36)$.\\
Vamos enseguida a mostrar el por qu\'e de los nombres que se le dan a
los operadores $a$ y $\ad$ . Para hacer esto comenzaremos calculando dos
conmutadores:
\be
[\hat{N},a]=[\ad a,a]=\ad[a,a]+[\ad,a]a=-a
\ee

Lo anterior se sigue de $[a,a]=0$ y $(43)$.Similarmente calculamos:
\be
[\hat{N},\ad]=[\ad a,\ad]=\ad[a,\ad]+[\ad,\ad]a=\ad
\ee
Con estos dos conmutadores podemos escribir:
\begin{eqnarray}
\hat{N}(\ad \m n>)&=&([\hat{N},\ad]+\ad\hat{N})\m n>\nonumber\\
&=&(\ad+\ad\hat{N})\m n>\\
&=&\ad(1+n)\m n>=(n+1)\ad\m n>\nonumber
\end{eqnarray}

Con un procedimiento similar se obtiene tambi\'en:
\be
\hat{N}(a\m n>)=([\hat{N},a]+a\hat{N})\m n>=(n-1)a\m n>
\ee
La expresi\'on $(54)$ implica que el
ket $\ad \m n>$ es un eigenket del operador de n\'umero , donde el
eigenvalor se ha incrementado por uno, esto es , se ha
creado un cuanto de energ\ii a al
actuar $\ad$ sobre el ket, de ah\ii \hspace{.1cm} su nombre de operador de
creaci\'on.\hspace{.3cm} Comentarios\hspace{.1cm} siguiendo\hspace{.1cm}
la\hspace{.1cm} misma\hspace{.1cm} l\ii nea\hspace{.1cm} de\hspace{.1cm} 
razonamiento son v\'alidos para el operador $a$, lo cual le da el nombre de
operador de\hspace{.1cm}
aniquilaci\'on ( un  cuanto\hspace{.1cm} de
energ\ii a  es\hspace{.1cm} disminu\ii do\hspace{.1cm}
al\hspace{.1cm} actuar\hspace{.1cm} este\hspace{.1cm} 
operador ).\\
La ecuaci\'on $(54)$ tambi\'en implica que el
ket $\ad\m n>$ y el ket $\m n+1>$ son proporcionales, podemos escribir
esta relaci\'on as\ii :
\be
\ad\m n>=c\m n+1>
\ee

Donde $c$ es una constante que hay que determinar. Considerando adem\'as que :
\be
(\ad\m n>)^{\dd}=<n\m a=c^*<n+1\m
\ee

Podemos entonces realizar el siguiente c\'alculo:
\be
<n\m a( \ad\m n>)=c^*<n+1\m (c\m n+1>)
\ee
\be
<n\m a\ad \m n>=c^*c<n+1\m n+1>
\ee
\be
<n\m a\ad \m n>=\m c\m^2
\ee

Pero de la relaci\'on de conmutaci\'on para los operadores $a$ y $\ad$ :
\be
[a,\ad]=a\ad-\ad a=a\ad-\hat{N}=1
\ee

Esto es :
\be
a\ad=\hat{N}+1
\ee

Sustituyendo en $(60)$:
\be
<n\m \hat{N}+1\m n>=<n\m n>+<n\m \hat{N}\m n>=n+1=\m c\m^2
\ee

Pidiendo que $c$ sea real y positiva ( por convenci\'on ), obtenemos su valor:
\be
c=\sqrt{n+1}
\ee

Con lo cual se tiene la relaci\'on:
\be
\ad \m n>=\sqrt{n+1}\m n+1>
\ee

Siguiendo el mismo camino se llega a una relaci\'on para el operador de 
aniquilaci\'on :
\be
a\m n>=\sqrt{n}\m n-1>
\ee

Vamos ahora a mostrar que los valores de $n$ deben ser enteros no negativos. 
Para esto, acudiremos al requerimiento de positividad de la  norma, aplicado 
en especial al vector de estado $a\m n>$. Este requerimiento nos dice que el 
producto interior de este vector con su adjunto($ (a\m n>)^\dd=<n\m \ad$) 
debe ser mayor  o igual que cero :
\be
( <n\m \ad)\cdot(a\m n>)\geq 0
\ee

Pero lo anterior no es m\'as que :
\be
<n\m \ad a\m n>=<n\m \hat{N}\m n>=n \geq 0
\ee

Por lo tanto $n$ nunca puede ser negativo. Y tiene que ser entero pues si no 
lo fuera al aplicar en repetidas ocasiones el operador de aniquilaci\'on nos 
llevar\ii a a valores negativos de $n$, lo cual est\'a en contraposici\'on con 
lo anterior.\\
Es posible expresar el estado $n$ $(\m n>)$ en t\'erminos del estado 
base $(\m 0>)$ usando el operador de creaci\'on, veamos como hacerlo:

\begin{eqnarray}
 \m 1>=\ad \m 0> \\
 \m 2>=[\frac{\ad}{\sqrt{2}}]\m 1>=[\frac{(\ad)^2}{\sqrt{2!}}]\m 0> \\
 \m 3>=[\frac{\ad}{\sqrt{3}}]\m 2>=[ \frac{ (\ad)^3}{\sqrt{3!}}]\m 0> 
\end{eqnarray}
%\begin{center}
\vdots
%\end{center}
\begin{eqnarray}
 \m n>=[ \frac{ (\ad)^n}{\sqrt{n!}}]\m 0>
\end{eqnarray}

Podemos tambi\'en aplicar este m\'etodo para encontar las eigenfunciones en el espacio de configuraciones. Para hacer esto, partiremos del estado base:
\be
a\m 0>=0
\ee
En la representaci\'on $x$ tenemos:
\be
\hat{ a}  \Psi_0(x)=\sqrt{\frac{m\omega}{2\hbar}} (x+\frac{ip}{m\omega}) \Psi_0(x)=0
\ee
Recordando la forma que toma el operador momento en la representaci\'on $x$, podemos llegar a una ecuaci\'on diferencial para la funci\'on de onda del estado base; introduciremos tambi\'en la definici\'on siguiente $x_0=\sqrt{\frac{\hbar}{m\omega}}$ , con
 esto :
\be
(x+x_0^2\frac{d}{dx})\Psi_0=0
\ee
Esta ecuaci\'on se puede resolver f\'acilmente, resolvi\'endola y normaliz\'andola ( su integral de $-\infty$ a $\infty$ debe ser la unidad ), llegamos a la funci\'on de onda del estado base:
\be
\Psi_0(x)=(\frac{1}{\sqrt{ \sqrt{\pi}x_0}})e^{ -\frac{1}{2}(\frac{x}{x_0})^2}
\ee
Las dem\'as eigenfunciones, esto es, las eigenfunciones para los estados excitados del \oa , se pueden obtener usando el operador de creaci\'on, el procedimiento es el siguiente:
\begin{eqnarray}
\Psi_1=\ad \Psi_0 =(\frac{1}{\sqrt{2}x_0})(x-x_0^2\frac{d}{dx})\Psi_0\\
\Psi_2=\frac{1}{\sqrt{2}}(\ad)^2\Psi_0=\frac{1}{\sqrt{2!}}(\frac{1}{\sqrt{2}x_0})^2(x-x_0^2\frac{d}{dx})^2\Psi_0
\end{eqnarray}
Siguiendo con este procedimiento, por inducci\'on se puede mostrar que:
\be
\Psi_n=\frac{1}{\sqrt{ \sqrt{\pi}2^nn!}}\hspace{.2cm}
\frac{1}{x_0^{n+\frac{1}{2}}}
\hspace{.2cm}(x-x_0^2\frac{d}{dx})^n
\hspace{.2cm}e^{-\frac{1}{2}(\frac{x}{x_0})^2}
\ee

%\newpage
%\\

\section*{Evoluci\'on temporal del oscilador}

En esta secci\'on vamos a ilustrar con  el \oa una manera en la cual se trabaja 
con la representaci\'on de Heisenberg, esto es, dejaremos que los estados 
est\'en fijos en el tiempo y haremos evolucionar a los operadores en \'el. 
Veremos a los operadores como funciones del tiempo, espec\ii ficamente, 
encontraremos como es que 
evolucionan los operadores posici\'on , momento, $a$ y $\ad$ en el tiempo, 
para el caso del \oa.
Las ecuaciones de movimiento de Heisenberg para $p$ y $x$ son :
\begin{eqnarray}
\frac{d\hat{p}}{dt}&=&-\frac{\partial}{\partial\hat{x}}V({\bf \hat{x})}\\
\nonumber\\
\frac{d\hat{x}}{dt}&=&\frac{\hat{p}}{m}
\end{eqnarray}

De aqu\ii \hspace{.1cm}  se sigue que las ecuaciones de movimiento para $x$ y $p$ en el caso del \oa son:
\begin{eqnarray}
\frac{d\hat{p}}{dt}&=&-m\omega^2\hat{x}\\
\nonumber\\
\frac{d\hat{x}}{dt}&=&\frac{\hat{p}}{m}
\end{eqnarray}

Se tiene un par de ecuaciones acopladas , estas son equivalentes a un par de ecuaciones para los operadores de creaci\'on y aniquilaci\'on, salvo que estas dos \'ultimas no est\'an acopladas, ve\'amoslas expl\ii citamente :
\begin{eqnarray}
\frac{d a}{dt}&=&\sqrt{\frac{m\omega}{2\hbar}}\frac{d}{dt}(\hat{x}+\frac{i\hat{p}}{m\omega})\\
\nonumber\\
\frac{da}{dt}&=&\sqrt{\frac{m\omega}{2\hbar}}(\frac{d\hat{x}}{dt}+\frac{i}{m\omega}\frac{d\hat{p}}{dt})
\end{eqnarray}
 
Sustituyendo $(82)$ y $(83)$ en $(85)$ :
\be
\frac{da}{dt}=\sqrt{\frac{m\omega}{2\hbar}}(\frac{\hat{p}}{m}-i\omega\hat{x})=-i\omega a
\ee
Similarmente se puede obtener una ecuaci\'on diferencial para el operador de creaci\'on, la cual no est\'a acoplada:
\be
\frac{d\ad}{dt}=i\omega\ad
\ee
Las ecuaciones diferenciales que hemos encontrado para la evoluci\'on temporal de los operadores de creaci\'on y aniquilaci\'on , pueden ser integradas inmediatamente, d\'andonos la evoluci\'on expl\ii cita de estos operadores en el tiempo :
\begin{eqnarray}
a(t)&=&a(0)e^{-i\omega t}\\
\ad (t)&=&\ad (0)e^{i\omega t}
\end{eqnarray}

Podemos observar de estos resultados y de las ecuaciones $(44)$ y $(47)$ , que tanto el hamiltoniano como el operador de n\'umero , no dependen del tiempo, tal y como podr\ii amos esperar.\\
Con los dos resultados anteriores , podemos encontrar los operadores de posici\'on y momento como funci\'on del tiempo, pues ellos est\'an dados en t\'erminos de los operadores de creaci\'on y aniquilaci\'on:
\begin{eqnarray}
\hat{x}&=&\sqrt{\frac{\hbar}{2m\omega}}(a+\ad)\\
\hat{p}&=&i\sqrt{ \frac{m\hbar\omega}{2}}(\ad-a)
\end{eqnarray}

Sustituyendo los operadores de creaci\'on y aniquilaci\'on se obtiene:
\begin{eqnarray}
\hat{x}(t)&=&\hat{x}(0)\cos{\omega t}+\frac{\hat{p}(0)}{m\omega}\sin{\omega t}\\
\nonumber\\
\hat{p}(t)&=&-m\omega\hat{x}(0)\sin{\omega t}+\hat{p}(0)\cos{\omega t}
\end{eqnarray}

La evoluci\'on temporal de los operadores de posici\'on y momento es la
misma que las ecuaciones cl\'asicas de movimiento.\\
Hemos finalizado  esta secci\'on, mostrando la forma expl\ii cita en 
que evolucionan cuatro operadores en el caso del \oa , reflejando de 
esta manera una forma de trabajar en la poco mencionada representaci\'on 
de Heisenberg.

%\newpage

\section*{El \oa tridimensional}

Al iniciar nuestro estudio  cu\'antico del \oa, hac\ii amos comentarios
acerca del por qu\'e la importacia del \oa . Si hici\'eramos un an\'alogo
tridimensional, considerar\ii amos entonces un desarrollo de Taylor en tres
variables\footnote{Es posible expresar el desarrollo de Taylor como un
operador exponencial como sigue:
\bc
$e^{[ (x-x_o)+(y-y_o)+(z-z_o)]
(\frac{\partial}{\partial x}+\frac{\partial}{\partial y}+
\frac{\partial}{\partial z})} \hspace{.1cm} f({\bf r_o})$\\
Esto es el desarrollo de Taylor en tres variables alrededor de ${\bf r_o}$.
\ec }
reteniendo t\'erminos s\'olo hasta segundo orden, lo que tenemos es una
forma cuadr\'atica (en el caso mas general), el problema de resolver para
esta aproximaci\'on no es sencillo, es decir, para el caso :
%\bc

\be
V(x,y,z)=ax^2+by^2+cz^2+dxy+exz+fyz
\ee

%\ec

Afortunadamente hay varios sistemas que se ajustan bien a la
simetr\ii a esf\'erica, esto es, para el caso:
%\bc

\be
V(x,y,z)=K(x^2+y^2+z^2)
\ee

%\ec
Esto \'ultimo, equivale a decir que las parciales
segundas ( no cruzadas ) toman todas el mismo valor ( en el caso
anterior representado por $K$), y podr\ii amos agregar que esta es una
buena aproximaci\'on en el caso en  que los valores de las parciales 
cruzadas sean peque\~nas comparadas con las parciales segundas no cruzadas.\\
Cuando se satisfacen los requerimientos anteriores, y tenemos un potencial
como el dado por $(95)$, entonces tenemos el
denominado {\em \oa tridimensional esf\'ericamente sim\'etrico}.\\
El hamiltoniano para este caso es de la forma:
%\bc

\be
\hat{H}=\frac{-\hbar^2}{2m}\bigtriangledown^2 + \frac{m\omega^2}{2}r^2
\ee

%\ec

Donde el laplaciano est\'a dado en coordenadas esf\'ericas y $r$ es la 
variable esf\'erica convencional.\\
Tenemos entonces que el potencial es independiente del tiempo, por tanto la 
energ\ii a se va a conservar; adem\'as dada la simetr\ii a esf\'erica , el 
momento angular se conservar\'a tambi\'en, se tienen por tanto dos cantidades 
conservadas, y puesto que 
a cada cantidad conservada le corresponde un n\'umero cu\'antico, podemos 
adelantar que nuestras funciones de onda depender\'an de dos n\'umeros 
cu\'anticos (aunque en este caso, como veremos surge otro ). Esto 
es , necesitamos solucionar la ecuaci\'on :
%\bc

\be
\hat{H}\Psi_{nl}=E_{nl}\Psi_{nl}
\ee

%\ec

El laplaciano en coordenadas esf\'ericas es :
%\bc

\be
\bigtriangledown^2
=\frac{\partial^2}{\partial r^2}+\frac{2}{r}\frac{\partial}{\partial r}
-\frac{\hat{L}^2}{\hbar^2r^2}
\ee

%\ec
Esto se sigue del hecho que :
%\bc

\be
\hat{L}^2=-\hbar^2[ \frac{1}{\sin{\theta}}
\frac{\partial}{\partial\theta}
( \sin{\theta}\frac{\partial}{\partial\theta})
+\frac{1}{\sin{\theta}^2}\frac{\partial^2}{\partial\varphi^2}]
\ee

%\ec

{\em Las eigenfunciones de $\hat{L}^2$ son los arm\'onicos esf\'ericos}, se
tiene:
%\bc

\be
\hat{L}^2Y_{lm}(\theta,\varphi)=-\hbar^2l(l+1)Y_{lm}(\theta,\varphi)
\ee

%\ec
Podemos observar que el hecho de que los arm\'onicos esf\'ericos
lleven el n\'umero cu\'antico $m$ , produce la intromisi\'on del mismo en
la funci\'on de onda, es decir tendremos $\Psi_{nlm}$.\\
Para separar la ecuaci\'on diferencial se propone la sustituci\'on:
%\bc

\be
\Psi_{nlm}(r, \theta,\varphi)=\frac{R_{nl}(r)}{r}Y_{lm}(\theta,\varphi)
\ee

%\ec
Esto al sustituirlo en la ecuaci\'on de Schr\"odinger, nos va a separar la
parte espacial de la parte angular; la parte angular , son las eigenfunciones
del operador momento angular ( al cuadrado ), en la parte espacial
llegamos a la ecuaci\'on :
%\bc

\be
R_{nl}^{\prime\prime}+(\frac{2mE_{nl}}{\hbar^2}
-\frac{m^2\omega^2}{\hbar^2}r^2-\frac{l(l+1)}{r^2})R_{nl}(r)=0
\ee

%\ec

Usando las definiciones $(7)$ y $(8)$ , la ecuaci\'on anterior toma
exactamente la misma forma que $(9)$, excepto por el t\'ermino del momento
angular, \'este t\'ermino usualmente es
llamado {\em la barrera de momento angular}.
%\bc

\be
R_{nl}^{\prime\prime}+(k^2-\lambda^2r^2-\frac{l(l+1)}{r^2})R_{nl}=0
\ee

%\ec

Para resolver esta ecuaci\'on ,
partiremos del an\'alisis asint\'otico de la misma. Si
consideramos primero el l\'{\i}m $r\rightarrow\infty$, observamos que
el t\'ermino del momento angular es despreciable, de manera que el
comportamiento asint\'otico en este l\ii mite es id\'entico a $(9)$ con lo
cual obtenemos:
%\bc

\be
R_{nl}(r)\sim\exp{\frac{-\lambda r^2}{2}}\hspace{2cm}\mbox{en}
\hspace{.3cm}\lim\hspace{.1cm}r\rightarrow\infty
\ee

%\ec

Si observamos ahora el comportamiento cerca de cero, vemos que el
comportamiento dominante est\'a dado por el t\'ermino de momento angular,
es decir, la ecuaci\'on
diferencial $(102)$ se convierte en este l\ii mite en :
%\bc

\be
R_{nl}^{\prime\prime}-\frac{l(l+1)}{r^2}R_{nl}=0
\ee

%\ec

Esta es una ecuaci\'on diferencial tipo
Euler \footnote{Una ecuaci\'on tipo Euler es :
%\bc

\[x^n y^{(n)}(x)+x^{n-1} y^{(n-1)}(x)+\cdots+x y^{\prime}(x)+y(x)=0\]

%\ec
La cual tiene soluciones del tipo $x^{\alpha}$, se sustituye y se encuentra un 
polinomio para $\alpha$.} , solucion\'andola encontramos dos soluciones 
independientes:
%\bc
\be
R_{nl}(r)\sim \hspace{.2cm}r^{l+1}\hspace{.2cm}\mbox{o}
\hspace{.4cm}r^{-l}\hspace{2cm}\mbox{en}\hspace{.4cm}\lim\hspace{.1cm}r
\rightarrow 0
\ee
%\ec

Lo anterior nos lleva a proponer la sustituci\'on :
%\bc
\be
R_{nl}(r)=r^{l+1}\exp{\frac{-\lambda r^2}{2}}\phi(r)
\ee
%\ec

Podr\ii amos hacer tambi\'en la sustituci\'on:
%\bc
\be
R_{nl}(r)=r^{-l}\exp{\frac{-\lambda r^2}{2}}v(r)
\ee
%\ec

Sin embargo esto lleva a las mismas soluciones
que $(107)$ ( mostrar esto es un buen ejercicio).
Sustituyendo $(107)$ en $(103)$ , se obtiene una
ecuaci\'on diferencial para $\phi$ :
%\bc
\be
\phi^{\prime\prime}+2(\frac{l+1}{r}-\lambda r)\phi^{\prime}
-[ \lambda (2l+3)-k^2]\phi=0
\ee
%\ec

Haciendo ahora la sustituci\'on de la variable $w=\lambda r^2$, obtenemos:
%\bc
\be
w\phi^{\prime\prime}+(l+\frac{3}{2}-w)\phi^{\prime}-[ \frac{1}{2}(l+\frac{3}{2})-\frac{\kappa}{2}]\phi=0
\ee
%\ec
Donde hemos introducido $\kappa =\frac {k^2}{2\lambda}=\frac{E}{\hbar\omega}$. Tenemos nuevamente una ecuaci\'on diferencial tipo hipergeom\'etrica confluente la cual tiene por soluciones ( v\'ease $(21)$ y $(22)$):
%\bc
\be
\phi(r)=A\hspace{.2cm}_1F_1[\frac{1}{2}(l+\frac{3}{2}-\kappa);l+\frac{3}{2},\lambda r^2]+B\hspace{.2cm}r^{-(2l+1)}\hspace{.3cm}_1F_1[\frac{1}{2}(-l+\frac{1}{2}-\kappa);-l+\frac{1}{2},\lambda r^2]
\ee
%\ec
 
La segunda soluci\'on particular no puede ser normalizada , pues diverge 
fuertemente en cero, de manera que tomamos $B=0$ y se tiene :
%\bc
\be
\phi(r)=A\hspace{.2cm}_1F_1[\frac{1}{2}(l+\frac{3}{2}-\kappa);l+\frac{3}{2},\lambda r^2]
\ee
%\ec
Empleando los mismos argumentos que para el oscilador unidimensional,
es decir, pedimos que las soluciones sean regulares en el infinito, nos
lleva a la condici\'on de truncamiento de la serie, lo cual nos lleva
nuevamente a la cuantizaci\'on de la energ\ii a ; imponiendo la condici\'on
de truncamiento:
\be
\frac{1}{2}(l+\frac{3}{2}-\kappa)=-n
\ee

Esto es , poniendo expl\ii citamente $\kappa$, obtenemos el espectro de
energ\ii a :
\be
E_{nl}=\hbar\omega(2n+l+\frac{3}{2})
\ee

Podemos observar que se tiene una energ\ii a de punto cero igual
a $\frac{3}{2}\hbar\omega$ para el \oa tridimensional esf\'ericamente
sim\'etrico.\\
Las eigenfunciones son (no normalizadas ):
\be
\Psi_{nlm}(r,\theta,\varphi)
=r^{l}e^{\frac{-\lambda r^2}{2}}\hspace{.2cm}_1F_1(-n;l
+\frac{3}{2},\lambda r^2)\hspace{.1cm}Y_{lm}(\theta,\varphi)
\ee

\newpage

\centerline{P r o b l e m a s}

\subsection*{Problema 5.1}

{\bf Encuentre los eigenvalores y eigenfunciones del \oa en el espacio de  
momentos}

El hamiltoniano mec\'anico cu\'antico para el \oa est\'a dado por:
\[
\hat{H}=\frac{\hat{p}^2}{2m}+\frac{1}{2}m\omega^2\hat{x}^2
\]
Ahora bien , en el espacio de momentos, los operadores  $\hat{x}$  
y $\hat{p}$ toman la siguiente forma :
\[
\hat{p}\rightarrow\hspace{.2cm}p
\]
\[
\hat{x}\rightarrow\hspace{.2cm}i\hbar\frac{\partial}{\partial p}
\]
Por tanto el hamiltoniano mec\'anico cu\'antico para el \oa en el espacio de 
los momentos es :
\[
\hat{H}=\frac{p^2}{2m}-\frac{1}{2}m\omega^2\hbar^2\frac{d^2}{dp^2}
\]
Tenemos entonces que resolver el problema de eigenvalores ( esto es, encontrar las eigenfunciones y los eigenvalores) dado por $(5)$ , lo cual nos lleva, con el hamiltoniano anterior, a la siguiente ecuaci\'on diferencial :
\be
\frac{d^2\Psi(p)}{dp^2}+( \frac{2E}{m\hbar^2\omega^2}-\frac{p^2}{m^2\hbar^2\omega^2})\Psi(p)=0
\ee
Se puede observar que la ecuaci\'on diferencial obtenida, es id\'entica, 
salvo constantes, con la ecuaci\'on diferencial que obtuvimos en el espacio de 
configuraciones ( ec. $(6)$ ). S\'olo que para ejemplificar otra forma de 
resolverla, no seguiremos exactamente el mismo camino que se sigui\'o para 
obtener la soluci\'on de aquella.\\
Definiremos dos par\'ametros, de manera similar a como hicimos en $(7)$ y $(8)$:
\be
k^2=\frac{2E}{m\hbar^2\omega^2} \hspace{1cm}\lambda=\frac{1}{m\hbar\omega}
\ee
Con estas definiciones, arribamos exactamente a la ecuaci\'on diferencial $(9)$, se sigue por tanto que la  soluci\'on buscada ( despu\'es de realizar el an\'alisis asint\'otico ) es de la forma:
\be
\Psi(y)=e^{-\frac{1}{2}y}\phi (y)
\ee
Donde $y$ est\'a dada por $y=\lambda p^2$ y $\lambda$ definida en $(117)$. 
Sustituiremos entonces $(118)$ en $(116)$, s\'olo que regresando $(118)$ a la 
variable $p$. Haciendo esta sustituci\'on se obtiene una ecuaci\'on 
diferencial para $ \phi$ :
\be
\frac{d^2\phi(p)}{dp^2}-2\lambda p\frac{d\phi (p)}{dp}+(k^2-\lambda)\phi (p)=0
\ee
Haremos  finalmente el cambio de variable $u=\sqrt{\lambda}p$, la ecuaci\'on 
anterior se transforma en la ecuaci\'on diferencial de Hermite:
\be
\frac{d^2\phi (u)}{du^2}-2u\frac{d\phi (u)}{du}+2n\phi(u)=0
\ee
Con $n$ un entero no negativo, y donde hemos hecho:
\[
\frac{k^2}{\lambda}-1=2n
\]
De aqu\ii \hspace{.1cm}  y de las definiciones dadas en  $(117)$  se sigue 
que los eigenvalores de la energ\ii a est\'an dados por:
\[
E_n=\hbar\omega(n+\frac{1}{2})
\]
Y las soluciones a  $(120)$  est\'an dadas por los polinomios de Hermite, de 
manera que  $\phi(u)=H_n(u)$,  con lo cual se tiene que las eigenfunciones  
(no normalizadas) est\'an dadas por:
\[
\Psi(p)=A e^{-\frac{\lambda}{2}p^2}H_n(\sqrt{\lambda}p)
\]

\vspace{1mm}
%\newpage

\subsection*{Problema 5.2}

{\bf Demuestre que los polinomios de Hermite pueden ser expresados con la 
siguiente representaci\'on integral:
\be
H_n(x)=\frac{2^n}{\sqrt{\pi}}\int_{-\infty}^{\infty} (x+iy)^n e^{-y^2}dy
\ee
}

La representaci\'on anterior de los polinomios de Hermite es una poco usual, 
pero que es muy \'util en ciertos casos. Lo que vamos a hacer para demostrar 
la igualdad, es desarrollar la integral que se presenta y mostrar que lo obtenido es id\'entico con 
la representaci\'on en serie de los polinomios de Hermite, la cual est\'a dada 
por:
\be
\sum_{k=0}^{[\frac{n}{2}]} \frac{ (-1)^k n!}{(n-2k)!k!}(2x)^{n-2k}
\ee

Donde el s\ii mbolo $[c]$ significa, el mayor entero menor o igual que $c$.\\
Lo primero que haremos es desarrollar el binomio que est\'a dentro de la integral usando el teorema del binomio:
\[
(x+y)^n = \sum_{m=0}^n \frac{n!}{(n-m)!m!}x^{n-m}y^m
\]
Usando esto, el binomio dentro de la integral, tiene el desarrollo siguiente:
\be
(x+iy)^n= \sum_{m=0}^n \frac{n!}{(n-m)!m!}i^mx^{n-m}y^m
\ee

Sustituyendo esto en la integral:
\be
\frac{2^n}{\sqrt{\pi}}\sum_{m=0}^n \frac{n!}{(n-m)!m!}i^m x^{n-m}\int_{-\infty}^{\infty} y^m e^{-y^2}dy
\ee

De la forma del integrando podemos ver que la integral es distinta de cero 
cuando $m$ es par, pues de lo contrario el integrando ser\ii a impar y la 
integral se anula. Haremos por tanto el cambio $m=2k$; con este cambio se tiene:

\be
\frac{2^n}{\sqrt{\pi}}\sum_{k=0}^{[\frac{n}{2}]}\frac{n!}{(n-2k)!(2k)!}i^{2k}x^{n-2k}\hspace{.2cm}2\int_{0}^{\infty} y^{2k}e^{-y^2}dy
\ee

Con el cambio de variable $u=y^2$ la integral se convierte en una funci\'on
gamma, haciendo el cambio de variable:
\be
\frac{2^n}{\sqrt{\pi}}\sum_{k=0}^{[\frac{n}{2}]}\frac{n!}{(n-2k)!(2k)!}i^{2k}x^{n-2k}\int_{0}^{\infty}u^{k-\frac{1}{2}}e^{-u}du
\ee

La integral es precisamente $\Gamma(k+\frac{1}{2})$ , la cual puede ser expresada en forma de factoriales ( para $k$ entero, desde luego) :
\[
\Gamma(k+\frac{1}{2})=\frac{(2k)!}{2^{2k}k!}\sqrt{\pi}
\]
Sustituyendo este valor en la sumatoria y usando el hecho de que $i^{2k}=(-1)^k$ 

\be
\sum_{k=0}^{[\frac{n}{2}]} \frac{ (-1)^k n!}{(n-2k)!k!}(2x)^{n-2k}
\ee

El cual es id\'entico con $(122)$, con lo cual se completa la demostraci\'on.

\vspace{1mm}
%\newpage

\subsection*{Problema 5.3}

{\bf Muestre que los eigenestados del \oa satisfacen la relaci\'on de 
incertidumbre}

Debemos mostrar que para cualesquier eigenestado $\Psi_n$ se satisface:
\be
<(\Delta p)^2(\Delta x)^2>\hspace{.3cm}\geq \frac{\hbar^2}{4}
\ee
Donde la notaci\'on $<>$ significa promedio.\\
Vamos a calcular por separado $<(\Delta p)^2>$ y $<(\Delta x)^2>$, donde cada 
una de estas expresiones es:

\[<(\Delta p)^2>=< (p-<p>)^2 >=< p^2 - 2p<p>+<p>^2>=<p^2>-<p>^2\]

\[<(\Delta x)^2>=< (x-<x>)^2 >=< x^2 - 2x<x>+<x>^2>=<x^2>-<x>^2 \]

Vamos primeramente a mostrar que tanto el promedio de $x$, como el de $p$ se 
anulan. Consideremos primeramente el promedio de $x$:

\[<x>=\int_{-\infty}^{\infty} x[\Psi_n(x)]^2 dx\]

Esta integral se anula pues $[\Psi_n(x)]^2$ es una funci\'on par, esto se 
puede ver considerando que la paridad est\'a dada por la parte polinomial 
( pues la exponencial involucrada es una funci\'on par ). Los polinomios de 
Hermite tienen paridad definida, y se tienen s\'olo dos casos, $n$ es par o es 
impar. Si $n$ es par se sigue de inmediato que $[\Psi_n(x)]^2$ lo es. Si $n$ 
es impar entonces tenemos que $H_n(-x)=(-1)^n H_n(x)$, e inmediatamente se ve 
que esta funci\'on al cuadrado es par tambi\'en ( esto es, cualquier 
polinomio par o impar, elevado al cuadrado es par ). Hemos mostrado 
que $[\Psi_n(x)]^2$ es una funci\'on par para $n$ cualquiera, por tanto al 
multiplicarla por $x$ se vuelve impar, de manera que la integral se anula. 
Tenemos entonces como resultado:
\be
<x>=0
\ee

Los mismos argumentos son v\'alidos para el promedio
de $p$, si calculamos este en el espacio de momentos con las
funciones encontradas en el problema 1, pues la forma funcional es
la misma. De manera que:
\be
<p>=0
\ee

Calculemos ahora el promedio de $x^2$. Para hacer esto usaremos el teorema del virial \footnote{ El teorema del virial en mec\'anica cu\'antica nos dice que:
\[ 2<T>=<{\bf r}\cdot\bigtriangledown V({\bf r})>\]
 Para un potencial de la forma $V=\lambda x^n$ se satisface:
\[
2<T>=n<V>
\]
Donde $T$ representa la energ\ii a cin\'etica y V la
energ\ii a potencial.}. Observemos primeramente que :
\[
<V>=\frac{1}{2}m\omega^2 <x^2>~.
\]
De manera que es posible relacionar el promedio de $x^2$ con el promedio del potencial ( y poder usar   el teorema del virial).
\be
<x^2>=\frac{2}{m\omega^2}<V>~.
\ee

Necesitamos considerar tambi\'en el promedio de la energ\ii a :
\[
<H>=<T>+<V>
\]

Usando el teorema del virial ( para $n=2$ ) se obtiene:
\be
<H>=2<V>
\ee

Con lo cual se obtiene:
\be
<x^2>=\frac{<H>}{m\omega^2}=\frac{\hbar\omega(n+\frac{1}{2})}{m\omega^2}
\ee

\be
<x^2>=\frac{\hbar}{m\omega}( n+\frac{1}{2})
\ee

De forma similar calculamos el promedio de $p^2$, expl\ii citamente:
\be
<p^2>=2m<\frac{p^2}{2m}>=2m<T>=m<H>=m\hbar\omega(n+\frac{1}{2})
\ee

Con $(133)$ y $(135)$ se tiene:
\be
<(\Delta p)^2(\Delta x)^2>=(n+\frac{1}{2})^2\hbar^2
\ee

De este resultado inmediatamente se puede ver que los eigenestados
satisfacen la relaci\'on de incertidumbre, con
el m\ii nimo valor precisamente para el estado base ( $n=0$ ).

\vspace{1mm}
%\newpage

\subsection*{Problema 5.4}

{\bf Obt\'enganse los elementos de
matriz de los operadores $a$, $\ad$, $\hat{x}$ y $\hat{p}$ }

Encontraremos primero los elementos de matriz de los operadores de 
creaci\'on y aniquilaci\'on, ya que estos nos ayudaran a encontrar los 
elementos de matriz para los otros dos operadores.\\
Usaremos las relaciones $(65)$ y $(66)$, con las cuales se tiene:
\be
<m \m a\m n>=\sqrt{n}<m\m n-1>=\sqrt{n}\delta_{m,n-1}
\ee

Similarmente para el operador de creaci\'on se tiene:
\be
<m\m \ad \m n>=\sqrt{n+1}<m\m n+1>=\sqrt{n+1}\delta_{m,n+1}~.
\ee

Pasaremos enseguida a calcular los elementos de matriz del operador
de posici\'on. Para hacer esto, expresaremos el operador de
posici\'on en t\'erminos de los operadores de creaci\'on y
aniquilaci\'on. Usando las definiciones $(39)$ y $(40)$, se comprueba
inmediatamente que el operador de posici\'on est\'a dado por:
\be
\hat{x}=\sqrt{ \frac{\hbar}{2m\omega}}(a+\ad)~.
\ee

Usando esto, los elementos de matriz del operador $\hat{x}$ pueden ser 
inmediatamente calculados:
\begin{eqnarray}
<m\m \hat{x}\m n>&=&<m\m \sqrt{ \frac{\hbar}{2m\omega}}(a+\ad)\m n>\nonumber\\
&=&\sqrt{ \frac{\hbar}{2m\omega}}[\sqrt{n}\delta_{m,n-1}+\sqrt{n+1}\delta_{m,n+1}]
\end{eqnarray}
Siguiendo el mismo procedimiento podemos calcular los 
elementos de matriz del operador momento, considerando que $\hat{p}$ est\'a 
dado en t\'erminos de los operadores de creaci\'on y aniquilaci\'on, de la 
forma que a continuaci\'on se muestra:
\be
\hat{p}= i\sqrt{ \frac{m\hbar\omega}{2}}(\ad -a)
\ee

Usando esto se tiene:
\be
<m\m\hat{p}\m n>= i\sqrt{ \frac{m\hbar\omega}{2}}[\sqrt{n+1}\delta_{m,n+1}
-\sqrt{n}\delta_{m,n-1}]
\ee

Hemos encontrado entonces , los elementos de matriz
de los cuatro operadores y se puede observar la
sencillez con la cual son calculados los mismos para los
operadores de posici\'on y momento con la ayuda de los operadores
de creaci\'on y aniquilaci\'on. 
Finalmente podemos hacer la observaci\'on acerca de  la no diagonalidad de
los elementos de matriz que hemos encontrado, lo cual era de esperarse
por el hecho de que la representaci\'on que estamos usando es la del
operador de n\'umero, y ninguno de los cuatro operadores conmuta con \'el.

\vspace{1mm}
%\newpage

\subsection*{Problema 5.5}

{\bf Encu\'entrense los valores esperados
de $\hat{x}^2$ y $\hat{p}^2$ para el \oa unidimensional y \'usense estos
para encontrar los valores esperados de la energ\ii a cin\'etica y la
energ\ii a potencial. Comp\'arese  este \'ultimo resultado con  el
teorema del virial}

Se encontrar\'a primeramente el valor esperado de $\hat{x}^2$. Para hacerlo
recurriremos a la expresi\'on $(139)$, de la cual se sigue:
\be
\hat{x}^2 = \frac{\hbar}{2m\omega} (a^2 + (\ad) ^2 +\ad a+a\ad)
\ee
Recu\'erdese que los operadores de creaci\'on y aniquilaci\'on no conmutan
entre s\ii . Con lo anterior podemos calcular el valor
esperado de $\hat{x}^2$:
\begin{eqnarray}
<\hat{x}^2>&=&<n\m \hat{x}^2\m n>\nonumber\\
&=&\frac{\hbar}{2m\omega}[ \sqrt{n(n-1)}\delta_{n,n-2}+\sqrt{(n+1)(n+2)}
\delta_{n,n+2}\nonumber\\
&+& \hspace{.2cm}n\hspace{.3cm}\delta_{n,n}\hspace{.2cm}+
\hspace{.2cm}(n+1)\hspace{.3cm}\delta_{n,n}]
\end{eqnarray}

Esto es, el valor esperado de $\hat{x}^2$ est\'a dado por:
\be
<\hat{x}^2>=<n\m \hat{x}^2\m n> = \frac{\hbar}{2m\omega}(2n+1)
\ee
Para calcular el valor esperado de $\hat{p}^2$ solo necesitamos expresar
este operador en t\'erminos de los operadores de creaci\'on y
aniquilaci\'on, lo cual lo podemos
realizar a partir de $(141)$, obteni\'endose:
\be
\hat{p}^2 = -\frac{m\hbar\omega}{2}(a^2+(\ad)^2-a\ad-\ad a)
\ee
Para el valor esperado del cuadrado del momento se obtiene:
\be
<\hat{p}^2>=<n\m \hat{p}^2\m n>=\frac{m\hbar\omega}{2}(2n+1)
\ee
Con este \'ultimo resultado podemos encontrar el valor esperado de la
energ\ii a cin\'etica :
\be
<\hat{T}>=<\frac{\hat{p}^2}{2m}>=\frac{1}{2m}<\hat{p}^2>
\ee
Usando el $(147)$, se obtiene el valor esperado de la energ\ii a cin\'etica:
\be
<\hat{T}>=<n\m \hat{T}\m n>=\frac{\hbar\omega}{4}(2n+1)~.
\ee

Nos falta obtener ahora el valor esperado de la energ\ii a potencial:
\be
<\hat{V}>=<\frac{1}{2}m\omega^2 \hat{x}^2>=\frac{1}{2}m\omega^2 <\hat{x}^2>~.
\ee
Con el resultado de $(145)$, inmediatamente
se sigue el valor esperado para la energ\ii a potencial:
\be
<\hat{V}>=<n\m \hat{V}\m n>=\frac{\hbar\omega}{4}(2n+1)~.
\ee

Observamos que los valores esperados de la energ\ii a cin\'etica y la
energ\ii a potencial coinciden para toda $n$, esto es, para cualquier
estado de energ\ii a, lo cual est\'a en correspondencia con el teorema del
virial el cual establece que para un potencial cuadr\'atico, como el
del \oa, los valores esperados de la energ\ii a cin\'etica y la
energ\ii a potencial deben coincidir, y mas a\'un estos deben ser iguales a un
medio del valor esperado de la energ\ii a total del sistema, lo cual
efectivamente se satisface.

\vspace{1mm}
%\newpage

\subsection*{Problema 5.6}

{\bf Una part\ii cula cargada (carga q) se mueve en la
direcci\'on $z$, en la presencia de un campo magn\'etico uniforme en la
misma direcci\'on ( $\vec{B}=B\hat{k}$). Comparando el hamiltoniano para
este sistema con el del \oa unidimensional, muestre que los eigenvalores de
la energ\ii a pueden ser inmediatamente escritos:
\[ E_{kn}=\frac{\hbar^2k^2}{2m}+\frac{\m qB\m\hbar}{mc}(n+\frac{1}{2})~,\]
donde $\hbar k$ es el eigenvalor continuo del operador $p_z$ y $n$ un
entero no negativo.}

El hamiltoniano para una part\ii cula de carga el\'ectrica $q$, la cual
se mueve en presencia de un campo electromagn\'etico est\'a dado por:
\be
H = \frac{1}{2m}( \vec{p} - \frac{q\vec{A}}{c})^2 +q\phi~,
\ee
donde $\vec{A}$ es el potencial vectorial que genera el
campo magn\'etico y $\phi$ es el potencial escalar que genera el
campo el\'ectrico.\\
En nuestro problema no tenemos campos el\'ectricos presentes, de
manera que el potencial escalar $\phi$  es igual a cero. Nuestro
hamiltoniano toma entonces la forma:
\be
H= \frac{{\bf p^2}}{2m}-\frac{q}{2mc}
({\bf p\cdot A+A\cdot p})+\frac{1}{2m}(\frac{q}{c})^2 {\bf A^2}~.
\ee
Esto dado que el potencial vectorial y el momento no conmutan, pues el
potencial vectorial es funci\'on de las coordenadas.\\
Dado que la part\ii cula se desplaza s\'olo en la
direcci\'on $z$, s\'olo tenemos momento lineal asociado a esta
coordenada, esto es :
\be
\vec{p}= (o,o,p_z)
\ee
Y por otro lado, el potencial vectorial $\vec{A}$ que genera el
campo magn\'etico en la direcci\'on $z$ es:
\be
\vec{A}=(-By,0,0)~.
\ee
Esto \'ultimo se puede comprobar sabiendo
que $\vec{B}=\bigtriangledown\times\vec{A}$.\\
Bajo estas circunstancias, el segundo t\'ermino del hamiltoniano
en $(153)$ se anula, de manera que el hamiltoniano a considerar es:
\be
H=\frac{p_z^2}{2m}+\frac{q^2B^2}{2mc^2}y^2~.
\ee
Observamos que el hamiltoniano es la suma de una parte de part\ii cula
libre y otra de \oa,
identificando $\omega=\frac{\m Bq\m}{mc}$ podemos
escribir la energ\ii a asociada a cada contribuci\'on
inmediatamente, con lo cual se obtiene:
\be
E_{kn}=\frac{\hbar^2k^2}{2m}+\frac{\m qB\m\hbar}{mc}(n+\frac{1}{2})
\ee
Obteniendo as\ii \hspace{.1cm}el resultado buscado.

%\end{document}

\newpage
%%%%%%%%%%%%%%%%%%%%%%%%%%%%%%%%%%%%%%%%%%%%%%%%%%%%%%%%%%%%%%%%%%%
%\documentstyle[aps,preprint,tighten]{revtex}
%\begin{document}
%\draft
\def\bi{bigskip}
\def\noi{noindent}
\def\ii{\'{\i}}
\begin{center}{\huge 6. EL \'ATOMO DE HIDR\'OGENO}
\end{center}
%\author{Edgar Alvarado Anell}
%\address{Universidad de Guanajuato,
%Guanajuato; M\'exico.}
%\maketitle
%\begin{abstract}
%\begin{center}
Se estudia el \'atomo de hidr\'ogeno resolviendo la ecuaci\'on de 
Schr\"odinger independiente del tiempo con un potencial debido a 
dos part\ii culas cargadas como lo son el electr\'on y el prot\'on, con el 
Laplaciano en coordenadas esf\'ericas, 
mediante separaci\'on de variables, dando una interpretaci\'on f\ii sica 
de la funci\'on de onda como una soluci\'on de la ecuaci\'on de Schr\"odinger 
para el \'atomo de hidr\'ogeno, adem\'as de las interpretaciones de 
los n\'umeros cu\'anticos y de las densidades de probabilidad.\\
%\end{center}
%\end{abstract}
%%%%%%%%%%%%%%%%%%%%%%%%%%%%%%%%%%%%%%%%%%%%%%%%%%%%%%%%%%%%%%%%%%%
%\setcounter{equation}
\section*{INTRODUCCI\'ON A LA MEC\'ANICA CU\'ANTICA}
\setcounter{equation}{0}
Como nuestro inter\'es es el de describir el 
\'atomo de hidr\'ogeno, el cual est\'a a una escala muy peque\~na, se har\'a
mediante el uso de la mec\'anica cu\'antica, la cual trata las relaciones
entre magnitudes observables, pero el principio de incertidumbre altera
radicalmente la definici\'on de ``magnitud observable" en el campo
at\'omico. De acuerdo con el principio de incertidumbre, la posici\'on y
el momento de una part\ii cula no se pueden medir simult\'aneamente con
precisi\'on. Las cantidades cuyas relaciones busca la mec\'anica
cu\'antica son probabilidades. En vez de afirmar, por ejemplo, que el
radio de la \'orbita del electr\'on en un estado fundamental del \'atomo
de hidr\'ogeno es siempre exactamente $5.3 \times 10^{-11}$ m, la mec\'anica
cu\'antica afirma que \'este es el radio m\'as probable; si realizamos un
experimento adecuado, la mayor parte de las pruebas dar\'an un valor
distinto, m\'as grande o m\'as peque\~no, pero el valor m\'as probable
ser\'a aproximadamente $5.3 \times 10^{-11}$ m. 

\section*{ECUACI\'ON DE ONDA}
Como ya se sabe, la cantidad
con que est\'a relacionada la mec\'anica cu\'antica es la funci\'on de onda
$\Psi$ de una part\ii cula. Aunque $\Psi$ no tiene interpretaci\'on f\ii 
sica, el
cuadrado de su valor absoluto $\mid \Psi \mid^{2}$ calculado para un
punto y en un instante determinado es proporcional a la cantidad de
encontrar experimentalmente a la part\ii cula ah\ii\  y en ese instante. El 
problema
de la mec\'anica cu\'antica es determinar $\Psi$ para una part\ii cula 
cuando su
libertad de movimiento est\'a limitada por la acci\'on de fuerzas
externas. 

Antes de considerar el c\'alculo real de $\Psi$, debemos
establecer algunos requisitos que siempre se deben cumplir. En primer
lugar, ya que $\mid \Psi \mid^{2}$ es proporcional a la probabilidad P de 
encontrar a la part\ii cula descrita por $\Psi$, la integral de $\mid \Psi 
\mid^{2}$ sobre todo el espacio debe ser finita, ya que la part\ii cula 
est\'a en alguna parte. Si tenemos que
\begin{equation} %1
\int_{-\infty}^{\infty} \mid \Psi \mid^{2} dV = 0
\end{equation}
la part\ii cula no existe y la integral evidentemente no puede ser $\infty$
y tener cierto significado; $\mid \Psi \mid^{2}$ no puede ser negativa o 
compleja a causa del camino seguido para definirla, y as\ii\   la \'unica
posibilidad dada es que su integral sea una cantidad finita para que 
$\Psi$ describa a\-pro\-pia\-da\-men\-te una part\ii cula real.
Generalmente es conveniente tener $\mid \Psi \mid^{2}$ igual a la 
probabilidad P de encontrar la part\ii cula descrita por $\Psi$, en lugar 
de ser simplemente proporcional a P. Para que $\mid \Psi \mid^{2}$ sea 
igual a P se tiene que cumplir la relaci\'on
\begin{equation} %2
\int_{-\infty}^{\infty}\mid \Psi \mid^{2} dV = 1
\end{equation}
ya que 
\begin{equation} %3
\int_{-\infty}^{\infty}P dV = 1
\end{equation}
es la afirmaci\'on matem\'atica de que la part\ii cula existe en alg\'un 
lugar en todo momento. Una funci\'on que obedezca a la ec. 2 se dice que 
est\'a normalizada. Adem\'as de ser normalizable, $\Psi$ debe tener un s\'olo 
valor, ya que P debe tener un valor \'unico en un tiempo y en un lugar 
determinados. Otra condici\'on que $\Psi$ debe obedecer es que ella y sus 
derivadas parciales $\frac{\partial \Psi}{\partial x}$, $\frac{\partial 
\Psi}{\partial y}$, $\frac{\partial \Psi}{\partial z}$  sean continuas en 
cualquier lugar.

La ecuaci\'on de Schr\"odinger, que es la ecuaci\'on fundamental de 
la mec\'anica cu\'antica, en el mismo sentido que la segunda ley del 
movimiento es la ecuaci\'on fundamental de la mec\'anica newtoniana, es 
una ecuaci\'on de onda en la variable $\Psi$. Antes de abordar la 
ecuaci\'on de Schr\"odinger repasemos la ecuaci\'on de onda general
\begin {equation} %4
\frac{\partial^{2}y}{\partial x^{2}} = \frac{1}{v^2} 
\frac{\partial^{2}y}{\partial t^{2}}
\end{equation}
que gobierna a una onda cuya cantidad variable es $y$ que se propaga en 
la direcci\'on de $x$ con la velocidad $v$. En el caso de una onda en una 
cuerda tensa, $y$ es el desplazamiento de la cuerda medido desde el eje 
$x$; en el caso de una onda sonora, $y$ es la diferencia de presi\'on; 
en el caso de una onda luminosa $y$ es la magnitud del campo el\'ectrico
o la del campo magn\'etico.

Las soluciones de la ecuaci\'on de onda pueden ser de varios tipos, como 
consecuencia de la variedad de ondas que puede haber (un pulso \'unico en 
desplazamiento, un tren de ondas de amplitud y longitud de onda 
constantes, un tren de ondas superpuestas de amplitudes y longitudes de 
onda id\'enticas, un tren de ondas superpuestas de amplitudes y longitudes de
onda diferentes, una onda estacionaria en una cuerda fija por ambos 
extremos, etc.). Todas las soluciones deben ser de la forma
\begin{equation} %5
y(x,t) = F\left(t \pm \frac{x}{v}\right)
\end{equation}
donde $F$ es cualquier funci\'on que pueda ser diferenciada. Las 
soluciones $F(t-x/v)$ re\-pre\-sen\-tan ondas que viajan en el sentido 
$+x$, y 
las soluciones $F(t+x/v)$ representan ondas que viajan en el sentido $-x$.
Aqu\ii\ nos interesa el equivalente ondulatorio de una part\ii cula 
``libre", es decir, una part\ii cula que no est\'e bajo la influencia de 
ninguna fuerza y que, por lo tanto, viaja en una trayectoria recta a 
velocidad constante. Este equivalente corresponde a la soluci\'on general 
de la ec. 4 para ondas arm\'onicas no amortiguadas ( es decir, de 
amplitud constante A), monocrom\'aticas (de frecuencia angular $\omega$ 
constante) en la direcci\'on $+x$,
\begin{equation} %6
y(x,t) = Ae^{-i \omega (t - x/v)}
\end{equation}
En esta f\'ormula, $y$ es una cantidad compleja, con parte real e imaginaria.
Como
\begin{equation} %7
e^{-i \theta} = cos \theta - isen \theta
\end{equation}
la ec. 6 se puede escribir en la forma
\begin{equation} %8
y(x,t) = Acos \omega (t - x/v) - iAsen \omega (t - x/v)
\end{equation}

\'Unicamente la parte real de la ec. 7 tiene significado en el caso de 
ondas en una cuerda en tensi\'on, donde $y$ representa el desplazamiento 
de la cuerda con respecto a su posici\'on normal, en este caso la parte 
imaginaria se descarta porque no se puede aplicar.

\section*{ECUACI\'ON DE SCHR\"ODINGER}
En mec\'anica cu\'antica, la funci\'on de onda $\Psi$ corresponde a la 
variable de onda $y$ del movimiento ondulatorio general. Sin embargo, 
$\Psi$, a diferencia de $y$, no es una cantidad mensurable en s\ii\ misma 
y puede, por tanto, ser compleja. Por esta raz\'on supondremos que $\Psi$ 
est\'a especificada en la direcci\'on $x$ por 
\begin{equation} %9
\Psi(x,t) = Ae^{-i \omega (t - x/v)}
\end{equation}
Cuando se sustituye en esta f\'ormula $\omega$ por 2$\pi \nu$ y $v$ por 
$\lambda \nu$, obtenemos
\begin{equation} %10
\Psi(x,t) = Ae^{-2 \pi i(\nu t - x/\lambda)}
\end{equation}
que es conveniente, ya que sabemos que $\nu$ y $\lambda$ est\'an en 
funci\'on de la energ\ii a total E y del momento $p$ de la part\ii cula 
descrita por $\Psi$. Ya que 
\begin{equation} %11
E = h \nu = 2 \pi \hbar \nu
\end{equation}
y
\begin{equation} %12 
\lambda = \frac{h}{p} = \frac{2 \pi \hbar}{p}
\end{equation}
tenemos
\begin{equation} %13
\Psi(x,t) = Ae^{-(i/\hbar)(Et - px)}
\end{equation}

La ec. 13 es una descripci\'on matem\'atica de la onda equivalente a una 
part\ii cula libre, de energ\ii a total E y momento $p$, que se mueve en 
la direcci\'on y sentido $+x$, del mismo modo que la ec. 6 es la 
descripci\'on matem\'atica de un desplazamiento de onda arm\'onica que se 
mueve libremente a lo largo de una cuerda en tensi\'on.

La expresi\'on de la funci\'on de onda $\Psi$, dada por la ec. 13, es 
correcta solamente para part\ii culas que se mueven libremente, pero 
estamos m\'as interesados en situaciones donde el movimiento de una 
part\ii cula est\'a sujeto a varias restricciones, como el caso de un 
electr\'on ligado a un \'atomo por el campo el\'ectrico de su n\'ucleo. Lo 
que debemos hacer ahora es obtener la ecuaci\'on diferencial fundamental 
para $\Psi$, la que se puede resolver en una situaci\'on espec\ii fica.

Comenzamos por la diferenciaci\'on de la ec. 13 dos veces con respecto a 
$x$, 
\begin{equation} %14
\frac{\partial^{2} \Psi}{\partial x^{2}} = -\frac{p^2}{\hbar^2}\Psi
\end{equation}
y una vez respecto a $t$
\begin{equation} %15
\frac{\partial \Psi}{\partial t} = -\frac{iE}{\hbar}\Psi
\end{equation}
A velocidades peque\~nas comparadas con la de la luz, la energ\ii a total 
E de una part\ii cula es la suma de su energ\ii a cin\'etica $p^{2}/2m$ y 
de su energ\ii a potencial $V$, donde $V$ es una funci\'on general de la 
posici\'on $x$ y del tiempo $t$:
\begin{equation} %16
E = \frac{p^{2}}{2m} + V
\end{equation}
Multiplicando ambos miembros de esta ecuaci\'on por la funci\'on de onda
\begin{equation} %17
E \Psi = \frac{p^{2} \Psi}{2m} + V \Psi
\end{equation}
De las ecs. 14 y 15 vemos que
\begin{equation} %18
E \Psi = -\frac{\hbar}{i} \frac{\partial \Psi}{\partial t}
\end{equation}
y
\begin{equation} %19
p^{2} \Psi = -\hbar^{2} \frac{\partial^{2} \Psi}{\partial x^{2}}
\end{equation}
Sustituyendo estas expresiones de E$\Psi$ y $p^{2}\Psi$ en la ec. 17 
obtenemos
\begin{equation} %20
i\hbar \frac{\partial \Psi}{\partial t} = -\frac{\hbar^{2}}{2m} 
\frac{\partial^{2} \Psi}{\partial x^{2}} + V \Psi
\end{equation}
La ec. 20 es la ecuaci\'on de Schr\"odinger dependiente del tiempo, 
donde la energ\ii a potencial $V$ es una funci\'on de $x,y,z,t$. En tres 
dimensiones, la ecuaci\'on de Schr\"odinger dependiente del tiempo 
es 
\begin{equation} %21
i\hbar \frac{\partial \Psi}{\partial t} = -\frac{\hbar^{2}}{2m}
\left(\frac{\partial^{2} \Psi}{\partial x^{2}} + \frac{\partial^{2} 
\Psi}{\partial y^{2}} + \frac{\partial^{2} \Psi}{\partial z^{2}}\right) + V 
\Psi \end{equation}

Una vez conocida $V$, se puede resolver la ecuaci\'on de Schr\"odinger 
para la funci\'on de onda $\Psi$ de la part\ii cula, cuya 
densidad de probabilidad $\mid \Psi \mid^{2}$ se puede determinar para 
$x,y,z,t$.
En muchas situaciones, la energ\ii a potencial de una part\ii cula no 
depende expl\ii citamente del tiempo; las fuerzas que act\'uan sobre ella 
y, por lo tanto, $V$, var\ii an solamente con la posici\'on de la part\ii 
cula. Cuando esto se cumple, la ecuaci\'on de Schr\"odinger se 
puede simplificar eliminando todo lo referente a $t$. Notemos que se puede 
escribir la funci\'on de onda unidimensional de una part\ii cula libre
\begin{eqnarray} %22
\Psi(x,t) & = &  Ae^{(-i/\hbar)(Et - px)} \nonumber \\
& = & Ae^{-(iE/\hbar)t}e^{(ip/\hbar)x} \nonumber\\
& = & \psi(x) e^{-(iE/\hbar)t}
\end{eqnarray}
Esto es, $\Psi(x,t)$ es el producto de una funci\'on dependiente del tiempo 
$e^{-(iE/\hbar)t}$ y una funci\'on dependiente de la posici\'on $\psi(x,t)$. 
Sucede que las variaciones con el tiempo de todas las funciones de 
part\ii culas, sobre las que act\'uan fuerzas estacionarias, tienen la 
misma forma que las de una part\ii cula libre. Sustituyendo la $\Psi$ de 
la ec. 21 en la ecuaci\'on de Schr\"odinger dependiente del tiempo, 
encontramos que 
\begin{equation} %23
E \psi e^{-(iE/\hbar)t} = -\frac{\hbar^{2}}{2m}e^{-(iE/\hbar)t}
\frac{\partial^{2} \psi}{\partial x^{2}} + V \psi e^{-(iE/\hbar)t}
\end{equation}
y, as\ii\, dividiendo ambos miembros entre el factor exponencial com\'un,
\begin{equation} %24
\frac{\partial^{2} \psi}{\partial x^{2}} + \frac{2m}{\hbar^{2}}(E-V)\psi = 0
\end{equation}
que es la ecuaci\'on de Schr\"odinger en estado estacionario. En 
tres dimensiones es
\begin{equation} %25
\frac{\partial^{2} \psi}{\partial x^{2}} + \frac{\partial^{2} 
\psi}{\partial y^{2}} + \frac{\partial^{2} \psi}{\partial z^{2}} + 
\frac{2m}{\hbar^{2}}(E-V)\psi = 0 
\end{equation}

En general, la ecuaci\'on de Schr\"odinger en estado estacionario 
se puede resolver \'unicamente para algunos valores de la energ\ii a E. 
Lo que queremos decir con esto no se refiere a las dificultades 
matem\'aticas que se pueden presentar, sino a algo m\'as fundamental. 
``Resolver" la ecuaci\'on de Schr\"odinger para un sistema dado 
significa obtener una funci\'on de onda $\psi$ que no s\'olo obedezca a la 
ecuaci\'on y a las condiciones en la frontera que existan, sino que 
tambi\'en cumplan las condiciones de una funci\'on de onda aceptable, es 
decir, que la funci\'on y su derivada sean continuas finitas y 
univaluadas. De esta manera, la cuantizaci\'on de energ\ii a aparece en 
la mec\'anica ondulatoria como un elemento natural de la teor\ii a. 
As\ii\, la cuantizaci\'on de la energ\ii a en el mundo f\ii sico se ha 
revelado como un fen\'omeno universal caracter\ii stico de todos los 
sistemas estables.

\section*{ECUACI\'ON DE SCHR\"ODINGER PARA EL \'ATOMO DE HIDR\'OGENO}
A continuaci\'on aplicaremos la ecuaci\'on de Schr\"odinger al 
\'atomo de hidr\'ogeno el cual est\'a formado por un prot\'on, part\ii cula 
con carga el\'ectrica +$e$, y un electr\'on, que tiene carga -$e$ y que 
es 1,836 veces m\'as ligero que el prot\'on. 

Ahora, si la interacci\'on entre dos part\ii culas es de tipo $u ( \mid 
\vec r_{1} - \vec r_{2} \mid)$, el problema de movimiento de tales 
part\ii culas en mec\'anica cu\'antica y tambi\'en en mec\'anica cl\'asica 
se reduce al movimiento de una sola part\ii cula en el campo de simetr\ii 
a esf\'erica, entonces tenemos el siguiente Lagrangiano:
\begin{equation} %26
L = \frac{1}{2} m_{1} \dot { \vec r_{1}^{2}} + \frac{1}{2} m_{2} \dot {\vec
r_{2}^{2} } - u ( \mid \vec r_{1} - \vec r_{2} \mid)
\end{equation}
Introduciendo las siguientes expresiones:
\begin{equation} %27
\vec r = \vec r_{1} - \vec r_{2}
\end{equation}
y
\begin{equation} %28
\vec R = \frac{m_{1} \vec r_{1} + m_{2} \vec r_{2}}{m_{1} + m_{2}}
\end{equation}
por lo tanto el Lagrangiano nos queda:
\begin{equation} %29
L = \frac{1}{2} M \dot { \vec R^{2}} + \frac{1}{2} \mu \dot {\vec
r^{2} } - u (r)
\end{equation}
donde
\begin{equation} %30
M = m_{1} + m_{2}
\end{equation}
y
\begin{equation} %31
\mu =\frac{m_{1} m_{2}}{m_{1} + m_{2}}
\end{equation}

Por otro lado la introducci\'on del impulso se hace con las f\'ormulas de 
Lagrange
\begin{equation} %32
\vec P = \frac{\partial L}{\partial \dot { \vec R}} = M \dot { \vec R}
\end{equation}
y
\begin{equation} %33
\vec p = \frac{\partial L}{\partial \dot { \vec r}} = m \dot { \vec r}
\end{equation}
lo que permite escribir la funci\'on cl\'asica de Hamilton
\begin{equation} %34
H = \frac{P^{2}}{2M} + \frac{p^{2}}{2m} + u(r)
\end{equation}

Entonces se puede obtener el operador Hamiltoniano del problema 
correspondiente cu\'antico con conmutadores de tipo
\begin{equation} %35
[P_{i},P_{k}] = -i \hbar \delta_{ik}
\end{equation}
y
\begin{equation} %36
[p_{i},p_{k}] = -i \hbar \delta_{ik}
\end{equation}
por lo tanto el operador Hamiltoniano es de la forma
\begin{equation} %37
\hat H = -\frac{\hbar^{2}}{2M}\nabla_{R}^{2} - 
\frac{\hbar^{2}}{2m}\nabla_{r}^{2} + u(r)
\end{equation}
Este operador Hamiltoniano es la parte fundamental de la ecuaci\'on de 
Schr\"odinger puesta en la forma
\begin{equation} %38
\hat H \psi = E \psi
\end{equation}
lo cual es una forma muy pr\'actica de escribirla, pero lo m\'as importante 
hasta ahora escrito en esta secci\'on es que se ha tratado al sistema 
formado por el prot\'on y el electr\'on como un sistema cl\'asico con 
part\ii culas de masa no despreciable, como lo demuestran las ecs. 24-29, 
ya que no se est\'an tomando en cuenta velocidades cercanas a la de la luz, 
por este motivo se puede aplicar perfectamente la ecuaci\'on de 
Schr\"odinger con resultados muy satisfactorios.

La ecuaci\'on de Schr\"odinger para el electr\'on en tres 
dimensiones, que es la que debemos emplear para el \'atomo de 
hidr\'ogeno, es la ec. 21. Utilizaremos esta ecuaci\'on de Schr\"odinger 
independiente del tiempo debido a que el potencial $V$ depende 
solamente de $r$ y no del tiempo. 

La energ\ii a potencial $V$, a causa de la 
energ\ii a potencial electrost\'atica de una carga -$e$ a una distancia 
$r$ de otra carga +$e$, es
\begin{equation} %39
V = -\frac{e^{2}}{4\pi \epsilon _{0} r}
\end{equation}

Puesto que $V$ es una funci\'on de $r$ en vez de serlo de $x,y,z$, no 
podemos sustituir la ec. 39 directamente en la ec. 21. Hay dos 
posibilidades: expresar $V$ en funci\'on de las coordenadas cartesianas 
$x,y,z$ sustituyendo a $r$ por $\sqrt{x^{2}+y^{2}+z^{2}}$, o expresar la 
ecuaci\'on de Schr\"odinger en funci\'on de las coordenadas polares 
esf\'ericas $r,\theta,\phi$. Haciendo esto \'ultimo debido a la simetr\ii 
a de la situaci\'on f\ii sica, el problema se simplifica considerablemente.

Por lo tanto, en coordenadas polares esf\'ericas, la ecuaci\'on de 
Schr\"odinger es
\begin{equation} %40
\frac{1}{r^{2}} \frac{\partial}{\partial r}\left(r^{2} \frac{\partial 
\psi}{\partial r}
\right) + \frac{1}{r^{2} sen\theta} \frac{\partial}{\partial 
\theta} \left(sen\theta \frac{\partial \psi}{\partial \theta}\right) + 
\frac{1}{r^{2}sen^{2}\theta} \frac{\partial^{2} \psi}{\partial \phi^{2}} 
+ \frac{2m}{\hbar^{2}}(E - V)\psi = 0
\end{equation}
Sustituyendo la energ\ii a potencial $V$ de la ec. 39 y multiplicando toda 
la ecuaci\'on por $r^{2}sen^{2}\theta$, se obtiene
\begin{equation} %41
sen^{2}\theta \frac{\partial}{\partial r}\left(r^{2}
\frac{\partial \psi}{\partial r}\right) + sen\theta \frac{\partial}{\partial
\theta}\left(sen\theta \frac{\partial \psi}{\partial \theta}\right) +
\frac{\partial^{2} \psi}{\partial \phi^{2}} +
\frac{2mr^{2}sen^{2}\theta}{\hbar^{2}} \left(\frac{e^{2}}{4\pi 
\epsilon_{0}r} + E\right)\psi = 0 
\end{equation} 
Esta ecuaci\'on, es la ecuaci\'on diferencial
parcial de la funci\'on de onda $\psi(r\theta,\phi)$ del electr\'on en un 
\'atomo de
hidr\'ogeno. Junto con las diversas condiciones que $\psi(r,\theta,\phi)$
debe cumplir
(por ejemplo, $\psi(r,\theta,\phi)$ tiene un s\'olo valor para cada punto
$r,\theta,\phi$),
esta ecuaci\'on especifica totalmente el comportamiento del electr\'on.
Para ver cual es este comportamiento, resolveremos la ec. 41 para 
$\psi(r,\theta,\phi)$ e interpretaremos los resultados obtenidos.

\section*{SEPARACI\'ON DE VARIABLES EN LA ECUACI\'ON DE SCHR\"ODINGER}
Lo verdaderamente valioso de escribir la ecuaci\'on de Schr\"odinger 
en coordenadas esf\'ericas para el problema del \'atomo de hidr\'ogeno 
est\'a en que de esta forma se puede separar f\'acilmente en tres 
ecuaciones independientes, cada una de ellas con una s\'ola coordenada. El 
procedimiento consiste en buscar las soluciones en que la funci\'on de 
onda $\psi(r, \theta, \phi)$ tiene la forma de un producto de tres 
funciones diferentes: $R(r)$, que depende solamente de $r$; 
$\Theta(\theta)$ que depende solamente de $\theta$; y $\Phi(\phi)$ que 
s\'olo depende de $\phi$. Esto es, suponemos que
\begin{equation} %42
\psi(r, \theta, \phi) = R(r)\Theta(\theta)\Phi(\phi)
\end{equation}
La funci\'on $R(r)$ describe la variaci\'on de la funci\'on de onda 
$\psi$ del electr\'on a lo largo de un radio vector desde el n\'ucleo, 
siendo $\theta$ y $\phi$ constantes. La variaci\'on de $\psi$ con el 
\'angulo cenital $\theta$ a lo largo de un meridiano de una esfera 
centrada sobre el n\'ucleo est\'a descrita por la funci\'on 
$\Theta(\theta)$  para $r$ y $\phi$ constantes. Finalmente, la funci\'on 
$\Phi(\phi)$ describe c\'omo var\ii a $\psi$ con el \'angulo azimutal 
$\phi$ a lo largo de un paralelo de una esfera centrada sobre el 
n\'ucleo, siendo $r$ y $\theta$ constantes.

La ec. 42 se puede escribir m\'as f\'acilmente como $\psi=R\Theta\Phi$ de 
donde vemos que
\begin{equation} %43
\frac{\partial \psi}{\partial r} = \Theta \Phi \frac{\partial 
R}{\partial r}  
\end{equation}
\begin{equation} %44
\frac{\partial \psi}{\partial \theta} = R\Phi \frac{\partial
\Theta}{\partial \theta} 
\end{equation} 
\begin{equation} %45 
\frac{\partial^{2} \psi}{\partial \phi^{2}} = R\Theta \frac{\partial^{2}
\Phi}{\partial \phi^{2}}  
\end{equation}
Al sustituir las ecs. 43-45 en la ec. 41, que es la ecuaci\'on de 
Schr\"odinger para el \'atomo de hidr\'ogeno, y se divide la 
ecuaci\'on total entre $R\Theta \Phi$ se tiene que
\begin{equation} %46
\frac{sen^{2}\theta}{R} \frac{\partial}{\partial r}\left(r^{2} \frac{\partial
R}{\partial r}\right)+\frac{sen\theta}{\Theta} \frac{\partial}{\partial 
\theta}\left(sen\theta
\frac{\partial \Theta}{\partial \theta}\right)+\frac{1}{\Phi}  
\frac{\partial^{2} \Phi}{\partial \phi^{2}} +
\frac{2mr^{2}sen^{2}\theta}{\hbar^{2}} \left(\frac{e^{2}}{4\pi 
\epsilon_{0}r} + E\right) = 0
\end{equation}
El tercer t\'ermino de esta ecuaci\'on s\'olo es funci\'on del \'angulo 
$\phi$, mientras que los otros dos son funci\'on de $r$ y $\theta$. 
Volviendo a escribir la ecuaci\'on anterior, tenemos
\begin{equation} %47
\frac{sen^{2}\theta}{R} \frac{\partial}{\partial r}\left(r^{2} \frac{\partial
R}{\partial r}\right)+\frac{sen\theta}{\Theta} \frac{\partial}{\partial
\theta}\left(sen\theta
\frac{\partial \Theta}{\partial \theta}\right)+
\frac{2mr^{2}sen^{2}\theta}{\hbar^{2}} \left(\frac{e^{2}}{4\pi 
\epsilon_{0}r} +
E\right) = -\frac{1}{\Phi}\frac{\partial^{2} \Phi}{\partial \phi^{2}}
\end{equation}
Esta ecuaci\'on solamente puede ser correcta si sus dos miembros son 
iguales a la misma constante, ya que son funciones de variables 
diferentes. A esta constante es conveniente llamarla $m_{l}^{2}$. La 
ecuaci\'on diferencial para la funci\'on $\Phi$ es
\begin{equation}  %48
-\frac{1}{\Phi}\frac{\partial^{2} \Phi}{\partial \phi^{2}} = m_{l}^{2}
\end{equation}
Si se sustituye $m_{l}^{2}$ en el segundo miembro de la ec. 47 , se divide 
la ecuaci\'on resultante entre $sen^{2}\theta$ y se reagrupan t\'erminos, 
se tiene
\begin{equation} %49
\frac{1}{R} \frac{\partial}{\partial r}\left(r^{2} \frac{\partial
R}{\partial r}\right) + 
\frac{2mr^{2}}{\hbar^{2}} \left(\frac{e^{2}}{4\pi \epsilon_{0}r} 
+ E\right) = \frac{m_{l}^{2}}{sen^{2}\theta} - \frac{1}{\Theta sen\theta} 
\frac{\partial}{\partial\theta}\left(sen\theta\frac{\partial 
\Theta}{\partial \theta}\right)
\end{equation}
Se tiene otra vez una ecuaci\'on en que aparecen variables diferentes en 
cada miembro, requiri\'endose que ambas sean iguales a la misma constante. 
A esta constante se le llamar\'a, por razones que veremos m\'as adelante, 
$l(l+1)$. Las ecuaciones para las funciones $\Theta(\theta)$ y $R(r)$ son
\begin{equation} %50
\frac{m_{l}^{2}}{sen^{2}\theta} - \frac{1}{\Theta 
sen\theta}\frac{d}{d\theta}\left(sen\theta \frac{d\Theta}{d\theta}\right) 
= l(l +1)
\end{equation}
y
\begin{equation} %51
\frac{1}{R}\frac{d}{dr}\left(r^{2}\frac{dR}{dr}\right) + 
\frac{2mr^{2}}{\hbar^{2}}\left(\frac{e^{2}}{4\pi \epsilon_{0}r} + 
E\right) = l(l+1)
\end{equation}
Las ecs. 48,50 y 51 se escriben normalmente como
\begin{equation} %52
\frac{d^{2}\Phi}{d\phi^{2}} + m_{l}^{2}\Phi = 0
\end{equation}
\begin{equation} %53
\frac{1}{sen\theta}\frac{d}{d\theta}\left(sen\theta 
\frac{d\Theta}{d\theta}\right) + 
\left[l(l+1)-\frac{m_{l}^{2}}{sen^{2}\theta}\right]\Theta = 0
\end{equation}
\begin{equation} %54
\frac{1}{r^{2}}\frac{d}{dr}\left(r^{2}\frac{dR}{dr}\right) + 
\left[\frac{2m}{\hbar^{2}}\left(\frac{e^{2}}{4\pi \epsilon_{0}r} + 
E\right) - \frac{l(l+1)}{r^{2}}\right]R = 0
\end{equation}

Cada una de estas ecuaciones es una ecuaci\'on diferencial ordinaria de 
una funci\'on con una sola variable. Con ello se ha conseguido 
simplificar la ecuaci\'on de Schr\"odinger para el \'atomo de 
hidr\'ogeno que, al principio, era una ecuaci\'on diferencial parcial de 
una funci\'on $\psi$ de tres variables.

\section*{LOS N\'UMEROS CU\'ANTICOS}

\subsection{Soluci\'on Para La Parte Azimutal}
La ec. 52 se resuelve f\'acilmente para encontrar que su soluci\'on es
\begin{equation} %55
\Phi(\phi) = A_{\phi}e^{im_{l}\phi}
\end{equation}
donde $A_{\phi}$ es la constante de integraci\'on. Una de las condiciones 
establecidas previamente que debe cumplir una funci\'on de onda (y por lo 
tanto $\Phi$, que es una componente de la funci\'on completa $\psi$) es 
que tenga un \'unico valor para cada punto del espacio. Por ejemplo se 
observa que $\phi$ y $\phi + 2\pi$ se identifican en el mismo plano 
meridiano. Por tanto, debe ser cierto que $\Phi(\phi)= \Phi(\phi + 
2\pi)$, o bien, que $Ae^{im_{l}\phi} = Ae^{im_{l}(\phi + 2\pi)}$, lo que 
solamente puede ser cuando $m_{l}$ sea 0 o un n\'umero entero positivo o 
negativo $(\pm 1, \pm 2, \pm 3,...)$. La constante $m_{l}$ se conoce como 
el n\'umero cu\'antico magn\'etico del \'atomo de hidr\'ogeno y gobierna 
a la direcci\'on del momento angular $L$. El n\'umero cu\'antico 
magn\'etico $m_{l}$ est\'a determinado por el n\'umero cu\'antico orbital 
$l$ que a su vez determina la magnitud del momento angular del electr\'on. 

La interpretaci\'on del n\'umero cu\'antico orbital $l$ no es tan 
evidente. Examinemos la ec. 54, que corresponde a la parte radial $R(r)$ 
de la funci\'on de onda $\psi$. Esta ecuaci\'on est\'a relacionada 
\'unicamente con el aspecto radial del movimiento de los electrones, es 
decir, con el movimiento de aproximaci\'on y alejamiento de los mismos al 
n\'ucleo; sin embargo, est\'a presente en ella la energ\ii a total del 
electr\'on $E$. Esta energ\ii a incluye la energ\ii a cin\'etica del 
electr\'on en su movimiento orbital que no tiene nada que ver con el 
movimiento radial. Esta contradicci\'on se puede eliminar con el 
siguiente razonamiento: la energ\ii a cin\'etica $T$ del electr\'on tiene 
dos partes, $T_{radial}$ debido a su movimiento de aproximaci\'on y 
alejamiento del n\'ucleo, y $T_{orbital}$ debida a su movimiento 
alrededor de \'el. La energ\ii a potencial $V$ del electr\'on es la 
energ\ii a electrost\'atica dada por la ec. 39. Por lo tanto, la energ\ii a 
total del electr\'on es
\begin{equation}  %56
E = T_{radial} + T_{orbital} - \frac{e^{2}}{4\pi \epsilon_{0}r}
\end{equation}
Sustituyendo esta expresi\'on de $E$ en la ec. 54 obtenemos, despu\'es de 
reagrupar t\'erminos,
\begin{equation}  %57
\frac{1}{r^{2}}\frac{d}{dr}\left(r^{2}\frac{dR}{dr}\right) + 
\frac{2m}{\hbar^{2}}\left[T_{radial} + T_{orbital} - 
\frac{\hbar^{2}l(l+1)}{2mr^{2}}\right]R=0
\end{equation}
Si los dos \'ultimos t\'erminos entre corchetes de esta ecuaci\'on se 
anulan entre s\ii\, tenemos lo que necesit\'abamos: una ecuaci\'on 
diferencial para $R(r)$ constituida exclusivamente por funciones del 
radio vector. Por lo tanto, necesitamos que
\begin{equation}  %58
T_{orbital} = \frac{\hbar^{2}l(l+1)}{2mr^{2}}
\end{equation}
La energ\ii a cin\'etica orbital del electr\'on es
\begin{equation}  %59
T_{orbital} = \frac{1}{2}mv_{orbital}^{2}
\end{equation}
Puesto que el momento angular $L$ del electr\'on es
\begin{equation}  %60
L = mv_{orbital}r
\end{equation}
podemos expresar la energ\ii a cin\'etica orbital
\begin{equation}  %61
T_{orbital} = \frac{L^{2}}{2mr^{2}}
\end{equation}
Por lo tanto, en la ec. 58 tenemos
\begin{equation}  %62
\frac{L^{2}}{2mr^{2}} = \frac{\hbar^{2}l(l+1)}{2mr^{2}}
\end{equation}
lo que nos da
\begin{equation}  %63
L = \sqrt{l(l+1)}\hbar
\end{equation}
La interpretaci\'on de este resultado es que, puesto que el n\'umero 
cu\'antico orbital $l$ est\'a limitado a los valores $l=0,1,2,...,(n-1)$, 
el electr\'on puede tener solamente los momentos angulares $L$ que se 
especifican mediante la ec. 63. Al igual que la energ\ii a total $E$, el 
momento angular se conserva y est\'a cuantizado. El termino 
$\hbar=h/2\pi=1.054 \times 10^{-34}$J-s es la unidad natural del momento 
angular.

En el movimiento planetario macrosc\'opico, una vez m\'as, el n\'umero 
cu\'antico que describe el momento angular es tan grande que la 
separaci\'on en estados discretos del momento angular no se puede 
observar experimentalmente. Por ejemplo, un electr\'on (o para este caso, 
cualquier otro cuerpo) cuyo n\'umero cu\'antico orbital sea 2, tiene un 
momento angular $L=2.6 \times 10^{-34}$J-s. Por el contrario, el momento 
angular orbital de la tierra es !`$2.7 \times 10^{40}$J-s!

Se acostumbra designar a los estados del momento angular con la letra $s$ 
para $l=0$, con la letra $p$ cuando $l=1$, y as\ii\ sucesivamente.
Este original c\'odigo se origin\'o en la clasificaci\'on emp\ii rica de 
los espectros en las llamadas series que recibieron los nombres de 
definida, principal, difusa y fundamental, nombres que se les di\'o desde 
antes de que se desarrollara la teor\ii a del \'atomo. As\ii\, un estado 
$s$ es el que no tiene momento angular, un estado $p$ tiene el momento 
angular $\sqrt{2}\hbar$, etc.

La combinaci\'on del n\'umero cu\'antico total con la letra que 
representa al momento angular orbital proporciona una notaci\'on 
apropiada, y que es muy com\'un para los estados at\'omicos. En esta 
notaci\'on, por ejemplo un estado en el que $n=2$, $l=0$ es un estado 
$2s$ y uno en el que $n=4$, $l=2$ es un estado $4d$.

Por otro lado para la interpretaci\'on del n\'umero cu\'antico magn\'etico, 
tenemos que, al igual que el momento lineal, el momento angular es un 
vector, de modo que para describirlo se requiere que se especifique su 
direcci\'on, su sentido y su magnitud. (El vector $L$ es perpendicular al 
plano en el que tiene lugar el movimiento de rotaci\'on, y su direcci\'on 
y sentido est\'an dados por la regla de la mano derecha: cuando los dedos 
apuntan en la direcci\'on del movimiento, el pulgar tiene la direcci\'on 
y el sentido de $L$.)

?`Qu\'e significado posible pueden tener una direcci\'on y un sentido en el 
espacio para un \'atomo de hidr\'ogeno ? La respuesta es sencilla si
pensamos que un electr\'on que gira alrededor de un n\'ucleo es un 
diminuto circuito que, como dipolo magn\'etico, tiene tambi\'en un campo 
magn\'etico. En consecuencia, un electr\'on at\'omico que posee momento 
angular interact\'ua con un campo magn\'etico externo $B$. El n\'umero 
cu\'antico magn\'etico $m_{l}$ especifica la direcci\'on de $L$, 
determinando la componente de $L$ en la direcci\'on del campo. Este 
fen\'omeno se conoce com\'unmente con el nombre de cuantizaci\'on espacial.

Si hacemos que la direcci\'on del campo magn\'etico sea paralela al eje 
$z$, la componente de $L$ en esta direcci\'on es
\begin{equation} %64
L_{z} = m_{l}\hbar
\end{equation}
Los valores posibles de $m_{l}$ para un valor dado de $l$, van desde $+l$ 
hasta $-l$, pasando por 0, de modo que las posibles orientaciones del 
vector momento angular $L$ en un campo magn\'etico son $2l+1$. Cuando 
$l=0$, $L_{z}$ puede tener solamente el valor cero; cuando $l=1$, $L_{z}$ 
puede ser $\hbar$, 0, \'o $-\hbar$; cuando $l=2$, $L_{z}$ puede ser 
$2\hbar$, $\hbar$, 0, $-\hbar$, \'o $-2\hbar$, y as\ii\  sucesivamente. 
Aclaremos que $L$ nunca puede estar alineado exactamente (paralela o 
antiparalelamente) con $B$, ya que $L_{z}$ es siempre m\'as pequen\~o que 
la magnitud $\sqrt{l(l+1)}\hbar$ del momento angular total.

La cuantizaci\'on espacial del momento angular orbital del \'atomo de 
hidr\'ogeno se muestra en la fig. 6.1.

%%%%%%%%%%%%%%
\vskip 2ex
\centerline{
\epsfxsize=280pt
\epsfbox{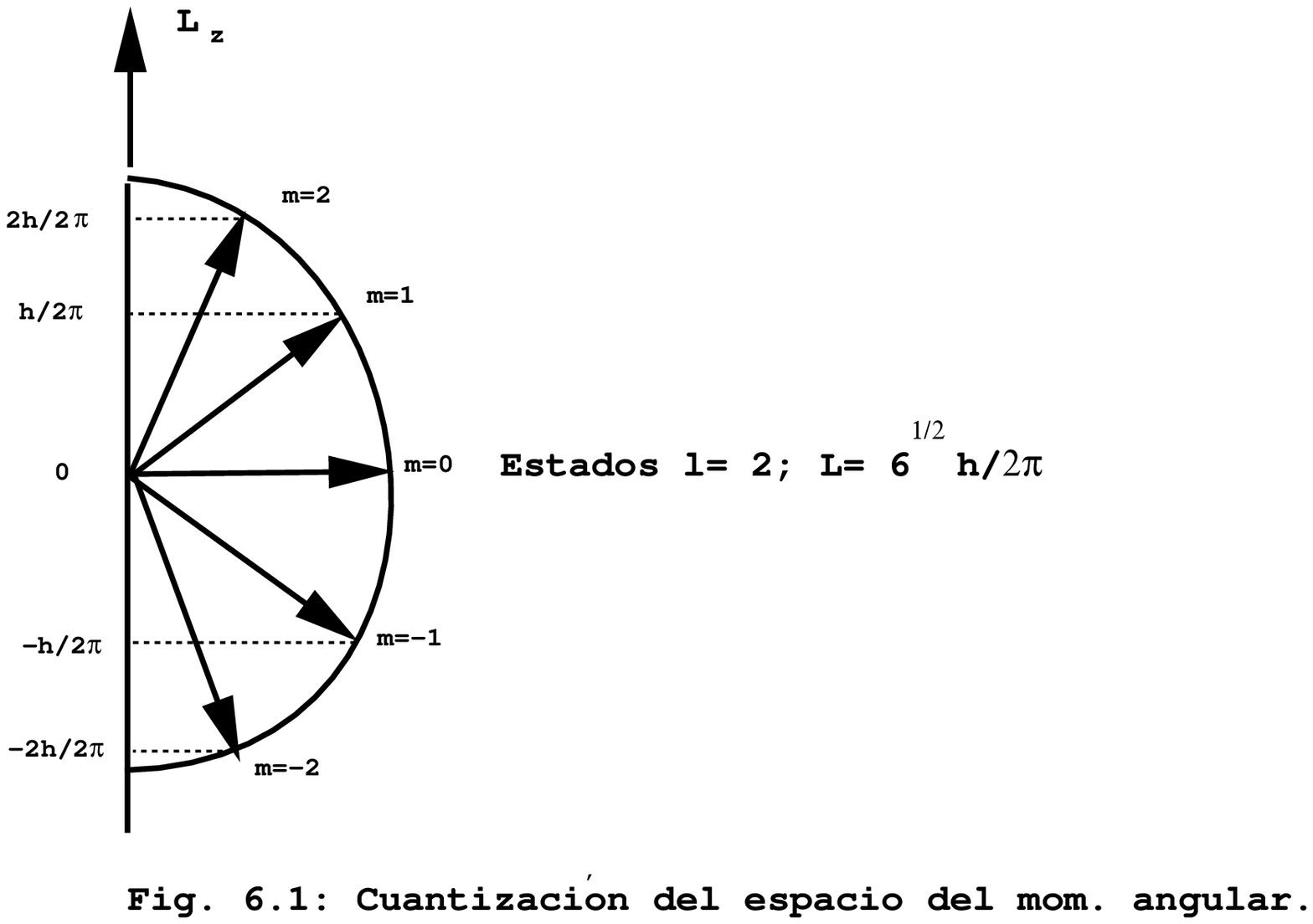}}
\vskip 4ex
%\begin{center}
%{\small{Fig. 1}\\
%}
%\end{center}
%%%%%%%%%%%%%%%%

 Debemos considerar al \'atomo 
caracterizado por un cierto valor de 
$m_{l}$ como preparado para tomar una determinada orientaci\'on de su 
momento angular $L$, relativo a un campo magn\'etico externo en el caso 
de encontrarse en \'el.

En ausencia de un campo magn\'etico externo, la direcci\'on del eje $z$ 
es completamente arbitraria. Por tanto, debe ser cierto que la componente 
de $L$ en cualquier direcci\'on que escojamos es $m_{l}\hbar$; el 
significado de un campo magn\'etico externo es que proporciona una 
direcci\'on de referencia importante experimentalmente. Un campo 
magn\'etico no es la \'unica direcci\'on de referencia posible. Por 
ejemplo, la l\ii nea entre los dos \'atomos $H$ en la mol\'ecula de 
hidr\'ogeno $H_{2}$ tiene tanto significado experimental como la 
direcci\'on de un campo magn\'etico y, a lo largo de esta l\ii nea, las 
componentes de los momentos angulares de los \'atomos de $H$ est\'an 
determinados por sus valores $m_{l}$.

?`Por qu\'e est\'a cuantizada \'unicamente la componente de $L$? La 
respuesta se relaciona estrechamente con el hecho de que $L$ nunca puede 
apuntar a cualquier direcci\'on $z$ espec\ii fica; en lugar de ello 
describe un cono en el espacio, de manera que su proyecci\'on $L_{z}$ es 
$m_{l}\hbar$. La raz\'on de este fen\'omeno es el principio de 
incertidumbre: si $L$ estuviera fijo en el espacio, de manera que 
$L_{x}$, $L_{y}$ y $L_{z}$ tuvieran valores definidos, el electr\'on 
estar\ii a confinado en un plano definido. Por ejemplo, si $L$ estuviera 
en la direcci\'on $z$, el electr\'on tendr\ii a que estar en el plano
$xy$ todo el tiempo (fig. 6.2a).

%%%%%%%%%%%%%%
\vskip 2ex
\centerline{
\epsfxsize=280pt
\epsfbox{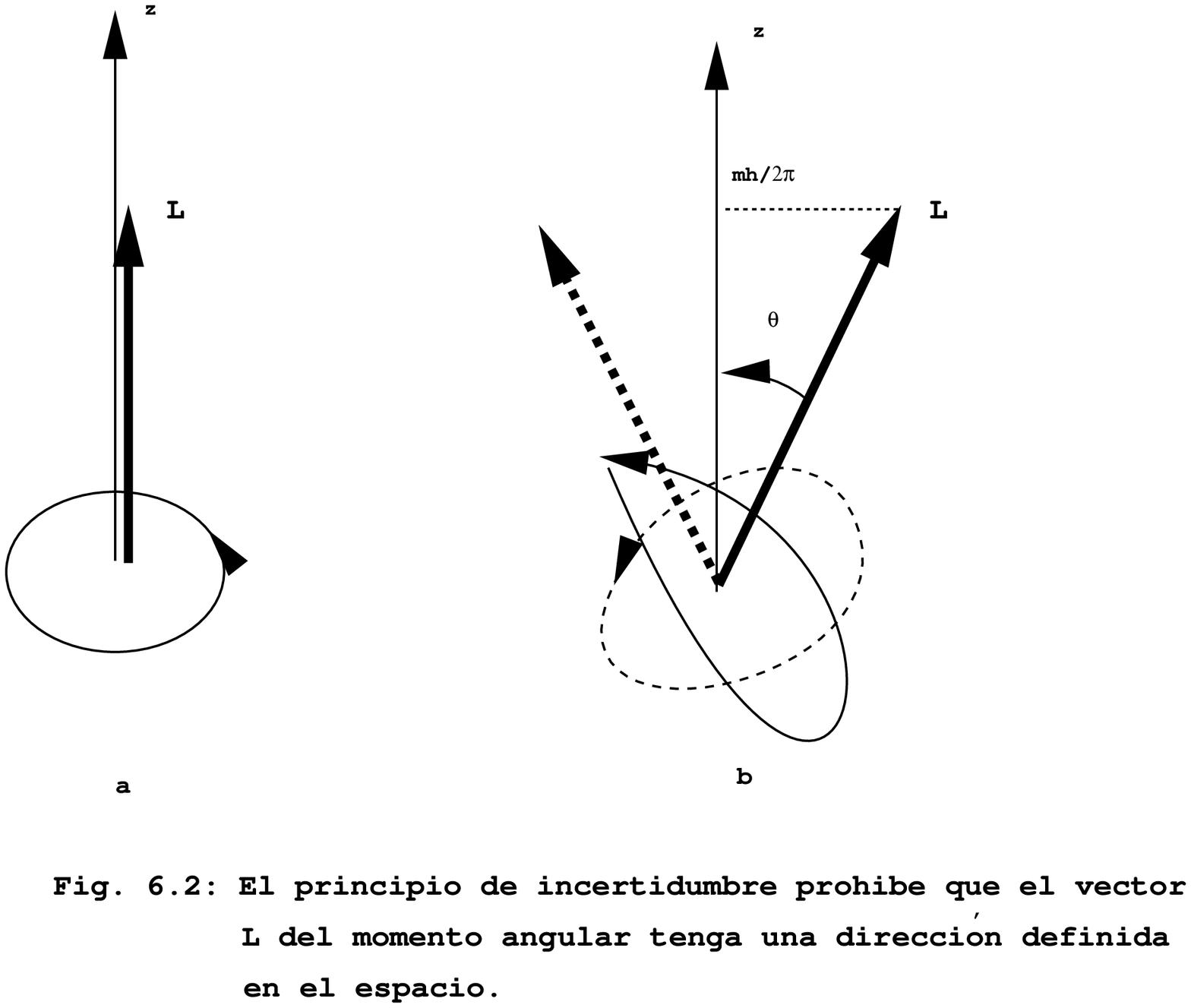}}
\vskip 4ex
%\begin{center}
%{\small{Fig. 1}\\
%}
%\end{center}
%%%%%%%%%%%%%%%%

 Esto \'unicamente puede ocurrir si la 
componente 
del momento del electr\'on $p_{z}$ en la direcci\'on $z$ es infinitamente 
incierta, lo que, por supuesto, es imposible si es parte de un \'atomo de 
hidr\'ogeno. Sin embargo, como en realidad \'unicamente una componente 
$L_{z}$  de $L$ junto con su magnitud $L$ tiene valores definidos y $\mid 
L \mid > \mid L_{z} \mid$, el electr\'on no est\'a limitado a un plano 
\'unico (fig. 6.2b), y si as\ii\ fuera, habr\ii a una fundada incertidumbre 
en la coordenada $z$ del electr\'on. La direcci\'on de $L$ cambia 
constantemente (fig. 6.3) y as\ii\ los valores promedio de $L_{x}$ y $L_{y}$ 
son 0, aunque $L_{z}$ tenga siempre el valor espec\ii fico $m_{l}\hbar$.

%%%%%%%%%%%%%%
\vskip 2ex
\centerline{
\epsfxsize=280pt
\epsfbox{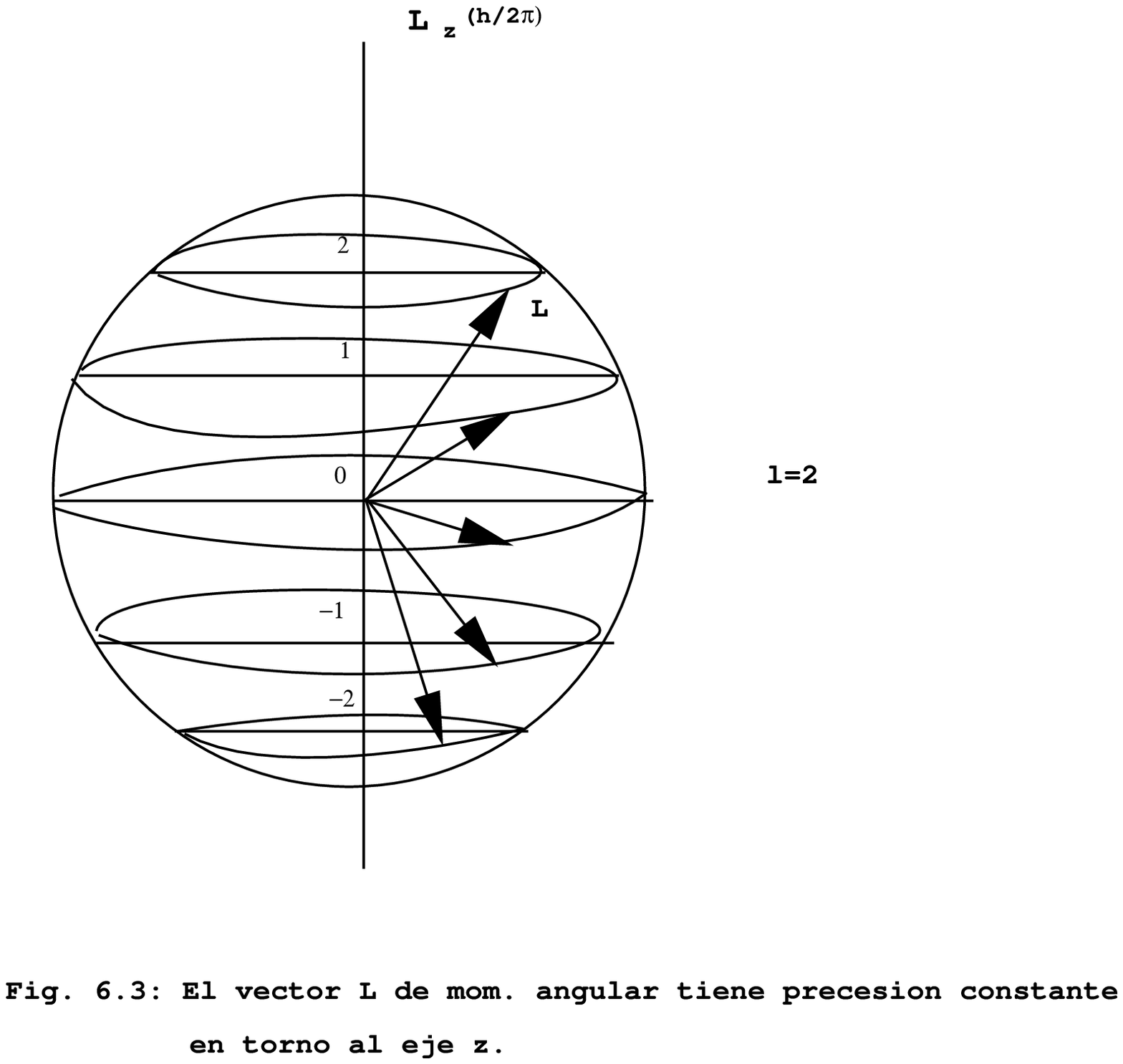}}
\vskip 4ex
%\begin{center}
%{\small{Fig. 1}\\
%}
%\end{center}
%%%%%%%%%%%%%%%%

La soluci\'on para $\Phi$ tambi\'en debe cumplir con la 
condici\'on de 
normalizaci\'on la cual est\'a dada por la ec. 2, entonces para $\Phi$ tenemos
\begin{equation} %65
\int_{0}^{2\pi} \mid \Phi \mid^{2}d\phi = 1
\end{equation}
al sustituir $\Phi$ se tiene
\begin{equation} %66
\int_{0}^{2\pi} A_{\phi}^{2}d\phi = 1
\end{equation}
con lo cual se tiene que $A_{\phi}=1/\sqrt{2\pi}$ y por lo tanto $\Phi$ ya 
normalizada est\'a dado por
\begin{equation} %67
\Phi(\phi) = \frac{1}{\sqrt{2\pi}}e^{im_{l}\phi}
\end{equation}

\subsection*{Soluci\'on Para La Parte Polar}
La ecuaci\'on diferencial 53 para $\Theta(\theta)$ tiene una soluci\'on 
m\'as complicada y est\'a dada por los polinomios asociados de Legendre
\begin{equation} %68
P_{l}^{m_{l}}(x) = 
(-1)^{m_{l}}(1-x^{2})^{m_{l}/2} 
\frac{d^{m_{l}}}{dx^{m_{l}}}P_{l}(x) = 
(-1)^{m_{l}}\frac{(1-x^{2})^{m_{l}/2}}{2^{l}l!}\frac{d^{m_{l} + 
l}}{dx^{{m_{l} + l}}}(x^{2} - 1)^{l}
\end{equation}
estas funciones cumplen con la relaci\'on de ortogonalidad
\begin{equation} %69
\int_{-1}^{1} [P_{l}^{m_{l}}(cos\theta)]^{2}dcos\theta = 
\frac{2}{2l+1}\frac{(l+m_{l})!}{(l-m_{l})!}
\end{equation}
Ahora, $\Theta(\theta)$, que es la soluci\'on para la ec. 53, est\'a dada 
por los polinomios de Legendre normalizados, esto es, si
\begin{equation} %70
\Theta(\theta) = A_{\theta}P_{l}^{m_{l}}(cos\theta)
\end{equation}
entonces la condici\'on de normalizaci\'on est\'a dada por
\begin{equation} %71
\int_{-1}^{1} A_{\theta}^{2}[P_{l}^{m_{l}}(cos\theta)]^{2}dcos\theta = 1
\end{equation}
por lo tanto la constante de normalizaci\'on para la parte polar es
\begin{equation} %72
A_{\theta} = \sqrt{\frac{2l+1}{2} \frac{(l-m_{l})!}{(l+m_{l})!}}
\end{equation}
y por consiguiente, la funci\'on $\Theta(\theta)$ ya normalizada es
\begin{equation} %73
\Theta(\theta) = 
\sqrt{\frac{2l+1}{2}\frac{(l-m_{l})!}{(l+m_{l})!}} P_{l}^{m_{l}}(cos\theta)
\end{equation}

Para nuestro prop\'osito, lo m\'as importante de estas funciones es que, 
como ya se dijo anteriormente, existen solamente cuando la constante $l$ 
es un n\'umero entero igual o 
mayor que $\mid m_{l}\mid$, que es el valor absoluto de $m_{l}$. Esta 
exigencia se puede expresar como una condici\'on de $m_{l}$ en la forma 
\begin{equation} %74
m_{l} = 0,\pm 1, \pm 2,...,\pm l 
\end{equation}

\subsection*{Arm\'onicos Esf\'ericos}
Las soluciones para la parte azimutal y polar se pueden juntar para 
formar los llamados arm\'onicos esf\'ericos, estos dependen de $\phi$ y 
$\theta$ y de alguna forma hacen m\'as f\'acil el manejo de la funci\'on de 
onda completa $\psi(r,\theta,\phi)$. Los arm\'onicos esf\'ericos est\'an dados 
de la siguiente manera:
\begin{equation} %75
Y_{l}^{m_{l}}(\theta,\phi) = (-1)^{m_{l}} \sqrt{\frac{2l+1}{4\pi} 
\frac{(l-m_{l})!}{(l+m_{l})!}} P_{l}^{m_{l}}(cos\theta)e^{im_{l}\phi}
\end{equation} 
El factor $(-1)^{m_{l}}$ que se ha introducido sin ningun problema debido a 
que la ecuaci\'on de Schr\"odinger es lineal y homog\'enea y adem\'as de que 
particularmente es conveniente para el estudio del momento angular. Este 
factor es un factor fase llamado fase Condon-Shortley y el efecto es para 
introducir una alternancia de signo.

\subsection*{Soluci\'on Para La Parte Radial}
La soluci\'on de la ecuaci\'on final, ec. 54, para la parte radial $R(r)$ 
de la funci\'on de onda $\psi$ del \'atomo de hidr\'ogeno tambi\'en es 
complicada, y viene dada por los polinomios asociados de Laguerre. La ec. 
54 s\'olo se puede resolver cuando E es positivo o tiene uno de los 
valores negativos $E_{n}$ (lo que significa que el electr\'on est\'a unido 
al \'atomo), dados por
\begin{equation} %76
E_{n} = 
-\frac{m 
e^{4}}{32\pi^{2}\epsilon_{0}^{2}\hbar^{2}}\left(\frac{1}{n^{2}}\right) 
\end{equation}
donde $n$ es un n\'umero entero y se conoce como n\'umero cu\'antico 
principal y describe la cuantizaci\'on de la energ\ii a del electr\'on en 
el \'atomo de hidr\'ogeno. Esta ecuaci\'on es la que 
obtuvo Bohr para los niveles de energ\ii a del \'atomo de hidr\'ogeno.

Otra condici\'on que se debe cumplir para resolver la ec. 54, es que $n$, 
conocido como n\'umero cu\'antico principal, sea igual o mayor que $l+1$. 
Esto se puede expresar como una condici\'on para $l$ en la forma
\begin{equation} %77
l = 0,1,2,...,(n-1) 
\end{equation}

La ec. 54 tambi\'en se puede poner de la siguiente forma
\begin{equation} %78
r^{2}\frac{d^{2}R}{dr^{2}} + 2r\frac{dR}{dr} + \left[\frac{2m 
E}{\hbar^{2}}r^{2} + \frac{2me^{2}}{4\pi \epsilon_{0} \hbar^{2}}r - 
l(l+1)\right]R = 0
\end{equation}
y su soluci\'on est\'a dada por los polinomios asociados de Laguerre los 
cuales cumplen con la siguiente condici\'on de normalizaci\'on
\begin{equation} %79
\int_{0}^{\infty}e^{-\rho}\rho^{2l}[L_{n+l}^{2l+1}(\rho)]^{2}\rho^{2}d\rho = 
\frac{2n[(n+l)!]^{3}}{(n-l-1)!}
\end{equation}
Por lo tanto la soluci\'on para la ec. 78, que corresponde a la parte 
radial, es:
\begin{equation} %80
R(r) = 
-\sqrt{\frac{(n-l-1)!}{2n[(n+l)!]^{3}}}
\left(\frac{2}{na_{0}}\right)^{3/2}
e^{-\rho /2}\rho^{l} L_{n+l}^{2l+1}(\rho)
\end{equation}
donde $\rho=2r/na_{0}$ y $a_{0}=\hbar^{2}/me^{2}$.

Ahora que ya tenemos las soluciones de cada una de las ecuaciones que 
s\'olo dependen de una variable, ya podemos construir nuestra funci\'on de 
onda para cada estado del electr\'on en el \'atomo de hidr\'ogeno, esto 
es si tenemos que $\psi(r,\theta,\phi)=R(r)\Theta(\theta)\Phi(\phi)$, 
entonces la funci\'on de onda completa es
$$%81
\psi(r,\theta,\phi)=-\sqrt{\frac{2l+1}{4\pi}\frac{(l-m_{l})!}{(l+m_{l})!} 
\frac{(n-l-1)!}{2n[(n+l)!]^{3}}} 
\left(\frac{2}{na_{0}}\right)^{3/2}\times
$$
\begin{equation} %81
(\alpha r)^{l} 
e^{-\alpha r/2} L_{n+l}^{2l+1}(\alpha r) 
P_{l}^{m_{l}}(cos\theta)e^{im_{l}\phi}
\end{equation}
donde $\alpha=2/na_{0}$.

Utilizando los arm\'onicos esf\'ericos nuestra soluci\'on queda de la 
siguiente manera
\begin{equation}  %82
\psi(r,\theta,\phi)=-\sqrt{\frac{(n-l-1)!}{2n[(n+l)!]^{3}}}
\left(\frac{2}{na_{0}}\right)^{3/2}(\alpha r)^{l}
e^{-\alpha r/2} L_{n+l}^{2l+1}(\alpha r)Y_{l}^{m_{l}}(\theta,\phi)
\end{equation}

Esta es la soluci\'on a la ecuaci\'on de Schr\"odinger para el 
\'atomo de hidr\'ogeno, la cual describe cada uno de los estados del 
electr\'on. Esta funci\'on de onda por si sola no tiene interpretaci\'on 
f\ii sica como se dijo anteriormente, pero el cuadrado de su valor absoluto 
$\mid \psi \mid^{2}$ calculado para un punto y en un instante determinado 
es proporcional a la probabilidad de encontrar experimentalmente al 
electr\'on ah\ii\ y en ese instante.
%En la tabla 1 se dan las funciones
%de onda normalizadas del \'atomo de hidr\'ogeno para $n=1,2,3$.

\section*{LA DENSIDAD DE PROBABILIDAD ELECTR\'ONICA}
En el modelo de Bohr del \'atomo de hidr\'ogeno, el electr\'on gira 
alrededor del n\'ucleo con una trayectoria circular. Si se realizara un 
experimento adecuado, se ver\ii a que el electr\'on estar\ii a siempre a 
una distancia del n\'ucleo $r=n^{2}a_{0}$ (donde $n$ es el n\'umero 
cu\'antico de la \'orbita y $a_{0}=0.53$ \AA es el radio de la \'orbita 
m\'as pr\'oxima al n\'ucleo) y en el plano ecuatorial $\theta=90$, 
mientras que el \'angulo azimutal $\phi$ var\ii a con el tiempo.

La teor\ii a cu\'antica del \'atomo de hidr\'ogeno modifica las 
conclusiones del modelo de Bohr en dos aspectos. En primer lugar, no se 
pueden dar valores correctos de $r,\theta,\phi$, sino \'unicamente 
probabilidades relativas de encontrar al electr\'on en un lugar dado. 
Esta imprecisi\'on es, por supuesto, una consecuencia de la 
naturaleza ondulatoria del electr\'on. En segundo lugar, no se puede 
pensar que el electr\'on se mueve alrededor del n\'ucleo en un sentido 
convencional, ya que la densidad de probabilidad $\mid \psi \mid^{2}$ es 
independiente del tiempo y puede variar considerablemente de un lugar a 
otro. 

La funci\'on de onda del electr\'on $\psi$ en un \'atomo de hidr\'ogeno 
viene dada por $\psi=R\Theta\Phi$ donde $R=R_{nl}(r)$ describe c\'omo 
var\ii a $\psi$ con $r$ cuando los n\'umeros cu\'anticos orbital y total 
tienen los valores $n$ y $l$; $\Theta=\Theta_{lm_{l}}(\theta)$ describe a 
su vez la variaci\'on de $\psi$ con $\theta$ cuando los n\'umeros 
cu\'anticos magn\'etico y orbital tienen los valores $l$ y $m_{l}$; y 
$\Phi=\Phi_{m_{l}}(\phi)$ que proporciona la variaci\'on de $\psi$ con 
$\phi$ cuando el n\'umero cu\'antico magn\'etico es $m_{l}$. Entonces, la 
densidad de probabilidad $\mid \psi \mid^{2}$ se puede escribir como
\begin{equation}  %83
\mid \psi \mid^{2} = \mid R \mid^{2} \mid \Theta \mid^{2} \mid \Phi \mid^{2}
\end{equation}
donde se comprende que si la funci\'on es compleja, hay que tener en 
cuenta que su cuadrado se debe sustituir por el producto de ella y su 
conjugada compleja.

La densidad de probabilidad $\mid \Phi \mid^{2}$, que mide la posibilidad 
de encontrar al electr\'on con un \'angulo azimutal $\phi$ dado, es una 
constante que no depende para nada de $\phi$. Por lo tanto, la densidad 
de probabilidad del electr\'on es sim\'etrica respecto al eje de las $z$, 
independientemente del estado cu\'antico, de manera que el electr\'on 
tiene igual oportunidad de encontrarse en un \'angulo $\phi$ como en 
otro.  

La parte radial $R$ de la funci\'on de onda, en contraste con $\Phi$, no 
solamente var\ii a con $r$, sino que lo hace de una manera diferente para 
cada combinaci\'on de n\'umeros cu\'anticos $n$ y $l$. La fig. 6.4 muestra 
gr\'aficas de $R$ en funci\'on de $r$ para los estados $1s, 2s$,
y $2p$ del \'atomo de hidr\'ogeno. Evidentemente, $R$ es m\'aximo al
ser $r=0$
(esto es, en el n\'ucleo mismo) para todos los estados $s$, mientras que 
es cero en $r=0$ para todos los estados que poseen momento angular.

%%%%%%%%%%%%%%
\vskip 2ex
\centerline{
\epsfxsize=280pt
\epsfbox{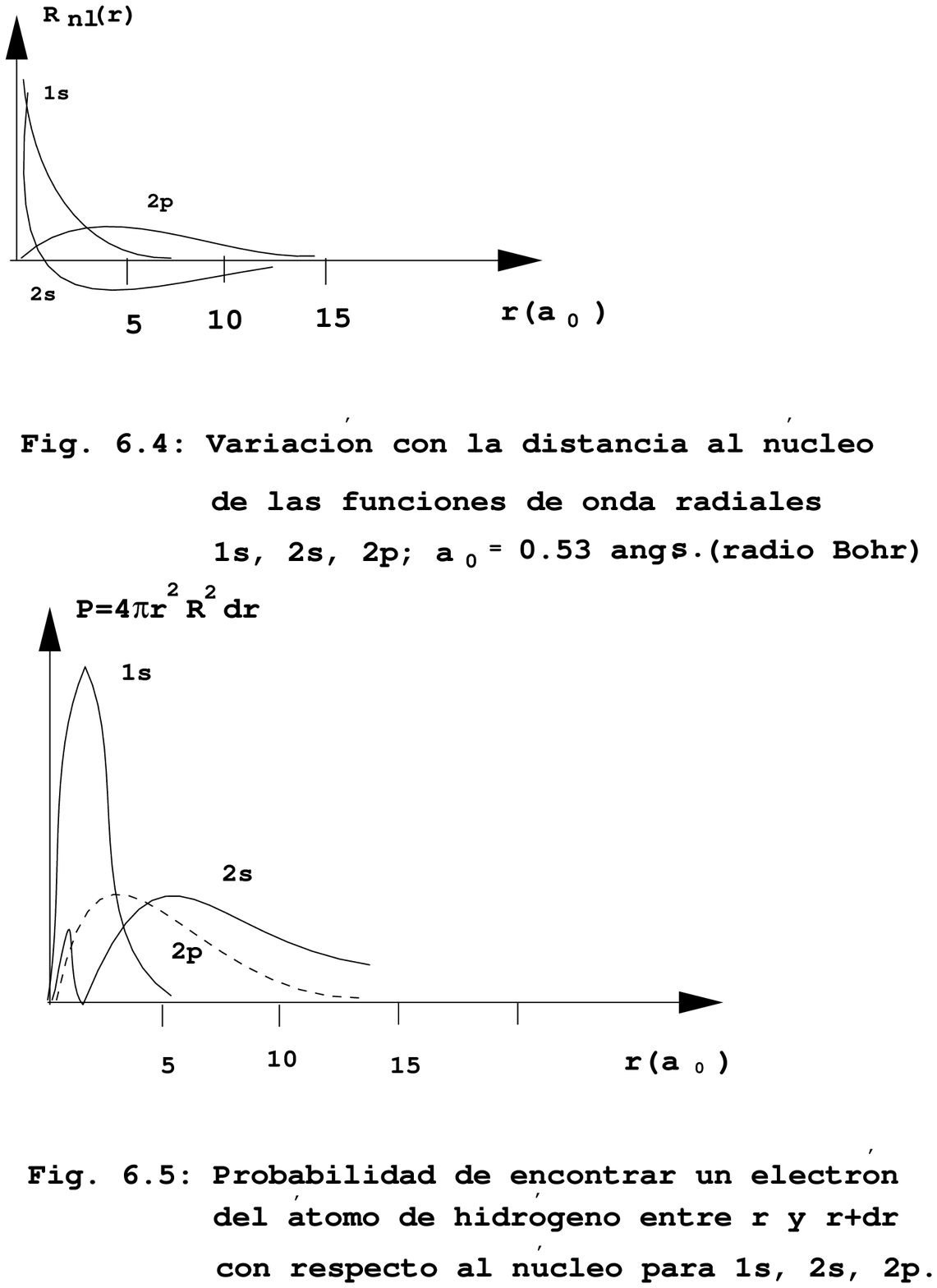}}
\vskip 4ex
%\begin{center}
%{\small{Fig. 1}\\
%}
%\end{center}
%%%%%%%%%%%%%%%%

La densidad de probabilidad del electr\'on en el punto $r,\theta,\phi$ es 
proporcional a $\mid \psi \mid^{2}$, pero la probabilidad real de 
encontrarlo en el elemento de volumen infinitesimal $dV$ es 
$\mid \psi \mid^{2}dV$. Ahora, en coordenadas polares esf\'ericas
\begin{equation}  %84
dV=r^{2}sen\theta dr d\theta d\phi
\end{equation}
de manera que, como $\Theta$ y $\Phi$ son funciones normalizadas, la 
probabilidad num\'erica real $P(r)dr$ de encontrar al electr\'on en el 
\'atomo de hidr\'ogeno, a una distancia comprendida entre $r$ y $r+dr$ 
del n\'ucleo, es
\begin{eqnarray}  %85
P(r)dr & = & r^{2}\mid R \mid^{2}dr \int_{0}^{\pi} 
\mid\ \Theta \mid^{2} sen\theta d\theta \int_{0}^{2\pi} 
\mid\ \Phi \mid^{2}d\phi \nonumber\\
& = & r^{2}\mid R \mid^{2}dr
\end{eqnarray}
Esta ecuaci\'on est\'a representada en la fig. 6.5 para los mismos estados 
cuyas funciones radiales $R$ aparecen en la fig. 6.4; en principio, las
curvas son completamente diferentes. Observamos de inmediato que $P$ no 
es m\'aximo en el n\'ucleo para los estados $s$, como lo es $R$, sino que 
tiene su m\'aximo a una distancia finita de \'el. El valor m\'as probable 
de $r$ para un electr\'on $1s$ es exactamente $a_{0}$, que es el radio de 
la \'orbita del electr\'on en estado fundamental en el modelo de Bohr. 
Sin embargo, el valor medio de $r$ para un electr\'on $1s$ es $1.5a_{0}$, 
lo cual parece enigm\'atico a primera vista, ya que los niveles de 
energ\ii a son los mismos en mec\'anica cu\'antica y en el modelo 
at\'omico de Bohr. Esta aparente discrepancia se elimina cuando se tiene 
en cuenta que la energ\ii a del electr\'on depende de $1/r$ y no 
directamente de $r$, y el valor medio de $1/r$ para un electr\'on $1s$ es 
precisamente $1/a_{0}$.

La funci\'on $\Theta$ var\ii a con el \'angulo polar $\theta$ para todos 
los n\'umeros cu\'anticos $l$ y $m_{l}$, excepto para $l=m_{l}=0$, que 
son estados $s$. La densidad de probabilidad $\mid\ \Theta \mid^{2}$ para 
un estado $s$, es una constante (1/2), lo que significa que, como 
$\mid \Phi \mid^{2}$ es tambi\'en constante, la densidad de probabilidad 
electr\'onica $\mid \psi \mid^{2}$ tiene el mismo valor, para un valor de 
$r$ dado, en todas las direcciones. Los electrones en otros estados 
tienen preferencias angulares que algunas veces llegan a ser muy 
complicadas. Esto se puede observar en la fig. 6.5, donde se muestran, para 
varios estados at\'omicos, las densidades de probabilidad electr\'onica 
en funci\'on de $r$ y $\theta$. (El t\'ermino que se representa es 
$\mid \psi \mid^{2}$, y no $\mid \psi \mid^{2}dV$.) Puesto que 
$\mid \psi \mid^{2}$ es independiente de $\phi$, una representaci\'on 
tridimensional de $\mid \psi \mid^{2}$ se obtiene haciendo girar una 
representaci\'on particular alrededor de un eje vertical. Al hacerlo, las 
densidades de probabilidad para los estados $s$ tienen evidentemente 
simetr\ii a esf\'erica, mientras que las otras no la tienen. Los tipos de 
l\'obulos pronunciados, caracter\ii sticos de muchos de los estados, 
tienen importancia en qu\ii mica ya que estos modelos ayudan a determinar 
la manera como interact\'uan en las mol\'eculas los \'atomos adyacentes.
\\
\\
{\bf Notas}:

1. E. Schr\"odinger consigi\'o el premio Nobel en 1933 (junto con Dirac)
por ``el descubrimiento de nuevas formas productivas de la teor\'{\i}a
\'atomica". Schr\"odinger escribi\'o una serie notable de cuatro
art\'{\i}culos intitulada ``Quantisierung als Eigenwertproblem" (I-IV,
recibidos por Annalen der Physik el 27 de Enero, el 23 de Febrero, el
10 de Mayo y el 21 de Junio de 1926).

\newpage
\centerline{P r o b l e m a s}
  
{\bf Problema 6.1} - Obtener la ecuaciones para las \'orbitas estables
y para los niveles de energ\ii a del electr\'on en el \'atomo de hidr\'ogeno.

{\bf Respuesta}: Tenemos que la longitud de onda del electr\'on est\'a dada por
\begin{eqnarray}
\lambda = \frac{h}{mv}\nonumber
\end{eqnarray}
mientras que por otro lado al igualar la fuerza el\'ectrica con la 
fuerza centr\ii peta, esto es
\begin{eqnarray}
\frac{mv^{2}}{r} = \frac{1}{4\pi \epsilon_{0}}\frac{e^{2}}{r^{2}}\nonumber
\end{eqnarray}
nosotros obtenemos que la velocidad del electr\'on est\'a dada por
\begin{eqnarray}
v = \frac{e}{\sqrt{4\pi \epsilon_{0} mr}}\nonumber
\end{eqnarray}
entonces la longitud de onda del electr\'on es
\begin{eqnarray}
\lambda = \frac{h}{e}\sqrt{\frac{4\pi \epsilon_{0}r}{m}}\nonumber
\end{eqnarray}
Ahora, si damos el valor $5.3 \times 10^{-11}$m al radio $r$ de la 
\'orbita electr\'onica, vemos que la longitud de onda del electr\'on es 
$\lambda=33 \times 10^{-11}$m. Esta longitud de onda tiene exactamente el 
mismo valor que la circunferencia de la \'orbita del electr\'on, $2\pi 
r=33 \times 10^{-11}$m. Como se puede ver, la \'orbita de un electr\'on 
en un \'atomo de hidr\'ogeno corresponde as\ii\ a una onda completa 
cerrada sobre s\ii\ misma. Esto se puede comparar con las vibraciones de 
un anillo de alambre, si las longitudes de onda est\'an en un n\'umero 
entero de veces en su circunferencia este podr\'a seguir vibrando 
indefinidamente, pero si un n\'umero no entero de longitudes de onda 
tiene lugar sobre el anillo se producir\'a una interferencia negativa a 
medida que las ondas se desplacen en torno a \'el, y las vibraciones 
desaparecer\'an r\'apidamente. Con esto se puede afirmar que un 
electr\'on puede girar indefinidamente alrededor de un n\'ucleo sin 
irradiar energ\ii a con tal que su \'orbita contenga un n\'umero entero 
de longitudes de la onda de De Broglie. Con esto tenemos que la 
condici\'on de estabilidad es
\begin{eqnarray}
n\lambda = 2\pi r_{n}\nonumber
\end{eqnarray}
donde $r_{n}$ designa el radio de la \'orbita que contiene $n$ longitudes 
de onda. Al sustituir $\lambda$ tenemos
\begin{eqnarray}
\frac{nh}{e}\sqrt{\frac{4\pi \epsilon_{0}r_{n}}{m}} = 2\pi r_{n}\nonumber
\end{eqnarray}
por lo tanto las \'orbitas estables del electr\'on son
\begin{eqnarray}
r_{n} = \frac{n^{2}\hbar^{2}\epsilon_{0}}{\pi me^{2}}\nonumber
\end{eqnarray}
\\

Para los niveles de energ\ii a tenemos que $E=T+V$, al sustituir las 
energ\ii as potencial y cin\'eticas obtenemos
\begin{eqnarray}
E = \frac{1}{2}mv^{2} - \frac{e^{2}}{4\pi \epsilon_{0}r}\nonumber
\end{eqnarray}
o lo que es igual
\begin{eqnarray}
E_{n} = -\frac{e^{2}}{8\pi \epsilon_{0}r_{n}}\nonumber
\end{eqnarray}
al sustituir el valor de $r_{n}$ en esta \'ultima ecuaci\'on obtenemos
\begin{eqnarray}
E_{n} = 
-\frac{me^{4}}{8\epsilon_{0}^{2}\hbar^{2}} 
\left(\frac{1}{n^{2}}\right)\nonumber 
\end{eqnarray}
lo que nos da los niveles de energ\ii a.
\\
\\

{\bf Problema 6.2} - El teorema de Uns\"old dice que, para cualquier valor
del n\'umero cu\'antico orbital $l$, las densidades de probabilidad, sumadas
para todos los estados posibles, desde $m_{l}=-l$ hasta $m_{l}=+l$ da una 
constante independiente de los \'angulos $\theta$ o $\phi$ esto es
\begin{eqnarray}
\sum_{m_{l}=-l}^{+l} \mid \Theta_{lm_{l}} \mid^{2} \mid \Phi_{m_{l}} 
\mid^{2} = cte.\nonumber 
\end{eqnarray}

Este teorema significa que todo \'atomo o ion con subcapa cerrada tiene 
una distribuci\'on sim\'etrica esf\'erica de carga el\'etrica. Comprobar 
el teorema de Uns\"old para $l=0$, $l=1$ y $l=2$.
%con ayuda de la tabla 1.

{\bf Respuesta}: Tenemos que para $l=0$, $\Theta_{00}=1/\sqrt{2}$ y
$\Phi_{0}=1/\sqrt{2\pi}$ por lo tanto del teorema de Uns\"old vemos que
\begin{eqnarray}
\mid \Theta_{0,0} \mid^{2} \mid \Phi_{0} \mid^{2} = \frac{1}{4\pi}\nonumber
\end{eqnarray}

Para $l=1$ tenemos que
\begin{eqnarray}
\sum_{m_{l}=-1}^{+1} \mid \Theta_{lm_{l}} \mid^{2} 
\mid \Phi_{m_{l}} \mid^{2} = \mid \Theta_{1,-1} \mid^{2} \mid \Phi_{-1} 
\mid^{2} + \mid \Theta_{1,0} \mid^{2} \mid \Phi_{0} \mid^{2} + \mid 
\Theta_{1,1} \mid^{2} \mid \Phi_{1} \mid^{2}\nonumber
\end{eqnarray}
por otro lado las funciones de onda son: 
$\Theta_{1,-1}=(\sqrt{3}/2)sen\theta$, 
$\Phi_{-1}=(1/\sqrt{2\pi})e^{-i\phi}$, 
$\Theta_{1,0}=(\sqrt{6}/2)cos\theta$, $\Phi_{0}=1/\sqrt{2\pi}$, 
$\Theta_{1,1}=(\sqrt{3}/2)sen\theta$, $\Phi_{1}=(1/\sqrt{2\pi})e^{i\phi}$
que al sustituirlas en la ecuaci\'on anterior nos queda
\begin{eqnarray}
\sum_{m_{l}=-1}^{+1} \mid \Theta_{lm_{l}} \mid^{2} 
\mid \Phi_{m_{l}} \mid^{2} = \frac{3}{8\pi}sen^{2}\theta + 
\frac{3}{4\pi}cos^{2}\theta + \frac{3}{8\pi}sen^{2}\theta = 
\frac{3}{4\pi}\nonumber 
\end{eqnarray}
lo que demuestra que tambi\'en es una constante.

Para $l=2$ tenemos que
\begin{eqnarray}
\sum_{m_{l}=-2}^{+2} \mid \Theta_{lm_{l}} \mid^{2}
\mid \Phi_{m_{l}} \mid^{2} =\nonumber
\end{eqnarray}
\begin{eqnarray}
\mid \Theta_{2,-2} \mid^{2} \mid \Phi_{-2} \mid^{2}
\mid \Theta_{2,-1} \mid^{2} \mid \Phi_{-1} \mid^{2} 
+ \mid \Theta_{2,0} \mid^{2} \mid \Phi_{0} \mid^{2} 
+ \mid \Theta_{2,1} \mid^{2} \mid \Phi_{1} \mid^{2}
+ \mid \Theta_{2,2} \mid^{2} \mid \Phi_{2} \mid^{2}\nonumber
\end{eqnarray}
y las funciones de onda son:
$\Theta_{2,-2}=(\sqrt{15}/4)sen^{2}\theta$, 
$\Phi_{-2}=(1/\sqrt{2\pi})e^{-2i\phi}$, 
$\Theta_{2,-1}=(\sqrt{15}/2)sen\theta cos\theta$,
$\Phi_{-1}=(1/\sqrt{2\pi})e^{-i\phi}$,
$\Theta_{2,0}=(\sqrt{10}/4)(3cos^{2}\theta-1)$,
$\Phi_{0}=1/\sqrt{2\pi}$,
$\Theta_{2,1}=(\sqrt{15}/2)sen\theta cos\theta$,
$\Phi_{1}=(1/\sqrt{2\pi})e^{i\phi}$,
$\Theta_{2,2}=(\sqrt{15}/4)sen^{2}\theta$,
$\Phi_{2}=(1/\sqrt{2\pi})e^{2i\phi}$,
sustituyendo en la ecuaci\'on anterior nos queda
\begin{eqnarray}
\sum_{m_{l}=-2}^{+2} \mid \Theta_{lm_{l}} \mid^{2}
\mid \Phi_{m_{l}} \mid^{2} = \frac{5}{4\pi}\nonumber
\end{eqnarray}
con lo que queda demostrado el teorema de Uns\"old.
\\
\\

{\bf Problema 6.3} - La probabilidad de encontrar un electr\'on at\'omico
cuya
funci\'on de onda radial sea $R(r)$, fuera de una esfera de radio $r_{0}$ 
centrada en el n\'ucleo, es
\begin{eqnarray}
\int_{r_{0}}^{\infty} \mid R(r) \mid^{2}r^{2}dr\nonumber 
\end{eqnarray}
La funci\'on de onda $R_{10}(r)$ corresponde al estado fundamental de un 
\'atomo de hidr\'ogeno, y $a_{0}$ es el radio de la \'orbita de Bohr 
correspondiente a este estado. Calcular la probabilidad de encontrar un 
electr\'on en estado fundamental en un \'atomo de hidr\'ogeno a una 
distancia del n\'ucleo mayor que $a_{0}$.

{\bf Respuesta}: Tenemos que la funci\'on de onda radial que corresponde al
estado fundamental es
\begin{eqnarray}
R_{10}(r) = \frac{2}{a_{0}^{3/2}}e^{-r/a_{0}}\nonumber
\end{eqnarray}
sustituyendo en la integral nos queda

\begin{eqnarray}
\int_{a_{0}}^{\infty} \mid R(r) \mid^{2}r^{2}dr = 
\frac{4}{a_{0}^{3}} \int_{a_{0}}^{\infty} r^{2} e^{-2r/a_{0}}dr\nonumber
\end{eqnarray}
o lo que es igual
\begin{eqnarray}
\int_{a_{0}}^{\infty} \mid R(r) \mid^{2}r^{2}dr =
\frac{4}{a_{0}^{3}}\left[-\frac{a_{0}}{2}r^{2}e^{-2r/a_{0}}  
-\frac{a_{0}^{2}}{2}re^{-2r/a_{0}}
-\frac{a_{0}^{3}}{4}e^{-2r/a_{0}}\right]_{a_{0}}^{\infty}\nonumber
\end{eqnarray}
esto nos da como resultado al evaluar
\begin{eqnarray}
\int_{a_{0}}^{\infty} \mid R(r) \mid^{2}r^{2}dr = \frac{5}{e^{2}}\nonumber
\end{eqnarray}
que es la probabilidad de encontrar al electr\'on.

%\end{document}

\newpage
%%%%%%%%%%%%%%%%%%%%%%%%%%%%%%%%%%%%%%%%%%%%%%%
%\protect
%\setcounter{equation}
\begin{center}{\huge 7. DISPERSI\'ON EN LA MC}
\end{center}
%\author{\it Daniel Jim\'enez Alvarez}
%\date{}
%\maketitle
%%%%%%%%%%%%%%%%%%%%%%%%%%%%%%%%%%%%%%%%%%%%%%%%%%%%%%%%%%%%%%%%%%%%

\section*{\bf Introducci\'on}
%%%%%%%%%%%%%%%%%%%%%%%%%%%%%%
\setcounter{equation}{0}
Para la teor\'{\i}a cu\'antica de dispersi\'on nos ayudaremos de los
resultados ya conocidos de la dispersi\'on en campos de fuerzas centrales, y
asumiremos ciertas situaciones que simplificar\'an los c\'alculos, si bien
no nos alejar\'an demasiado del problema ``real". Sabemos que en el
an\'alisis experimental de una colisi\'on podemos obtener datos que nos
ayuden a entender el estado de la materia ``blanco", o bien, la
interacci\'on entre el haz incidente y el ``blanco".  Las hip\'otesis que
asumiremos son:

i) Asumiremos que las part\'{\i}culas no tienen esp\'{\i}n, lo cual 
{\em no} implica que \'este no sea importante en la dispersi\'on.

ii) Nos ocuparemos s\'olo de la dispersi\'on el\'astica, en
la cual no consideramos la posible estructura interna de las
part\'{\i}culas.

iii) Asumiremos que el blanco es lo suficientemente
delgado como para no permitir la dispersi\'on m\'ultiple.

iv) Asumiremos que las interacciones son descritas por un potencial que
depende s\'olo de la posici\'on relativa de las part\'{\i}culas.

As\'{\i}, usaremos los resultados ya conocidos de la teor\'{\i}a de
dispersi\'on de la mec\'anica cl\'asica, ah\'{\i} se define:

\begin{equation}
\frac{d\sigma}{d\Omega}=\frac{I(\theta,\varphi)}{I_{0}}
\end{equation}

\noindent
donde $d\sigma$ es el elemento de \'angulo s\'olido, $I_{0}$ es
el n\'umero de part\'{\i}culas incidentes por unidad de \'area, e
$I{}d\Omega$ el n\'umero de part\'{\i}culas dispersadas en el elemento de
\'angulo s\'olido. 

Con esto, llegamos a que:

\begin{equation}
\frac{d\sigma}{d\Omega}=\frac{\rho}{\sin\theta}\vert 
\frac{d\rho}{d\theta}\vert
\end{equation}

Si deseamos conocer en t\'erminos de mec\'anica cu\'antica lo que sucede,
debemos estudiar la evoluci\'on en el tiempo de un paquete de ondas. 
Sea ahora $F_{i}$ el flujo de part\'{\i}culas del haz incidente, es 
decir, el n\'umero de part\'{\i}culas por unidad de tiempo que atraviesan 
una superficie unitaria perpendicular al eje. Colocamos un detector lejos
de la regi\'on de influencia del potencial, y que subtiende un \'angulo 
s\'olido $d\Omega$; con esto, podemos contar el n\'umero de 
part\'{\i}culas $dn$ dispersadas por unidad de tiempo en $d\Omega$ en la 
direcci\'on $(\theta,\varphi)$.

%%%%%%%%%%%%%%
\vskip 2ex
\centerline{
\epsfxsize=280pt
\epsfbox{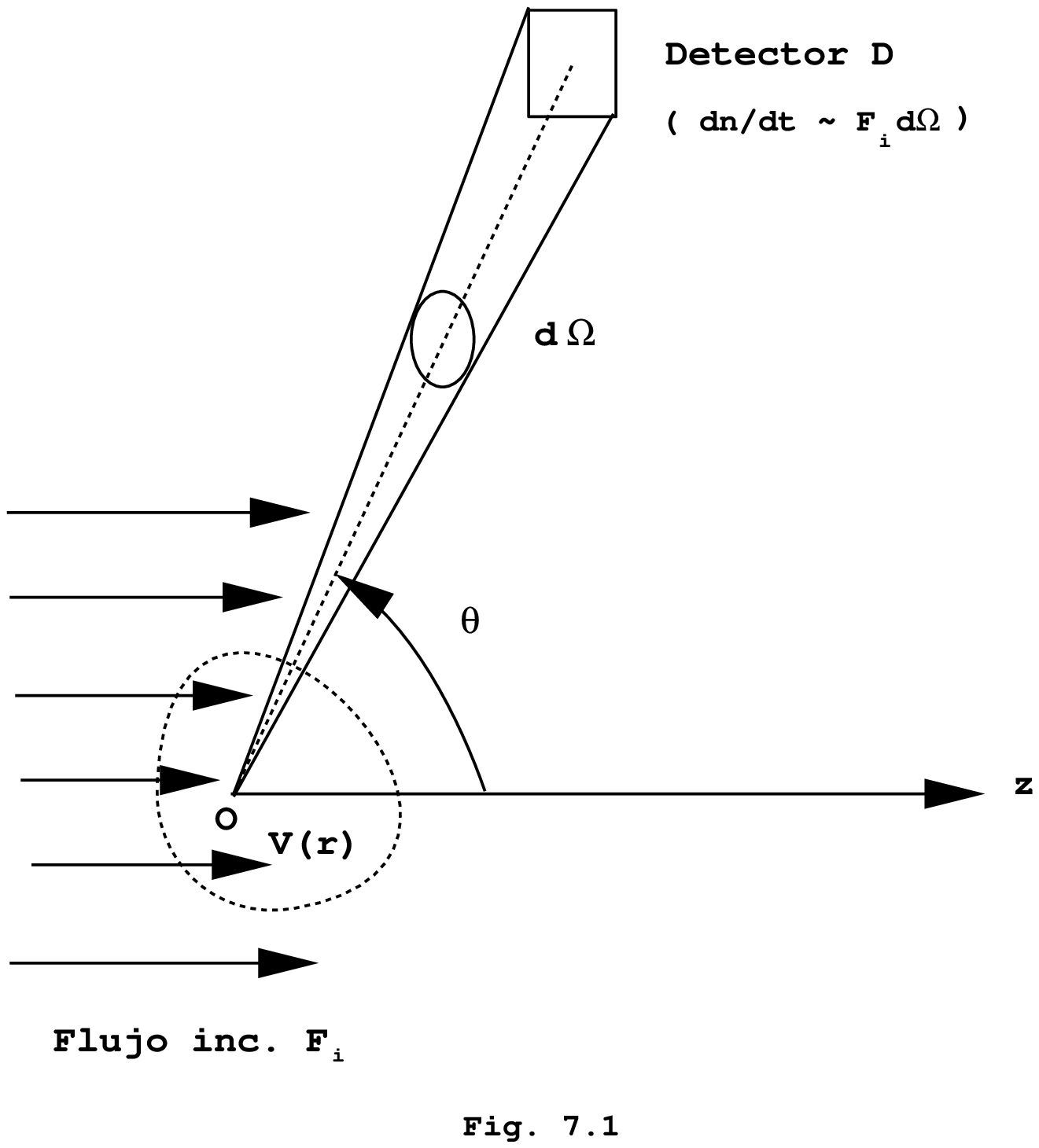}}
\vskip 4ex
%\begin{center}
%{\small{Fig. x}\\
%}
%\end{center}
%%%%%%%%%%%%%%%%

$dn$ es proporcional a $d\Omega$ y a
$F_{i}$. Llamaremos $\sigma (\theta,\varphi)$ al coeficiente de 
proporcionalidad entre $dn$ y $F_{i} d\Omega$:
\begin{equation}
 dn=\sigma (\theta,\varphi)F_{i} d\Omega~,
\end{equation}

esta es la llamada secci\'on diferencial transversal.

El n\'umero de part\'{\i}culas por unidad de tiempo que alcanzan el 
detector es igual al n\'umero de part\'{\i}culas que cruzar\'{\i}an una 
superficie $\sigma (\theta,\varphi) d\Omega$ colocada perpendicular al eje 
del haz. La secci\'on total de dispersi\'on se define por:
\begin{equation}
 \sigma=\int \sigma (\theta,\varphi) d\Omega 
\end{equation}

Ya que podemos orientar los ejes de coordenadas a nuestra elecci\'on, lo 
haremos de tal forma que el eje del haz incidente de part\'{\i}culas 
coincida con, digamos, el eje z (por simplicidad en los c\'alculos, donde 
usaremos coordenadas esf\'ericas). \\
En la regi\'on negativa del eje, para $t$ negativo grande, la 
part\'{\i}cula ser\'a pr\'acticamente libre; no se ver\'a afectada por 
$V({\bf r})$, y su estado se puede representar por ondas planas, por lo 
tanto la funci\'on de onda debe contener t\'erminos de la forma 
$e^{ikz}$, donde $k$ es la constante que aparece en la ecuaci\'on de 
Helmholtz. Por analog\'{\i}a con \'optica, la forma de la onda dispersada 
es:
\begin{equation}
 f(r)= \frac{e^{ikr}}{r} 
\end{equation}

en efecto, pues:
\begin{equation}
 (\nabla ^{2} + k^{2})e^{ikr} \neq 0
\end{equation} 

y
\begin{equation} 
(\nabla ^{2} + k^{2}) \frac{e^{ikr}}{r}=0
\end{equation}
para $r>r_{0}$, donde $r_{0}$ es cualquier n\'umero positivo.

Asumamos que el movimiento de la part\'{\i}cula est\'a descrito por el 
hamiltoniano:\\

\begin{equation}
H=\frac{{\rm \bf p^2}}{2\mu}+V=H_{0}+V
\end{equation}

V es diferente de cero s\'olo para una peque\~na vecindad del origen. 
Sabemos la evoluci\'on y descripci\'on de un paquete de ondas en $t=0$:

\begin{equation}
 \psi({\bf{r}},0)=\frac{1}{(2\pi)^\frac{3}{2}}\int 
\varphi({\rm \bf k})\exp[i{\rm \bf k\cdot (r-r_{0})}]
{\rm {\bf d^{3}k}}
\end{equation}

\noindent
donde, como ya se sabe, $\psi$ es una funci\'on de ancho 
$\Delta {\rm \bf k}$,
centrada alrededor de ${\rm \bf k_{0}}$. Asumimos tambi\'en
que ${\rm \bf k_{0}}$ es
paralela a ${\rm \bf r_{0}}$ pero en direcci\'on opuesta.
Para averiguar qu\'e sucede con el paquete de onda--que es lo que nos 
interesa--cuando en un tiempo posterior el paquete ha chocado con el 
blanco y ha sido dispersado por \'este, podemos valernos de la 
expansi\'on de $\psi({\rm \bf r},0)$ en t\'erminos de las eigenfunciones
$\psi_{n}({\rm \bf r})$, de $H$. As\'{\i}, podemos expandir:
$\psi({\bf{r}},0)=\sum_{n}c_{n}\psi_{n}(\bf{r})$. De esta forma, el paquete
de onda al tiempo $t$ es:
\begin{equation}
\psi({\bf 
r},t)=\sum_{n}c_{n}\varphi _{n}({\bf r})\exp(-\frac{i}{\hbar}E_{n}t)~.
\end{equation}

Esta es una eigenfunci\'on del operador $H_{0}$ y no de $H$, pero
podemos sustituir estas eigenfunciones por eigenfunciones 
particulares de $H$, que designaremos por $\psi_{k}^{(+)}(\bf{r})$. La 
forma asint\'otica de estas \'ultimas son de la forma:
\begin{equation}
\psi _{k}^{(+)}(\bf{r})\simeq e^{i\bf{k\cdot r}} +
f({\rm \bf r})\frac{e^{ikr}}{r}~,
\end{equation}

donde, como es usual,
${\rm \bf p}=\hbar {\rm \bf k}$ y  $E=\frac{\hbar ^{2}k^{2}}{2m}$.

Esto corresponde a una onda plana para el haz 
incidente, y otra onda esf\'erica divergente, como resultado de la 
interacci\'on entre el haz y el blanco. Estas soluciones de la ecuaci\'on
de Schr\"odinger existen en realidad, y podemos expandir
$\psi ({\rm \bf r},0)$ en
t\'erminos de ondas planas o en t\'erminos de $\psi _{k}({\rm \bf r})$:
\begin{equation}
\psi({\rm \bf r},0)=\int \varphi ({\rm \bf k})\exp(-i{\rm \bf k\cdot
r_{0}})\psi _{{\rm \bf k}}({\rm \bf r}) d^{3}k
\end{equation}

donde $ \hbar\omega= \frac{\hbar^{2}k^{2}}{2m}$.

Esto puede verse entonces como que la onda esf\'erica divergente no tiene 
contribuci\'on alguna al paquete de ondas inicial.

\section*{\bf Dispersi\'on}
%%%%%%%%%%%%%%%%%%%%%%%%%%%

Al viajar la onda, necesariamente sufre una dispersi\'on, de manera que 
aqu\'{\i} no podemos ya pasar por alto el efecto de la onda divergente, 
pero podemos usar lo siguiente:

\begin{equation} 
\omega= 
\frac{\hbar}{2m}k^{2}= 
\frac{\hbar}{2m}[{\bf k_{0}+(k-k_{0}})]^{2}= 
\frac{\hbar}{2m}[2{\bf k_{0}\cdot k - k_{0}^{2}+ (k-k_{0})^{2}}]~,
\end{equation}
  
\noindent para poder despreciar el \'ultimo t\'ermino del par\'entesis, al 
sustituir $\omega$ en la relaci\'on para $\psi$, requerimos que: 
\( \frac{\hbar}{2m}({\bf k-k_{0}})^{2}T \ll 1 \),
donde $T \simeq \frac{2mr_{0}}{\hbar k_{0}}$, con lo que obtenemos:

\begin{equation}
\frac{(\Delta k)^{2}r_{0}}{k_{0}} \ll 1 
\end{equation}

\noindent Esta condici\'on implica que el paquete de onda no se dispersa de 
manera apreciable al desplazarse una distancia macrosc\'opica $r_{0}$.

Escogiendo la direcci\'on del vector $\bf{k}$ de la onda incidente alineado
con el sistema de coordenadas esf\'ericas, podemos escribir:

\( \psi_{k}(r,\theta,\varphi) \simeq e^{ikz} + 
\frac{f(k,\theta,\varphi)e^{ikr}}{r} \) \\

Ya que $H$, el operador hamiltoniano (que hemos considerado hasta ahora 
no como operador, pero cuyos resultados son los mismos) es invariante 
ante rotaciones en el eje z, podemos escoger condiciones de frontera que 
tambi\'en sean invariantes, de manera que:

\( \psi_{k}(r,\theta,\varphi)\simeq e^{ikz}+\frac{f(\theta)e^{ikr}}{r}\)\\

\noindent estas se conocen como funciones de onda de choque. El coeficiente 
$f(\theta)$ se conoce como amplitud de choque.\\

\section*{\bf Amplitud de probabilidad}
%%%%%%%%%%%%%%%%%%%%%%%%%%%%%%%%%%%%%%%%

Recordemos ahora la ecuaci\'on de Schr\"odinger a resolver:
\begin{equation}
 i\hbar \frac{\partial\psi}{\partial t}= - \frac{\hbar^{2}}{2m} 
\nabla^{2}\psi + V({\bf r},t)\psi 
\end{equation}

Y, tenemos
\begin{equation}
P({\bf r},t)= \psi^{*}({\bf r},t)\psi ({\bf r},t)=\vert \psi ({\bf 
r},t) \vert ^{2} 
\end{equation}

\noindent se interpreta, de acuerdo a Max Born, como una densidad de 
probabilidad. Esta funci\'on de onda debe estar normalizada de manera tal
que:
\begin{equation}
\int \vert \psi ({\rm \bf r},t) \vert ^{2}  d^{3}r = 1~.
\end{equation}

El coeficiente de normalizaci\'on de $\psi$, y la integral de 
normalizaci\'on deben ser independientes 
del tiempo, si ha de cumplir con la ecuaci\'on de Schr$\ddot{o}$dinger. 
Esto lo podemos notar de la siguiente manera:
\begin{equation}
I= \frac{\partial}{\partial t} \int _{\Omega} P({\rm \bf r},t) d^{3}r=
\int_{\Omega} (\psi^{*}\frac{\partial\psi}{\partial t}
+\frac{\partial\psi^{*}}{\partial t}\psi) d^{3}r 
\end{equation}

\noindent de la ecuaci\'on de Schr\"odinger:
\begin{equation}
\frac{\partial\psi}{\partial t}= \frac{i\hbar}{2m}
\nabla ^{2}\psi-\frac{i}{\hbar}V({\bf r},t)\psi
\end{equation} 

\noindent entonces:
$$
I=\frac{i\hbar}{2m} \int_{\Omega}
[\psi^{*}\nabla^{2}-(\nabla^{2}\psi^{*})\psi]d^{3}r = \frac{i\hbar}{2m} 
\int_{\Omega} \nabla \cdot 
[\psi^{*}\nabla\psi-(\nabla\psi^{*})\psi]d^{3}r=
$$
\begin{equation}
=\frac{i\hbar}{2m} \int_{A}[\psi^{*}\nabla\psi-(\nabla\psi^{*})\psi]_{n}
dA 
\end{equation}

\noindent donde se ha usado el teorema de Green para evaluar la integral 
sobre el 
volumen. $A$ es la superficie que limita la regi\'on de integraci\'on y 
$[\quad]_{n}$ denota la componente en la direcci\'on normal a la superficie 
$dA$.

Definiendo:
\begin{equation}
 {\bf 
S}({\bf r},t)=\frac{\hbar}{2im} [\psi^{*}\nabla\psi-(\nabla\psi^{*})\psi] 
\end{equation}

obtenemos:
\begin{equation}
 I= \frac{\partial}{\partial t} \int_{\Omega} P({\bf r},t) d^{3}r= - 
\int _{\Omega} \nabla\cdot {\bf S} d^{3}r = -\int_{A} S_{n} dA
\end{equation}

\noindent para paquetes de onda en los que $\psi$ se hace cero a grandes 
distancias 
y la integral de normalizaci\'on converge, la integral de superficie es 
cero cuando $\Omega$ es todo el espacio. Se puede demostrar (v\'ease P. 
Dennery y A. Krzywicki, {\it Mathematical methods for physicists}) que la
integral de 
superficie es cero, de manera que la integral de normalizaci\'on es 
constante en el tiempo, y se cumple el requerimiento inicial. De la misma 
ecuaci\'on para ${\bf S}$ obtenemos:
\begin{equation}
 \frac{\partial P({\bf r},t)}{\partial t} + \nabla \cdot {\bf S}({\bf 
r},t)= 0 
\end{equation}

\noindent esto nos recuerda la ecuaci\'on de continuidad de 
electrodin\'amica, en 
este caso con un flujo de densidad $P$ y corriente de densidad ${\bf S}$, 
sin fuentes o sumideros. As\'{\i}, es razonable interpretar ${\bf S}$ como 
una densidad de corriente de probabilidad. Por semejanza con la 
electrodin\'amica, $\frac{\hbar}{im}\nabla$ es el operador velocidad y:
\begin{equation}
{\bf S}({\bf r}, t)= Re(\psi ^{*}\frac{\hbar}{im}\nabla\psi)
\end{equation}

\section*{\bf Funci\'on de Green en teor\'{\i}a de dispersi\'on}
%%%%%%%%%%%%%%%%%%%%%%%%%%%%%%%%%%%%%%%%%%%%%%%%%%%%%%%%%%%%%%%%%%%

Otra forma de escribir la ecuaci\'on de Schr\"odinger a resolver es
$(-\frac{\hbar^{2}}{2m} \nabla^{2} + V)\psi = E\psi $ o
$(\nabla^{2} + k^{2})\psi = U\psi $ donde:
$ k^{2}=\frac{2mE}{\hbar^{2}}$, y $U=\frac{2mV}{\hbar^{2}}$.

Resulta mas conveniente transformar esta ecuaci\'on a una de forma 
integral. Esto podemos hacerlo si consideramos a $U\psi$ del lado derecho 
de la ecuaci\'on como una inhomogeneidad, y de esta manera una soluci\'on 
de la ecuaci\'on se construye con las funciones de Green, por medio de:
\begin{equation} \label{eq:e1}
(\nabla^{2}+k^{2})G(\bf{r,r'}) =  
\delta(\bf{r-r'}) 
\end{equation} 

\noindent entonces, una soluci\'on a la ecuaci\'on de Schr\"odinger, se
obtendr\'a de la suma de la soluci\'on a la ecuaci\'on homog\'enea y la 
soluci\'on a la parte inhomog\'enea:
\begin{equation}
\psi(\bf{r})=\lambda(\bf{r})-\int
G(\bf{r,r'})U(\bf{r'})\psi(\bf{r'})d^{3}r'
\end{equation}

Buscamos una funci\'on $G$ que sea producto de funciones linealmente
independientes, como lo son las ondas planas:
\begin{equation}
G({\bf r,r'}=\int A({\bf q})e^{i{\bf q\cdot (r-r')}}dq
\end{equation}

\noindent usando la ecuaci\'on \ref{eq:e1}:
\begin{equation}
\int A({\bf q})(k^{2}-q^{2})e^{i{\bf q\cdot(r-r')}}dq=
\delta{\bf(r-r')}
\end{equation}
 
\noindent lo cual se cumple como identidad si:
\begin{equation}
A({\bf q})= (2\pi)^{-3}(k^{2}-q^{2})^{-1}
\end{equation}

\noindent de lo que resulta:

\begin{equation}
G({\bf r,r'})=\frac{1}{(2\pi)^{3}} \int 
\frac{e^{iqR}}{k^{2}-q^{2}}d^{3}q~,
\end{equation}
con $R=\vert {\bf r-r'} \vert$.
Tras un c\'alculo usando los m\'etodos de variable
compleja \footnote{V\'ease el problema 7.1}, llegamos a:
\begin{equation}
G(r)= - \frac{1}{4\pi} \frac{e^{ikr}}{r}
\end{equation}

Esta funci\'on no est\'a determinada de manera un\'{\i}voca; la 
funci\'on de
Green puede ser cualquier soluci\'on de la ecuaci\'on \ref{eq:e1} y existen
una infinidad de ellas; por lo tanto, la elecci\'on de una en particular
impone condiciones a la frontera sobre las eigenfunciones $\psi_{k}({\bf
r})$.

La funci\'on de Green obtenida es por lo tanto de la forma:
\begin{equation}
G({\bf r,r'})= -\left( \frac{e^{ik \vert {\bf r-r'} \vert}}{4\pi
\vert{\bf r-r'}\vert }\right)
\end{equation}

De esta forma, llegamos a la ecuaci\'on integral para la funci\'on de
onda de choque:
\begin{equation}
\psi (k,{\bf r})= \varphi (k,{\bf r}) - \frac{m}{2 \pi \hbar^{2}} \int
\frac{e^{ik \vert {\bf r-r'} \vert}}{ {\bf r-r'} } U({\bf r'}) \psi (k,{\bf 
r})d{\bf r}~,
\end{equation}
donde $\varphi$ es una soluci\'on de la ecuaci\'on de Helmholtz. Haciendo:
$ \vert {\bf r-r'} \vert = R $,
\begin{equation}
(\nabla^{2}+k^{2})\psi=(\nabla^{2}+k^{2})[\varphi + \int G({\bf r,r'}) 
U({\bf r'}) \psi({\bf r'}) d^{3}r']
\end{equation}

\noindent asumiendo que podemos intercambiar el orden y poner el operador 
$\nabla$ dentro de la integral: 
\begin{equation}
(\nabla^{2}+k^{2})\psi= \int (\nabla^{2}+k^{2}) G ({\bf r,r'}) U({\bf
r'}) \psi({\bf r'}) d^{3}r'= U({\bf r}) \psi ({\bf r})
\end{equation}
y se verifica entonces que
$G(R)= \frac{1}{4\pi} \frac{e^{ikR}}{R}$ es soluci\'on.

\section*{\bf Teorema \'optico}
%%%%%%%%%%%%%%%%%%%%%%%%%%%%%%%%
 
La secci\'on diferencial total est\'a dada por:
\begin{equation}
\sigma_{tot}(k)= \int \frac{d\sigma}{d\Omega} d\Omega
\end{equation}

Expresamos ahora $f(\theta)$ expresada en t\'erminos del corrimiento de 
fase $S_{l}(k)=e^{2i\delta_{l}(k)}$ de forma que:
\begin{equation}
f(\theta)=  \frac{1}{k} \sum_{l=0}^{\infty} (2l+1) e^{i\delta_{i}(k)} 
\sin \delta_{l}(k) P_{l}(\cos \theta)
\end{equation}

entonces
\begin{equation}
\sigma_{tot} = \int [\frac{1}{k} \sum_{l=0}^{\infty} (2l+1)
e^{i\delta_{l}(k)}\sin \delta_{l}(k) P_{l}(\cos \theta)][\int 
[\frac{1}{k} \sum_{l'=0}^{\infty} (2l'+1)e^{i\delta_{l'}(k)}\sin 
\delta_{l'}(k) P_{l'}(\cos \theta)]~.
\end{equation}
Usando ahora $\int P_{l}(\cos\theta)P_{l'}(\cos\theta)= \frac{4\pi}{2l+1}
\delta_{ll'}$ obtenemos
\begin{equation}
\sigma_{tot}= \frac{4\pi}{k^{2}} \sum_{l=0}^{\infty} (2l+1)\sin
\delta_{l}(k)^{2}~.
\end{equation}
Lo que aqu\'{\i} nos interesa es que:
\begin{equation}
{\rm Im} f(0)=\frac{1}{k} \sum_{l=0}^{\infty} (2l+1)
{\rm Im}[e^{i\delta_{l}(k)}\sin \delta_{l}(k)]P_{l}(1) =
\frac{1}{k} \sum _{l=0}^{\infty} (2l+1) \sin
\delta_{l}(k)^{2}=\frac{k}{4\pi} \sigma_{tot}
\end{equation}

Esta relaci\'on es conocida como el {\em teorema \'optico}. Su significado
f\'{\i}sico es que la interferencia de la onda incidente con la onda 
dispersada en \'angulo cero produce la salida de la part\'{\i}cula de la 
onda incidente, lo que permite la conservaci\'on de la probabilidad.

\section*{\bf Aproximaci\'on de Born} 
%%%%%%%%%%%%%%%%%%%%%%%%%%%%%%%%%%%%%%%%%%%

Consideremos la siguiente situaci\'on (Fig. 7.2).

%%%%%%%%%%%%%%
\vskip 2ex
\centerline{
\epsfxsize=120pt
\epsfbox{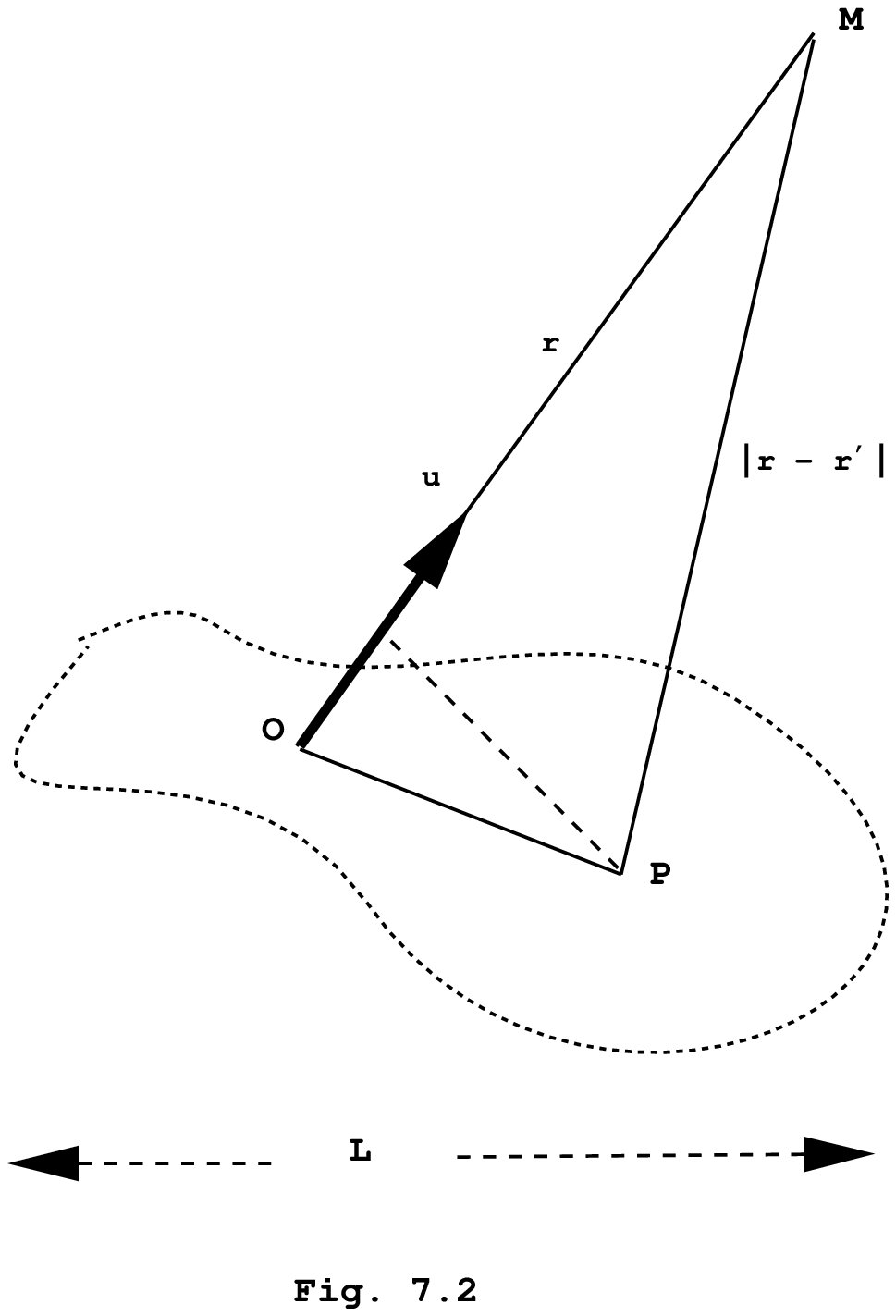}}
\vskip 2ex
%\begin{center}
%{\small{Fig. x}\\
%}
%\end{center}
%%%%%%%%%%%%%%%%
Nos situamos en un punto muy alejado de P, que corresponde a la regi\'on
de influencia del potencial $U$, y $r\gg L \quad r'\ll l$.
La longitud MP, que corresponde a $\vert {\bf r-r'} \vert$ es
aproximadamente igual a la proyecci\'on de MP en MO:
\begin{equation}
\vert {\bf r-r'} \vert \simeq r-{\bf u \cdot r'}
\end{equation}
\noindent donde ${\bf u}$ es el vector unitario en la direcci\'on ${\bf
r}$. Entonces, para $r$ grande:
\begin{equation}
G=- \frac{1}{4\pi} \frac{e^{ik \vert {\bf r-r'} \vert}}{\vert {\bf 
r-r'}\vert} \simeq_{r \rightarrow \infty}  -\frac{1}{4 \pi} 
\frac{e^{ikr}}{r} e^{-ik {\bf u \cdot r}}~.
\end{equation}

\noindent Sustituimos ahora en la expresi\'on integral para la funci\'on de
onda de choque, y obtenemos:
\begin{equation}
\psi({\bf r})= e^{ikz} - \frac{1}{4\pi} \frac 
{e^{ikr}}{r} 
\int e^{-ik {\bf u \cdot r}}U({\bf r'})\psi ({\bf r'}) 
d^{3}r'~.
\end{equation}

\noindent Esta ya no es una funci\'on de la distancia $r=OM$, sino
solamente de $\theta$ y $\psi$, y entonces:
\begin{equation}
f(\theta, \psi)= - \frac{1}{4\pi} \int e^{-ik {\bf u\ 
\cdot r}} U({\bf r'}) \psi ({\bf r'}) d^{3}r'
\end{equation}
Definimos ahora el vector de onda incidente ${\bf k_{i}}$
como un vector de m\'odulo $k$ dirigido a lo largo del 
eje polar del haz tal que:
$ e^{ikz}=e^{i {\bf k_{i} \cdot r}}$;
de manera similar, ${\bf k_{d}}$, con m\'odulo
$k$ y con direcci\'on fijada por $\theta$ y $\varphi$ se 
llama vector de onda desplazada en la direcci\'on 
$(\theta, \varphi)$:
$ {\bf k_{d}}= k{\bf u} $

El vector de onda transferido en la direcci\'on $(\theta, 
\varphi)$ se define como: ${\bf K}= {\bf k_{d}-k_{i}}$

%%%%%%%%%%%%%%
\vskip 1ex
\centerline{
\epsfxsize=80pt
\epsfbox{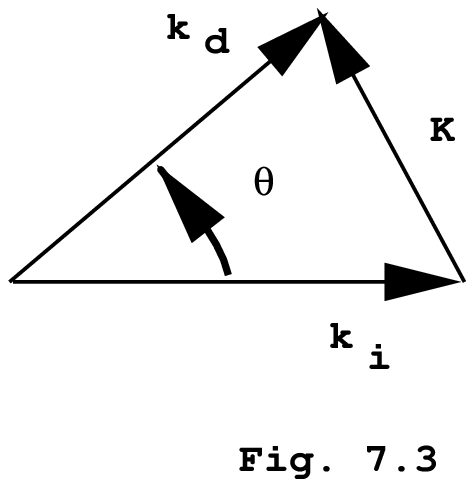}}
\vskip 2ex
%\begin{center}
%{\small{Fig. x}\\
%}
%\end{center}
%%%%%%%%%%%%%%%%

Con esto, podemos escribir la ecuaci\'on integral de dispersi\'on como:
\begin{equation}
\label{eq:e3}
\psi ({\bf r})= e^{i{\bf k_{i}\cdot r}} + \int
G({\bf r,r'}) U({\bf r'}) \psi({\bf r'}) d^{3}r'
\end{equation}

Ahora podemos intentar resolver esta ecuaci\'on por iteraci\'on. Hacemos 
el cambio ${\bf r} \rightarrow {\bf r'}; {\bf r'} \rightarrow {\bf r''}$ y 
con esto escribimos:
\begin{equation}
\psi ({\bf r'})= e^{i{\bf k_{i}\cdot r'}} + \int G({\bf r',r''}) U({\bf
r''}) \psi({\bf r''}) d^{3}r''
\end{equation}

Sustituyendo esta expresi\'on en \ref{eq:e3} obtenemos:
\begin{equation}
\label{eq:e4}
\psi({\bf r})= e^{i{\bf k}_{i}\cdot r} + \int G({\bf r,r'})U({\bf
r'})e^{i{\bf k_{i} \cdot r'}}d^{3}r'
+ \int \int G({\bf r,r'})U({\bf
r'})G({\bf r',r''})U({\bf r''}) \psi({\bf r''})d^{3}r'' d^{3}r'
\end{equation}

Los dos primeros t\'erminos de lado derecho de la ecuaci\'on son 
conocidos y solamente el tercero contiene la funci\'on desconocida 
$\psi({\bf r})$. Podemos repetir este procedimiento: cambiando ${\bf r}$ 
por ${\bf r''}$ y ${\bf r'}$ por ${\bf r'''}$ nos da $\psi ({\bf r''})$ 
la cual podemos reinsertar en la ecuaci\'on \ref{eq:e4} con esto obtenemos:
$$
\psi({\bf r}) = e^{i {\bf k_{i} \cdot r}} + \int G({\bf r,r'})U({\bf 
r'}) e^{i {\bf k_{i} \cdot r'}}
+\int \int G({\bf r,r'})U({\bf r'}) G({\bf r',r''})U({\bf r''})e^{i {\bf
k_{i} \cdot r''}}d^{3}r'd^{3}r''$$
\begin{equation}
+ \int \int \int  G({\bf r,r'})U({\bf r'}) G({\bf r',r''})U({\bf
r''})e^{i {\bf k_{i}\cdot r''}} G({\bf r'',r'''})U({\bf r'''}) \psi ({\bf 
r'''})~.
\end{equation}

\noindent Los primeros tres t\'erminos son conocidos; la funci\'on
desconocida 
$\psi({\bf r})$ se ha ido hasta el cuarto t\'ermino. De esta forma, por 
iteraci\'on construimos la funci\'on de onda de dispersi\'on 
estacionaria. N\'otese que cada t\'ermino de la expansi\'on lleva una 
potencia mayor del potencial que la que le precede. Podemos continuar de 
esta forma hasta obtener una expresi\'on despreciable del lado derecho, y 
obtenemos $\psi({\bf r})$ en t\'erminos de cantidades conocidas.

Sustituyendo la expresi\'on de $\psi({\bf r})$ en $f(\theta, \varphi)$ 
obtenemos la expansi\'on de Born de la amplitud de 
dispersi\'on. Limit\'andonos al primer orden en $U$, todo lo que hay que
hacer es sustituir $\psi({\bf r'})$ por $e^{i{\bf k_{i}\cdot r'}}$ en el 
lado derecho de la ecuaci\'on, con esto obtenemos:
%\begin{equation}
$$
f^{(B)}(\theta, \varphi)= \frac{-1}{4\pi}  \int e^{i{\bf k_{i}\cdot
r'}} U({\bf r'}) e^{-ik {\bf u\cdot r'}} d^{3}r'=
\frac{-1}{4\pi} \int e^{-i{\bf (k_{d}-k_{i})\cdot r'}} U({\bf r'})
d^{3}r'=
$$
\begin{equation}
 \frac{-1}{4\pi} \int e^{-i{\bf K \cdot r'}}U({\bf r'})d^{3}r'
 \end{equation}

${\bf K}$ es el vector de onda dispersada definido anteriormente. La
secci\'on de dispersi\'on se relaciona entonces de manera muy sencilla a
la transformada de Fourier del potencial, recordando:
$V({\bf r})= \frac{\hbar^{2}}{2m} U({\bf r})$, y
$\sigma (\theta,\varphi)= \vert f(\theta, \varphi) \vert^{2} \quad \)
tenemos que:
\begin{equation}
\sigma^{(B)} (\theta,\varphi)=\frac{m^{2}}{4\pi^{2}\hbar^{4}} \vert
\int e^{-i{\bf K \cdot r}} V({\bf r})d^{3}r \vert^{2}
\end{equation}

La direcci\'on y m\'odulo del vector de onda dispersada ${\bf K}$ depende 
del m\'odulo $k$ de ${\bf k_{i}}$ y ${\bf k_{d}}$ y de la direcci\'on de 
dispersi\'on $(\theta,\varphi)$. Para un $\theta$ y $\varphi$, la 
secci\'on eficaz var\'{\i}a con $k$, la energ\'{\i}a del haz incidente, y 
de manera an\'aloga, con una energ\'{\i}a dada $\sigma^{(B)}$ var\'{\i}a 
con $\theta$ y $\varphi$. Con esta aproximaci\'on de Born estudiando la 
variaci\'on de la secci\'on diferencial eficaz en t\'erminos de la 
direcci\'on de dispersi\'on y la energ\'{\i}a incidente nos d\'a 
informaci\'on del potencial $V({\bf r})$. \\

%atencion: para imprimir es: dvips blabla.dvi, para pasar a otro archivo
%que sea ps entonces es: dvips -o bla.dvi bla.ps, para escoger impresora 
%es la opcion -p %

\noindent {\bf Notas}: Uno de los primeros trabajos de dispersi\'on
cu\'antica es:

\noindent
M. Born, ``Quantenmechanik der Stossvorg\"ange",
Zf. f. Physik {\bf 37}, 863-867 (1926)

%%%%%%%%%%%%%%%%%%%%%%%%%%%%%%%
%\documentclass{article}
%\begin{document}
%%%%%%%%%%%%%%%%%%%%%%%%%%%%%%
\newpage
\centerline{P r o b l e m a s}

{\bf Problema 7.1}

\noindent{\bf C\'alculo por medio de variable compleja de la funci\'on de 
Green}

Recu\'erdese que obtuvimos:

\( G({\bf r,r'})=\frac{1}{(2\pi)^{3}} \int
\frac{e^{iqR}}{k^{2}-q^{2}}d^{3}q~, \)
con $R=\vert {\bf r-r'} \vert$.
Ya que $d^{3}q=q^{2} \sin\theta dq d\theta d\phi$, llegamos,
despu\'es de la integraci\'on angular, a que:

\( G({\bf r,r'})= \frac{i}{4\pi^{2}R}\int_{-\infty} ^{\infty}
\frac{(e^{-iqR}-e^{iqR})}{k^{2}-q^{2}} q dq~. \)

\noindent Hacemos:
$C=\frac{i}{4\pi^{2}R}$; y dividimos la integral en dos partes:

\( C(\int _{-\infty} ^{\infty} \frac{e^{-iqR}}{k^{2}-q^{2}} q dq -
\int _{-\infty} ^{\infty}\frac{e^{iqR}}{k^{2}-q^{2}} q dq)~. \)

\noindent Hacemos ahora $q \rightarrow -q$ en la primera integral:

\(  \int _{-\infty} ^{\infty} \frac{e^{-i(-q)R}}{k^{2}-(-q)^{2}} (-q)
d(-q)= \int _{\infty} ^{-\infty} \frac{e^{iqR}}{k^{2}-q^{2}} q dq
= -\int _{-\infty} ^{\infty} \frac{e^{iqR}}{k^{2}-q^{2}} q dq \)

\noindent por lo que:

\( G({\bf r,r'})= -2C ( \int_{-\infty} ^{\infty} \frac{qe^{iqR}}{k^{2}-q^{2}}
dq)\)

\noindent sustituyendo $C$,
%\( C= \frac{i}{2\pi^{2}}R \) \\
%\noindent 
obtenemos:

\( G({\bf r,r'})= \frac{-i}{2\pi^{2}R}\int_{-\infty} ^{\infty}
\frac{qe^{iqR}}{k^{2}-q^{2}}dq \)

Ahora, esta integral podemos evaluarla por los residuos que posee, usando
los m\'etodos de variable compleja. Notamos que existen polos simples en
$q=_{-}^{+}k$.

%%%%%%%%%%%%%%
\vskip 2ex
\centerline{
\epsfxsize=280pt
\epsfbox{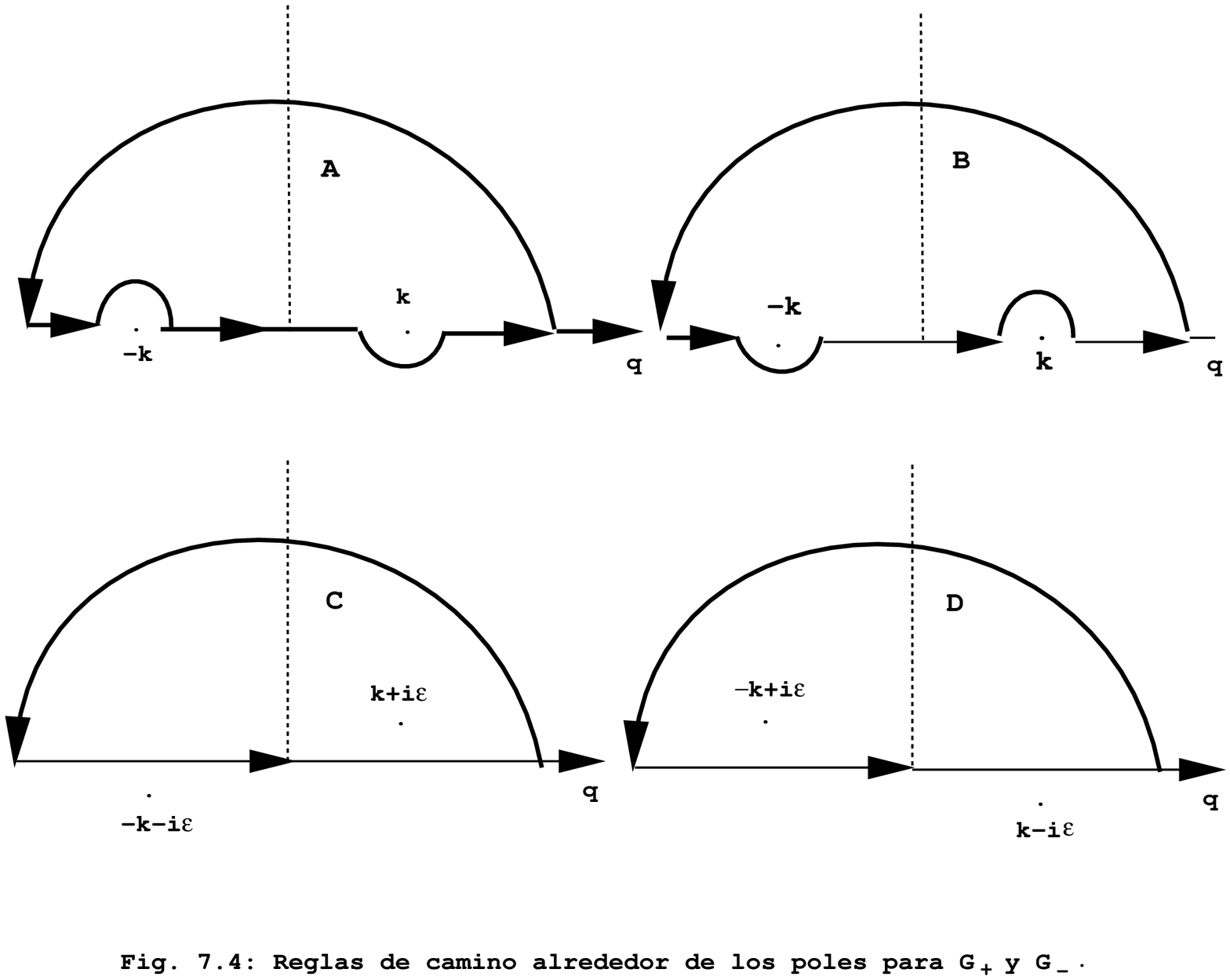}}
\vskip 4ex
%\begin{center}
%{\small{Fig. 1}\\
%}
%\end{center}
%%%%%%%%%%%%%%%%

 Usamos el contorno mostrado en la figura 7.4, el cual rodee a
%AQUI VA UNA IMAGEN DEL CONTORNO DE INTEGRACION 
los polos de la manera se\~nalada, lo cual dar\'a el efecto f\'{\i}sico
buscado, pues notemos que, de acuerdo al teorema del residuo,

\( G(r)= - \frac{1}{4\pi} \frac{e^{ikr}}{r}\quad({\rm Im} k > 0) \) ,

\(G(r)= - \frac{1}{4\pi} \frac{e^{-ikr}}{r}\quad ({\rm Im} k < 0) \)

La soluci\'on que nos interesa es la primera, porque da ondas dispersadas
{\em divergentes}, mientras que la otra soluci\'on representa ondas
dispersadas convergentes, y a\'un m\'as, la combinaci\'on lineal

\( \frac{1}{2} \lim_{\epsilon\rightarrow 0} [G_{k+i\epsilon} +
G_{k-i\epsilon}] = - {\frac{1}{4\pi}} \frac{\cos kr}{r} \)

\noindent corresponde a ondas estacionarias.

La evaluaci\'on formal de la integral se puede hacer tomando
$k^{2}-q^{2}\rightarrow k^{2}+i\epsilon-q^{2}$ , de tal manera
que:
\(\int _{-\infty} ^{\infty}\frac{qe^{iqR}}{k^{2}-q^{2}}dq \rightarrow
\int _{-\infty} ^{\infty}\frac{qe^{iqR}}{(k^{2}+i\epsilon)-q^{2}}dq~. \)

\noindent Esto es posible porque $R>0$, de manera que el contorno a
calcular estar\'a en el semiplano complejo superior. As\'{\i}, los polos
del integrando est\'an en:
$q=_{-}^{+}\sqrt{k^{2}+i\epsilon} \simeq
^{+}_{-}(k+\frac{i\epsilon}{2k})$.
El proceso l\'{\i}mite cuando $\epsilon \rightarrow 0$ debe
hacerse {\em despu\'es} de la evaluaci\'on de la integral.\\

{\bf Problema 7.2}

\noindent {\bf Forma asint\'otica de la expresi\'on radial}

Como ya se vi\'o antes\footnote{V\'ease {\it El \'atomo de hidr\'ogeno}}, 
la parte radial de la ecuaci\'on de Schr\"odinger
se puede escribir como:

\( ( \frac {d^2}{dr^{2}} + \frac{2}{r} \frac{d}{dr} ) 
R_{nlm}(r)-\frac{2m}{\hbar^{2}}[V(r)+\frac{l(l+1) 
\hbar^{2}}{2mr^{2}}]R_{nlm}(r)+\frac{2mE}{\hbar^{2}}R_{nlm}(r)=0 \)

\noindent $n,l,m$ son los n\'umeros cu\'anticos. De aqu\'{\i} en adelante 
se eliminar\'an por comodidad, y donde se sabe que $R$ depende s\'olo de 
$r$. Asumiremos que los potenciales decaen a cero mas r\'apido que $1/r$, y:
$\lim_{r \rightarrow 0} r^{2}V(r)=0$.

Usamos ahora $u(r)=rR$, y como:
$(\frac{d^{2}}{dr^{2}} + \frac{2}{r} \frac{d}{dr})\frac{u}{r} =
\frac{1}{r} \frac{d^{2}}{dr^{2}}u$ tenemos que

\( \frac{d^{2}}{dr^{2}}u + 
\frac{2m}{\hbar^{2}}[E-V(r)-\frac{l(l+1)\hbar^{2}}{2mr^{2}}] u=0~. \)

\noindent N\'otese que el potencial ahora tiene un t\'ermino mas:

\( V(r)\rightarrow V(r)+\frac{l(l+1)\hbar^{2}}{2mr^{2}}\)

\noindent que corresponde a una barrera repulsiva centr\'{\i}fuga. Para una
part\'{\i}cula libre $V(r)=0$ y la ecuaci\'on toma la forma

\( [\frac{d^{2}}{dr^{2}} + \frac{2}{r} 
\frac{d}{dr})-\frac{l(l+1)}{r^{2}}]R + k^{2}R=0~. \)

\noindent Introduciendo la variable $\rho=kr$, obtenemos

\( \frac{d^{2}R}{d\rho^{2}} + \frac{2}{\rho} \frac{dR}{d\rho} - 
\frac{l(l+1)}{\rho^{2}}R + R=0~. \)

Las soluciones a esta ecuaci\'on son las llamadas {\it funciones 
esf\'ericas de Bessel}. La soluci\'on regular es:

\( j_{l}(\rho)=(-\rho)^{l} (\frac{1}{\rho} \frac{d}{d\rho})^{l} (\frac{\sin 
\rho}{\rho}) \)

\noindent y la soluci\'on irregular:

\( n_{l}(\rho)= - (-\rho)^{l} (\frac{1}{\rho} \frac{d}{d\rho})^{l} 
(\frac{\cos \rho}{\rho}) \)

Para $\rho$ grande, las funciones de inter\'es son las funciones 
esf\'ericas de Hankel:

\( h_{l}^{(1)}(\rho)=j_{l}(\rho)+ in_{l}(\rho) \)
y

\( h_{l}^{(2)}(\rho)=[ h_{l}^{(1)}(\rho)]^{*} \)

De especial inter\'es es el comportamiento para $\rho \gg l$, cuyo 
comportamiento asint\'otico es:
$$
\label{eq:P1}
j_{l}(\rho) \simeq \frac{1}{\rho} \sin
(\rho-\frac{l\pi}{2})
$$  
\begin{equation}
\label{eq:P1}
 n_{l}(\rho) \simeq - \frac{1}{\rho} \cos(\rho-\frac{l\pi}{2}) 
\end{equation}

\noindent y entonces

\( h_{l}^{1} \simeq  -\frac{i}{\rho} e^{i(\rho - l\pi/2)} \)

La soluci\'on regular en el origen es:
$R_{l}(r)=j_{l}(kr)$

La forma asint\'otica es, usando la ecuaci\'on \ref{eq:P1}

\( R_{l}(r) \simeq \frac{1}{2ikr}[e^{-ikr-l\pi/2}-e^{ikr-l\pi/2}]~.\)

{\bf Problema 7.3}

\noindent {\bf La aproximaci\'on de Born para potencial de Yukawa}

Consideramos un potencial de la forma:

\begin{equation} 
V({\bf r})= V_{0} \frac{e^{-\alpha r}}{r} 
\end{equation}

\noindent con $V_{0}$ y $\alpha$ constantes reales y $\alpha$ positiva. El 
potencial 
es atractivo o repulsivo dependiendo de si $V_{0}$ es negativo o 
positivo; entre m\'as grande $\vert V_{0} \vert$, m\'as intenso el potencial.
Asumimos que $\vert V_{0} \vert$ es suficientemente peque\~no para que la 
aproximaci\'on de Born sea v\'alida. De acuerdo a la f\'ormula ya 
obtenida antes en el apartado de amplitud de probabilidad, la amplitud de 
dispersi\'on est\'a dada por:\\

\( f^{(B)}(\theta, \varphi)= - \frac{1}{4\pi} \frac{2mV_{0}}{\hbar^{2}} 
\int e^{-i {\bf K \cdot r}} \frac{e^{-\alpha r}}{r} d^{3}r~. \)

Como este potencial depende s\'olo de $r$ las integraciones angulares 
pueden hacerse f\'acilmente, y llegamos a la forma:

\( f^{(B)}(\theta, \varphi)=  \frac{1}{4\pi} \frac{2mV_{0}}{\hbar^{2}} 
\frac{4\pi}{\vert {\bf K} \vert} \int_{0}^{\infty} \sin \vert {\bf K} 
\vert r \frac{e^{-\alpha r}}{r} r dr~. \)

Con esto llegamos a:

\( f^{(B)}(\theta, \varphi)=  -\frac{2mV_{0}}{\hbar^{2}} \frac 
{1}{\alpha^{2} + \vert {\bf K}\vert^{2}}~.\)

%AQUI VA UNA FIGURA 

De la figura se observa que:
$\vert {\bf K} \vert = 2k \sin \frac{\theta}{2}$;
por lo tanto:

\( \sigma^{(B)}(\theta)=\frac{4m^{2}V_{0}^{2}}{\hbar^{4}} 
\frac{1}{[\alpha^{2} + 4k^{2} \sin \frac{\theta}{2}^{2}]^{2}}~. \)

La secci\'on de dispersi\'on total se obtiene por integraci\'on:

\( \sigma^{(B)} = \int \sigma^{(B})(\theta) d\Omega= 
\frac{4m^{2}V_{0}^{2}}{\hbar^{4}} \frac{4\pi}{\alpha^{2}(\alpha^{2}+4k^{2})}
~. \)

% \end{document}

\newpage
%%%%%%%%%%%%%%%%%%%%%%%%%%%%%%%%%%%%%%%%%%%%%
%\documentstyle[aps,preprint,tighten]{revtex}
%\begin{document}
%\draft
\def\bi{bigskip}
\def\noi{noindent}
%\protect
%\setcounter{equation}
%%%%%%%%%%%%%%%%%%%%%%%%%%%%%%%%%
\begin{center}
{\huge 8. LAS ONDAS PARCIALES}
\end{center}
%\author{Pedro Basilio Espinoza Padilla}
%\address{Universidad de Guanajuato, Instituto de F\'isica, \\ Le\'on,
%Guanajuato; M\'exico.}
%\maketitle
%\begin{abstract}
\begin{center}
%En el presente trabajo,
Explicamos brevemente en que consiste el m\'etodo de ondas
parciales en el estudio de problemas de dispersi\'on.
\end{center}
%\end{abstract}

\section*{Introducci\'on.}
\setcounter{equation}{0}
El problema de dispersi\'on desde el punto de vista cu\'antico, consiste en 
tratar a una part\'{\i}cula que interacciona con otra llamada dispersor (en la
presente exposici\'on supondremos que el dispersor siempre se encuentra fijo)
en una regi\'on muy peque\~na del espacio. Fuera de esta regi\'on, la
interacci\'on
entre ambas part\'{\i}culas es despreciable. De esta manera es posible
describir a la part\'{\i}cula dispersada por el siguiente Hamiltoniano:
\begin{equation}
H=H_0+V
\end{equation}

Donde $H_0$ corresponde al hamiltoniano para la part\'{\i}cula libre.
Entonces nuestro 
problema consiste en resolver la siguiente ecuaci\'on:
\begin{equation}
(H_0+V) \mid \psi \rangle = E \mid \psi \rangle
\end{equation}

Es evidente que el espectro ser\'a cont\'{\i}nuo (estamos tratando el caso de
dispersi\'on el\'astica). La soluci\'on a la
ecuaci\'on anterior est\'a dada por: 
\begin{equation}
\mid \psi \rangle = \frac {1}{E-H_0} V\mid \psi \rangle + \mid \phi \rangle
\end{equation}

De un ligero an\'alisis podemos ver que en el caso que $V=0$ obtenemos la 
soluci\'on: $\mid \phi \rangle $, es decir, la soluci\'on correspondiente al 
caso de la part\'{\i}cula libre. Hay que notar que el operador $\frac{1}{E-H_0}$
en cierto sentido es an\'omalo, pues tiene un continuo de polos en el eje
real que coinciden con los valores propios de $H_0$, para ``librarnos" de ese
problema induzcamos un peque\~no desplazamiento del corte que yace sobre
el eje real, de esta manera tenemos:
\begin{equation}
\mid \psi^{\pm} \rangle = \frac {1}{E-H_0 \pm i\varepsilon} V\mid \psi^{\pm} 
\rangle + \mid \phi \rangle
\end{equation}

La ecuaci\'on anterior es conocida como la ecuaci\'on de Lippmann-Schwinger.
Al final el desplazamiento de los polos ser\'a en el sentido positivo de
el eje imaginario (para que el principio de causalidad no se viole [seg\'un 
Feynman]). Tomemos la  x-representaci\'on:
\begin{equation}
\langle {\bf {x}}\mid \psi^{\pm} \rangle =\langle {\bf {x}}\mid \phi \rangle + 
\int d^{3} x^{'}\left \langle {\bf {x}} \vert \frac {1}{E-H_0 \pm i\varepsilon 
}\vert {\bf {x^{'}}} \right \rangle \langle {\bf {x^{'}}} \mid V\mid 
\psi^{\pm}\rangle
\end{equation}

En el primer t\'ermino del lado derecho de la ecuaci\'on anterior vemos que 
corresponde a una part\'{\i}cula libre y el segundo t\'ermino se interpreta
como una 
onda esf\'erica que emerge del dispersor. El kernel de la integral anterior
lo podemos asociar con una funci\'on de Green o propagador y es muy sencillo
calcular:
\begin{equation}
G_{\pm}({\bf {x}},{\bf {x^{'}}})=\frac{\hbar^{2}}{2m}\left \langle {\bf {x}} \vert 
\frac {1}{E-H_0 \pm i\varepsilon}\vert {\bf {x^{'}}} \right \rangle = 
-\frac{1}{4\pi} \frac{e^{\pm ik\mid {\bf {x}}-{\bf {x^{'}}}\mid}}{\mid {\bf 
{x}}-{\bf {x^{'}}}\mid}~,
\end{equation}

donde $E={\hbar^{2}}{k^2}/2m$.
Como ve\'{\i}amos anteriormente la funci\'on de onda la podemos escribir
como una 
onda plana m\'as una onda esf\'erica que emana del dispersor (salvo un factor 
constante):
\begin{equation}
\langle {\bf {x}}\mid \psi^{+} \rangle =e^{{\bf {k}}\cdot {\bf {x}}} + 
\frac{e^{ikr}}{r} f({\bf {k}},{\bf {k^{'}}})
\end{equation}

A la cantidad $f({\bf {k}},{\bf {k^{'}}})$ que aparece en la ecuaci\'on 7
se le conoce como amplitud de dispersi\'on y expl\'{\i}citamente la
podemos escribir como:
\begin{equation}
f({\bf {k}},{\bf {k^{'}}})=-\frac{1}{4\pi} {(2\pi )^3}\frac{2m}{\hbar^2}\langle 
{\bf {k^{'}}}\mid V \mid \psi^{+} \rangle 
\end{equation}

Definamos ahora un operador T tal que:
\begin{equation}
T\mid \phi \rangle = V\mid \psi^{+} \rangle
\end{equation}

Si multiplicamos la ecuaci\'on de Lippman-Schwinger por V y a partir de la
definici\'on anterior obtenemos:
\begin{equation}
T\mid \phi \rangle = V\mid \phi \rangle + V\frac{1}{E-H_0+
i\varepsilon}T\mid \phi 
\rangle 
\end{equation}

As\'{\i} iterando la ecuaci\'on anterior (como en teor\'{\i}a de
perturbaciones)
podemos obtener la aproximaci\'on de Born y sus correcciones de orden superior.

\section*{El m\'etodo de ondas parciales.}
%%%%%%%%%%%%%%%%%%%%%%%%%%%%%%%%%%%%%%%%%%

Ahora consideremos el caso de un potencial central no nulo. Entonces de la 
definici\'on (9) del operador $T$ se deduce que conmuta con $\vec {L}^{2} $ y
con $\vec {L}$ de ah\'{\i} que se diga que $T$ es un operador escalar. De esta
manera 
para facilitar los c\'alculos es conveniente utilizar coordenadas esf\'ericas, 
puesto que dada la simetr\'{\i}a, el operador $T$ ser\'a diagonal. Ahora
veamos que 
forma adquiere la expresi\'on (8) para la amplitud de dispersi\'on:
$$
f({\bf {k}},{\bf {k^{'}}})=-\frac{1}{4\pi}\frac{2m}{\hbar^2} {(2\pi 
)^3}
$$
\begin{equation}
\sum_{l}\sum_{m} \sum_{l^{'}} \sum_{m^{'}} \int dE\int 
dE^{'}\langle {\bf {k^{'}}}\mid E^{'} l^{'} m^{'} \rangle \langle E^{'} 
l^{'} m^{'}\mid T\mid Elm\rangle \langle Elm\mid \bf {k} \rangle
\end{equation}

despu\'es de algunos c\'alculos podemos obtener:
\begin{equation}
f({\bf {k}},{\bf {k^{'}}})=-\frac{4\pi^2}{k}\sum_{l}\sum_{m} T_{l} (E) 
Y^{m}_{l} ({\bf {k^{'}}})Y^{m^{*}}_{l}(\bf {k})
\end{equation}

Escogiendo el sistema de coordenadas tal que $\bf {k}$ tenga la misma direcci\'on 
que el eje orientado z, de esta manera a la amplitud de dispersi\'on 
\'unicamente contribuir\'an los arm\'onicos esf\'ericos con m igual a cero; si 
definimos que $\theta$ sea el \'angulo entre ${\bf {k}}$ y ${\bf {k^{'}}}$ 
tenemos:
\begin{equation}
Y^{0}_{l} ({\bf {k^{'}}})=\sqrt {\frac{2l+1}{4\pi}} P_{l}(cos\theta)
\end{equation}

hagamos la siguiente definici\'on:
\begin{equation}
f_{l}(k)\equiv-\frac{\pi T_{l} (E)}{k}
\end{equation}

as\'{\i} la ecn. (12) se puede escribir de la siguiente manera:
\begin{equation}
f({\bf {k}},{\bf {k^{'}}})=f(\theta)=\sum^{\infty}_{l=0} 
(2l+1)f_{l}(k)P_{l}(cos\theta)
\end{equation}

A la cantidad $f_{l}(k)$ le podemos dar una interpretaci\'on sencilla a 
partir del desarrollo de una onda plana en ondas esf\'ericas, veamos el 
comportamiento de la funci\'on $\langle {\bf {x}}\mid \psi^{+} \rangle$ para 
grandes valores de r, que como ya hab\'{\i}amos establecido previamente debe
tener la forma:
$$%\begin{displaymath}
\langle {\bf {x}}\mid \psi^{+} \rangle = \frac{1}{{(2\pi )^{3/2}}}\left[ 
{e^{ikz}}+f(\theta ) \frac{{e^{ikr}}}{r}\right] =
$$%\end{displaymath}
$$%\begin{displaymath}
\frac{1}{{(2\pi)^{3/2}}}\left[ \sum_{l} (2l+1)P_{l}(\cos\theta 
)\left(\frac{{e^{ikr}}-{e^{i(kr-l\pi )}}}{2ikr} \right) 
+\sum_{l}(2l+1)f_{l}(k)P_{l}(\cos\theta )\frac{{e^{ikr}}}{r}\right]=
$$%\end{displaymath}
\begin{equation}
\frac{1}{{(2\pi )^{3/2}}}\sum_{l} 
(2l+1)\frac {P_{l}(\cos\theta )}{2ik}\left[ \left[ 
1+2ikf_{l}(k)\right]\frac{{e^{ikr}}}{r}-\frac{{e^{i(kr-l\pi 
)}}}{r} \right]
\end{equation}

La ecuaci\'on anterior la podemos interpretar de la manera siguiente. Los 
dos t\'erminos exponenciales corresponden a ondas esf\'ericas, el primero a 
una onda emergente y el segundo a una onda convergente; y el efecto de la 
dispersi\'on se ve en el coeficiente de la onda emergente, y es igual a uno 
cuando no existe un dispersor.

\section*{Corrimientos de fase}
%%%%%%%%%%%%%%%%%%%%%%%%%%%%%%%%%%

Imaginemos ahora una superficie cerrada centrada en el dispersor, si 
asumimos que no hay creaci\'on ni aniquilaci\'on de part\'{\i}culas se
verifica:
\begin{equation}
\int {\bf {j}}\cdot d{\bf {S}}=0
\end{equation}

Donde la regi\'on  de integraci\'on es evidentemente la superficie 
anteriormente definida y ${\bf {j}}$ es la densidad de corriente de 
probabilidad. A\'un m\'as, debido a la conservaci\'on del momento 
angular la ecuaci\'on anterior debe verificarse para cada onda parcial 
(en otras palabras, todas las ondas parciales tienen diferentes valores de 
las proyecciones del momento angular, lo cual las hace en esencia 
diferentes y la formulaci\'on ser\'{\i}a equivalente si trat\'aramos al
paquete de ondas como un 
haz de part\'{\i}culas que no interact\'uan entre s\'{\i}; m\'as a\'un, debido a
que el potencial de nuestro problema es central el momento angular de cada 
``part\'{\i}cula'' se conservar\'a de tal manera que podr\'{\i}amos decir
que las 
part\'{\i}culas contin\'uan siendo las mismas). Por las consideraciones
anteriores, podemos decir que tanto la onda divergente como la emergente 
difieren a lo mucho en un factor de fase, es decir, si definimos:
\begin{equation}
S_{l}(k)\equiv 1+ 2ikf_{l}(k)
\end{equation}

deber\'a suceder que
\begin{equation}
\mid S_{l}(k)\mid =1
\end{equation}

Los resultados anteriores los podemos interpretar en vista de la 
conservaci\'on de las probabilidades, y era de esperarse pues hemos supuesto 
que no existe creaci\'on ni aniquilaci\'on de part\'{\i}culas, as\'{\i} que
la 
influencia del dispersor consiste en agregar simplemente un factor de fase 
en las componentes de la onda emergente, en virtud de la unitariedad del 
factor de fase lo podemos escribir como:
\begin{equation}
S_{l}=e^{2i\delta_{l}}
\end{equation}
 
Donde $\delta_{l}$ es real y es funci\'on de k. A partir de la definici\'on 
(18) podemos escribir:
\begin{equation}
f_{l}=\frac{{e^{2i\delta_{l}}}-1}{2ik}=\frac{{e^{i\delta_{l}}}\sin 
(\delta_{l})}{k}=\frac{1}{k\cot (\delta_{l})-ik} 
\end{equation}

La secci\'on total de dispersi\'on adquiere la siguiente forma:
$$%\begin{displaymath}
\sigma_{total}=\int \mid f(\theta){\mid ^2}d\Omega =
$$%\end{displaymath}
$$%\begin{displaymath}
\frac{1}{{k^2}}{\int _{0} ^{2\pi}}d\phi {\int _{-1} ^{1}}d(\cos (\theta 
))\sum_{l} \sum_{{l^{'}}}(2l+1)(2{l^{'}}+1){e^{i\delta_{l}}}\sin 
(\delta_{l}){e^{i\delta_{{l^{'}}}}} \sin (\delta_{{l^{'}}})P_{l}P_{{l^{'}}}
$$%\end{displaymath}
\begin{equation}
=\frac{4\pi }{{k^2}}\sum_{l} (2l+1)\sin {^2}(\delta_{{l^{'}}})
\end{equation}

\section*{Determinaci\'on de los corrimientos de fase.}
%%%%%%%%%%%%%%%%%%%%%%%%%%%%%%%%%%%%%%%%%%%%%%%%%%%%%%%%
Consideremos ahora un potencial V tal que V se anula para $r>R$, el 
par\'ametro R se le conoce como el ``alcance del potencial'', as\'{\i} que en
la regi\'on $r>R$ evidentemente debe corresponder a una onda esf\'erica 
libre. Por otro lado la forma m\'as general de la expansi\'on de una onda 
plana en ondas esf\'ericas es de la forma:
\begin{equation}
\langle {\bf {x}}\mid \psi^{+} \rangle =\frac{1}{{(2\pi )^{3/2}}}\sum_{l} 
{i^{l}} (2l+1)A_{l}(r)P_{l}(\cos \theta ) \quad (r>R)
\end{equation}

Donde el coeficiente $A_{l}$ est\'a definido como:
\begin{equation}
A_{l}={c_{l} ^{(1)}}{h_{l} ^{(1)}}(kr)+{c_{l} ^{(2)}}{h_{l} ^{(2)}}(kr)
\end{equation}

donde ${h_{l} ^{(1)}}$ y ${h_{l} ^{(2)}}$ son las funciones de Hankel 
esf\'ericas y sus formas asint\'oticas est\'an dadas por:
%\begin{displaymath}
$$ 
{h_{l} ^{(1)}} \sim \frac{{e^{i(kr-l\pi /2)}}}{ikr}
$$
%\end{displaymath}
$$%\begin{displaymath}
{h_{l} ^{(2)}} \sim - \frac{{e^{-i(kr-l\pi /2)}}}{ikr}
$$%\end{displaymath}

Al ver la forma asint\'otica de la expresi\'on (23):
\begin{equation}
\frac{1}{{(2\pi )^{3/2}}}\sum_{l}(2l+1)P_{l}\left[ \frac{{e^{ikr}}}{2ikr}-
\frac{{e^{-i(kr-l\pi)}}}{2ikr} \right]
\end{equation}

De esta manera vemos que:
\begin{equation}
{c_{l} ^{(1)}}=\frac{1}{2} e^{2i\delta_{l}} \qquad {c_{l} ^{(2)}}=\frac{1}{2}
\end{equation}

Ahora podemos ver que la funci\'on de onda radial para $r>R$ se escribe como:
\begin{equation}
A_{l}=e^{2i\delta _{l}}\left[ \cos \delta _{l} j_{l} (kr)
- \sin \delta _{l}n_{l} 
(kr)\right] \end{equation}

A partir de ecuaci\'on anterior podemos evaluar su derivada logar\'{\i}tmica
en r=R, i.e., justo afuera del alcance del potencial:
\begin{equation}
\beta _{l}\equiv \left( \frac{r}{A_{l}}\frac{dA_{l}}{dr}\right)_{r=R}=
kR\left[
\frac{{j_{l}^{'}}\cos \delta _{l}-{n_{l}^{'}}(kR)\sin \delta _{l}}{j_{l}\cos
\delta _{l}-{n_{l}}(kR)\sin \delta_{l}}\right]
\end{equation}

Donde $j_{l}^{'}$ es la derivada de $j_{l}$ con respecto a $kr$ y evaluada en 
$r=R$. Otro resultado importante que podemos obtener conociendo el resultado 
anterior es el corrimiento de fase:
\begin{equation}
\tan \delta _{l}=\frac{kR{j_{l}^{'}}(kR)-\beta _{l}
j_{l}(kR)}{kR{n_{l}^{'}}(kR)-
\beta_{l} n_{l}(kR)}
\end{equation}

Para encontrar la soluci\'on completa a nuestro problema a\'un hay que hacer los 
c\'alculos para cuando $r<R$, es decir, dentro del rango del alcance del 
potencial. Para el caso de un potencial central la soluci\'on a la ecuaci\'on de 
Schr\"odinger en tres dimensiones se reduce a resolver la
ecuaci\'on:
\begin{equation}
\frac{{d^{2}}u_{l}}{d{r^{2}}}+\left( {k^{2}}-\frac{2m}{{\hbar ^{2}}} 
V-\frac{l(l+1)}{{r^{2}}} \right) u_{l}=0
\end{equation}

donde $u_{l}=rA_{l}(r)$ y est\'a sujeto a la condici\'on de 
frontera $u_{l}\mid _{r=0} \quad =0$ de esta manera ya podemos calcular 
la derivada logar\'{\i}tmica, la cual por la propiedad de continuidad de 
la derivada logar\'{\i}tmica (que es
equivalente a la continuidad de la derivada en un punto de discontinuidad):
\begin{equation}
\beta_{l} \mid_{dentro}=\beta_{l}\mid_{fuera}
\end{equation}

\section*{Un ejemplo: dispersi\'on por una esfera s\'olida.}
%%%%%%%%%%%%%%%%%%%%%%%%%%%%%%%%%%%%%%%%%%%%%%%%%%%%%%%%%%%%%%
Ahora tratemos un caso espec\'{\i}fico.
Consideremos un potencial definido como:
\begin{equation}
V=\left\{ 
\begin{array} {ll}
\infty & \mbox{ $r<R$} \\
0      & \mbox {$r>R$} 
\end{array}
 \right.
\end{equation}

Es sabido que una part\'{\i}cula no puede penetrar en una regi\'on donde el
potencial sea infinito, as\'{\i} que la funci\'on de onda se deber\'a anular
en $r=R$; de que la esfera es impenetrable tambi\'en se deduce que:
\begin{equation}
A_{l}(r)\mid_{r=R} =0
\end{equation}

As\'{\i} de la ecuaci\'on (27) tenemos:
\begin{equation}
\tan \delta_{l} = \frac{j_{l} (kR)}{n_{l} (kR)}
\end{equation}

Vemos que se puede calcular f\'acilmente el corrimiento de fase para
cualquier l.
Consideremos ahora el caso $l=0$ (dispersi\'on de una onda S) de esta forma 
tenemos:
$$%\begin{displaymath}
\delta_{l} = -kR
$$%\end{displaymath}

y de la ecuaci\'on (27):

\begin{equation}
A_{l=0}(r)\sim \frac{\sin kr}{kr}\cos\delta_{0}+\frac{\cos 
kr}{kr}\sin\delta_{0}=\frac{1}{kr}\sin (kr+\delta_{0})
\end{equation}

Vemos que la contribuci\'on al movimiento libre es la adici\'on de una fase. 
Claro que en el caso m\'as general diferentes ondas tendr\'an diferentes 
corrimientos de fase ocasionando una distorsi\'on transitoria en el paquete de 
ondas dispersado.
Estudiemos ahora el caso de energ\'{\i}as peque\~nas,
i.e., $kR<<1$ en este caso 
las expresiones para las funciones de Bessel (usadas para escribir las funciones 
de Hankel esf\'ericas) son las siguientes:
\begin{equation}
j_{l} (kr)\sim \frac{(kr)^{l}}{(2l+1)!!}
\end{equation}
\begin{equation}
n_{l} (kr)\sim -\frac{(2l-1)!!}{(kr)^{l+1}}
\end{equation}

obteniendo as\'{\i} la expresi\'on:
\begin{equation}
\tan\delta_{l} = \frac{-(kR)^{2l+1}}{(2l+1)[(2l-1)!!]^{2}}~.
\end{equation}

De la f\'ormula anterior podemos ver que la contribuci\'on apreciable al 
corrimiento de fase es principalmente de ondas con $l=0$, pero como 
$\delta_{0}=-kR$ tenemos para la secci\'on eficaz de dispersi\'on:
\begin{equation}
\sigma_{total}=\int\frac{d\sigma}{d\Omega}d\Omega=4\pi R^{2}
\end{equation}

Vemos que la secci\'on eficaz cu\'antica es cuatro veces mayor que
la secci\'on 
eficaz cl\'asica, y coincide con el \'area total de la esfera. Para grandes 
valores de la energ\'{\i}a del paquete incidente conjeturemos que todos
los valores de $l$ hasta un valor
m\'aximo $l_{max}\sim kR$ contribuyen a la secci\'on eficaz total.
\begin{equation}
\sigma_{total}
=\frac{4\pi}{k^{2}}{\sum_{l=0} ^{l\sim kR}}(2l+1){\sin}^{2}\delta_{l}~.
\end{equation}

De esta forma a partir de la ecuaci\'on (34) tenemos:
\begin{equation}
{\sin}^{2}\delta_{l}=\frac{\tan^{2}\delta_{l}}{1+\tan^{2}\delta_{l}}=
\frac{[j_{l} (kR)]^{2}}{[j_{l} (kR)]^{2}+[n_{l} (kR)]^{2}}\sim\sin^{2}\left( 
kR-\frac{l\pi}{2}\right)~.
\end{equation}

Aqu\'{\i} hemos utilizado las expresiones:
$$%\begin{displaymath}
j_{l} (kr)\sim\frac{1}{kr}\sin\left( kr-\frac{l\pi}{2}\right)
$$%\end{displaymath}
$$%\begin{displaymath}
n_{l} (kr)\sim -\frac{1}{kr}\cos\left( kr-\frac{l\pi}{2}\right)
$$%\end{displaymath}

vemos que $\delta_{l}$ decrece en $\frac{\pi}{2}$ cada vez que $l$ se 
incrementa 
en una unidad, as\'{\i} es evidente que se cumple ${\sin}^{2}\delta_{l}+
{\sin}^{2}\delta_{l+1}=1$, y aproximando ${\sin}^{2}\delta_{l}$ por su valor 
promedio $\frac{1}{2}$, as\'{\i} es sumamente sencillo obtener el resultado de la 
suma anterior, y la f\'ormula para la secci\'on eficaz total es:
\begin{equation}
\sigma_{total}=\frac{4\pi}{k^{2}}(kR)^{2}\frac{1}{2}=2\pi R^{2}
\end{equation}

Una vez m\'as el resultado del c\'alculo utilizando mec\'anica cu\'antica 
difiere del resultado cl\'asico, pero veamos cu\'al es el origen del factor 2; 
primero a la ecuaci\'on (15) la vamos a separar en dos partes:
\begin{equation}
f(\theta )=\frac{1}{2ik}{\sum_{l=0} ^{l=kR}}(2l+1){e^{2i\delta_{l}}}P_{l}
\cos (\theta )+\frac{i}{2k}{\sum_{l=0} ^{l=
kR}}(2l+1) P_{l}\cos (\theta )=f_{\mbox{reflexi\'on}}+f_{\mbox{sombra}}
\end{equation}

evaluando $\int |f_{\mbox{ reflexi\'on}}|^{2}d\Omega$:
\begin{equation}
\int |f_{\mbox{ reflexi\'on}}|^{2}d\Omega=\frac{2\pi}{4k^2}{\sum_{l=0} 
^{l_{max}}}{{\int_{-1}}^{1}}(2l+1)^{2}[P_{l}\cos (\theta )]^{2} d(\cos \theta 
)=\frac{\pi {l_{max}}^{2}}{k^{2}}=\pi R^{2}
\end{equation}

Ahora analizando $f_{\mbox{ sombra}}$ para \'angulos peque\~nos tenemos:
\begin{equation}
f_{\mbox{ sombra}}\sim\frac{i}{2k}\sum (2l+1)J_{0}(l\theta )\sim ik{\int_{0} 
^{R}}bJ_{0}(kb\theta )db=\frac{iRJ_{1}(kR\theta )}{\theta}
\end{equation}

Esta f\'ormula es bastante conocida en \'optica, es la f\'ormula de 
difracci\'on de Fraunhofer; haciendo el cambio de variable $z=kR\theta$ podemos 
evaluar la integral $\int |f_{{\rm sombra}}|^{2}d\Omega$:
\begin{equation}
\int |f_{\mbox{ sombra}}|^{2}d\Omega \sim 2\pi R^{2}{\int_{0} 
^{\infty}}\frac{[J_{1}(z)]^{2}}{z} dz\sim\pi R^{2}
\end{equation}

Finalmente despreciando la interferencia entre $f_{\mbox{reflexi\'on}}$ 
y $f_{\mbox{ sombra}}$ 
(porque la fase oscila entre $2\delta_{l+1}=2\delta_{l}-\pi$). De esta manera 
obtenemos el resultado (42). Hemos etiquetado un t\'ermino con el nombre de 
sombra, el origen de esta contribuci\'on se explica f\'acilmente al apelar al 
comportamiento ondulatorio de la part\'{\i}cula dispersada (en este punto da lo
mismo [f\'{\i}sicamente] un paquete de ondas que una part\'{\i}cula), 
tiene su origen
en las componentes del paquete dispersadas hacia atr\'as, entonces tendr\'an
una diferencia de fase con las ondas incidentes y habr\'a una interferencia
destructiva.

\section*{Dispersi\'on en un campo de Coulomb}
%%%%%%%%%%%%%%%%%%%%%%%%%%%%%%%%%%%%%%%%%%%%%%
Ahora consideremos un ejemplo cl\'asico y algo complicado: la dispersi\'on de 
part\'{\i}culas en un campo coulombiano, para este caso la ecuaci\'on de
Schr\"odinger es:
\begin{equation}
\left( -\frac{\hbar ^{2}}{2m}\nabla ^{2} - \frac{Z_{1}Z_{2}e^{2}}{r}\right)\psi 
({\bf {r}})=E\psi ({\bf {r}}), \qquad E>0
\end{equation}

donde m es la masa reducida de las part\'{\i}culas que interaccionan y
evidentemente $E>0$ debido a que tratamos el caso de dispersi\'on sin
creaci\'on 
de estados ligados. La ecuaci\'on anterior es equivalente a la siguiente 
expresi\'on (para valores adecuados de las constantes $k$ y $\gamma$ :
\begin{equation}
\left( \nabla ^{2} +{k^{2}} +\frac{2\gamma k}{r}\right)\psi ({\bf {r}})=0
\end{equation}

Si no consideramos la parte centr\'{\i}fuga del potencial efectivo, es decir,
que
nos quedamos \'unicamente con la interacci\'on de un campo de Coulomb puro, por 
eso podemos proponer una soluci\'on de la forma:
\begin{equation}
\psi ({\bf {r}})={e^{i{\bf {k\cdot r}}}}\chi (u)
\end{equation}

con
$$%\begin{displaymath}
u=ikr(1-\cos\theta )=ik(r-z)=ikw
$$%\end{displaymath}
$$%\begin{displaymath}
{\bf {k\cdot r}}=kz
$$%\end{displaymath}

$\psi ({\bf {r}})$ es la soluci\'on completa a la ecuaci\'on de
Schr\"odinger, y 
es de esperarse que su comportamiento asint\'otico conste de dos partes: una 
tipo onda plana ${e^{i{\bf {k\cdot r}}}}$; y otra tipo onda esf\'erica 
${r^{-1}e^{ikr}}$. Definiendo nuevas variables:
\begin{displaymath}
z=z \qquad w=r-z \qquad \lambda =\phi
\end{displaymath}

y con ayuda de las relaciones anteriores, la ecuaci\'on (48) adquiere la forma:
\begin{equation}
\left[ u \frac{d^{2}}{du^{2}}+(1-u)\frac{d}{du}-i\gamma\right]\chi (u)=0
\end{equation}

Para resolver la ecuaci\'on anterior, primero hay que estudiar sus 
comportamientos asint\'oticos, pero como todo eso ya est\'a hecho, la funci\'on 
de onda asint\'otica normalizada que se obtiene al final de todos los 
c\'alculos anteriores es:
\begin{equation}
\psi_{\bf k} ({\bf {r}})=\frac{1}{(2\pi )^{3/2}}\left( {e^{i[{\bf {k\cdot 
r}}-\gamma ln(kr-{\bf {k\cdot r}})]}}+
\frac{f_{c}(k,\theta){e^{i[kr+\gamma 
ln2kr]}}}{r}\right)
\end{equation}

Como vemos, la funci\'on de onda anterior tiene algunos t\'erminos que la hacen 
diferir de manera apreciable a nuestra ecuaci\'on (7) y eso se debe a que una 
fuerza de tipo de Coulomb es de largo alcance. Hacer el c\'alculo exacto para la 
amplitud de dispersi\'on de Coulomb es bastante dif\'{\i}cil de hacer (de hecho
casi todos los c\'alculos de este problema). Quien logre hacer todos los 
c\'alculos para obtener la funci\'on de onda normalizada encontrar\'a:
\begin{equation}
\psi_{\bf k} ({\bf {r}})=\frac{1}{(2\pi )^{3/2}}\left( {e^{i[{\bf {k\cdot r}}-
\gamma ln(kr-{\bf {k\cdot 
r}})]}}+\frac{g_{1}^{*}(\gamma )}{g_{1}(\gamma )}\frac{\gamma}{2k\sin 
(\theta /2) ^{2}}\frac{e^{i[kr+\gamma ln2kr]}}{r}\right)
\end{equation}

donde $g_{1}(\gamma )=\frac{1}{\Gamma (1-i\gamma )}$.

El an\'alisis de ondas parciales lo reduciremos a mostrar resultados ya 
obtenidos de una manera lo m\'as clara posible.
Primero escribamos la funci\'on de onda (49) $\psi ({\bf {r}})$ de la siguiente 
manera:
\begin{equation}
\psi ({\bf {r}})={e^{i{\bf {k\cdot r}}}}\chi (u)=A{e^{i{\bf {k\cdot 
r}}}}\int_{C}{e^{ut}}{t^{i\gamma -1}}(1-t)^{-i\gamma}dt
\end{equation}

donde $A$ es una constante de normalizaci\'on y toda la parte integral es la 
transformada de Laplace inversa de la transformada directa de la ecuaci\'on 
(50). La ecuaci\'on anterior se puede escribir de la siguiente forma:
\begin{equation}
\psi ({\bf {r}})=A\int_{C}{e^{i{\bf {k\cdot r}}}(1-t)}{e^{ikrt}}(1-t)
d(t,\gamma )dt
\end{equation}

con
\begin{equation}
d(t,\gamma )={t^{i\gamma -1}}(1-t)^{-i\gamma -1}
\end{equation}

Ya en el an\'alisis de ondas parciales procedemos a escribir:
\begin{equation}
\psi ({\bf {r}})={\sum_{l=0} ^{\infty}}(2l+1){i^{l}}P_{l}(\cos\theta )A_{l}(kr)
\end{equation}

donde
\begin{equation}
A_{l}(kr)=A\int_{C}{e^{ikrt}}j_{l}[kr(1-t)](1-t)d(t,\gamma )
\end{equation}

Y dada la relaci\'on que hay entre las funciones de Bessel esf\'ericas con las 
funciones esf\'ericas de Hankel tenemos:
\begin{equation}
A_{l}(kr)=A_{l}^{(1)}(kr)+A_{l}^{(2)}(kr)
\end{equation}

La evaluaci\'on de los coeficientes anteriores no la vamos a hacer aqui (dada su 
extensi\'on), nos basta con saber que:
\begin{equation}
A_{l}^{(1)}(kr)=0
\end{equation}
\begin{equation}
A_{l}^{(2)}(kr)\sim -\frac{Ae^{\pi\gamma /2}}{2ikr}[2\pi ig_{1}(\gamma)]
\left( 
e^{-i[kr-(l\pi /2)+\gamma \ln 2kr]}-{e^{2i\eta_{l} (k)}}
e^{i[kr-(l\pi /2)+\gamma 
\ln 2kr]}\right)
\end{equation}

donde
\begin{equation}
{e^{2i\eta_{l} (k)}}=\frac{\Gamma (1+l-i\gamma )}{\Gamma (1+l+i\gamma )}
\end{equation}

\section*{C\'alculo de la amplitud de dispersi\'on de Coulomb}
%%%%%%%%%%%%%%%%%%%%%%%%%%%%%%%%%%%%%%%%%%%%%%%%%%%%%%%%%%%%%%%

Si tomamos la transformada de Laplace de la ecuaci\'on (50) obtenemos:
\begin{equation}
\chi (u)=A\int_{C} e^{ut}t^{i\gamma-1}(1-t)^{-i\gamma}dt
\end{equation}

Donde el contorno C va de $-\infty $ a $\infty$ y se cierra por arriba del
eje
real, vemos que hay dos polos: cuando $t=0$ y $t=1$. Haciendo el cambio de 
variables $s=ut$ obtenemos:
\begin{equation}
\chi (u)=A\int_{C_{1}}e^{s}s^{i\gamma -1}(u-s)^{-i\gamma}
\end{equation}

$\chi (u)$ debe ser regular en cero y en efecto:
\begin{equation}
\chi (0)=(-1)^{-i\gamma}A\int_{C_{1}}\frac{e^{s}}{s}ds
=(-1)^{-i\gamma}A2\pi i
\end{equation}

Ahora tomando el l\'{\i}mite $u\to\infty$ haciendo un desplazamiento
infinitesimal 
(para quitarse el problema de que los polos est\'an sobre la trayectoria) y un 
cambio de variable tal que $\frac{s}{u}=
-\frac{(s_{0}\pm i\varepsilon)}{i\kappa}$, 
pero como este t\'ermino tiende a cero cuando $u\to\infty$ entonces podemos 
expander $(u-s)$ en series de potencias de $\frac{s}{u}$ para el polo con $s=0$. 
Pero el desarrollo anterior no es bueno en $s=1$, porque aqu\'{\i}
$s=-s_{0}+i(\kappa\pm\varepsilon$; y de ah\'{\i} que $\frac{s}{u}=
1-\frac{(s_{0}\pm 
i\varepsilon )}{\kappa}$ tiende a $1$ cuando $\kappa\to\infty$; pero si hacemos 
el cambio de variable $s^{'}=s-u$ ya no hay problema:
\begin{equation}
\chi (u)=A\int_{\rm C_{2}}\left([e^{s}s^{i\gamma
-1}(u-s)^{-i\gamma}]ds+[e^{s^{'}+u}(-s^{'})^{i\gamma}(u+s^{'})^{i\gamma-1}]
ds^{'}\right)
\end{equation}

Desarrollando las series de potencias es sencillo calcular las integrales 
anteriores, pero del resultado a\'un hay que tomar el
l\'{\i}mite $\frac{s}{u}\to 0$ para obtener las formas asint\'oticas
correctas para la dispersi\'on de Coulomb:
$$%\begin{displaymath}
\chi (u)\sim 2\pi iA\left[u^{-i\gamma}g_{1}(\gamma )-(-u)^{i\gamma 
-1}e^{u}g_{2}(\gamma )\right]
$$%\end{displaymath}
$$%\begin{displaymath}
2\pi g_{1}(\gamma )=i\int_{\rm C_{2}}e^{s}s^{i\gamma -1}ds
$$%\end{displaymath}
\begin{equation}
2\pi g_{2}(\gamma )=i\int_{\rm C_{2}}e^{s}s^{-i\gamma}ds
\end{equation}

Y despu\'es de tanto cambio de variables, regresamos a las originales y vemos:
$$%\begin{displaymath}
(u^{*})^{i\gamma}=(-i)^{i\gamma}[k(r-z)]^{i\gamma}
=e^{\gamma\pi /2}e^{i\gamma \ln k(r-z)}
$$%\end{displaymath}
\begin{equation}
(u)^{-i\gamma}=(i)^{-i\gamma}[k(r-z)]^{-i\gamma}=e^{\gamma\pi /2}e^{-i\gamma 
\ln k(r-z)}
\end{equation}

Ya hemos calculado $\chi$, lo cual equivale a tener $\psi_{\bf k} ({\bf {r}})$ a 
partir de (49).

\section*{Aproximaci\'on eikonal}

Hacemos una breve exposici\'on a cerca de la aproximaci\'on
eikonal. La filosof\'{\i}a de la aproximaci\'on eikonal en este caso es el
mismo que cuando hacemos el paso de la \'optica ondulatoria a la \'optica
geom\'etrica, por eso es v\'alida cuando el potencial var\'{\i}a m\'as 
lentamente 
que la longitud de onda del paquete de ondas dispersado, es decir, para el caso 
$E>>|V|$. As\'{\i} esta aproximaci\'on puede ser considerada como
una aproximaci\'on cuasicl\'asica. Primero propongamos que la funci\'on de onda 
cuasicl\'asica es:
\begin{equation}
\psi\sim e^{iS({\bf r})/\hbar}
\end{equation}

Donde S satisface la ecuaci\'on de Hamilton-Jacobi, y su soluci\'on es:
\begin{equation}
\frac{S}{\hbar}=\int_{-\infty}^{z}\left[ k^{2}-\frac{2m}{\hbar ^{2}}V\left( \sqrt 
{b^{2}+z'^{2}}\right)\right]^{1/2}dz'+ {\rm constante}
\end{equation}

La constante aditiva se escoje de tal forma que:
\begin{equation}
\frac{S}{\hbar}\to kz\qquad {\rm cuando} \qquad V\to 0
\end{equation}

El t\'ermino que multiplica el potencial lo podemos interpretar como un cambio 
de fase del paquete de ondas, su forma espec\'{\i}fica es la siguiente
\begin{equation}
\Delta (b)\equiv \frac{-m}{2k\hbar^{2}}\int_{-\infty}^{\infty} V\left( 
\sqrt{b^{2}+z^{2}}\right)dz
\end{equation}

La anterior aproximaci\'on dentro del m\'etodo de ondas parciales tiene la 
siguiente aplicaci\'on. Sabemos que la aproximaci\'on eikonal es v\'alida para 
grandes energ\'{\i}as, pero para altas energ\'{\i}as hay muchas ondas
parciales que 
contribuyen a la dispersi\'on, as\'{\i} que podemos tratar a $l$ como una
variable continua y por analog\'{\i}a con la mec\'anica cl\'asica: $l=bk$.
Adem\'as como
anteriormente ya hab\'{\i}amos mencionado $l_{max}=kR$, sustituyendo en la
expresi\'on (15) se obtiene:
\begin{equation}
f(\theta )=-ik\int bJ_{0}(kb\theta )[e^{2i\Delta (b)}-1]db
\end{equation}

}

%Cap. 7
%AQUI VA UN DIAGRAMA DEL ARREGLO EXPERIMENTAL 

%\end{document}

\end{document}